\documentclass[12pt, draftclsnofoot, onecolumn]{IEEEtran}
\newtheorem{Theorem}{Theorem}
\newtheorem{Lemma}[Theorem]{Lemma}
\newtheorem{prop}[Theorem]{Proposition}
\newtheorem{defi}{Definition}
\newtheorem{Corollary}[Theorem]{Corollary}
\newtheorem{Remark}{Remark}
\usepackage{algorithm}
\usepackage{algorithmic}
\usepackage{graphicx}
\usepackage{textcomp}
\usepackage{stfloats}
\usepackage{subfig}
\usepackage{amssymb,amsmath,bm}
\usepackage{cite}
\usepackage{times}
\usepackage{latexsym}
\usepackage{array}
\usepackage{fancyhdr}
\usepackage{psfrag}
\usepackage{multirow}
\usepackage{xcolor}
\usepackage{color}
\usepackage{makecell}
\usepackage{threeparttable}
\usepackage{hyperref}
\begin{document}
\title{Energy Efficiency of Massive Random Access in MIMO Quasi-Static Rayleigh Fading Channels with Finite Blocklength\\}
%\author{\IEEEauthorblockN{Junyuan Gao,
%Yongpeng Wu,~\IEEEmembership{Senior~Member,~IEEE,}
%Shuo Shao,~\IEEEmembership{Member,~IEEE,}
%Wei Yang,~\IEEEmembership{Member,~IEEE,}
%H. Vincent Poor,~\IEEEmembership{Life Fellow,~IEEE,}
%and~Wenjun~Zhang,~\IEEEmembership{Fellow,~IEEE}}
%\thanks{J. Gao, Y. Wu, S. Shao, and W. Zhang are with the Department of Electronic Engineering, Shanghai Jiao Tong University, Minhang 200240, China (e-mail: \{sunflower0515, yongpeng.wu, shuoshao, zhangwenjun\}@sjtu.edu.cn) (Corresponding author: Yongpeng Wu).}
%\thanks{W. Yang is with Qualcomm Technologies, Inc., San Diego, CA 92121, USA (e-mail: weiyang@qti.qualcomm.com).}
%\thanks{H. Vincent Poor is with the Department of Electrical and Computer Engineering, Princeton University, Princeton, NJ 08544, USA (email: poor@princeton.edu).}
%}

\author{\IEEEauthorblockN{Junyuan~Gao,
Yongpeng~Wu,
Shuo~Shao,
Wei~Yang,
and~H.~Vincent~Poor}
\thanks{J. Gao, Y. Wu, and S. Shao are with the Department of Electronic Engineering, Shanghai Jiao Tong University, Minhang 200240, China (e-mail: \{sunflower0515, yongpeng.wu, shuoshao\}@sjtu.edu.cn) (Corresponding author: Yongpeng Wu).}
\thanks{W. Yang is with Qualcomm Technologies, Inc., San Diego, CA 92121, USA (e-mail: weiyang@qti.qualcomm.com).}
\thanks{H. V. Poor is with the Department of Electrical and Computer Engineering, Princeton University, Princeton, NJ 08544, USA (email: poor@princeton.edu).}
}

%\author{\IEEEauthorblockN{Junyuan Gao, Yongpeng Wu, and Shuo Shao}
%\thanks{J. Gao, Y. Wu, and S. Shao are with the Department of Electronic Engineering, Shanghai Jiao Tong University, Minhang 200240, China (e-mail: sunflower0515@sjtu.edu.cn; yongpeng.wu@sjtu.edu.cn; shuoshao@sjtu.edu.cn) (Corresponding author: Yongpeng Wu)
%.} }
\maketitle

\begin{abstract}
  This paper considers the massive random access problem in multiple-input multiple-output quasi-static Rayleigh fading channels.
  Specifically, we derive achievability and converse bounds on the minimum energy-per-bit required for each active user to transmit $J$~bits with blocklength~$n$ and power~$P$ under a per-user probability of error~(PUPE) constraint, in the cases with and without \emph{a priori} channel state information at the receiver~(CSIR and no-CSI).
  In the case of no-CSI, we consider both the settings with and without the knowledge of the number $K_a$ of active users at the receiver.
  % we consider both the settings with and without known the number $K_a$ of active users.}
  % We further extend the achievability result in the no-CSI case with known $K_a$ to a general setting where $K_a$ is unknown in advance.
  The achievability bounds rely on the design of an appropriate ``good region''.
  Numerical evaluation shows the gap between achievability and converse bounds is less than $2.5$~dB for the CSIR case and less than $4$~dB for the no-CSI case in most considered regimes.
  % The no-CSI achievability bounds with and without known $K_a$ show that the extra required energy-per-bit due to the lack of knowledge of $K_a$ is about $5$~dB when $K_a$ is small, and this gap reduces to less than $1.5$~dB as $K_a$ increases.
  % Under the condition that the distribution of the number $K_a$ of active users is known in advance,
  Under the condition that the distribution of $K_a$ is known in advance,
  % the performance loss caused by the uncertainty of the exact value of $K_a$ is small.
  the performance gap between the cases with and without the knowledge of the exact value of $K_a$ is small.
%  Compared to the case where $K_a$ is known, the performance loss caused by unknown $K_a$ is small, under the condition that the distribution of $K_a$ is known in advance.
  For example, in the setup with blocklength $n=1000$, payload $J = 100$~bits, error requirement $\epsilon = 0.001$, and $L=128$ receive antennas,
  compared to the case with known $K_a$,
  the extra required energy-per-bit in the case where $K_a$ is unknown and distributed as $K_a\sim \text{Binom}(K,0.4)$ is less than $0.3$~dB on the converse side and less than $1.1$~dB on the achievability side.
%  Under the condition that the distribution of $K_a$ is known in advance, the performance loss caused by the uncertainty of the exact value of $K_a$ is small, which is less than $0.3$~dB on the converse side and less than $1.1$~dB on the achievability side.
%
%  Numerical results indicate that once the distribution $K_a\sim \text{Binom}(K,p_a)$ is known in advance, the uncertainty of the exact value of $K_a$ entails only a small penalty in terms of energy efficiency.
  The spectral efficiency grows approximately linearly with the number~$L$ of receive antennas with CSIR, whereas the growth rate decreases with no-CSI.
  Moreover, in the case of no-CSI, we study the performance of a pilot-assisted scheme, and numerical evaluation shows that it is suboptimal, especially when there exist many users.
  Building on non-asymptotic results, when all users are active and $J=\Theta(1)$, we obtain scaling laws of the number of supported users as follows:
  when $L = \Theta \left(n^2\right)$ and $P=\Theta\left(\frac{1}{n^2}\right)$, one can reliably serve $K = \mathcal{O}(n^2)$ users with no-CSI;
  under mild conditions with CSIR, the PUPE requirement is satisfied if and only if $\frac{nL\ln KP}{K}=\Omega\left(1\right)$.

%  {\red Moreover, from the achievability and converse bounds with and without known $K_a$, we can observe that once the distribution $K_a\sim \text{Binom}(K,p_a)$ is known in advance,
%  % there is a small penalty when only the distribution of $K_a$ rather than its exact value is known in advance.
%  % there is a slight penalty when the exact number $K_a$ of active users is unknown, under the condition that its distribution is known in advance,
%  the uncertainty of the exact value of $K_a$ entails only a small penalty in terms of energy efficiency, with the extra required energy-per-bit less than $0.3$~dB in the converse side and less than $1.1$~dB in the achievability side.
%  % the performance loss caused by the uncertainty of user activities is small once the distribution of $K_a$ is known in advance
%  }

\end{abstract}


\section{Introduction} \label{section1}

%% 背景 + 引出 Massive random access + 特点 + 引出研究问题
  The design of uplink communication systems in many contemporary wireless networks is influenced by four issues: the rapidly expanding number of users with random activity patterns; the relatively small quantity of information bits to transmit; the strict requirement in communication latency; and the stringent demand on communication energy efficiency.
  Notably, these issues are present in many Internet-of-Things (IoT) applications, in which a very large number of sensors are deployed, but only a fraction of them are active at any given time.
  Active sensors often transmit hundreds of bits describing the parameters they have sensed to the base station~(BS) within latency and energy constraints.
  To address these issues, massive random access technologies have been proposed recently, the study of which includes the information-theoretic analysis and the development of transmission strategies for massive numbers of users with sporadic activity patterns in the regime of finite blocklength.

%% 前人工作
\subsection{Previous work}
% 经典 + GUO (多用户-->大规模多用户)
  Some subsets of these issues have been discussed in recent years in the information-theoretic literature.
  The classical multiuser information theory in~\cite{classical_MAC_Ahlswede,classical_MAC_Liao,elements_IT} studied the fundamental limits of the conventional multiple access channel~(MAC), where the number of users is fixed and the blocklength is taken to infinity.
  To characterize the massive user population in IoT applications, a new model called the many-access channel~(MnAC) was proposed in~\cite{GuoDN}, which allows the number of users to grow unboundedly with the blocklength.
  Based on this model, a new notion of capacity was introduced and characterized with random user activity~\cite{GuoDN}.
  Since the publication of~\cite{GuoDN}, MnACs have been studied in various works in different settings, where a common assumption is that the number of users grows linearly and unboundedly with the blocklength~\cite{WeiF,improved_bound,finite_payloads_fading,A_perspective_on,GaoJY}.
  However, the work in~\cite{GuoDN} relies on the assumption of infinite payload size and infinite blocklength, which cannot capture stringent energy requirements in massive access systems.

% finite payload
  In addition to the massive user population, finite payload size and even finite blocklength should be taken into consideration to make the setting more relevant in practice.
  On this topic, Polyanskiy introduced the per-user probability of error (PUPE) criterion to measure the fraction of transmitted messages that are missing from the list of decoded messages, instead of utilizing the traditional joint error probability criterion, which results in another crucial departure from the classical MAC model~\cite{A_perspective_on}.

  Under the PUPE criterion, some works considered the regime with finite payload size, finite energy-per-bit, and infinite blocklength~\cite{improved_bound,finite_payloads_fading,GaoJY}.
  In particular, based on the MnAC model with the linear scaling mentioned above, under the assumptions of individual codebooks\footnote{It should be noted that individual codebook and common codebook assumptions correspond to different massive access models in practice \cite{wuyp}. In essence, the detection problem under these two assumptions reduces to the block sparse support recovery problem and the sparse support recovery problem, respectively.} and a single BS antenna, Zadik et al.~\cite{improved_bound} and Kowshik et al. \cite{finite_payloads_fading} presented bounds on the tradeoff between user density and energy-per-bit for reliable transmission in additive white Gaussian noise (AWGN)  channels and quasi-static fading channels, respectively.
  In both models, it was observed that in the low user density regime, the multi-user interference~(MUI) can be almost perfectly canceled with good coded access schemes.

  % finite blocklength
  Finite blocklength considerations have also been studied to address transmission within latency constraints.
  For point-to-point channels, Polyanskiy et al.~\cite{Channel_coding_rate} developed a tight approximation to the maximal achievable rate for various channels with positive Shannon capacity, and this approximation was extended to quasi-static fading channels by Yang et al.~\cite{TDMA_yangwei}.
  For the $K$-user Gaussian MAC, achievability bounds and normal approximations with a joint error probability criterion were studied in~\cite{K_MAC}.
  Yavas et al.~\cite{K_MAC2,indivicommon} improved the achievable third-order term in~\cite{K_MAC} for the Gaussian MAC model, and extended this result to Gaussian random access channels under the assumption that the number~$K$ of users does not grow with the blocklength~$n$.
  For the massive random access problem with finite blocklength, the works in~\cite{A_perspective_on} and~\cite{RAC_fading} derived non-asymptotic bounds for Gaussian and Rayleigh fading channels, respectively, under the PUPE criterion and the assumption that the number $K_a$ of active users is known \emph{a priori}.
  It was pointed out in~\cite{estimate_Ka} that the number $K_a$ of active users can be detected with high success probability in Rayleigh fading channels when both uplink and downlink transmissions are exploited to mitigate fading uncertainty, which supports the assumption of known $K_a$ in~\cite{A_perspective_on} and \cite{RAC_fading}.

  When only the uplink transmission is utilized, the success probability of detecting $K_a$ can be reduced~\cite{estimate_Ka}.
  The performance penalty in uplink Gaussian channels, suffering from the lack of knowledge of $K_a$, was analysed in~\cite{noKa,On_joint,GuoDN}.
  Specifically, in the asymptotic regime with infinite number of users, it was pointed out in~\cite{GuoDN} that the message-length capacity penalty due to unknown user activity on each of the $K_a$ active users is $H_2(p_a)/p_a$ under the joint error probability criterion and the assumption that each user becomes active independently with probability $p_a$.
  Moreover, in~\cite{On_joint}, Lancho et al. derived non-asymptotic achievability and converse bounds for the single-user random access scenario, and numerical results for the binary-input Gaussian channel indicated that the bound with unknown user activity approaches the one with known $K_a$ as the blocklength and the signal-to-noise ratio~(SNR) increase.
  Following from the maximum likelihood~(ML) principle, a non-asymptotic achievability bound was derived in~\cite{noKa} for the massive random access problem with unknown $K_a$, whereas a matching converse bound was not provided.
  As a result, it is of great significance to construct tight non-asymptotic bounds in both achievability and converse sides to characterize the performance loss caused by unknown $K_a$ in massive random access channels, which is an important goal of this paper.

  It should be noted that the above-mentioned non-asymptotic works on the massive random access communication problem~\cite{A_perspective_on,RAC_fading,noKa,On_joint} rely on the assumption of a single BS antenna.
  In practice, equipping multiple antennas at the BS can bring great benefits in massive random access systems.
  Specifically, for the user activity detection problem, it was demonstrated in~\cite{Caire1} that, with $n$ channel uses and a sufficiently large number $L$ of BS antennas satisfying $K_a/L=o(1)$, up to $K_a = \mathcal{O}\left(n^2\right)$ active users can be identified among $K$ potential users when $\frac{K_a}{K}=\Theta(1)$;
  it overcomes the fundamental limitation of the single-receive-antenna system, in which the number $K_a$ of active users that can be identified is at most linear with the blocklength~$n$.
  Given the great potential of multiple receive antennas for the activity detection problem as revealed by the scaling law in~\cite{Caire1}, it is natural to conjecture that multiple receive antennas could bring similar benefits for the joint activity and data detection problem in massive random access channels. An important goal of this paper is to characterize the impact of multiple BS antennas on the performance of joint activity and data detection in both the non-asymptotic regime and the asymptotic regime.

  From the perspective of channel state information~(CSI) availability, the above mentioned works can be divided into two categories: the case in which CSI is known at the receiver in advance~(CSIR)~\cite{A_perspective_on,improved_bound,finite_payloads_fading,GaoJY,K_MAC,K_MAC2,indivicommon,noKa} (the AWGN channel without fading is a special case of CSIR), and the case in which there is no \emph{a priori} CSI at the receiver~(no-CSI)~\cite{finite_payloads_fading,TDMA_yangwei,Caire1}.
  In the no-CSI case~(i.e. the so called noncoherent setting), the communication scheme suggested by the capacity result makes no effort to estimate channel coefficients~\cite{Zheng}.
  Thus, the scheme without explicit channel estimation is adopted in many works, such as~\cite{finite_payloads_fading,TDMA_yangwei,Caire1}.
  In addition, in the no-CSI case, the receiver is also allowed to gain channel knowledge, where channel estimation can be simply viewed as a specific form of coding~\cite{pilot_coding1,pilot_coding2}.
  In practical wireless systems, the pilot-assisted scheme is widely adopted, in which users first send pilots for explicit channel estimation, and then the estimated channels are utilized to decode the signals for each user.
  The performance of this scheme has been investigated in some works.
  In the single-user case, it was proved in~\cite{Zheng} that the pilot-assisted scheme is optimal at a high SNR in terms of degrees of freedom for block-fading channels, and non-asymptotic bounds on the maximum coding rate with finite blocklength were derived in~\cite{pilot_Durisi1}.
  For the scenario with multiple users, the large-antenna limit of the pilot-assisted scheme was studied in~\cite{pilot_Durisi2}, where the achievable error probability was derived at finite blocklength, assuming channels were estimated based on the minimum mean-square error~(MMSE) criterion and both the MMSE and maximum ratio criteria were utilized for mismatched combining.
  After combining, the complicated problem of jointly detecting $K$ transmitted codewords based on the received signals among $L$ BS antennas, is converted to the problem of separately detecting $K$ codewords in the single-receive-antenna fading channel, which, however, can result in a performance loss.
  %Thus, in order to assess the non-asymptotic performance of the pilot-assisted scheme for massive access in MIMO fading channels, we consider the mismatched nearest neighbor decoder~\cite{mismatch_Shamai,mismatch_Asyhari}, which yields the minimal error probability in the CSIR case~\cite{Gallager}.

\subsection{Our contributions}
  %% 本文内容
  In this paper, we consider the joint activity and data detection problem for massive random access in multiple-input multiple-output~(MIMO) quasi-static Rayleigh fading channels with stringent latency and energy constraints.
  Specifically, in both cases of CSIR and no-CSI, we derive achievability and converse bounds on the minimum energy-per-bit required for each active user to transmit $J=\log_2 M$ information bits with blocklength $n$, power $P$, and PUPE less than a constant,
  under the assumption that the number $K_a$ of active users is known \emph{a priori}.
  To characterize the performance loss caused by the uncertainty of user activities in the non-asymptotic regime, we further extend the achievability and converse results in the no-CSI case with known $K_a$ to a general setting where $K_a$ is random and unknown but its distribution $K_a\sim \text{Binom}(K,p_a)$ is known at the receiver in advance.
  Indeed, knowing the distribution of $K_a$ is a common assumption in many works such as~\cite{noKa,Yuwei_active,Gao_active}.
  Moreover, we study the performance of a pilot-assisted scheme in the no-CSI case.
  The derived non-asymptotic bounds provide theoretical benchmarks to evaluate practical transmission schemes.
  Building on these non-asymptotic bounds, we obtain scaling laws of the number of reliably served users in a special case where all users are assumed to be active. These results reveal the great potential of multiple receive antennas for the massive access problem.
  Meanwhile, they show a significant difference in the required number of BS antennas between utilizing the PUPE criterion and the joint error probability criterion.
%  It should be noted that we consider individual codebooks for joint activity and data detection, but the results can be extended to the framework of a common codebook. A similar extension with AWGN channels can be found in~\cite{A_perspective_on,indivicommon}.

  %% 难点
  \emph{Non-asymptotic analysis:} There are some twists in deriving non-asymptotic achievability bounds for massive random access in MIMO quasi-static Rayleigh fading channels.
  Specifically, compared with traditional MAC, the number of users is greatly increased in massive random access channels, leading to a considerable increase in the number of error events.
  As a consequence, the simple union bound can be substantially loosened if not applied with care, and we need to resort to more efficient tools.
  Moreover, in the case of no-CSI, the projection decoder was used in~\cite{finite_payloads_fading} to derive an achievability bound for the single-receive-antenna setting.
  When we employ this decoder to our considered massive random access problem in MIMO fading channels with individual codebooks and known $K_a$, the output is given by
  \begin{equation} \label{eq:intro_projection_decoder}
    \left[ \hat{\mathcal{K}}_a , \{\hat{W}_{k} : k \in \hat{\mathcal{K}}_a \} \right]
    = \underset{ \hat{\mathcal{K}}_a \subset [K], | \hat{\mathcal{K}}_a| = K_a }{\operatorname{argmax}}
    \max_{ \{\hat{W}_{k} \in [M] : k\in \hat{\mathcal{K}}_a  \} }
    \max_{\mathbf{H}}
    \mathbb{P} \left[ \mathbf{Y} \left| \mathbf{X},
    \{\hat{W}_{k} : k \in \hat{\mathcal{K}}_a\}, \mathbf{H} \right.\right] ,
  \end{equation}
  where $\mathbf{X}\in\mathbb{C}^{n\times MK}$ denotes the concatenation of codebooks of the $K$ users, $\mathbf{H}$ contains the channel fading coefficients, $\mathbf{Y}$ denotes the received signal,
  $\hat{\mathcal{K}}_a$ denotes the estimated set of active users,
  and $\hat{W}_{k}$ denotes the decoded message for user $k$.
  As we can see from~\eqref{eq:intro_projection_decoder}, an advantage of the projection decoder lies in that it requires no knowledge of the fading distribution.
  However, when the projection decoder is applied to the framework with multiple BS antennas, it can be ineffectual in two specific cases.
  First, the use of large antenna arrays allows the number of reliably served active users to be much larger than the blocklength. As a result, the dimension of the subspace spanned by the transmitted codewords of active users is limited by the blocklength. In this case, the subspace spanned by $K_a$ transmitted codewords can be the same as that spanned by another set of $K_a$ codewords, which prevents the projection decoder from distinguishing the two sets.
  Second, the signals received over different BS antennas share the same sparse support since they are linear combinations of the same $K_a$ codewords corrupted by different noise processes.
  Thus, it is ineffectual to apply the projection decoder to $L$ antennas separately.
  Moreover, it is challenging (although not impossible) to jointly deal with the signals received over $L$ BS antennas based on the projection decoder, because the analysis of the angle between the subspace spanned by $L$ received signals and the subspace spanned by $K_a$ transmitted codewords is quite involved.

  %% 技术路线，解决方案
  To alleviate the problems mentioned above, for massive random access in MIMO quasi-static Rayleigh fading channels, some techniques are utilized in this paper to derive non-asymptotic achievability bounds on the minimum required energy-per-bit.
  Specifically, in both cases of CSIR and no-CSI, we leverage the ML-based decoder when $K_a$ is known \emph{a priori}.
  Note that, in the no-CSI case with known $K_a$, in contrast to the projection decoder mentioned above, the ML decoder is applicable regardless of whether $K_a$ is less than the blocklength or not, but at the price of requiring \emph{a priori} distribution on $\mathbf{H}$.
  This can be observed from the ML decoding criterion given by
  \begin{equation} \label{eq:intro_ML_decoder}
    \left[ \hat{\mathcal{K}}_a , \{\hat{W}_{k} : k\in \hat{\mathcal{K}}_a \} \right]
    = \underset{ \hat{\mathcal{K}}_a \subset [K], |\hat{\mathcal{K}}_a| = K_a }
    {\operatorname{argmax}}
    \max_{ \{\hat{W}_k \in [M] : k \in \hat{\mathcal{K}}_a  \} }
    \mathbb{P} \left[ \mathbf{Y} \left| \mathbf{X},
    \{\hat{W}_{k} : k\in \hat{\mathcal{K}}_a\} \right.\right],
  \end{equation}
  \begin{equation}\label{eq:intro_ML_decoder2}
    \mathbb{P} \left[ \mathbf{Y} \left| \mathbf{X},
    \{\hat{W}_{k} : k\in \hat{\mathcal{K}}_a\} \right.\right]
    =\mathbb{E}_{\mathbf{H}} \left\{
    \mathbb{P} \left[ \mathbf{Y} \left| \mathbf{X},
    \{\hat{W}_{k} : k\in \hat{\mathcal{K}}_a\} , \mathbf{H} \right.\right]
    \right\} .
  \end{equation}
  Moreover, when $K_a$ is unknown, we first obtain an estimate of $K_a$ via an energy-based estimator; then, we output a set of decoded messages following the maximum \emph{a posteriori}~(MAP) principle, which incorporates prior distributions in users' messages of various sizes.
  For the pilot-assisted coded access scheme, in a special case where all users are active, we leverage the MMSE criterion to estimate channels in the first stage, and utilize the mismatched nearest neighbor criterion~\cite{mismatch_Shamai,mismatch_Asyhari} to decode in the second stage.
  The signals received over $L$ BS antennas can be jointly dealt with easily in aforementioned cases.

  To address the probability of the union of extremely many error events, we resort to standard bounding techniques proposed by Fano~\cite{1961} and by Gallager~\cite{1965}.
  Gallager's $\rho$-trick bound is only used for a special case in which both the user activity and CSI are known at the receiver, considering that this bound is difficult to evaluate by the Monte Carlo method when random access is taken into consideration.
  The Fano's bound is used to establish non-asymptotic achievability bounds in massive random access channels for the case of CSIR and no-CSI. Its performance relies on the choice of a region around the linear combination of the transmitted signals, which is interpreted as the ``good region''~\cite{goodregion}.
  In this work, we design an appropriate ``good region'' for massive random access channels, which is parameterized by two parameters $\omega$ and $\nu$.
  Our ``good region'' reduces to the one used in~\cite{finite_payloads_fading} if the parameter $\nu$ is set to 0.
  In the CSIR case with $0\leq \omega < 1$, our ``good region'' is essentially a sphere, where its center is determined by $\omega$ and its radius is controlled by both $\omega$ and $\nu$.
  However, for the region in~\cite{finite_payloads_fading}, both the center and the radius are controlled by $\omega$. The value of the radius depends on the position of the center for the region in~\cite{finite_payloads_fading}, whereas the radius of our region can be flexibly changed by adjusting $\nu$.
  As a result, we have better control of the ``good region''.

  %% gap
  Numerical results demonstrate the tightness of our bounds. Specifically, the gap between the achievability bound and the converse bound is less than $2.5$~dB for the CSIR case and less than $4$~dB for the no-CSI case in most considered regimes (the Fano type converse bound for the no-CSI case relies on the assumption of i.i.d. Gaussian codebooks).
  Compared to the case where the number $K_a$ of active users is known, the performance loss caused by unknown $K_a$ is small.
  For example, in the setup with blocklength $n=1000$, payload $J = 100$~bits, active probability $p_a=0.4$, error requirement $\epsilon = 0.001$, and $L=128$ receive antennas, the extra required energy-per-bit due to the uncertainty of the exact value of $K_a$ is less than $0.3$~dB on the converse side and less than $1.1$~dB on the achievability side.
  % The no-CSI achievability and converse bounds with and without the knowledge of $K_a$ at the receiver show that, the extra required energy-per-bit due to the uncertainty of the exact value of $K_a$ is small, which is less than $0.3$~dB on the converse side and less than $1.1$~dB on the achievability side.
%  {\red Moreover, from the achievability and converse bounds with and without known $K_a$, we can observe that once the distribution $K_a\sim \text{Binom}(K,p_a)$ is known in advance,
%  % there is a small penalty when only the distribution of $K_a$ rather than its exact value is known in advance.
%  % there is a slight penalty when the exact number $K_a$ of active users is unknown, under the condition that its distribution is known in advance,
%  the uncertainty of the exact value of $K_a$ entails only a small penalty in terms of energy efficiency, with the extra required energy-per-bit less than $0.3$~dB in the converse side and less than $1.1$~dB in the achievability side.
%  % the performance loss caused by the uncertainty of user activities is small once the distribution of $K_a$ is known in advance
%  }
  %% MUI
  Similar to AWGN channels~\cite{improved_bound} and single-receive-antenna quasi-static fading channels~\cite{finite_payloads_fading}, the MUI can be almost perfectly cancelled in multiple-receive-antenna quasi-static fading channels when the number of active users is below a critical threshold.
  %% SE linear nonlinear
  Additionally, in our considered regime, the spectral efficiency grows approximately linearly with the number of BS antennas for the CSIR case, but the lack of CSI at the receiver causes a slowdown in the growth rate.
  %% noncoherent pilot-assisted
  Furthermore, our results for the no-CSI case reveal that the orthogonal-pilot-assisted coded access scheme is suboptimal, especially when the number of active users is large, even if the power allocation between pilot and data symbols is optimized.
  Overall, we believe our non-asymptotic bounds provide theoretical benchmarks to evaluate practical transmission schemes, which are of considerable importance in massive random access systems.

  \emph{Asymptotic analysis:} Building on these non-asymptotic results, in a special case where all users are assumed to be active, we obtain scaling laws of the number of reliably served users under the PUPE criterion.
  For the CSIR case, assuming $ n, K \to \infty$, $M=\Theta(1)$, $\ln K = o(n)$, and $KP=\Omega\left(1\right)$ ($P$ denotes the transmitting power per channel use), the PUPE requirement is satisfied if and only if  $\frac{nL\ln KP}{K}=\Omega\left(1\right)$.
  It can be divided into the following two regimes: 1)~$\frac{nL}{K}=\Omega\left(1\right)$ and $KP =\Theta\left(1\right)$; 2)~$\frac{nL\ln KP}{K}=\Omega\left(1\right)$ and $KP \to \infty$.
  The first regime is power-limited, where the number of degrees of freedom grows linearly with the number of users.
  As a result, by allocating orthogonal resources to users, the minimum received energy-per-user can be $nLP=\Theta\left(1\right)$, which is as low as that in the single-user case~\cite{Yang_scalinglaw_singleuser}.
  %%%%%%%%%%%%%%%%
  The second regime is degrees-of-freedom-limited, where the number of degrees of freedom, i.e. $nL$, is far less than the number of users, and the minimum received energy-per-user $nLP\to\infty$.
  Two special scaling laws in the CSIR case are presented in Table~\ref{table:scalinglaw}.
  \begin{table}[!t]
  \scriptsize
    \caption{Comparison of scaling laws for massive access in quasi-static Rayleigh fading channels} \label{table:scalinglaw}
    \renewcommand{\arraystretch}{1.1}
    \centering
    \begin{threeparttable}          %这行要添加
    \begin{tabular}{|c|c|c|c|c|c|c|c|c|}
      \hline
      result                                & $K$           & $L$                                         & $P$                              & $M$        & CSIR/no-CSI       & error criterion         & achievability/converse  \\  \hline
      Theorem \ref{Theorem_scalinglaw_CSIR} & $\mathcal{O}(n^2)$ & $\Theta\!\left(n\right)$ & $\Theta\!\left(\frac{1}{n^2}\right)$ & $\Theta(1)$ &CSIR  & PUPE                    & both \\  \hline
      Theorem \ref{Theorem_scalinglaw_CSIR} & $\mathcal{O}(n^2)$ & $\Theta\!\left(\frac{n}{\ln n}\right)$ & $\Theta\!\left(\frac{1}{n}\right)$ & $\Theta(1)$ &CSIR  & PUPE                    & both \\  \hline
      Theorem \ref{Theorem_scalinglaw_noCSI}\tnote{1}& $\mathcal{O}(n^2)$ & $\Theta\left( n^2 \right)$                    & $\Theta\!\left(\frac{1}{n^2}\right)$  & $\Theta(1)$ &no-CSI& PUPE                    &
      both \\  \hline
      extended from~\cite{Caire1}          & $\mathcal{O}(n^2)$ & $ \Theta\!\left(n^2\! \ln \!n\right)\! $                & $\Theta\!\left(\frac{1}{n^2}\right)$                    & $\Theta(1)$ &no-CSI& { {joint error probability} }  & achievability             \\  \hline
      \cite{finite_payloads_fading}           & $\mathcal{O}(n)$   &$ 1 $                &  $\Theta\left(\frac{1}{n}\right)$                     & $\Theta(1)$ &both& PUPE & both             \\  \hline
      \cite{Ravi}          &  $o(n)$    & $ 1 $               &  $\Theta\left(\frac{1}{n}\right)$                  & $\Theta(1)$  &  CSIR (AWGN)  &   PUPE (vanish)\tnote{2}   &  both             \\  \hline
    \end{tabular}
    \begin{tablenotes}    %这行要添加， 从这开始
        \footnotesize               %这行要添加
        \item[1] {In the case of no-CSI, the converse bound relies on the assumption of i.i.d. Gaussian codebooks.}
        \item[2] {The PUPE is required to vanish for the scaling law in~\cite{Ravi}, and a positive constant PUPE is acceptable for other cases in Table~\ref{table:scalinglaw}.}
    \end{tablenotes}            %这行要添加
    \end{threeparttable}       %这行要添加，到这里结束
  \end{table}
  We can observe that, in order to reliably serve $K = \mathcal{O}(n^2)$ users, when the number of BS antennas is increased from $L=\Theta\left( \frac{n}{\ln n} \right)$ to $L=\Theta\left( n \right)$, the minimum required power can be considerably decreased from $P =\Theta\left(\frac{1}{n}\right)$ to $P =\Theta\left(\frac{1}{n^2}\right)$, which indicates the great potential of multiple receive antennas for the data detection problem.
  Moreover, our scaling laws reveal the tightness of the derived bounds in asymptotic cases since they are proved from both the achievability side and the converse side.
  The scaling law for the scenario with a single BS antenna is also presented in Table~\ref{table:scalinglaw} for comparison:
  one can reliably serve $K=\mathcal{O}(n)$ users when a positive constant PUPE is acceptable~\cite{finite_payloads_fading};
  however, the number of users is only allowed to grow sublinearly with $n$ even in AWGN channels when the PUPE is required to vanish~\cite{Ravi}.

  For the no-CSI case, the scaling law from our result is shown in Table~\ref{table:scalinglaw}, together with the result extended from~\cite{Caire1}, which is based on the joint error probability criterion.
  We observe a significant difference in the number of BS antennas to reliably serve $K$ users between utilizing the PUPE criterion and the joint error probability criterion.
  Specifically, in order to obtain the scaling law on the achievability side, both the activity detection problem considered in~\cite{Caire1} and the data detection problem of interest in this work can be formulated as sparse support recovery problems.
  Thus, the scaling law of the activity detection problem in~\cite{Caire1} can be extended to that of the data detection problem as follows: under the joint error probability criterion, with a coherence block of dimension $n\to \infty$ and a sufficient number of BS antennas $L = \Theta\left(n^2 \ln n\right)$, one can reliably serve up to $K = \mathcal{O}\left(n^2\right)$ users when the payload $J=\Theta(1)$ and the power $P=\Theta\left(\frac{1}{n^2}\right)$ in the case of no-CSI. %(see Appendix~\ref{Proof_scalinglaw_noCSI} for more details).
  In this work, we consider the PUPE criterion, which is more appropriate for the consideration of massive access~\cite{A_perspective_on}.
  Our result shows that the required number of BS antennas can be reduced from $L = \Theta\left(n^2 \ln n\right)$ to $L = \Theta\left(n^2\right)$ when we change from the joint error probability criterion to the PUPE criterion.
  %Moreover, the result in~\cite{Caire1} is on the achievability side; Theorem~\ref{Theorem_scalinglaw_noCSI} is proved from both the achievability side and the converse side, in which the converse result relies on the assumption that the codebooks have i.i.d. Gaussian entries.
  In addition, it should be noted that, the case of $nP = \Theta\left(1\right)$ and the case of $n^2P = \Theta\left(1\right)$ in Table~\ref{table:scalinglaw} imply that the energy-per-bit is finite and goes to $0$, respectively, which are crucial in practical communication systems with stringent energy constraints.

  The remainder of this paper is organized as follows.
  Section \ref{section2} introduces the system model.
  In Section \ref{section3}, we introduce a key proof technique used to derive non-asymptotic achievability bounds, where an appropriate ``good region'' is designed for massive random access channels.
  We also provide our main results in Section~\ref{section3}, including achievability and converse bounds in both cases of CSIR and no-CSI, respectively, and corresponding scaling laws.
  Section~\ref{sec_simulation} presents numerical results.
  Conclusions are drawn in Section~\ref{section7}.

  \emph{Notation:} Throughout this paper, uppercase and lowercase boldface letters denote matrices and column vectors, respectively.
  We use $\left[\mathbf{x} \right]_{m}$ to denote the $m$-th element of a vector $\mathbf{x}$, and use $\left[\mathbf{A} \right]_{m,n}$, $\left[\mathbf{A} \right]_{m,:}$, and $\left[\mathbf{A} \right]_{:,n}$ to denote the $\left( m,n \right)$-th element, the $m$-th row vector, and the $n$-th column vector of a matrix $\mathbf{A}$, respectively.
  The notation $\mathbf{I}_{n}$ denotes an $n\times n$ identity matrix, and $\mathbf{I}_{(t)} \in \left\{ 0,1 \right\}^{n\times n}$ denotes a diagonal matrix with the first $t\leq n$ diagonal entries being ones and all of the rest being 0.
  We use $\left(\cdot \right)^{T}$, $\left(\cdot \right)^{H}$, $\operatorname{vec}\left(\mathbf{X}\right)$, $\left|\mathbf{X}\right|$, $\left\|\mathbf{x} \right\|_{p}$, and $\left\|\mathbf{X} \right\|_{F}$ to denote transpose, conjugate transpose, vectorization of a matrix $\mathbf{X}$, determinant of a matrix $\mathbf{X}$, ${\ell}_p$-norm of a vector $\mathbf{x}$, and Frobenius norm of a matrix $\mathbf{X}$, respectively.
  The notations $\lceil \cdot \rceil$ and $k!$ depict the ceiling function and factorial function, respectively.
  Given any complex variable, vector or matrix, the notations $\Re\left(\cdot\right)$ and $\Im\left(\cdot\right)$ return its real and imaginary parts, respectively.
  We use $\operatorname{diag} \left\{ \mathbf{x} \right\}$ to denote a diagonal matrix with vector $\mathbf{x}$ comprising its diagonal elements, and $\operatorname{diag} \left\{ \mathbf{A}, \mathbf{B} \right\}$ to denote a block diagonal matrix with $\mathbf{A}$ and $\mathbf{B}$ in diagonal blocks.
  We use $\cdot\backslash\cdot$ and $\left| \mathcal{A} \right|$ to denote set subtraction and the cardinality of a set $\mathcal{A}$, respectively.
  We use $\mathbf{b}_{[\mathcal{A}]}=\left\{\mathbf{b}_i: i \in \mathcal{A} \right\}$ to denote a set of vectors.
  %We use $\mathbb{N}$ and $\mathbb{N}_{+}$ to denote the set of natural numbers and the set of nonnegative natural numbers, respectively.
  We denote the set of nonnegative natural numbers by $\mathbb{N}_{+}$.
  For an integer $k>0$, the notation $[k]$ denotes $\left\{1,2,\ldots,k \right\}$;
  for integers $k_2 \geq k_1>0$, the notation $[k_1:k_2]$ denotes $\left\{k_1,k_1+1,\ldots,k_2\right\}$.
  We denote $x^{+} = \max\{x, 0\}$.
  We denote the projection matrix onto the subspace spanned by $S\subset\mathbb{C}^{n}$ and its orthogonal complement as $\mathcal{P}_{S}$ and $\mathcal{P}_{S}^{\bot}$, respectively.
  The notation $\mathcal{G}^c$ denotes the complement of the event $\mathcal{G}$.
  We use $\mathcal{N}(\cdot ,\cdot)$, $\mathcal{CN}(\cdot ,\cdot)$, $\chi^2(d)$, $\chi^2(d,\lambda)$, and $\mathcal{W}_{m}(n, \mathbf{A})$ to denote the standard Gaussian distribution, circularly symmetric complex Gaussian distribution, central chi-squared distribution with $d$ degrees of freedom, non-central chi-squared distribution with $d$ degrees of freedom and noncentrality parameter $\lambda$, and Wishart distribution with $n$ degrees of freedom and covariance matrix $\mathbf{A}$ of size $m\times m$, respectively.
  The functions $\gamma\left(\cdot, \cdot\right)$ and $\Gamma\left(\cdot\right)$ denote the lower incomplete gamma function and gamma function, respectively, with the assumption that $\gamma\left(\cdot, a\right)=0$ if $a\leq0$.
  For $0 \leq  p \leq 1$, we denote $h(p) = -p\ln(p)-(1-p)\ln(1-p)$ and $h_2(p) = h(p)/\ln 2$ with $0\ln 0$ defined to be $0$.
  Let $f(x)$ and $g(x)$ be positive. The notation $f(x) = o\left( g(x)\right)$ means that $\lim_{x\to\infty} f(x)/g(x) = 0$,
  $f(x) = \mathcal{O} \left( g(x)\right)$ means that $\limsup_{x\to\infty} f(x)/g(x) < \infty$, $f(x) = \Theta \left( g(x)\right)$ means that $f(x) = \mathcal{O} \left( g(x)\right)$ and $g(x) = \mathcal{O} \left( f(x)\right)$,
  and $f(x) = \Omega \left( g(x)\right)$ means that $g(x) = \mathcal{O} \left( f(x)\right)$.

\section{System Model} \label{section2}
  We consider a massive random access system consisting of a BS equipped with $L$ receive antennas and $K$ potential users each equipped with a single transmit antenna.
  We assume that the user traffic is sporadic, i.e., only $K_a\leq K$ users are active at any given time.
  %Denote the ratio of the number of active users to the number of potential users as $p_a=\frac{K_a}{K}$.
  Each active user transmits $J$ information bits with blocklength $n$.
  The user set and active user set are denoted as $\mathcal{K}$ and $\mathcal{K}_a$, respectively.

  We assume each user has an individual codebook of size $M=2^J$ and blocklength $n$.
  The matrix $\mathbf{X}_k = \left[\mathbf{x}_{k,1}, \mathbf{x}_{k,2},\ldots, \mathbf{x}_{k,M}\right] \in \mathbb{C}^{n\times M}$ consists of the codewords of the $k$-th user and the matrix $\mathbf{X} = \left[ \mathbf{X}_1, \mathbf{X}_2, \ldots, \mathbf{X}_K\right] \in \mathbb{C}^{n\times MK}$ is obtained by concatenating all codebooks.

\begin{figure}
  \centering
  \includegraphics[width=0.73\linewidth]{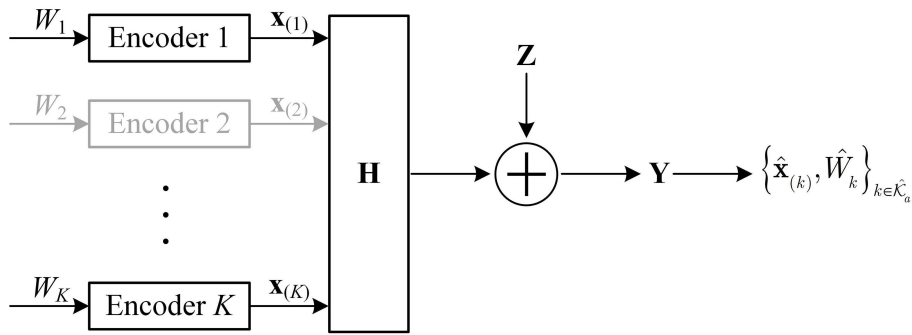}\\
  \caption{Massive random access in MIMO quasi-static Rayleigh fading channels.}
  \label{fig:system}
  \end{figure}

  We consider a quasi-static Rayleigh fading channel model, where the channel stays constant during the transmission of a codeword. We assume synchronous transmission.
  The $l$-th antenna of the BS observes $\mathbf{y}_l\in \mathbb{C}^{n}$ given by
  \begin{equation} \label{eq_yl}
    \mathbf{y}_l = \sum_{k\in{\mathcal{K}}}{h}_{k,l}\mathbf{x}_{(k)}+\mathbf{z}_l ,
  \end{equation}
  where ${h}_{k,l} \sim \mathcal{CN}(0,1)$ denotes the fading coefficient between the $k$-th user and the $l$-th antenna of the BS, which is i.i.d. across different users and different BS antennas;
  the noise vector $\mathbf{z}_l$ is distributed as $\mathcal{CN}(\mathbf{0},\mathbf{I}_n)$, which is i.i.d. across $L$ BS antennas;
  the transmitted codeword of the $k$-th user is denoted as $\mathbf{x}_{(k)} = \mathbf{x}_{k,W_k}$.
  Here, if the $k$-th user is active, its message $W_{k} \in [M]$ is chosen uniformly at random; if it is inactive, we denote $W_{k}=0$ and $\mathbf{x}_{(k)}=\boldsymbol{0}$.
  Denote $\boldsymbol{\Phi}\in\{0,1\}^{MK\times K}$ the binary selection matrix that satisfies $\left[\boldsymbol{\Phi}\right]_{(k-1)M + W_k,k} = 1$ if the $k$-th user is active and the $W_{k}$-th codeword is transmitted by this user, and $\left[\boldsymbol{\Phi}\right]_{(k-1)M + W_k,k} = 0$ otherwise.
  As presented in Fig.~\ref{fig:system}, the received signal over $L$ antennas of the BS can be written as
  \begin{equation} \label{eq_y}
    \mathbf{Y} =\mathbf{X}\boldsymbol{\Phi}\mathbf{H}+\mathbf{Z}  ,
  \end{equation}
  where $\mathbf{Y} = \left[ \mathbf{y}_1, \mathbf{y}_2, \ldots, \mathbf{y}_L \right] \in \mathbb{C}^{n\times L}$,
  $\mathbf{H} = \left[\mathbf{h}_1, \mathbf{h}_2, \ldots, \mathbf{h}_L \right] \in \mathbb{C}^{K\times L} $,
  $\mathbf{h}_l = \left[ {h}_{1,l}, {h}_{2,l},\ldots,{h}_{K,l} \right]^T \in \mathbb{C}^{K}$,
  and $\mathbf{Z} = \left[\mathbf{z}_1, \mathbf{z}_2, \ldots, \mathbf{z}_L \right] \in \mathbb{C}^{n\times L} $.

  The decoder aims to find the estimated set $\hat{\mathcal{K}}_a$ of active users, and find the estimate $\hat{\mathbf{x}}_{(k)}$ of ${\mathbf{x}}_{(k)}$ and corresponding message $\hat{W}_k$ of ${W}_k$ for $ k \in \hat{\mathcal{K}}_a $.
  We denote $\hat{W}_k = 0$ and $\hat{\mathbf{x}}_{(k)}=\boldsymbol{0}$ for $ k \notin \hat{\mathcal{K}}_a $.
  As noted previously, in this work, we consider two scenarios: CSIR (the decoder knows the realization of the fading channel beforehand) and no-CSI (the decoder does not have \emph{a priori} knowledge of the realization of the fading channel but it knows its distribution in advance).
  In the case of CSIR, we assume the number $K_a$ of active users is fixed and known to the receiver in advance as in~\cite{A_perspective_on};
  in the case of no-CSI, we consider two settings: 1)~$K_a$ is fixed and known to the receiver \emph{a priori}; 2)~$K_a$ is random and unknown to the receiver, but its distribution is known in advance.

  Based on the PUPE criterion in~\cite{A_perspective_on,finite_payloads_fading}, we introduce the notion of a massive random access code for the case of CSIR and no-CSI with known $K_a$ as follows:
  \begin{defi}[Massive random access code with CSIR and known $K_a$] \label{defi1}
    Let $\mathcal{X}_k$, $\mathcal{H}_k$, and $\mathcal{Y}$ denote the input alphabet of user $k$, the channel fading coefficient alphabet of user $k$, and the output alphabet, respectively.
    An $(n,M,\epsilon,P)_{\text{CSIR},K_a}$ massive random access code consists of
    \begin{enumerate}
      \item An encoder $\emph{f}_{\text{en},k}: [M] \mapsto \mathcal{X}_k$ that maps the message $W_k \in [M]$ to a codeword $\mathbf{x}_{(k)} \in \mathcal{X}_k$ for $k\in\mathcal{K}_a$.
          The codewords in $\left\{ \mathcal{X}_k: k\in\mathcal{K} \right\}$ satisfy the power constraint
          \begin{equation}\label{power_constraint}
            \left\|\mathbf{x}_{k,m}\right\|_{2}^{2} \leq nP,\;\;\;\;k\in\mathcal{K},\;\; m\in[M].
          \end{equation}
          We assume that $W_k$ is equiprobable on $[M]$ for $k\in\mathcal{K}_a$.
      \item A decoder $\emph{g}_{\text{de,CSIR},K_a}: \mathcal{Y} \times \prod_{k\in\mathcal{K}}\mathcal{H}_k \mapsto [M]^{K_a}$ that satisfies the PUPE constraint
          \begin{equation} \label{PUPE}
            P_{e} = \frac{1}{K_a} \sum_{k\in {\mathcal{K}_a}} \mathbb{P}\left[ W_{k} \neq \hat{W}_{k} \right] \leq \epsilon,
          \end{equation}
          where $\hat{W}_{k} = \left( \emph{g}_{\text{de,CSIR},K_a}\left( \mathbf{Y} , \mathbf{H} \right) \right)_{k}$ denotes the decoded message for user $k$ in the case of CSIR with known $K_a$ to the receiver in advance.
    \end{enumerate}
  \end{defi}
  \begin{defi}[Massive random access code with no-CSI and known $K_a$] \label{defi2}
    Let $\mathcal{X}_k$ and $\mathcal{Y}$ denote the input alphabet of user $k$ and the output alphabet, respectively. An $(n,M,\epsilon,P)_{\text{no-CSI},K_a}$ massive random access code consists of
    \begin{enumerate}
      \item An encoder $\emph{f}_{\text{en},k}: [M] \mapsto \mathcal{X}_k$ that maps the message $W_k\in[M]$ to a codeword $\mathbf{x}_{(k)} \in \mathcal{X}_k$ for $k\in\mathcal{K}_a$.
          The codewords satisfy the power constraint in \eqref{power_constraint}.
          We assume that $W_k$ is equiprobable on $[M]$ for $k\in\mathcal{K}_a$.
      \item A decoder $\emph{g}_{\text{de,no-CSI},K_a}: \mathcal{Y} \mapsto [M]^{K_a} $ that satisfies the PUPE constraint in~\eqref{PUPE} for the case of no-CSI with known $K_a$ to the receiver in advance.
          The decoded message for user $k$ is denoted as $\hat{W}_{k} = \left( \emph{g}_{\text{de,no-CSI},K_a} \left( \mathbf{Y}  \right) \right)_{k}$.
    \end{enumerate}
  \end{defi}

  In the following, we introduce the notion of a massive random access code for the no-CSI case when the number $K_a$ of active users is random and unknown.
  Specifically, we assume that each user becomes active independently with identical probability $p_a$ during any given block.
  In this case, the number $K_a$ of active users is random and distributed as $K_a\sim \text{Binom}(K,p_a)$, which is assumed to be known to the receiver as in~\cite{Yuwei_active,Gao_active}. % with mean $\bar{K}_a = p_a K$.
  The probability of the event that $K_a = {\rm{K}}_a$, i.e., there are exactly ${\rm{K}}_a\in\{0,1,\ldots,K\}$ active users among $K$ potential users, is given by
% If the $k$-th user is active, its message $W_{k} \in [M]$ is chosen uniformly at random; if it is inactive, we denote $W_{k}=0$ and $\mathbf{x}_{(k)}=\boldsymbol{0}$, i.e., we have
%  \begin{equation} \label{eq:active_prob}
%     \mathbb{P}\left[ W_k = w \right] = \left\{
%        \begin{array}{ll}
%           1 - p_a ,      &  w = 0 \\
%           \frac{p_a}{M}, &  w \in [M]
%        \end{array}\right. .
%  \end{equation}
  \begin{equation}\label{eq_conv_pKa_noKa}
    P_{K_a}({\rm{K}}_a) = \binom{K}{{\rm{K}}_a} {p_a}^{{\rm{K}}_a} {(1-p_a)}^{K-{\rm{K}}_a} .
  \end{equation}
  Based on the per-user probability of misdetection/false-alarm in~\cite{noKa}, we introduce the notion of a massive random access code for the no-CSI case with random and unknown $K_a$ as follows:
  \begin{defi}[Massive random access code with no-CSI and unknown $K_a$] \label{defi3}
    Let $\mathcal{X}_k$ and $\mathcal{Y}$ denote the input alphabet of user $k$ and output alphabet, respectively. An $(n,\!M,\!\epsilon_{\mathrm{MD}},\!\epsilon_{\mathrm{FA}},\!P)_{\text{no-CSI,no-}K_{\!a}}$ massive random access code consists of
    \begin{enumerate}
      \item An encoder $\emph{f}_{\text{en},k}: [M] \mapsto \mathcal{X}_k$ that maps the message $W_k\in[M]$ to a codeword $\mathbf{x}_{(k)} \in \mathcal{X}_k$ for $k\in\mathcal{K}_a$.
          The codewords satisfy the power constraint in \eqref{power_constraint}.
          We assume that $W_k$ is equiprobable on $[M]$ for $k\in\mathcal{K}_a$.
      \item A decoder $\emph{g}_{\text{de,no-CSI,no-}K_a}: \mathcal{Y} \mapsto [M]^{|\hat{\mathcal{K}}_a|} $ that satisfies the per-user probability of misdetection constraint in~\eqref{eq:MD} and the per-user probability of false-alarm constraint in~\eqref{eq:FA}:
          \begin{equation}\label{eq:MD}
            P_{e,\mathrm{MD}} = \mathbb{E}_{K_a} \left[ 1 \left[ K_a > 0 \right]
            \cdot \frac{1}{K_a} \sum_{k\in {\mathcal{K}_a}} \mathbb{P} \left[ W_{k} \neq \hat{W}_{k} \right]\right] \leq \epsilon_{\mathrm{MD}},
          \end{equation}
          \begin{equation}\label{eq:FA}
            P_{e,\mathrm{FA}} = \mathbb{E}_{|\hat{\mathcal{K}}_a|} \left[ 1 \left[ |\hat{\mathcal{K}}_a| > 0 \right]
            \cdot  \frac{1}{ |\hat{\mathcal{K}}_a| } \sum_{k \in \hat{\mathcal{K}}_a } \mathbb{P} \left[ \hat{W}_{k} \neq W_{k} \right]\right] \leq \epsilon_{\mathrm{FA}},
          \end{equation}
          where the decoded message $\hat{W}_{k}$ for user $k$ is given by $\hat{W}_{k} = \left( \emph{g}_{\text{de,no-CSI,no-}K_a} \left( \mathbf{Y}  \right) \right)_{k}$ in the case of no-CSI with unknown $K_a$ at the decoder.
    \end{enumerate}
  \end{defi}

  Let $S_e=\frac{K_a J}{n}$ denote the spectral efficiency and $E_b =\frac{nP}{J}$ denote the energy-per-bit. The minimum energy-per-bit in the case of CSIR and no-CSI with known $K_a$ is defined as
  \begin{equation}\label{minimum_energyperbit}
   E^{*}_{b,i}(n,M,\epsilon) \triangleq \inf \left\{E_b : \exists (n,M,\epsilon,P)_{i} \text { code } \right\} , \; i \in \{ \{\text{CSIR,}K_a\} ,  \{\text{no-CSI,}K_a\} \}.
  \end{equation}
  The minimum energy-per-bit in the case of no-CSI with unknown $K_a$ is defined as
  \begin{equation}\label{minimum_energyperbit_noKa}
   E^{*}_{b,\text{no-CSI,no-}K_a}(n,M,\epsilon_{\mathrm{MD}},\epsilon_{\mathrm{FA}}) \triangleq \inf \left\{E_b : \exists (n,M,\epsilon_{\mathrm{MD}},\epsilon_{\mathrm{FA}},P)_{\text{no-CSI,no-}K_a} \text { code } \right\} .
  \end{equation}

%  {\red
%
%  We consider a MAC in which a random number $\mathrm{K}_{a}$ of users transmit their messages to a receiver
%
%  $\mathrm{K}_{a}$ follows a distribution with probability mass function (PMF) $P_{\mathrm{K}_{a}}$
%
%  We further assume that the receiver does not know $\mathrm{K}_{a}$ a priori, but can choose to estimate it
%
%  However, as opposed to [4], where the number of active users is assumed to be fixed and known, we assume that $\mathrm{K}_{a}$ is random and unknown. We therefore need to account for both MD and FA probabilities.
%
%  Consider the $\mathrm{K}_{a}$-user Gaussian MAC with $\mathrm{K}_{a} \sim P_{\mathrm{K}_{a}}$
%}

\section{Main Results} \label{section3}
  In this section, we aim to bound the minimum energy-per-bit for ensuring reliable communication in MIMO quasi-static Rayleigh fading massive random access channels with finite blocklength and finite payload size, and to provide corresponding scaling laws.
  In Section~\ref{section3_sub_goodregion}, we first introduce the main proof technique used to derive non-asymptotic achievability bounds for both the CSIR and no-CSI cases, where an appropriate ``good region'' is designed for massive random access channels.
  Next, we provide non-asymptotic bounds and scaling laws for the case of CSIR in Section~\ref{section3_sub1} and for the case of no-CSI in Section~\ref{section3_sub2}, respectively.
  Then, in Section~\ref{section3_nocsi_pilot}, we derive a non-asymptotic achievability bound for a pilot-assisted scheme.
  Several possible generalizations of our results are provided in Section~\ref{section3_generalizations}.

\subsection{``Good region'' for massive random access channels} \label{section3_sub_goodregion}

\begin{figure*}[!t]
\centering
\subfloat[]{\includegraphics[width=0.45\linewidth]{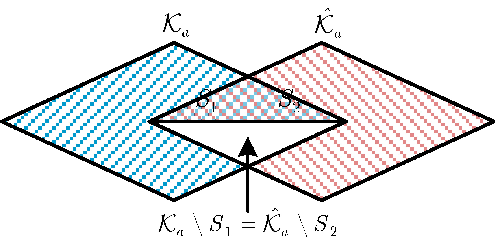} \label{fig_set:a}}
\hfil
\subfloat[]{\includegraphics[width=0.48\linewidth]{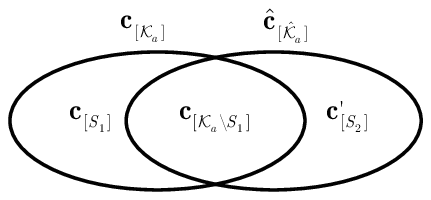}\label{fig_set:b}}
\caption{
The set relationship:
(a) users:
% $\mathcal{K}_a$ and $\hat{\mathcal{K}}_a$ denote the set of active users and identified users, respectively,
$S_1$ (in blue) denotes the set of active users whose transmitted codewords are misdecoded,
$S_2$ (in red) denotes the set of identified users with false alarm codewords,
$S_1 \cap S_2$ includes users that are correctly identified but incorrectly decoded,
and $\mathcal{K}_a \backslash S_1 = \hat{\mathcal{K}}_a \backslash S_2$ (in white) includes users that are correctly identified and correctly decoded;
(b)~codewords: for $S \in \left\{ S_1, \mathcal{K}_a, \mathcal{K}_a\backslash S_1 \right\}$, the set $\mathbf{c}_{[S]}$ includes codewords transmitted by users in the set $S$,
$\hat{\mathbf{c}}_{[\hat{\mathcal{K}}_a]}$ denotes the set of decoded codewords for users in the set $\hat{\mathcal{K}}_a$,
and the set $\mathbf{c}'_{[S_2]}$ includes false alarm codewords corresponding to users in the set $S_2$.}
\label{fig_set}
\end{figure*}

  In this subsection, we consider a special case where the number $K_a$ of active users is known \emph{a priori}.
  A crucial step to derive an achievability bound on the minimum required energy-per-bit is to establish an upper bound on the probability $\mathbb{P}\left[\mathcal{F}_{t,S_1}\right]$ with fixed blocklength~$n$, payload $J$, and power $P$.
  Here, $\mathcal{F}_{t,S_1}$ denotes the event that there are exactly $t$ misdecoded codewords transmitted by users in the set $S_1\subset\mathcal{K}_a$.
  In massive random access channels, a major challenge lies in that the event $\mathcal{F}_{t,S_1}$ is the union of a massive number of error events and most of them are not disjoint.
  Specifically, we have $\mathcal{F}_{t,S_1} = \bigcup_{S_2} \bigcup_{\mathbf{c}_{[S_2]}^{'}} \mathcal{F}_{t,S_1,S_2,\mathbf{c}_{[S_2]}^{'}}$.
  Here, the set $S_2\subset \mathcal{K} \backslash \mathcal{K}_a \cup S_1$ of size $t$ includes identified users with false alarm codewords, and it is worth noting that $S_2$ can also take values in $S_1$ because some users that are correctly identified can still be incorrectly decoded;
  the set $\mathbf{c}_{[S_2]}^{'}$ includes $t$ false alarm codewords corresponding to users in the set $S_2$.
  As a result, the event $\mathcal{F}_{t,S_1}$ is the union of about ${\binom {K-K_{a}+t} {t}} M^t$ events, which is considerably large for the massive random access communication problem.
  The set relationship is presented in Fig.~\ref{fig_set}.

  A classical method of upper-bounding $\mathbb{P}\left[\mathcal{F}_{t,S_1}\right]$ is applying the union bound, which yields $\mathbb{P}\left[ \mathcal{F}_{t,S_1} \right] \leq \sum_{S_2} \sum_{\mathbf{c}_{[S_2]}^{'}} \mathbb{P}\left[ \mathcal{F}_{t,S_1,S_2,\mathbf{c}_{[S_2]}^{'}} \right]$.
  However, it may be very loose when the number of terms in the summation is large, as in the massive random access scenario considered in this paper.
  In order to tightly upper-bound the probability of the union of extremely many events, a standard bounding technique was proposed by Fano~\cite{1961}, which upper-bounds $\mathbb{P}\left[\mathcal{F}_{t,S_1}\right]$ as~follows:
  \begin{equation}
    \mathbb{P}\left[\mathcal{F}_{t,S_1}\right]
    \leq \mathbb{P}\left[\mathcal{F}_{t,S_1}, \mathbf{Y} \in \mathcal{R}_{t,S_1}\right]
         +\mathbb{P}\left[\mathbf{Y} \notin \mathcal{R}_{t,S_1}\right],
         \label{eq_good region_1961}
  \end{equation}
  where $\mathbf{Y}$ denotes the received signal and $\mathcal{R}_{t,S_1}$ represents a region around the linear combination of the transmitted signals, also known as the ``good region''~\cite{goodregion}.
  The union bound is only applied on the first term on the right-hand side (RHS) of~\eqref{eq_good region_1961}, and the second term on the RHS of~\eqref{eq_good region_1961} can be tightly bounded and even accurately computed if $\mathcal{R}_{t,S_1}$ is chosen appropriately.
  With this technique, the probability of the union of many events can be tightly bounded.

  To get a tight non-asymptotic achievability bound in massive random access channels, we design an appropriate ``good region'' $\mathcal{R}_{t,S_1}$ in the remainder of this subsection.
  Assuming there is no power constraint, let $\mathbf{c}_{(k)}$ denote the transmitted codeword of the $k$-th user, which is chosen uniformly at random from its codebook $\mathcal{C}_k$.
  For a given received signal $\mathbf{Y}$, the decoder searches for the estimated set of active users, i.e. ${ \hat{\mathcal{K}}_{a } \subset \mathcal{K}}$ of size $K_a$, and the estimated set of transmitted codewords, i.e. $\hat{\mathbf{c}}_{[\hat{\mathcal{K}}_a]} = \left\{ \hat{\mathbf{c}}_{(k)} \in \mathcal{C}_k:  k\in\hat{\mathcal{K}}_a \right\}$, to minimize the decoding metric $g\left(  \mathbf{Y} ,  \hat{\mathbf{c}}_{[\hat{\mathcal{K}}_a]}  \right)$.
  % Here, $g\left( \cdot , \cdot \right)$ is a non-negative function (see Appendix \ref{Proof_achiCSIR}, \ref{section5}, and \ref{section6} for the expression of $g\left( \cdot , \cdot \right)$ in the case of CSIR, noncoherent transmission, and pilot-assisted transmission, respectively).
  An error event $\mathcal{F}_{t,S_1}$ occurs, if there exists a set of codewords $\mathbf{c}_{[\mathcal{K}_a \backslash S_1]} \cup \mathbf{c}^{'}_{[S_2]}$ satisfying $g\left(  \mathbf{Y}, \mathbf{c}_{[\mathcal{K}_a \backslash S_1]} \cup \mathbf{c}^{'}_{[S_2]} \right) \leq g\left(  \mathbf{Y}, \mathbf{c}_{[\mathcal{K}_a]}   \right)$,
  where $\mathbf{c}_{[S]} = \left\{ \mathbf{c}_{(k)} \in \mathcal{C}_k:  k\in S \right\}$ and $\mathbf{c}_{[S]}^{'} = \left\{ \mathbf{c}'_{(k)} \in \mathcal{C}_k: k \in S, \mathbf{c}'_{(k)} \neq \mathbf{c}_{(k)} \right\}$ for the set $S\subset \mathcal{K}$.
  Roughly speaking, the more similar the ``good region'' $\mathcal{R}_{t,S_1}$ is to the Voronoi region $\mathcal{V}_{t,S_1}$, the tighter the upper bound on the RHS of~\eqref{eq_good region_1961} is but the higher the complexity is to compute this bound~\cite{goodregion}, where $\mathcal{V}_{t,S_1}$ is given by
  \begin{equation}
    \mathcal{V}_{t,S_1} = \left\{ \mathbf{Y}: g\left(  \mathbf{Y}, \mathbf{c}_{[\mathcal{K}_a]} \right)
    \leq  g\left(  \mathbf{Y} , \mathbf{c}_{[\mathcal{K}_a \backslash S_1]} \cup \mathbf{c}^{'}_{[S_2]} \right) , \forall S_2\subset \mathcal{K} \backslash \mathcal{K}_a \cup S_1
    , \forall \mathbf{c}^{'}_{[S_2]}   \right\}.
  \end{equation}
  For massive random access in MIMO fading channels, the \emph{``good region''} $\mathcal{R}_{t,S_1}$ used for deriving a tight upper bound on the probability $\mathbb{P}\left[\mathcal{F}_{t,S_1}\right]$ in~\eqref{eq_good region_1961} is selected as follows:
  \begin{equation} \label{eq_good region_selection}
    \mathcal{R}_{t,S_1} = \left\{ \mathbf{Y}: g\left(  \mathbf{Y}, \mathbf{c}_{[\mathcal{K}_a]}  \right)
    \leq \omega g\left(  \mathbf{Y} , \mathbf{c}_{[\mathcal{K}_a \backslash S_1]}  \right) + \nu nL  \right\},
  \end{equation}
  where $0\leq \omega \leq 1$ and $\nu\geq 0$.
  % in the case of CSIR and noncoherent no-CSI, and choose $\omega=0$ and $\nu\geq 0$ to derive the achievability bound for the pilot-assisted coded access scheme for simplicity.
  By adjusting $\omega$ and $\nu$, we can find a ``good region'' $\mathcal{R}_{t,S_1}$ similar to the Voronoi region $\mathcal{V}_{t,S_1}$.
  As a result, when the received signal $\mathbf{Y}$ falls inside $\mathcal{R}_{t,S_1}$, $K_a$ transmitted codewords are likely to be correctly decoded rather than with $t$ misdecoded codewords corresponding to users in the set~$S_1$.

\begin{figure*}[!t]
\centering
\subfloat[]{\includegraphics[width=0.44\linewidth]{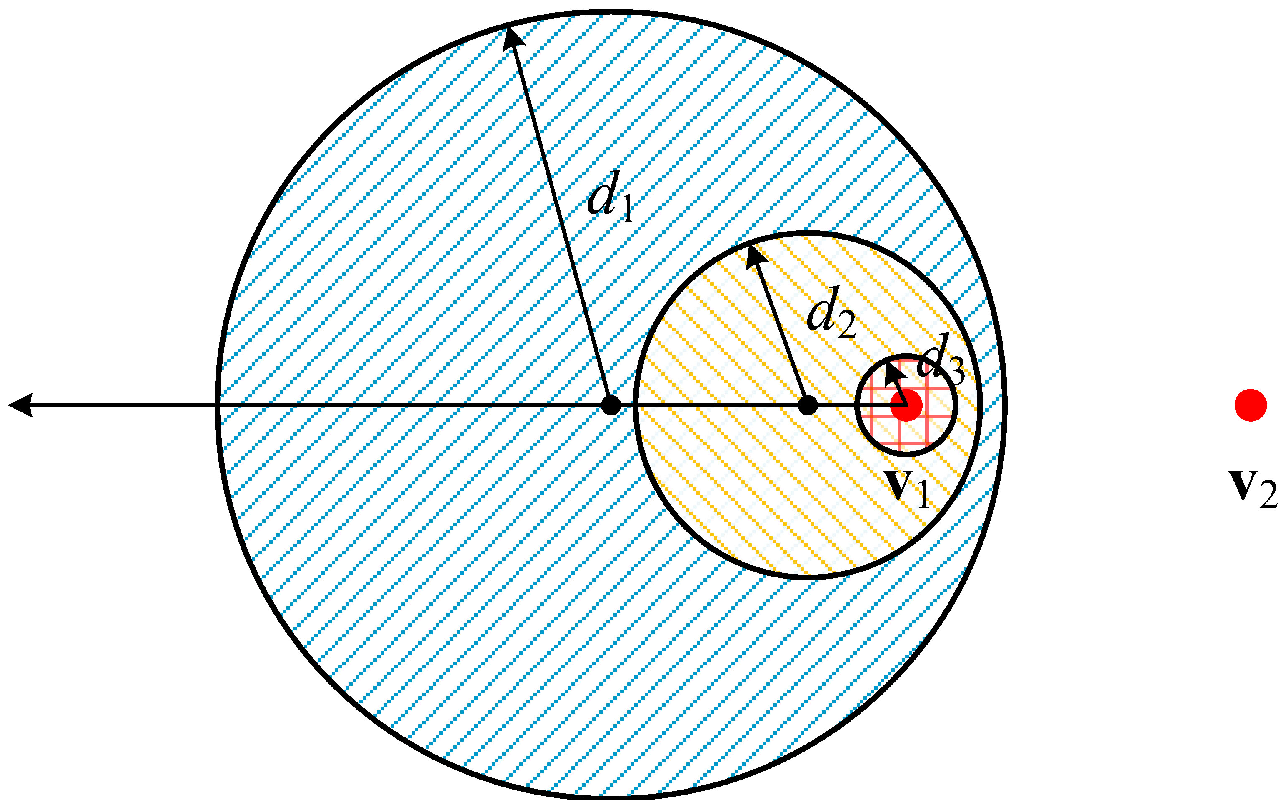} \label{fig_goodregion:a}}
\hfil
\subfloat[]{\includegraphics[width=0.35\linewidth]{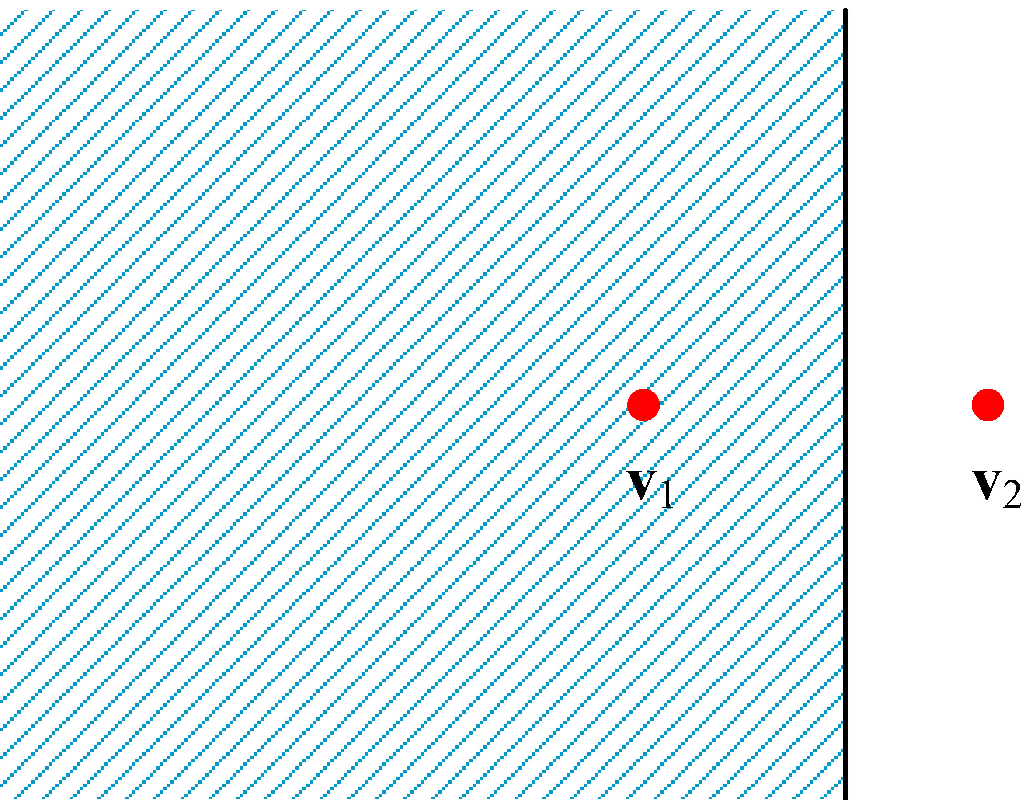}\label{fig_goodregion:b}}
\caption{A geometric illustration of the cross section of the ``good region'' $\mathcal{R}_{t,S_1}$ in the CSIR case: (a) $0= \omega_3 < \omega_2 < \omega_1 < 1$, $\nu>0$; (b) $\omega=1$, $\nu>0$.}
\label{fig_goodregion}
\end{figure*}
  In the following, we take the case of CSIR as an example to clearly illustrate the ``good region'' $\mathcal{R}_{t,S_1}$. Based on the ML decoding metric, the region $\mathcal{R}_{t,S_1}$
  in~\eqref{eq_good region_selection} can be expressed as
\begin{equation} \label{eq_CSIR_goodregion_R1}
  \mathcal{R}_{t,S_1} = \left\{ \mathbf{Y}: \sum_{l = 1}^{L} \left\| \mathbf{y}_l - \sum_{k\in {\mathcal{K}}_{a}} h_{k,l} \mathbf{c}_{(k)} \right\|_2^2
\leq  \omega \sum_{l = 1}^{L} \left\| \mathbf{y}_l - \sum_{k\in \mathcal{K}_{a} \backslash {S}_{1}} h_{k,l} \mathbf{c}_{(k)} \right\|_2^2 + \nu nL \right\}.
\end{equation}
  In the special case of $0\leq \omega < 1$, by straightforward manipulations, the ``good region'' $\mathcal{R}_{t,S_1}$ in~\eqref{eq_CSIR_goodregion_R1} can be rewritten as
\begin{align} \label{eq_CSIR_goodregion_R1_2}
  \mathcal{R}_{t,S_1} &= \left\{ \mathbf{Y}: \sum_{l = 1}^{L}  \left\| \mathbf{y}_l - \frac{\sum_{k\in {\mathcal{K}}_{a}} h_{k,l} \mathbf{c}_{(k)} - \omega \sum_{k\in {\mathcal{K}}_{a}\backslash {S}_{1}} h_{k,l} \mathbf{c}_{(k)} }{1-\omega} \right\|_2^2 \right. \notag \\
  & \;\;\;\;\;\;\;\;\;\;\;\;\;\;  \left.\leq \frac{\omega}{(1-\omega)^2} \sum_{l = 1}^{L} \left\|\sum_{k \in  {S}_{1}} h_{k,l} \mathbf{c}_{(k)} \right\|_2^2  + \frac{\nu nL}{1-\omega}\right\}.
\end{align}
  It can be regarded as a sphere with flexible center and radius for different fading coefficients and codewords.
  For convenience, we denote $\mathbf{v}_1 = \left[ \sum_{k\in {\mathcal{K}}_{a}} h_{k,1} \mathbf{c}_{(k)}, \ldots,  \sum_{k\in {\mathcal{K}}_{a}} h_{k,L} \mathbf{c}_{(k)} \right] \in \mathbb{C}^{n\times L}$ and $\mathbf{v}_2 = \left[ \sum_{k\in {\mathcal{K}}_{a}\backslash {S}_{1}} h_{k,l} \mathbf{c}_{(k)} , \ldots,  \sum_{k\in {\mathcal{K}}_{a}\backslash {S}_{1}} h_{k,L} \mathbf{c}_{(k)} \right] \in \mathbb{C}^{n\times L}$.
  The center of this sphere can be any point in the ray with endpoint $\mathbf{v}_1$ and direction $ \mathbf{v}_1-\mathbf{v}_2 $; the radius of this sphere is  $\sqrt{ \frac{\omega}{(1-\omega)^2}  \left\| \mathbf{v}_1 - \mathbf{v}_2 \right\|_F^2  + \frac{\nu nL}{1-\omega} }$.
  When $\omega = 0$, the region $\mathcal{R}_{t,S_1}$ becomes a sphere with center $\mathbf{v}_1$ and radius $\sqrt{\nu nL}$.
  We illustrate the cross section of the region $\mathcal{R}_{t,S_1}$ with $0\leq \omega < 1$ in Fig.~\ref{fig_goodregion:a}.
  As mentioned above, the region $\mathcal{R}_{t,S_1}$ (the shaded area) is around the sum of the faded codewords transmitted from active users, i.e., around $\mathbf{v}_1$.
  As in Fig.~\ref{fig_goodregion:a}, for a given $\nu$, as $\omega$ increases, the radius of the sphere gradually increases and its center (located in the ray with endpoint $\mathbf{v}_1$ and direction $ \mathbf{v}_1-\mathbf{v}_2 $) gradually moves away from $\mathbf{v}_1$.
  In the special case of $\omega = 1$, the region $\mathcal{R}_{t,S_1}$ becomes a halfspace as shown in Fig. \ref{fig_goodregion:b}.
  In other words, the upper bound based on the region $\mathcal{R}_{t,S_1}$ reduces to the commonly used sphere bound~\cite{SB} and tangential bound~\cite{TB} in some special~cases.

  In general, the ``good region'' $\mathcal{R}_{t,S_1}$ in~\eqref{eq_good region_selection} has some properties as follows:
  \begin{itemize}
    \item {When the ``good region'' $\mathcal{R}_{t,S_1}$ in~\eqref{eq_good region_selection} is the whole observation space, such as in the case of $\omega = 0$ and $\nu = \infty$, the upper bound on $\mathbb{P}\left[\mathcal{F}_{t,S_1}\right]$ in~\eqref{eq_good region_1961} based on $\mathcal{R}_{t,S_1}$ reduces to that obtained by straightforwardly applying the union bound to $\mathbb{P}\left[\mathcal{F}_{t,S_1}\right]$ as provided above.}
    \item {In the special case of $\omega=0$, the ``good region'' $\mathcal{R}_{t,S_1}$ is independent of the set $S_1$ and reduces to $\mathcal{R} = \left\{ \mathbf{Y}: g\left(  \mathbf{Y}, \mathbf{c}_{[{\mathcal{K}}_a]} \right) \leq \nu nL \right\}$.
        In essence, the transmitted signals from active users are treated as a whole for $\mathcal{R}$ with $\omega=0$, which is equivalent to the case of a single user. However, in the case of $0<\omega\leq1$, the region $\mathcal{R}_{t,S_1}$ relies on the set of misdecoded users, which incorporates more details of the massive access model.}
    \item {Our ``good region'' in~\eqref{eq_good region_selection} is parameterized by two parameters $\omega$ and $\nu$, which reduces to the one used in~\cite{finite_payloads_fading} if $\nu$ is set to 0.
        In general, in order to derive non-asymptotic achievability bounds, using our ``good region'' in~\eqref{eq_good region_selection} is better than using the one in~\cite{finite_payloads_fading} for two reasons:
        \begin{itemize}
          \item We have better control of the ``good region'' when taking both $\omega$ and $\nu$ into consideration. Thus, the upper bound based on $\mathcal{R}_{t,S_1}$ is tighter than that based on the region in~\cite{finite_payloads_fading}.
              Specifically, when $\omega=0$, the region
              in~\eqref{eq_good region_selection} reduces to $\mathcal{R}$ as explained above, but the upper bound in~\cite{finite_payloads_fading} diverges in this case.
              Moreover, as in~\eqref{eq_CSIR_goodregion_R1_2}, in the CSIR case with $0\leq \omega < 1$, the region $\mathcal{R}_{t,S_1}$ is essentially a sphere, where its center is~determined by $\omega$ and its radius is controlled by both $\omega$ and $\nu$.
              However, for the region with $\nu=0$, both the center and the radius are controlled by $\omega$.
              Thus, the value of the radius depends on the position of the center for the region in~\cite{finite_payloads_fading}, whereas the radius of our ``good region'' can be flexibly changed by adjusting $\nu$.
              As a result, it is more likely to find a ``good region'' similar to the Voronoi region by simultaneously adjusting $\omega$ and $\nu$.
          \item The problem of searching for an appropriate ``good region'' $\mathcal{R}_{t,S_1}$ can be expressed as $\arg \min_{\omega,\nu}   f_{t,S_1}(\omega,\nu)$, where $f_{t,S_1}(\omega,\nu)$ denotes an upper bound on $\mathbb{P}[\mathcal{F}_{t,S_1}]$.
              Since it is difficult to obtain closed-form solutions for the optimal values of $\omega$ and $\nu$, we resort to numerical evaluations with exhaustive search to find $\omega$ and $\nu$ that yield a tight bound.
              Note that since the dependency of $f_{t,S_1}(\omega,\nu)$ on $\omega$ is more complicated than its dependency on $\nu$, there is a much higher complexity when searching for $\omega$ than $\nu$ (see Theorem~\ref{Theorem_noCSI_achi} for instance).
              In contrast to the case of $\nu=0$, the feasible region of $\omega$, in which the error requirement is satisfied, is enlarged when both $\omega$ and $\nu$ are taken into consideration.
              As a result, by introducing $\nu$ in~\eqref{eq_good region_selection}, we can reduce the number of sampling points when searching for $\omega$, thereby reducing the complexity of finding an appropriate ``good region''.
        \end{itemize}
        }
  \end{itemize}

\subsection{CSIR} \label{section3_sub1}
  In this subsection, we consider the case where CSI and the number of active users are available at the receiver, and establish non-asymptotic bounds for the massive random access model described in Section~\ref{section2}.
  Specifically, we establish an upper bound on the PUPE in Theorem~\ref{Theorem_CSIR_achi}.
  On the basis of it, Corollary~\ref{Theorem_CSIR_achi_energyperbit} is obtained, which presents an achievability bound (upper bound) on the minimum required energy-per-bit for massive random access.
  In a special case where all users are assumed to be active, we obtain a simplified achievability bound in Corollary \ref{Theorem_CSIR_achi_knownUE}.
  Then, in Theorem~\ref{prop_converse_CSIR}, we establish a converse bound (lower bound) on the minimum required energy-per-bit assuming user activity is known, and thus it can also be regarded as a converse bound for massive random access.
  Finally, on the basis of Corollary~\ref{Theorem_CSIR_achi_knownUE} and Theorem~\ref{prop_converse_CSIR}, we establish scaling laws in Theorem~\ref{Theorem_scalinglaw_CSIR} for a special case where all users are assumed to be active.

\subsubsection{Achievability bound} \label{section3_sub1_subsub1}
  An upper bound on the PUPE for massive random access in MIMO quasi-static Rayleigh fading channels with CSIR and known $K_a$ is given in Theorem~\ref{Theorem_CSIR_achi}.
  \begin{Theorem} \label{Theorem_CSIR_achi}
    Assume that there are $K_a$ active users among $K$ potential users each equipped with a single antenna and the number of BS antennas is $L$. Each user has an individual codebook with size $M=2^J$ and length $n$ satisfying the maximum power constraint in \eqref{power_constraint}.
    For massive random access in MIMO quasi-static Rayleigh fading channels with CSIR and known $K_a$, the PUPE can be upper-bounded as
    \begin{equation} \label{eq_CSIR_PUPE}
      P_e \leq \min_{0< P'< P} \left\{ p_0 + \sum_{t=1}^{K_{a }} \frac{ t}{K_a} \min \left\{ 1, p_{1,t}, p_{2,t} \right\} \right\} ,
    \end{equation}
    where
    \begin{equation} \label{eq_CSIR_p0}
      p_0=K_a \left( 1 - \frac{\gamma\left(n, \frac{nP}{P'}\right)}{\Gamma\left(n\right)}\right),
    \end{equation}
    \begin{equation} \label{eq_CSIR_p2t}
      p_{1,t} = \min_{ 0 \leq \omega \leq 1, 0 \leq \nu }
       \left\{ q_{1,t}\!\left( \omega ,\nu \right)
       + q_{2,t}\!\left( \omega ,\nu \right) \right\}  ,
    \end{equation}
    \begin{equation} \label{eq_CSIR_q1t}
       q_{1,t}\!\left( \omega ,\nu \right) = \sum_{t_0=0}^{t} C_{t_0,t} \; \!\mathbb{E}_{ \tilde{\mathbf{A}}_{S_1}, \tilde{\mathbf{A}}^{'}_{S_2}  } \!\! \left[ \min_{\substack{ {u\geq 0,r\geq 0,} \\ { \lambda_{\min}\left({\tilde{\mathbf{B}}}\right) >-1 } } }  \!\!\!\!
       \exp\!\left\{ \!-L \!\left( n \ln\!\left( 1+r \left( 1-\omega \right) \right) + \ln \!\left| \mathbf{I}_{K} + \tilde{\mathbf{B}} \right| - rn\nu \right) \right\} \right] \!,
    \end{equation}
    \begin{equation} \label{eq_CSIR_Ct}
       C_{t_0,t} = {\binom {K_a}{t}}  {\binom {t}{t_0}}  {\binom {K-K_a}{t-t_0}} (M-1)^{t_0}  M^{t-t_0},
    \end{equation}
    \begin{equation} \label{eq_CSIR_q1t_B}
        \tilde{\mathbf{B}} = \frac{ (1+r-u)(u-r\omega) }{ 1+r\left( 1-\omega \right) }  \left( \tilde{\mathbf{A}}_{S_1} -\! \frac{u}{u-r\omega} \tilde{\mathbf{A}}^{'}_{S_2} \!\right)^H   \!\left( \tilde{\mathbf{A}}_{S_1} \!- \frac{u}{u-r\omega} \tilde{\mathbf{A}}^{'}_{S_2} \right)
      - \frac{r\omega u}{u-r\omega}  \left(\tilde{\mathbf{A}}^{'}_{S_2}\right)^{\!H}  \!\tilde{\mathbf{A}}^{'}_{S_2},
    \end{equation}
    \begin{equation} \label{eq_CSIR_q2t}
        q_{2,t}\!\left( \omega ,\nu \right)   \!= \!\left\{
        \begin{array}{ll}
            \!\!\min\limits_{ {\substack{{\eta\geq 0}\\{\delta\geq 0}}} } \!{\binom {K_a} {t}}
\mathbb{E}_{ \tilde{\mathbf{A}}_{S_1}  } \!\!  \left[ \frac{\gamma\left( tL, L { \left(t(1+\eta) - n\nu + n(1+\delta)(1-\omega) \right)  {\left| \mathbf{I}_n + \omega \tilde{\mathbf{A}}_{S_1} \tilde{\mathbf{A}}_{S_1}^{H} \right|^{-\frac{1}{t}}}   } \right)}{\Gamma\left( tL \right)} \right] \\
\;\;\;\;\;\;\;\;\;  +  {\binom {K_a} {t}} \left( 2 - \frac{\gamma\left( tL, tL \left( 1+\eta \right)\right)}{\Gamma\left( tL \right)}
- \frac{\gamma\left( nL, nL \left( 1+\delta \right)\right)}{\Gamma\left( nL \right)}  \right),
&  \!\!\!t < n , \omega \in \!(0,1] \\
           \!\!\min\limits_{ \eta\geq0 } {\binom {K_a} {t}} \mathbb{E}_{ \tilde{\mathbf{A}}_{S_1}  } \! \!\left[ \frac{\gamma\left( nL, \frac{nL \left( { 1+\eta - \nu  }  \right)}{ \omega\left| \mathbf{I}_n + \tilde{\mathbf{A}}_{S_1} \tilde{\mathbf{A}}_{S_1}^{H} \right|^{1/n} } \right)}{\Gamma\left( nL \right)} \right]
+   1 - \frac{\gamma\left( nL, nL \left( 1+\eta \right)\right)}{\Gamma\left( nL \right)}, & \!\!\!t \geq n , \omega \in\! (0,1] \\
           \!1 - \frac{\gamma\left( nL, nL \nu \right)}{\Gamma\left( nL \right)}, & \!\!\!\omega = 0
        \end{array}\right. \!\!,\!
    \end{equation}
    \begin{align}\label{eq_CSIR_p1t}
      p_{2,t} \!= \! \min_{ 0\leq\rho\leq 1,0\leq\beta<\frac{1}{\rho} }
      \sum_{t_0=0}^{t} {\binom {K_a}{t}}  {\binom {t}{t_0}}  {\binom {K-K_a}{t-t_0}} M^{\rho t}\;
      \mathbb{E}_{ \mathbf{H}_1, \mathbf{H}_2 }\! & \left[ \exp \! \left\{ (1-\rho) n \ln \left| \mathbf{I}_{ L}+ \beta P' \mathbf{H}_2^H \mathbf{H}_2 \right|  \right. \right.  \notag\\
      & \!\!\!\!\!\!\!\!\!\!\!\!\!\!\!\!\!\!\!\!\!\!\!\!\!\!
      \!\!\!\!\!\!\!\!\!\!\!\!\!\!\!\!\!\!\!\!
      \left.\left.- n \ln\! \left| \mathbf{I}_{ L} +  \beta \left(1-\rho\beta \right) \!P' \! \left(\rho \mathbf{H}_1^H \mathbf{H}_1  + \mathbf{H}_2^H \mathbf{H}_2 \right) \right| \right\} \right] .
    \end{align}
  Here, $\mathbf{H}_1$ and $\mathbf{H}_2$ are $t\times L$ submatrices of $\mathbf{H}\in\mathbb{C}^{K\times L}$ formed by rows corresponding to the support of $S_1$ and $S_2$, respectively;
  $\mathbf{H}\in\mathbb{C}^{K\times L}$ has i.i.d. $\mathcal{CN}(0,1)$ entries;
  $S_{1}$ is an arbitrary $t$-subset of $\mathcal{K}_a$;
  $S_{2} = S_{2,1}\cup S_{2,2}$, where $S_{2,1}$ is an arbitrary $t_0$-subset of $S_1$ and $S_{2,2}$ is an arbitrary $(t-t_0)$-subset of $\mathcal{K} \backslash \mathcal{K}_a$;
  $\tilde{\mathbf{A}}_{S_1} = {\mathbf{A}} \boldsymbol{\Phi}_{ {S}_1} $ and $\tilde{\mathbf{A}}^{'}_{S_2} = {\mathbf{A}} \boldsymbol{\Phi}^{'}_{ {S}_2} $;
  the matrix $\mathbf{A}\in\mathbb{C}^{n\times MK}$ is the concatenation of codebooks of the $K$ users without power constraint, which has i.i.d. $\mathcal{CN}\left(0, P'\right)$ entries;
  the binary selection matrix $\boldsymbol{\Phi}_{ {S}_1} \in\{0,1\}^{MK\times K}$ indicates which codewords are transmitted by users in the set $ {S}_1$, where $\left[\boldsymbol{\Phi}_{ {S}_1}\right]_{(k-1)M + W_k,k} = 1$ if user $k\in S_1$ is active and transmits the $W_{k}$-th codeword, and $\left[\boldsymbol{\Phi}_{ {S}_1}\right]_{(k-1)M + W_k,k} = 0$ otherwise;
  and similarly, $\boldsymbol{\Phi}^{'}_{ {S}_2} \in\{0,1\}^{MK\times K} $ indicates which codewords are not transmitted but decoded for users in the set ${S}_2$.
  % {\blue $\mathbf{A}_{S_1}$ is an $n\times t$ submatrix of $\mathbf{A}$ formed by columns corresponding to the codewords transmitted by users in the set $ {S}_1$.}
  \begin{IEEEproof}[Proof sketch]
    We use a random coding scheme and an ML decoder, which searches for all possible support sets and finds the one that maximizes the likelihood function.
    As in~\eqref{eq_CSIR_PUPE}, the upper bound on the PUPE comprises of two terms: the first term $p_0$ upper-bounds the total variation distance between the measure with power constraint and the one without power constraint, whose expression is given in~\eqref{eq_CSIR_p0} relying on a straightforward utilization of the union bound;
    the second term $ \sum_{t=1}^{K_{a }} \frac{ t}{K_a} \min \left\{ 1, p_{1,t}, p_{2,t} \right\} $ upper-bounds the PUPE assuming there is no power constraint.
    Here, $p_{1,t}$ and $p_{2,t}$ denote two upper bounds on $\mathbb{P}\left[ \mathcal{F}_t \right]$, which indicates the probability of the event that there are exactly $t$ misdecoded users.
    We have $\mathbb{P}\left[\mathcal{F}_t\right] \leq \binom{K_a}{t} \mathbb{P}\left[ \mathcal{F}_{t, {S}_1} \right]$.
    As mentioned in Section~\ref{section3_sub_goodregion}, upper-bounding $\mathbb{P}\left[ \mathcal{F}_{t, {S}_1} \right]$ is involved since $\mathcal{F}_{t, {S}_1}$ is the union of a massive number of events.
    Two upper bounds on $\mathbb{P}\left[ \mathcal{F}_t \right]$, i.e. $p_{1,t}$ and $p_{2,t}$, are obtained as follows:
  \begin{itemize}
    \item In Appendix~\ref{Proof_achi_CSIR_noCSI}, we derive a general upper bound on the PUPE based on Fano's bounding technique~\cite{1961}.
        We obtain $p_{1,t}$ by particularizing this general bound to the CSIR case and performing additional manipulations as introduced in Appendix~\ref{Proof_achiCSIR_2}.
        Specifically, we upper-bound $\mathbb{P}\left[ \mathcal{F}_t \right]$ by the sum of two terms as presented in~\eqref{eq_CSIR_p2t}.
        The first term $q_{1,t}\!\left( \omega ,\nu \right)$ denotes an upper bound on the probability of the joint event that the decoder yields exactly $t$ misdecoded users and the received signal falls inside the ``good region''.
        The expression of $q_{1,t}\!\left( \omega ,\nu \right)$ is given in~\eqref{eq_CSIR_q1t}, which is obtained by applying the union bound, Chernoff bound, and moment generating function of quadratic forms~\cite{VP}.
        The second term $q_{2,t}\!\left( \omega ,\nu \right)$ upper-bounds the probability of the event that the received signal falls outside this region, whose expression is given in~\eqref{eq_CSIR_q2t}.
    \item The expression of $p_{2,t}$ is given in \eqref{eq_CSIR_p1t}, which is derived relying on Gallager's $\rho$-trick~\cite{1965} as introduced in Appendix~\ref{Proof_achiCSIR_1}.
        % with the proof provided in Appendix~\ref{Proof_achiCSIR_1}, which mainly relies on Gallager's $\rho$-trick~\cite{1965}.
        Specifically, given a set $ {S}_1$ including $t$ misdecoded users and a set $ {S}_2$ including $t$ detected users with false alarm codewords,  Gallager's $\rho$-trick is applied to the union of about $M^t$ events, corresponding to different sets of false alarm codewords.
  \end{itemize}
    See Appendix \ref{Proof_achiCSIR} for the complete proof.
  \end{IEEEproof}
\end{Theorem}

  The following corollary of Theorem~\ref{Theorem_CSIR_achi} provides an achievability bound on the minimum required energy-per-bit for the massive random access problem with CSIR and known $K_a$.
  \begin{Corollary} \label{Theorem_CSIR_achi_energyperbit}
      Assume that there are $K_a$ active users among $K$ potential users each equipped with a single antenna and the number of BS antennas is $L$. Each user has an individual codebook with size $M=2^J$ and length $n$ satisfying the maximum power constraint in \eqref{power_constraint}.
      For massive random access in MIMO quasi-static Rayleigh fading channels with CSIR and known $K_a$, the minimum energy-per-bit $E^{*}_{b,\text{CSIR},K_a}(n,M,\epsilon)$ for satisfying the PUPE requirement in \eqref{PUPE} can be upper-bounded as
      \begin{equation}
        E^{*}_{b,\text{CSIR},K_a}(n,M,\epsilon) \leq \inf \frac{n {P}}{J},
      \end{equation}
      where the $\inf$ is taken over all $P > 0$ satisfying that
      \begin{equation} \label{eq_CSIR_epsilon}
        \epsilon \geq \min_{0< P'< P} \left\{ p_0 + \sum_{t=1}^{K_{a }} \frac{ t}{K_a} \min \left\{ 1, p_{1,t}, p_{2,t} \right\} \right\}.
      \end{equation}
      Here, $p_0$, $p_{1,t}$, and $p_{2,t}$ are the same as those in Theorem~\ref{Theorem_CSIR_achi}.
  \end{Corollary}

  In a special case where all users are assumed to be active, Corollary~\ref{Theorem_CSIR_achi_energyperbit} reduces to the following Corollary~\ref{Theorem_CSIR_achi_knownUE}.
  % , which explores the effect of multiple BS antennas on the massive access communication problem.
  In essence, the achievability bound for the case where all users are active is equivalent to that with knowledge of the active user set.
  \begin{Corollary} \label{Theorem_CSIR_achi_knownUE}
    Assume that all users are active, i.e. $K_a=K$.
    Suppose each user is equipped with a single antenna and the number of BS antennas is $L$.
    Each user has an individual codebook with size $M=2^J$ and length $n$ satisfying the maximum power constraint in \eqref{power_constraint}.
    In MIMO quasi-static Rayleigh fading channels with CSIR, the minimum energy-per-bit $E^{*}_{b,\text{CSIR},K_a}(n,M,\epsilon)$ for satisfying the PUPE requirement in \eqref{PUPE} can be upper-bounded as
      \begin{equation}
        E^{*}_{b,\text{CSIR},K_a}(n,M,\epsilon) \leq \inf \frac{n {P}}{J},
      \end{equation}
    where the $\inf$ is taken over all $P > 0$ satisfying that
      \begin{equation} \label{eq_CSIR_epsilon_knownUE}
        \epsilon \geq \min_{0< P'< P} \left\{ \tilde{p}_0 + \sum_{t=1}^{K } \frac{ t}{K } \min\left\{ 1, \tilde{p}_{1,t}, \tilde{p}_{2,t} \right\} \right\}.
      \end{equation}
    Here, $\tilde{p}_0$ follows from $p_0$ in~\eqref{eq_CSIR_p0} by allowing $K_a=K$;
    $\tilde{p}_{1,t}$ is obtained by assuming $ {S}_1 = {S}_2$ and $K_a=K$ in \eqref{eq_CSIR_p2t}, \eqref{eq_CSIR_q1t}, \eqref{eq_CSIR_Ct}, \eqref{eq_CSIR_q1t_B}, and \eqref{eq_CSIR_q2t};
    and $\tilde{p}_{2,t}$ is given by
    \begin{equation}
      \tilde{p}_{2,t}
      = \min_{ 0\leq\rho\leq 1, \rho n \in \mathbb{N}_{+} } {\binom {K } {t}}
      M^{\rho t}\;
      \mathbb{E}_{ \mathbf{G} } \! \left[ \left| \mathbf{I}_{t} + \frac{P'}{1+\rho} {\mathbf{G}} {\mathbf{G}}^H \right|^{-L} \right], \label{eq_CSIR_p1t_knownUE}
    \end{equation}
    where each element of $\mathbf{G} \in \mathbb{C}^{t\times \rho n}$ is i.i.d. $\mathcal{CN}\left( 0,1 \right)$ distributed.
    To simplify simulation complexities, $\tilde{p}_{2,t}$ can be further upper-bounded as
    \begin{equation}\label{eq_CSIR_p1t_knownUE_bound1}
        \tilde{p}_{2,t} \leq \tilde{p}_{2,t}^{\rm{u}} = \min_{ 0\leq\rho\leq 1, \rho n \in \mathbb{N}_{+} } q(\rho),
    \end{equation}
    \begin{equation}\label{eq_CSIR_p1t_knownUE_bound}
        q(\rho) = \left\{
        \begin{array}{ll}
            {\binom {K } {t}}
            M^{\rho t}  \left( \frac{P'}{ 1+\rho } \right)^{-Lt}
            \prod_{i=\rho n - t + 1}^{\rho n}  \frac{\Gamma(i-L)}{\Gamma(i)},
            &  \rho n \geq t + L   \\
            {\binom {K } {t}}
            M^{\rho t}  \left( \frac{P'}{1+\rho} \right)^{-L\rho n }
            \prod_{i=t - \rho n + 1}^{t}  \frac{\Gamma(i-L)}{\Gamma(i)},
            &  \rho n \leq t - L   \\
            1,
            &  t - L < \rho n < t + L    \\
        \end{array}\right. .
    \end{equation}
  \begin{IEEEproof}
    See Appendix \ref{proof_eq_CSIR_p1t_knownUE}.
  \end{IEEEproof}
  \end{Corollary}

  When the BS is equipped with a single antenna, assuming all users are active and the number of users grows linearly and unboundedly with the blocklength, an achievability bound on the minimum required energy-per-bit was derived in~\cite[Theorem IV.4]{finite_payloads_fading} for the case of CSIR.
  In contrast, we consider a more practical communication system with random access, multiple BS antennas, and finite blocklength.
  % To be specific, in the case of CSIR, for massive random access in MIMO quasi-static Rayleigh fading channels, Theorem~\ref{Theorem_CSIR_achi} and Corollary~\ref{Theorem_CSIR_achi_energyperbit} present non-asymptotic achievability bounds on the PUPE and the minimum required energy-per-bit, respectively.
  % In a special case where all users are active, Corollary~\ref{Theorem_CSIR_achi_energyperbit} reduces to Corollary~\ref{Theorem_CSIR_achi_knownUE}.
  In general, there are two major differences in the proof ideas of our achievability bounds and the result   in~\cite[Theorem IV.4]{finite_payloads_fading}.
  First, we utilize standard bounding techniques proposed by Fano~\cite{1961} and by Gallager~\cite{1965} (corresponding to~\eqref{eq_CSIR_p2t} and~\eqref{eq_CSIR_p1t} in Theorem~\ref{Theorem_CSIR_achi}, respectively), whereas only the latter one, namely Gallager's $\rho$-trick, is used in~\cite[Theorem IV.4]{finite_payloads_fading}.
  When random access is taken into consideration, in contrast to the ``good region''-based bound~\eqref{eq_CSIR_p2t}, more samples are required by Gallager's $\rho$-trick bound~\eqref{eq_CSIR_p1t} to obtain a good estimate, which can be observed from numerical simulation.
  Gallager's $\rho$-trick bound~\eqref{eq_CSIR_p1t} is easy-to-evaluate only for a special case with knowledge of the active user set.
  Thus, for the massive random access problem, we resort to the bounding technique proposed by Fano~\cite{1961} and the ``good region'' designed in~\eqref{eq_good region_selection}.
  Second, when the BS is equipped with a single antenna, a key idea used in~\cite[Theorem IV.4]{finite_payloads_fading} is to drop a subset of users (less than $\epsilon K_a$) with very bad channel gains and decode the rest~\cite{Shamai_Bettesh}.
  However, this idea is not applicable in our regime for two reasons:
  1) as introduced in Section~\ref{sec_simulation}, $\epsilon K_a$ is very small and even less than $1$ in most of our considered settings, thereby making this decoding technique useless;
  2) the channel quality imbalance between different users is greatly reduced when multiple antennas are equipped at the BS, and it is not necessary to drop some users.

\subsubsection{Converse bound} \label{section3_sub1_subsub2}
    Apart from the achievability bound, we provide a converse bound on the minimum required energy-per-bit for massive random access in MIMO quasi-static Rayleigh fading channels with CSIR in the following theorem.
    \begin{Theorem}\label{prop_converse_CSIR}
       Assume that there are $K_a$ active users among $K$ potential users each equipped with a single antenna and the number of BS antennas is $L$.
       Let $M=2^J$ be the codebook size and $n$ be the blocklength.
       For massive random access in MIMO quasi-static Rayleigh fading channels with CSIR, the minimum energy-per-bit $E^{*}_{b,\text{CSIR},K_a}(n,M,\epsilon)$ required for satisfying the PUPE requirement in~\eqref{PUPE} can be lower-bounded as
       \begin{equation}\label{eq:EbN0_conv_CSIR}
         E^{*}_{b,\text{CSIR},K_a}(n,M,\epsilon) \geq \inf \frac{nP}{J}.
       \end{equation}
       The $\inf$ is taken over all $P>0$ satisfying that
       \begin{equation}\label{P_tot_conv_CSIR}
        \left( \frac{t}{K_a} - \epsilon \right) \!J - h_2 \left( \epsilon \right)
        \leq \!\frac{n}{K_a}
        \mathbb{E}_{ \mathbf{H}_t } \!  \left[ \log_2 \left| \mathbf{I}_L + P \mathbf{H}_t^H  \mathbf{H}_t \right| \right],   \forall t \in [K_a],
       \end{equation}
       where $\mathbf{H}_t \in \mathbb{C}^{t\times L}$ has i.i.d. $\mathcal{CN}(0,1)$ entries.
       The condition in~\eqref{P_tot_conv_CSIR} can be loosened to:
      \begin{equation}\label{P_tot_conv_CSIR_bound}
        \left( \frac{t}{K_a} - \epsilon \right) J - h_2 \left( \epsilon \right) \leq
        \frac{n}{K_a}  \min\left\{ L \log_2\left( 1+Pt \right), t \log_2\left( 1+PL \right) \right\},   \forall t \in [K_a].
      \end{equation}
      Note that the minimum required energy-per-bit $E^{*}_{b,\text{CSIR},K_a}(n,M,\epsilon)$ should also satisfy the meta-converse bound for the single-user multiple-receive-antenna channel with CSIR~\cite[Theorem~1]{Beta_dis}.
      \begin{IEEEproof}[Proof sketch]
        For the converse bound with multiple users, we first utilize the Fano inequality and then bound the mutual information therein under the assumption of CSIR, which contributes to~\eqref{P_tot_conv_CSIR}.
        In order to simplify calculations, we further upper-bound the RHS of~\eqref{P_tot_conv_CSIR} and obtain~\eqref{P_tot_conv_CSIR_bound} by applying the concavity of the $\log_2\left| \cdot \right|$ function.
        Moreover, the minimum required energy-per-bit $E^{*}_{b,\text{CSIR},K_a}(n,M,\epsilon)$ should also satisfy the converse bound for the single-user multiple-antenna channels in the CSIR case~\cite[Theorem 1]{Beta_dis}, which is based on the meta-converse theorem in~\cite{Channel_coding_rate}.
        See Appendix~\ref{proof_converse_CSIR} for the complete proof.
      \end{IEEEproof}
    \end{Theorem}

\subsubsection{Asymptotic analysis} \label{section3_sub1_subsub3}
  On the basis of the achievability bound in Corollary~\ref{Theorem_CSIR_achi_knownUE} and the converse bound in Theorem~\ref{prop_converse_CSIR}, we establish scaling laws of the number of reliably served users in Theorem~\ref{Theorem_scalinglaw_CSIR} for a special case where all users are assumed to be active.

  \begin{Theorem}\label{Theorem_scalinglaw_CSIR}
    Assume that all users are active, i.e. $K_a=K$.
    Each user is equipped with a single antenna and the number of BS antennas is $L$.
    The channel is assumed to be Rayleigh distributed.
    Each user has an individual codebook with size $M$ and length $n$ satisfying the maximum power constraint in~\eqref{power_constraint}.
    Let $ n, K\to \infty$, $M=\Theta(1)$, $\ln K = o(n)$, and $KP=\Omega\left(1\right)$.
    In the case of CSIR, the PUPE requirement in~\eqref{PUPE} is satisfied if and only if $\frac{nL\ln KP}{K}=\Omega\left(1\right)$.
    \begin{IEEEproof}
        See Appendix \ref{Proof_scalinglaw_CSIR}.
    \end{IEEEproof}
  \end{Theorem}

  \begin{Remark}\label{Theorem_scalinglaw_CSIR_Remark1}
    In the case of CSIR, under the assumptions in Theorem~\ref{Theorem_scalinglaw_CSIR}, the sufficient and necessary condition $\frac{nL\ln KP}{K}=\Omega\left(1\right)$ for satisfying the PUPE requirement can be divided into the following two regimes:
    1)~$\frac{nL}{K}=\Omega\left(1\right)$ and $KP =\Theta\left(1\right)$; 2)~$\frac{nL\ln KP}{K}=\Omega\left(1\right)$ and $KP \to \infty$.
    The first regime is power-limited, where the number of degrees of freedom, i.e., $n\min\left\{ K,L\right\}=nL$, grows linearly with the number of users.
    It was pointed out in~\cite{Yang_scalinglaw_singleuser} that, in the single-user case, the minimum received energy-per-user required to transmit a finite number of information bits is given by $nLP=\Theta\left(1\right)$.
    By allocating orthogonal resources to $K$ users, the minimum required energy-per-user $nLP$ in the first regime can be as low as that in the single-user case.
    %In other words, when $nLP$ and $J$ are both finite, the number of users that can be reliably served is in the order of $K=\mathcal{O}\left(nL\right)$.
    The second regime is degrees-of-freedom-limited, where the number of degrees of freedom, i.e. $nL$, is far less than the number of users, and the minimum received energy-per-user $nLP\to\infty$.
  \end{Remark}

  \begin{Remark}\label{Theorem_scalinglaw_CSIR_Remark2}
    In the case of CSIR, under the maximum power constraint in~\eqref{power_constraint} and the PUPE requirement in~\eqref{PUPE}, the number of reliably served users is in the order of $K = \mathcal{O}(n^2)$ in two regimes: 1)~the number of BS antennas is $L=\Theta\left( n \right)$ and the power satisfies $P =\Theta\left(\frac{1}{n^2}\right)$;
    2)~the number of BS antennas is $L=\Theta\left( \frac{n}{\ln n} \right)$ and the power satisfies $P =\Theta\left(\frac{1}{n}\right)$.
    %when the number of BS antennas is $L=\Theta\left( n \right)$ (resp. $L=\Theta\left( \frac{n}{\ln n} \right)$) and the power satisfies $P =\Theta\left(\frac{1}{n^2}\right)$ (resp. $P =\Theta\left(\frac{1}{n}\right)$), the number of reliably served users is in the order of $K = \mathcal{O}(n^2)$.
    \begin{IEEEproof}
        See Appendix \ref{Proof_scalinglaw_CSIR}.
    \end{IEEEproof}
  \end{Remark}

  Our scaling law in Theorem~\ref{Theorem_scalinglaw_CSIR} is proved from both the achievability side and the converse side, which reveals the tightness of our bounds in Corollary~\ref{Theorem_CSIR_achi_knownUE} and Theorem~\ref{prop_converse_CSIR} in asymptotic cases.
  Moreover, it indicates the great potential of multiple receive antennas for the data detection problem.
  Specifically, we can observe from the condition $\frac{nL\ln KP}{K}=\Omega\left(1\right)$ that, when the number $L$ of BS antennas is increased, the maximum number $K$ of reliably served users can be greatly increased and the required blocklength $n$ and power $P$ can be greatly decreased.
  As in Remark~\ref{Theorem_scalinglaw_CSIR_Remark2}, in order to reliably serve $K = \mathcal{O}(n^2)$ users, when the number of BS antennas is increased from $L=\Theta\left( \frac{n}{\ln n} \right)$ to $L=\Theta\left( n \right)$, the minimum required power can be considerably decreased from $P =\Theta\left(\frac{1}{n}\right)$ to $P =\Theta\left(\frac{1}{n^2}\right)$.
  Notably, the case of $P = \Theta\left(\frac{1}{n}\right)$ and the case of $P = \Theta\left(\frac{1}{n^2}\right)$ imply that the energy-per-bit is finite and goes to $0$, respectively, which are crucial in practical communication systems with stringent energy constraints.

\subsection{No-CSI} \label{section3_sub2}
  In this subsection, we consider the case where neither the transmitters nor the decoder knows the realization of fading coefficients, but they both know the distribution.
  In this noncoherent setting, we establish non-asymptotic bounds for the massive random access model described in Section~\ref{section2},
  where both the cases with known $K_a$ and unknown $K_a$ are considered.
  Specifically, in Theorem~\ref{Theorem_noCSI_achi}, we establish an upper bound on the PUPE for massive random access with known $K_a$.
  On the basis of it, Corollary~\ref{Theorem_noCSI_achi_energyperbit} is established, which presents an achievability bound (upper bound) on the minimum required energy-per-bit.
  % By allowing $K_a=K$, it reduces to the achievability bound for a special case in which all users are active.
  For a general setting where the number of active users is random and unknown at the receiver, we establish an achievability bound on the minimum required energy-per-bit in Theorem~\ref{Theorem_noCSI_noKa_achi_energyperbit}.
  Then, we present the converse bounds (lower bounds) on the minimum required energy-per-bit in the cases with and without the knowledge of the number $K_a$ of active users at the receiver in Theorem~\ref{prop_converse_noCSI} and Theorem~\ref{prop_converse_noCSI_noKa}, respectively, where the multiple-user Fano type bounds are established under the assumption of i.i.d. Gaussian codebooks.
  Finally, on the basis of Corollary~\ref{Theorem_noCSI_achi_energyperbit} and Theorem~\ref{prop_converse_noCSI}, we establish scaling laws in Theorem~\ref{Theorem_scalinglaw_noCSI} for a special case where all users are assumed to be active.

\subsubsection{\texorpdfstring{Achievability bound with known $K_a$}{Achievability bound with known Ka}} \label{section3_sub2_subsub1}
  An upper bound on the PUPE for massive random access in MIMO quasi-static Rayleigh fading channels in the case with no-CSI and known $K_a$ is given in Theorem~\ref{Theorem_noCSI_achi}.
  \begin{Theorem} \label{Theorem_noCSI_achi}
    Assume that there are $K_a$ active users among $K$ potential users each equipped with a single antenna and the number of BS antennas is $L$. Each user has an individual codebook with size $M=2^J$ and length $n$ satisfying the maximum power constraint in~\eqref{power_constraint}.
    For massive random access in MIMO quasi-static Rayleigh fading channels with known $K_a$ but unknown CSI at the receiver, the PUPE is upper-bounded as
    \begin{equation} \label{eq_noCSI_PUPE}
      P_e \leq \min_{0< P'< P} \left\{ p_0 + \sum_{t=1}^{K_{a }} \frac{ t}{K_a} \min \left\{ 1, p_{t} \right\} \right\} ,
    \end{equation}
    where
    \begin{equation} \label{eq_noCSI_p0}
      p_0=K_a \left( 1 - \frac{\gamma\left(n, \frac{nP}{P'}\right)}{\Gamma\left(n\right)}\right),
    \end{equation}
    \begin{equation} \label{eq_noCSI_pt}
      p_t = \min_{ 0 \leq \omega \leq 1, 0 \leq \nu }
       \left\{ q_{1,t}\left(\omega,\nu\right)
       + q_{2,t}\left(\omega,\nu\right) \right\}  ,
    \end{equation}
    \begin{align}
      q_{1,t} \left(\omega,\nu\right)
      = {\binom {K_a} {t}} {\binom {K-K_{a}+t} {t}}  M^t\;
      & \mathbb{E}_{ {\mathbf{A}}_{ \mathcal{K}_a },  {\mathbf{A}}_{ \mathcal{K}_a \backslash S_1 },   {\mathbf{A}}'_{  S_2} } \! \left[
      \min_{ {u\geq 0,r\geq 0, \lambda_{\min}\left(\mathbf{B}\right) > 0}}
      \exp \left\{  L rn\nu  \right\}
      \right. \notag \\
      & \!\!\!\!\!\!\!\!\!\!\!\!\!\!\!\!\!\!\!\!\!\!\!\!\!\!\! \!\!\!
      \cdot \exp \left\{  L  \left( (u-r) \ln \left|\mathbf{F}\right|
      - u \ln \left| {\mathbf{F}'} \right|
      + r\omega \ln \left| \mathbf{F}_{1} \right|
      - \ln \left| \mathbf{B} \right| \right)
      \right\} \bigg] , \label{eq_noCSI_q1t}
    \end{align}
    \begin{equation}
      \mathbf{B} = (1-u+r) \mathbf{I}_n
      + u \left( \mathbf{F}' \right)^{-1} \mathbf{F}
      - r\omega \mathbf{F}_{ 1}^{-1} \mathbf{F},
    \end{equation}
    \begin{equation} \label{eq_noCSI_F}
      \mathbf{F} = \mathbf{I}_n + \mathbf{A}_{\mathcal{K}_a} \mathbf{A}_{\mathcal{K}_a}^H,
    \end{equation}
    \begin{equation}\label{eq_noCSI_Fprime}
      \mathbf{F}'  = \mathbf{I}_n +
      \mathbf{A}_{ \mathcal{K}_a \backslash  {S}_1 }  \mathbf{A}_{ \mathcal{K}_a \backslash  {S}_1 }^H
      + \mathbf{A}'_{ {S}_2}  \left(\mathbf{A}'_{ {S}_2}\right)^H,
    \end{equation}
    \begin{equation}\label{eq_noCSI_F1}
      \mathbf{F}_1 = \mathbf{I}_n+ \mathbf{A}_{ \mathcal{K}_a \backslash  {S}_1 }  \mathbf{A}_{ \mathcal{K}_a \backslash  {S}_1 }^H,
    \end{equation}
    \begin{align}
      q_{2,t}\left(\omega,\nu\right) =  \min_{ \delta\geq0 } & \left\{ \binom {K_a} {t} \mathbb{E}_{ {\mathbf{A}}_{ \mathcal{K}_a },  {\mathbf{A}}_{ \mathcal{K}_a \backslash S_1 } } \!  \left[ \frac{\gamma\left( Lm, L \prod_{i=1}^{m}\lambda_i^{-\frac{1}{m}} \frac{ n(1+\delta)(1-\omega) - \omega\ln\left|\mathbf{F}_1\right| + \ln\left|\mathbf{F}\right|  - n\nu }
      {\omega}  \right)}{\Gamma\left( Lm \right)}  \right] \right. \notag\\
      &  \;\;\;\; + \binom {K_a} {t} \left( 1 - \frac{\gamma\left( nL, nL \left( 1+\delta \right)\right)}{\Gamma\left( nL \right)} \right) \Bigg\} . \label{eq_noCSI_q2t}
    \end{align}
  Here, ${S}_{1}$ is an arbitrary $t$-subset of $\mathcal{K}_a$;
  ${S}_{2}$ is an arbitrary $t$-subset of $\mathcal{K} \backslash \mathcal{K}_a\cup  {S}_{1}$;
  $\mathbf{A}_{S}$ denotes an $n \times |S|$ submatrix of $\mathbf{A}$ including transmitted codewords of active users in the set $S\subset\mathcal{K}_a$;
  $\mathbf{A}'_{S_2}$ denotes an $n \times |S_2 |$ submatrix of $\mathbf{A}$ including false-alarm codewords for users in the set $S_2$;
  the matrix $\mathbf{A}\in\mathbb{C}^{n\times MK}$ is the concatenation of codebooks of the $K$ users without power constraint, which has i.i.d. $\mathcal{CN}\left(0, P'\right)$ entries;
  and $\lambda_1,\ldots, \lambda_m$ denote non-zero eigenvalues of $\mathbf{F}_1^{-1} \mathbf{A}_{{S}_1}\mathbf{A}_{{S}_1}^H$ with $m=\min\left\{ n,t \right\}$.
  \begin{IEEEproof}[Proof sketch]
    Similar to the CSIR case, we use the random coding scheme and the ML decoder in the no-CSI case with known $K_a$.
    The PUPE can be upper-bounded as the sum of two terms as in~\eqref{eq_noCSI_PUPE}:
    the first term $p_0$ upper-bounds the total variation distance between the measures with and without power constraint;
    the second term $ \sum_{t=1}^{K_{a }} \frac{ t}{K_a} \min \left\{ 1, p_{ t} \right\} $ upper-bounds the PUPE assuming there is no power constraint.
    Here, $p_{t}$ denotes an upper bound on $\mathbb{P}\left[ \mathcal{F}_t \right]$, which indicates the probability of the event that there are exactly $t$ misdecoded users.
    There are some differences in bounding $\mathbb{P}\left[ \mathcal{F}_t \right]$ between in the case of CSIR and no-CSI.
    First, Gallager's $\rho$-trick is difficult to apply in the no-CSI case.
    Specifically, for a set ${ {S}_2}$ including $t$ detected users with false alarm codewords, there are about $M^t$ (extremely large) events corresponding to different sets of false alarm codewords.
    Gallager's $\rho$-trick can be useful only if it is applied to the union of these events at first.
    However, the probability over false alarm codewords conditioned on other variables is difficult to handle in the no-CSI case because they exist in many terms including $\ln \left| \cdot \right|$ and $\operatorname{tr}\left( \cdot \right)$.
    Therefore, in this case, we only utilize the bounding technique proposed by Fano~\cite{1961} and apply the ``good region'' designed in~\eqref{eq_good region_selection}.
    Second, different from the CSIR case, both the channel and noise are unknown to the receiver in the no-CSI case. As a result, the effects due to noise and channel are coupled together, and it is difficult to separate these two effects in the analysis.
    Fortunately, when channels are Rayleigh distributed, conditioned on $\mathbf{X}$, the received signal $\mathbf{Y}$ is Gaussian distributed, making the analysis easier.
    The main techniques used for bounding $\mathbb{P}\left[ \mathcal{F}_t \right]$ in the CSIR case, such as the Chernoff bound and moment generating function of quadratic forms, are applied in the no-CSI case.
    See Appendix~\ref{section5} for the complete proof.
  \end{IEEEproof}
  \end{Theorem}

  The following corollary of Theorem~\ref{Theorem_noCSI_achi} provides an achievability bound on the minimum~required energy-per-bit for massive random access with known $K_a$ and no-CSI at the
  BS.\begin{Corollary} \label{Theorem_noCSI_achi_energyperbit}
      Assume that there are $K_a$ active users among $K$ potential users each equipped with a single antenna and the number of BS antennas is $L$. Each user has an individual codebook with size $M=2^J$ and length $n$ satisfying the maximum power constraint in \eqref{power_constraint}.
      For massive random access in MIMO quasi-static Rayleigh fading channels with known $K_a$ but unknown CSI at the receiver, the minimum energy-per-bit $E^{*}_{b,\text{no-CSI},K_a}(n,M,\epsilon)$ for satisfying the PUPE requirement in~\eqref{PUPE} can be upper-bounded~as
      \begin{equation}
        E^{*}_{b,\text{no-CSI},K_a}(n,M,\epsilon) \leq \inf \frac{n {P}}{J},
      \end{equation}
      where the $\inf$ is taken over all $P > 0$ satisfying that
      \begin{equation} \label{eq_noCSI_epsilon}
        \epsilon \geq \min_{0< P'< P} \left\{ p_0 + \sum_{t=1}^{K_{a }} \frac{ t}{K_a} \min \left\{ 1, p_{t} \right\} \right\}.
      \end{equation}
      Here, $p_0$ and $p_{t}$ are the same as those in Theorem~\ref{Theorem_noCSI_achi}.
  \end{Corollary}

  In the single-receive-antenna setting with known active user set, an asymptotic achievability bound on the minimum required energy-per-bit was derived in~\cite[Theorem IV.1]{finite_payloads_fading} for the no-CSI case.
  In the multiple-receive-antenna setting, a non-asymptotic achievability bound is provided in  Corollary~\ref{Theorem_noCSI_achi_energyperbit}.
  There are some differences between the proof ideas of Theorem~IV.1 in~\cite{finite_payloads_fading} and Theorem~\ref{Theorem_noCSI_achi} in our work.
  Specifically, we utilize the ``good region'' designed in~\eqref{eq_good region_selection}, which is better than the one used in~\cite[Theorem IV.1]{finite_payloads_fading} and reduces to it if $\nu$ is set to 0 as mentioned in Section~\ref{section3_sub_goodregion}.
  %%%%%%%%%%%%%%
  Moreover, the projection decoder is used in~\cite[Theorem IV.1]{finite_payloads_fading} for the single-receive-antenna model, but we leverage the ML decoder in the multiple-receive-antenna setting.
  As mentioned in the introduction, the projection decoder has the advantage of requiring no knowledge of the fading distribution, but can be ineffectual in two specific cases when applied to the framework with multiple BS antennas:
  1) it is ineffectual when the number of active users is larger than the blocklength;
  2) it is ineffectual to apply the projection decoder to $L$ BS antennas separately because the signals received over different BS antennas share the same sparse support.
  Meanwhile, it is challenging (although not impossible) to jointly deal with the signals received over $L$ antennas based on the projection decoder, because the analysis of the angle between the subspace spanned by $L$ received signals and the one spanned by $K_a$ codewords is quite involved.
  Thus, we leverage the ML decoder, which is efficient in the multiple-receive-antenna model no matter whether $K_a$ is less than $n$ or not, at the price of requiring \emph{a priori} distribution on $\mathbf{H}$.

%% Ka 未知
\subsubsection{\texorpdfstring{Achievability bound with random and unknown $K_a$}{Achievability bound with random and unknown Ka}} \label{section3_sub2_subsub12}
  In Theorem~\ref{Theorem_noCSI_achi} and Corollary~\ref{Theorem_noCSI_achi_energyperbit}, we assume $K_a$ is known at the receiver in advance and the decoder outputs $K_a$ messages.
  In such a setup, a misdetection for a user implies a false-alarm for another user, and vice versa.
  Next, we consider a general case in which the number of active users is random and unknown to the receiver.
  In this case, we need to account for both the per-user probability of misdetection and the per-user probability of false-alarm.
  The following theorem provides an achievability bound on the minimum required energy-per-bit for the no-CSI case with random and unknown $K_a$.
  \begin{Theorem} \label{Theorem_noCSI_noKa_achi_energyperbit}
      Assume that there are $K$ potential users each equipped with a single antenna and the number of BS antennas is $L$.
      The number of active users is random and unknown, which is distributed as $K_a \sim \text{Binom}(K,p_a)$.
      Each user has an individual codebook with size $M=2^J$ and length $n$ satisfying the maximum power constraint in \eqref{power_constraint}.
      For massive random access in MIMO quasi-static Rayleigh fading channels with no-CSI, the minimum energy-per-bit $E^{*}_{b,{\text{no-CSI,no-}K_a}}(n,M,\epsilon_{\mathrm{MD}},\epsilon_{\mathrm{FA}})$ for satisfying the per-user probability of misdetection and the per-user probability of false-alarm requirements in~\eqref{eq:MD} and~\eqref{eq:FA} can be upper-bounded as
      \begin{equation}\label{eq_noCSI_noKa_EbN0}
        E^{*}_{b,\text{no-CSI,no-}K_a}(n,M,\epsilon_{\mathrm{MD}},\epsilon_{\mathrm{FA}}) \leq \inf \frac{n {P}}{J}.
      \end{equation}
      The $\inf$ is taken over all $P > 0$ satisfying
      \begin{equation} \label{eq_noCSI_noKa_epsilonMD}
        \epsilon_{\mathrm{MD}} \!\geq \!\! \min_{0< P'< P} \!\!\left\{ \! p_0 +\!\!
        \sum_{{\rm{K}}_a = 1}^{K}  \!\!P_{K_a}\!({\rm{K}}_a)\!
        \sum_{{\rm{K}}'_a = 0}^{K}
        \sum_{t\in\mathcal{T}_{{\rm{K}}'_a}}
        \!\! \frac{t\!+\!({\rm{K}}_a\!-\!{\rm{K}}'_{a,u} )^{+}\!}{{\rm{K}}_a}
        \min\!\left\{ \!1, \!\sum_{t' \in \bar{\mathcal{T}}_{{\rm{K}}'_a,t}} \!\!\!p_{{\rm{K}}'_a,t,t'}, p_{{\rm{K}}_a\to {\rm{K}}'_a} \!\right\}
        \!\right\} \!,
      \end{equation}
      \begin{equation} \label{eq_noCSI_noKa_epsilonFA}
        \epsilon_{\mathrm{FA}} \geq \! \min_{0< P'< P} \!\left\{ p_0 + \!
        \sum_{{\rm{K}}_a = 0 }^{K} \! P_{K_a}({\rm{K}}_a)\!
        \sum_{{\rm{K}}'_a=0}^{K}
        \sum_{t\in\mathcal{T}_{{\rm{K}}'_a}}
        \sum_{t' \in \mathcal{T}_{{\rm{K}}'_a,t}}
         \!\!\frac{t' + ( {\rm{K}}'_{a,l} - {\rm{K}}_a )^{+}}{ \hat{{\rm{K}}}_a }
        \min\!\left\{ 1, p_{{\rm{K}}'_a,t,t'}, p_{{\rm{K}}_a\to {\rm{K}}'_a} \right\} \!\right\} ,
      \end{equation}
  where
      \begin{equation}\label{eq_noCSI_noKa_p0}
        p_0 = % \mathbb{E}[K_a]
        p_a K \left( 1 - \frac{\gamma\left(n, \frac{nP}{{P}^{\prime}} \right)}{\Gamma\left(n\right)} \right) ,
      \end{equation}
      \begin{equation}\label{eq_noCSI_noKa_estimate_Tset1}
        \mathcal{T}_{{\rm{K}}'_a } = \left[ 0  :
        \min\{{\rm{K}}_a,{\rm{K}}'_{a,u}\} \right] ,
      \end{equation}
      \begin{equation}\label{eq_noCSI_noKa_estimate_Tset2bar}
        \bar{\mathcal{T}}_{{\rm{K}}'_a,t} = \left[ \left( ({\rm{K}}_a - {\rm{K}}'_{a,u} )^{+} - ({\rm{K}}_a-{\rm{K}}'_{a,l} )^{+} + t \right)^{+}  :
        ({\rm{K}}'_{a,u}-{\rm{K}}_a)^{+} - ({\rm{K}}'_{a,l}-{\rm{K}}_a)^{+} + t \right] ,
      \end{equation}
      \begin{equation}\label{eq_noCSI_noKa_estimate_Tset2}
        \mathcal{T}_{{\rm{K}}'_a,t} \!=\!\! \left[ \left( ({\rm{K}}_a \!-\! {\rm{K}}'_{a,u} )^{+} \!- ({\rm{K}}'_{a,l}\!-\!{\rm{K}}_a)^{+} \!+\! \max\{{\rm{K}}'_{a,l},1\} \!-\! {\rm{K}}_a \!+\!  t \right)^{+}  \!:
        ({\rm{K}}'_{a,u}\!\!-\!{\rm{K}}_a)^{+} \!-\! ({\rm{K}}'_{a,l}\!-\!{\rm{K}}_a)^{+} \!+\! t \right] \!,
      \end{equation}
      \begin{equation}\label{eq_noCSI_noKa_estimate_Kahat}
        \hat{{\rm{K}}}_a = {\rm{K}}_a - t - ({\rm{K}}_a-{\rm{K}}'_{a,u} )^{+} + t' + ( {\rm{K}}'_{a,l}  - {\rm{K}}_a )^{+} ,
      \end{equation}
      \begin{equation}\label{eq_noCSI_noKa_estimate_Kal}
        {\rm{K}}'_{a,l}  = \max\left\{ 0 , {\rm{K}}'_{a} -r' \right\} ,
      \end{equation}
      \begin{equation}\label{eq_noCSI_noKa_estimate_Kau}
        {\rm{K}}'_{a,u}  = \min\left\{ K , {\rm{K}}'_{a} +r' \right\} ,
      \end{equation}
    \begin{align} \label{eq_noCSI_noKa_estimate_ptt}
      p_{{\rm K}'_a,t,t'} & = \min_{ 0 \leq \omega \leq 1, 0 \leq \nu }
      \left\{ q_{1,{\rm K}'_a,t,t'}\left(\omega,\nu\right)
      + 1 \left[ t+({\rm K}_a-{\rm K}'_{a,u})^{+} > 0 \right]
      q_{2,{\rm K}'_a,t }\left(\omega,\nu\right) \right. \notag \\
      & \;\;\;\;\;\;\;\;\;\;\;\;\;\;\;\;\;\;\;\;\;\;\;\;\;\;\;\;\;\;\;\;\;\;\;\;\;\;\;\;\;\;\!
      \left. +\; 1 \left[ t+({\rm K}_a-{\rm K}'_{a,u})^{+} = 0 \right] q_{2,{\rm K}'_a,t,0 }\left(\omega,\nu\right) \right\}  ,
    \end{align}
    \begin{align}
      q_{1,{\rm K}'_a,t,t'} \left(\omega,\nu\right)
      = C_{{\rm K}'_a,t,t'}\;
      & \mathbb{E}_{ {\mathbf{A}}_{ \mathcal{K}_a },  {\mathbf{A}}_{ \mathcal{K}_a \backslash S_1 },  {\mathbf{A}}_{ \mathcal{K}_a \backslash S_{1,1} },   {\mathbf{A}}'_{ S_2},   {\mathbf{A}}'_{ S_{2,1}}  } \! \left[
      \min_{ {u\geq 0,r\geq 0, \lambda_{\min}\left(\mathbf{B}\right) > 0}}
      \exp \left\{  L rn\nu + b_{u,r} \right\}
      \right. \notag \\
      & \;\;\;
      \cdot \exp \! \left\{  L  \left( u \ln \!\left|\mathbf{F}''\right|
      - r \ln \!\left|\mathbf{F}\right|
      - u \ln \!\left| {\mathbf{F}'} \right|
      + r\omega \ln \!\left| \mathbf{F}_{1} \right|
      - \ln \!\left| \mathbf{B} \right| \right)
      \right\} \!\bigg] , \label{eq_noCSI_noKa_estimation_q1t}
    \end{align}
      \begin{equation}\label{eq_noCSI_noKa_estimation_Ctt}
        C_{{\rm K}'_a,t,t'} = \binom{{\rm K}_a}{ t+({\rm K}_a-{\rm K}'_{a,u})^{+} } \binom{K-\min\{{\rm K}_a,{\rm K}'_{a,u}\}+t}{t'+({\rm K}'_{a,l}-{\rm K}_a)^{+}} M^{t'+({\rm K}'_{a,l}-{\rm K}_a)^{+}} ,
      \end{equation}
      \begin{equation}\label{eq_noCSI_noKa_estimation_Fprime2}
        \mathbf{F}''  = \mathbf{I}_n + \mathbf{A}_{ \mathcal{K}_a \backslash {S}_{1,1} }  \mathbf{A}_{ \mathcal{K}_a \backslash  {S}_{1,1} }^H  + \mathbf{A}'_{  {S}_{2,1}} (\mathbf{A}'_{ {S}_{2,1}})^H ,
      \end{equation}
      \begin{equation}\label{eq_noCSI_noKa_estimation_B}
        \mathbf{B} = (1+r) \mathbf{I}_n - u \left( \mathbf{F}'' \right)^{-1} \mathbf{F}
        + u \left( \mathbf{F}' \right)^{-1} \mathbf{F}
        - r\omega \mathbf{F}_{ 1}^{-1} \mathbf{F} ,
      \end{equation}
    \begin{equation}\label{eq_noCSI_noKa_estimation_bur}
      b_{u,r} = -u b'' + r b + u b' -r\omega b_1,
    \end{equation}
    \begin{equation}\label{eq_noCSI_noKa_estimation_b}
      b = \ln \left( P_{K_a}\!\left({\rm{K}}_a\right) \right) - {\rm{K}}_a \ln M,
    \end{equation}
    \begin{equation}\label{eq_noCSI_noKa_estimation_b1}
      b_1 = \ln \left( P_{K_a}\!\left( {\rm{K}}_a - t - ({\rm K}_a-{\rm K}'_{a,u})^{+} \right) \right)
      - \left( {\rm{K}}_a - t - ({\rm K}_a-{\rm K}'_{a,u})^{+} \right) \ln M,
    \end{equation}
    \begin{equation}\label{eq_noCSI_noKa_estimation_bprime}
      b' = \ln \left( P_{K_a}\!\left({\hat{\rm{K}}}_a\right) \right) - {\hat{\rm{K}}}_a \ln M,
    \end{equation}
    \begin{equation}\label{eq_noCSI_noKa_estimation_bprime2}
      b'' = \ln \left( P_{K_a}\!\left({\rm{K}}_a - ({\rm{K}}_a\!-\!{\rm{K}}'_{a,u} )^{+} + ( {\rm{K}}'_{a,l}  \!-\! {\rm{K}}_a )^{+}   \right) \right) - \left({\rm{K}}_a  -  ({\rm{K}}_a\!-\!{\rm{K}}'_{a,u} )^{+} + ( {\rm{K}}'_{a,l}  \!-\! {\rm{K}}_a )^{+}   \right) \ln M,
    \end{equation}
    \begin{align}
      q_{2,{\rm K}'_a,t}\left(\omega,\nu\right) =  & \binom{{\rm K}_a}{t+({\rm K}_a-{\rm K}'_{a,u})^{+}} \cdot
      \min_{ \delta\geq0 }
      \Bigg\{ 1 - \frac{\gamma\left( nL, nL \left( 1+\delta \right)\right)}{\Gamma\left( nL \right)} \notag\\
      &  \left. +\;\!   \mathbb{E}_{{\mathbf{A}}_{ \mathcal{K}_a },  {\mathbf{A}}_{ \mathcal{K}_a \!\backslash S_1 } } \!\!\!  \left[ \! \frac{\gamma\!\left(\! Lm, \prod_{i=1}^{m} \! \lambda_i^{\!-\frac{1}{m}}  \frac{ nL(1+\delta)(1-\omega) - \omega\left( L\!\ln\left|\mathbf{F}_1\right| - b_1 \right) + L\! \ln\left|\mathbf{F}\right| - b - nL\nu }
      {\omega  } \! \right)}{\Gamma\left( Lm \right)} \!\right]
      \!\right\} \! , \label{eq_noCSI_noKa_estimation_q2t}
    \end{align}
    \begin{equation}\label{eq_noCSI_noKa_estimation_q2t0}
      q_{2,{\rm K}'_a,t,0} = \mathbb{E}_{\mathbf{A}_{\mathcal{K}_a}} \left[ 1 - \frac{\gamma\left( nL, \frac{nL\nu}{1-\omega} - L \ln \left| \mathbf{F} \right| + b \right)}{\Gamma\left( nL \right)}  \right] ,
    \end{equation}
    \begin{equation}\label{eq_noCSI_noKa_pKa_Kahat}
       p_{{\rm{K}}_a\to {\rm{K}}'_a}
       =  \min_{ \tilde{{\rm{K}}}_a \in \left[0 : K\right],
       \tilde{{\rm{K}}}_a \neq {\rm{K}}'_a }
       \left\{ 1 \left[ {\rm{K}}'_a < \tilde{{\rm{K}}}_a \right]
       p_{{\rm{K}}_a\to {\rm{K}}'_{a},1}
       +  1 \left[ {\rm{K}}'_a > \tilde{{\rm{K}}}_a \right]p_{{\rm{K}}_a\to {\rm{K}}'_a,2} \right\} ,
    \end{equation}
    \begin{align}\label{eq_noCSI_noKa_pKa_Kahat1}
      p_{{\rm{K}}_a\to {\rm{K}}'_a,1} \!
      & = \min\!\left\{\!
      \min_{\eta>0}\!
      \left\{\! \mathbb{E}_{\mathbf{A}_{\mathcal{K}_a}} \!\!\! \left[
      \frac{\gamma\!\left( Lm', \prod_{i=1}^{m'}\!{(\lambda^{'}_i)}^{-\frac{1}{m'}}
      nL\!\left( 1 \!+ \frac{{\rm{K}}'_a+\tilde{{\rm{K}}}_a}{2} P^{\prime} \!- \eta \right)
      \right)}{\Gamma\left( Lm' \right)}  \right]
      \!+ \frac{\gamma\left( nL, nL\eta \right)}{\Gamma\left( nL \right)} \right\} \!, \right. \notag \\
      & \;\;\;\;\;\;\;\;\;\;\;\;\;  \left.
      \mathbb{E}_{\mathbf{A}_{\mathcal{K}_a}} \!\left[ \min_{\rho\geq0} \exp\left\{ \rho nL\left(1+\frac{{\rm{K}}'_a+\tilde{{\rm{K}}}_a}{2} P^{\prime}\right)
      - L \ln \left| \mathbf{I}_n + \rho \mathbf{F} \right| \right\} \right]  \right\},
    \end{align}
    \begin{align}\label{eq_noCSI_noKa_pKa_Kahat2}
      p_{{\rm{K}}_a\to {\rm{K}}'_a,2} \!
      & = \min\!\left\{ \min_{\eta>0} \left\{ 2 - \mathbb{E}_{\mathbf{A}_{\mathcal{K}_a}} \!\! \left[
      \frac{\gamma\left( Lm',
      \frac{nL}{\lambda'_{1}}\left( 1 + \frac{{\rm{K}}'_a+\tilde{\rm{K}}_a}{2} P^{\prime} - \eta \right)
      \right)}{\Gamma\left( Lm' \right)}  \right]
      - \frac{\gamma\left( nL, nL\eta \right)}{\Gamma\left( nL \right)}  \right\} , \right. \notag \\
      & \;\;\;\;\;\;\;\;\;\;\;\;\; \left.
      \mathbb{E}_{\mathbf{A}_{\mathcal{K}_a}} \!\! \left[
      \min_{ 0\leq \rho < \frac{1}{1+\lambda'_{1}  }}
      \! \exp\left\{ -\rho nL\left(\!1+\frac{{\rm{K}}'_a+\tilde{\rm{K}}_a}{2} P^{\prime}\!\right)
      -L \ln \left|  \mathbf{I}_n - \rho\mathbf{F} \right|  \right\} \right]  \right\} .
    \end{align}
    Here, $\mathbf{F}$, $\mathbf{F}'$, and $\mathbf{F}_1$ are defined in \eqref{eq_noCSI_F}, \eqref{eq_noCSI_Fprime}, and \eqref{eq_noCSI_F1}, respectively;
    $r'$ denotes a nonnegative integer referred to as the decoding radius;
    $S_1$ is an arbitrary subset of $\mathcal{K}_a$ of size $t+({\rm{K}}_a-{\rm{K}}'_{a,u})^{+}$, which denotes the set of users whose codewords are misdecoded and can be divided into two subsets $S_{1,1}$ and $S_{1,2}$ of size $({\rm{K}}_a-{\rm{K}}'_{a,u})^{+}$ and $t$, respectively;
    $S_2$ is an arbitrary subset of $\mathcal{K} \backslash \mathcal{K}_a\cup  {S}_{1}$ of size $t'+({\rm{K}}'_{a,l}-{\rm{K}}_a)^{+}$, which denotes the set of detected users with false-alarm codewords;
    $S_{2,1}$ is an arbitrary subset of $S_2$ of size $({\rm{K}}'_{a,l}-{\rm{K}}_a)^{+}$;
    $\mathbf{A}_{S}$ denotes an $n \times |S|$ submatrix of $\mathbf{A}$ including transmitted codewords of users in the set $S\subset\mathcal{K}_a$;
    $\mathbf{A}'_{S}$ denotes an $n \times |S|$ submatrix of $\mathbf{A}$ including false-alarm codewords for users in the set $S\subset\mathcal{K}$;
    the matrix $\mathbf{A}\in\mathbb{C}^{n\times MK}$ is the concatenation of codebooks of all users without power constraint, which has i.i.d. $\mathcal{CN}\left(0, P'\right)$ entries;
    $\lambda_1^{'},\ldots, \lambda_{m'}^{'}$ are non-zero eigenvalues of $\mathbf{A}_{\mathcal{K}_a} \mathbf{A}_{\mathcal{K}_a}^H$ in decreasing order with $m' = \min\left\{ n,{\rm{K}}_a \right\}$;
    and $\lambda_1,\ldots, \lambda_m$ denote non-zero eigenvalues of $\mathbf{F}_1^{-1} \mathbf{A}_{{S}_1}\mathbf{A}_{{S}_1}^H$ with $m=\min\left\{ n, t+({\rm{K}}_a-{\rm{K}}'_{a,u})^{+} \right\}$.

  \begin{IEEEproof}[Proof sketch]
     The receiver first estimates the number of active users via an energy-based estimator, which is denoted as ${\rm{K}}'_a$, and then outputs a set of decoded messages of size $\hat{\rm{K}}_a \in [{\rm{K}}'_{a,l} : {\rm{K}}'_{a,u}]$ via an MAP-based decoder.
     The quantity $p_0$ upper-bounds the total variation distance between the measures with and without power constraint.
     When there is no power constraint, $p_{{\rm{K}}_a\to {\rm{K}}'_a}$ upper-bounds the probability of the event that the estimation of ${\rm{K}}_a$ is ${\rm{K}}'_a$, which is obtained based on the Chernoff bound and moment generating function of quadratic forms.
     Moreover, $p_{{\rm{K}}'_a,t,t'}$ upper-bounds the probability of the event that there are exactly $t+({\rm{K}}_a-{\rm{K}}'_{a,u})^{+}$ misdetected codewords and $t'+({\rm{K}}'_{a,l}-{\rm{K}}_a)^{+}$ false-alarm codewords,
     which is derived along similar lines as in the case of known $K_a$.
     See Appendix~\ref{proof_achi_noCSI_noKa_weight} for the complete proof.
    \end{IEEEproof}
  \end{Theorem}

  Theorem~\ref{Theorem_noCSI_noKa_achi_energyperbit} presents an achievability bound on the minimum required energy-per-bit for the case in which the number $K_a$ of active users is random and unknown.
  Specifically, we first estimate the number of active users via an energy-based estimator, which is denoted as ${\rm{K}}'_a$;
  then, we obtain a set of decoded messages of size $\hat{\rm{K}}_a$ via an MAP-based decoder, where $\hat{\rm{K}}_a$ is selected from the interval $[{\rm{K}}'_{a,l} : {\rm{K}}'_{a,u}]$ determined by ${\rm{K}}'_a$ and $r'$.
  The decoding radius $r'$ can be optimized according to the target misdetection and false-alarm probabilities. In general, a large decoding radius $r'$ can reduce the error probabilities suffering from inaccurate estimation of the number of active users; however, increasing $r'$ may increase the chance that the decoder returns a set of codewords whose posterior probability is larger than that of the transmitted codewords, especially when $P$ is small~\cite{noKa}.

  Compared with~\cite{noKa}, where a random-coding achievability bound was derived for Gaussian massive random access channels assuming $K_a$ is unknown \emph{a priori}, there are two main changes in this work.
  First, we employ the MAP-based decoder rather than the ML-based decoder used in~\cite{noKa}.
  When $K_a$ is unknown, the number of decoded messages is not given in advance.
  % the prior distributions in users' messages of various sizes can be different,
  In this case, it is more advantageous to use the MAP-based decoder since it incorporates prior distributions in users' messages of various sizes, at the price of requiring the knowledge of the distribution of $K_a$.
  Indeed, knowing the distribution of $K_a$ is a common assumption in many works such as~\cite{noKa,Yuwei_active,Gao_active}.
  % As opposed to the case of known $K_a$ where the ML-based decoder is used, the number of decoded messages is not given in advance in the case of unknown $K_a$, and thus it is advantageous to apply the MAP-based decoder since it incorporates prior distributions in users' messages of various sizes.
  Second, compared with Gaussian channels considered in~\cite{noKa},
  we further consider the massive random access problem in MIMO quasi-static Rayleigh fading channels,
  which increases the difficulties of upper-bounding the error probabilities.
  For example, the probability of the event that the number of active users is estimated as ${\rm{K}}'_a$ is obtained by straightforward manipulation in~\cite{noKa}, whereas more techniques, such as the Chernoff bound, ``good region''-trick, and moment generating function of quadratic forms, are employed in quasi-static Rayleigh fading channels.

\subsubsection{\texorpdfstring{Converse bound with known $K_a$}{Converse bound with known Ka}} \label{section3_sub2_subsub2}
  In Theorem~\ref{prop_converse_noCSI}, we provide a converse bound on the minimum required energy-per-bit for massive random access in MIMO quasi-static Rayleigh fading channels with no-CSI and known $K_a$.
  This converse bound contains two parts, namely the multiple-user Fano type bound and the single-user bound, where the former relies on the assumption of i.i.d Gaussian codebooks (i.e., the converse is a weaker ensemble converse), but the single-user bound holds for all codes.
  \begin{Theorem}\label{prop_converse_noCSI}
    Assume that there are $K_a$ active users among $K$ potential users each equipped with a single antenna and the number of BS antennas is $L$.
    Each user has an individual codebook with size $M=2^J$ and length $n$.
    For massive random access in MIMO quasi-static Rayleigh fading channels with no-CSI and known $K_a$, the minimum energy-per-bit required for satisfying the PUPE requirement in~\eqref{PUPE} can be lower-bounded as
       \begin{equation} \label{eq:EbN0_conv_noCSI}
         E^{*}_{b,\text{no-CSI},K_a}(n,M,\epsilon) \geq \inf \frac{nP}{J}.
       \end{equation}
    The $\inf$ is taken over all $P>0$ satisfying the following two conditions:
    \begin{enumerate}
      \item
      {Under the assumption that codewords have i.i.d. Gaussian entries, it should be satisfied~that
      \begin{equation}\label{P_tot_conv_noCSI}
        b_1 \leq  \frac{LC}{K_a} - \frac{L}{K_a} \mathbb{E}_{ {\mathbf{X}}_{\!K_a} } \! \left[ \log_2  \left| \mathbf{I}_{K_a} + {\mathbf{X}}_{K_a}^{ H}  {\mathbf{X}}_{K_a} \right|  \right] ,
      \end{equation}
      \begin{equation}\label{P_tot_conv_noCSI_C}
        C = \min \left\{ n \log_2 \left( 1 + K_aP \right),  K_a M \log_2 \left( 1 + \frac{1}{M}nP \right) \right\},
      \end{equation}
      where $b_1 = J{\left(1 - \epsilon\right)} - h_2 \left( \epsilon \right)$ and $ {\mathbf{X}}_{K_a}$ is an $n\times K_{a}$ matrix with each entry i.i.d. from $\mathcal{CN}(0,P)$.
      The condition in~\eqref{P_tot_conv_noCSI} can be loosened to
    \begin{equation} \label{P_tot_conv_noCSI_bound}
        b_1 \leq \left\{
        \begin{array}{ll}
            \frac{LC}{K_a}
    - \frac{L}{K_a}  \sum\limits_{i=0}^{K_a-1} \left( \psi(n-i) \log_2 e +  \log_2  \left( {P + \frac{1}{ n -i }} \right) \right),
            &  1\leq K_a \leq n   \\
            \frac{LC}{K_a}
    - \frac{L}{K_a}  \sum\limits_{i=0}^{n-1}  \left( \psi(K_a-i) \log_2 e +  \log_2  \left( {P + \frac{1}{ K_a -i }} \right) \right),
            &  K_a > n
        \end{array}\right.  ,
    \end{equation}
    where $\psi(\cdot)$ denotes Euler's digamma function.
      }
      \item
      {
      The single-user finite-blocklength bound shows that
      \begin{equation}
        M \leq
        \frac{ 1 }{ \mathbb{P}\left[ \chi^2(2L) \geq (1+(n+1)P)r
        \right] } ,
      \end{equation}
      where $r$ is the solution of
      \begin{equation}
        \mathbb{P} \left[ \chi^2(2L) \leq r \right] = \epsilon .
      \end{equation}
      }
    \end{enumerate}
    \begin{IEEEproof}[Proof sketch]
      Similar to the CSIR case, we first utilize Fano's inequality; then, we follow the idea in~\cite{sparsity_pattern} to deal with the mutual information therein. %, which is more involved than that in the CSIR case.
      Under the assumption of i.i.d. Gaussian codebooks, we obtain~\eqref{P_tot_conv_noCSI} for the scenario with multiple BS antennas and finite blocklength, which reduces to an easy-to-evaluate bound in~\eqref{P_tot_conv_noCSI_bound}.
      Moreover, the minimum required energy-per-bit $E^{*}_{b,\text{no-CSI},K_a}(n,M,\epsilon)$ should also satisfy the single-user meta-converse bound in~\cite[Theorem~3]{noCSI_conv} % for single-user multiple-receive-antenna channels
      with three changes as follows:
      1)~both the number of transmitting antennas and the number of subcodewords are set to be $1$;
      2)~the blocklength is changed from $n$ to $n+1$ because we consider the maximum power constraint in~\eqref{power_constraint}, which can be replaced by the equal power constraint in~\cite{noCSI_conv} following from the standard $n \to n+ 1$ trick~\cite[Lemma~39]{Channel_coding_rate};
      3)~to reduce the simulation complexity of the meta-converse bound in the single-user case, we choose the auxiliary distribution as $Q_{Y^{(n+1)\times L}} = \prod_{l=1}^{L} \mathcal{CN}(0,\mathbf{I}_{n+1})$, rather than the output distribution induced by the input distribution as considered in~\cite{noCSI_conv}.
      %The reason for the second change lies in that we consider the maximum power constraint in~\eqref{power_constraint}, which can be replaced by the equal power constraint in~\cite{noCSI_conv} following from the standard $n \to n+ 1$ trick~\cite[Lemma~39]{Channel_coding_rate}.
      See Appendix~\ref{proof_converse_noCSI} for the complete proof of the Fano type bound.
    \end{IEEEproof}
  \end{Theorem}

  Under the assumption that the entries of codebooks are i.i.d. with mean zero and variance~$P$, a converse bound was established in~\cite{finite_payloads_fading,sparsity_pattern}, in which the number of users is assumed to grow linearly and unboundedly with the blocklength and the BS is assumed to be equipped with a single antenna.
  In the scenario with multiple BS antennas and finite blocklength, some useful techniques used in~\cite{finite_payloads_fading,sparsity_pattern}, such as some results from random matrix theory, are not applicable, and it becomes more involved to obtain an easy-to-evaluate converse bound.
  Instead, in Theorem~\ref{prop_converse_noCSI}, we make stronger assumptions, i.e., we assume codebooks have i.i.d. $\mathcal{CN}(0,P)$ entries, which makes the analysis easier.
  This raises an interesting open question of whether an easy-to-evaluate non-asymptotic converse bound can be obtained for the massive access problem in the multiple-receive-antenna setting under more general assumptions on the codebooks.

\subsubsection{\texorpdfstring{Converse bound with random and unknown $K_a$}{Converse bound with random and unknown Ka}} \label{section3_sub2_subsub22}

  In Theorem~\ref{prop_converse_noCSI_noKa}, we provide a converse bound on the minimum required energy-per-bit for massive random access in MIMO quasi-static Rayleigh fading channels with no-CSI and unknown number of active users.
  Similar to the case of known $K_a$, the converse bound in Theorem~\ref{prop_converse_noCSI_noKa} contains two parts, namely the multiple-user Fano type bound and the single-user bound, where the former relies on the assumption of i.i.d Gaussian codebooks (i.e., the converse is a weaker ensemble converse), but the single-user bound holds for all codes.
  \begin{Theorem}\label{prop_converse_noCSI_noKa}
    Assume that there are $K$ potential users each equipped with a single antenna and the number of BS antennas is $L$.
    The number of active users is random and unknown, which is distributed as $K_a \sim \text{Binom}(K,p_a)$.
    Each user has an individual codebook with size $M=2^J$ and length $n$.
    % Let $K$ users generate their codebooks independently with each entry i.i.d. from $\mathcal{CN}(0,P)$.
    % Let $M=2^J$ be the codebook size and $n$ be the blocklength.
    For massive random access in MIMO quasi-static Rayleigh fading channels with no-CSI, the minimum energy-per-bit required for satisfying the error requirements in~\eqref{eq:MD} and~\eqref{eq:FA} can be lower-bounded as
       \begin{equation} \label{eq:EbN0_conv_noCSI_noKa}
         E^{*}_{b,\text{no-CSI},\text{no-}K_a}(n,M,\epsilon_{\rm MD}, \epsilon_{\rm FA}) \geq \inf \frac{nP}{J}.
       \end{equation}
    The $\inf$ is taken over all $P>0$ satisfying the following two conditions:
    \begin{enumerate}
      \item
      {Under the assumptions that each codebook has i.i.d. $\mathcal{CN}(0,P)$ entries and $\epsilon_{\rm MD} + \epsilon_{\rm FA} \leq 1-\frac{1}{1+2^{h_2(p_a) + p_a J}}$, it should be satisfied that
      \begin{equation}\label{P_tot_conv_noCSI_noKa}
        b_1 \leq  \frac{LC}{K }  - \frac{L}{K } \sum_{ {\rm{K}}_a = 0 }^{K} P_{K_a}({\rm{K}}_a)
        \;\mathbb{E}_{{\mathbf{X}}_{{\rm{K}}_a}} \! \left[ \log_2  \left| \mathbf{I}_{n} +  {\mathbf{X}}_{{\rm{K}}_a}  {\mathbf{X}}_{{\rm{K}}_a}^{ H} \right| \right] ,
      \end{equation}
    \begin{equation}\label{P_tot_conv_noCSI_b1_noKa}
      b_1 = (1-\epsilon_{\rm MD} - \epsilon_{\rm FA}) \left( h_2(p_a) + p_a J \right)
      -  h_2( \epsilon_{\rm MD} + \epsilon_{\rm FA} ) ,
    \end{equation}
    \begin{equation}\label{P_tot_conv_noCSI_C_noKa}
      C = \min \left\{ n \log_2 \left( 1 + p_aKP \right),  K M  \log_2 \left( 1 + \frac{p_a}{M}nP \right) \right\},
    \end{equation}
    where $P_{K_a}({\rm{K}}_a)$ denotes the probability of the event that there are exactly ${\rm{K}}_a$ active users given in~\eqref{eq_conv_pKa_noKa} and ${\mathbf{X}}_{{\rm{K}}_a}$ denotes an $n\times {\rm{K}}_a$ matrix with each entry i.i.d. from $\mathcal{CN}(0,P)$.
    }
      \item
      {The single-user finite-blocklength bound shows that
      \begin{equation} \label{eqR:beta_conv_AWGN_singleUE_1e}
        M \leq % \sup_{ {\epsilon_{1},\epsilon_{2} \in [0,1]} }
        \frac{ \epsilon_{1} }{ \mathbb{P}\left[ \chi^2(2L) \geq (1+(n+1)P)r
        \right] } ,
      \end{equation}
      where $r$ is the solution of
      \begin{equation}\label{eqR:beta_conv_AWGN_singleUE_e_constraint1}
        \mathbb{P} \left[ \chi^2(2L) \leq r \right] = \epsilon_{2} ,% \frac{\epsilon_{\rm{MD}}}{p_a} .
      \end{equation}
      \begin{equation}\label{eqR:beta_conv_epsilon1}
        \epsilon_{1} = \min\left\{ 1, \frac{ \epsilon_{\rm{FA}} }{1-p_a} \right\} ,
      \end{equation}
      \begin{equation}\label{eqR:beta_conv_epsilon2}
        \epsilon_{2} = \min\left\{ 1, \frac{ \epsilon_{\rm{MD}} }{p_a} \right\} .
      \end{equation}
      }
    \end{enumerate}
    \begin{IEEEproof}[Proof sketch]
      Both Condition~1 and Condition~2 take the uncertainty of user activities into consideration.
      Inspired by~\cite{GuoDN}, condition~1 is established for the massive random access problem applying Fano's inequality, under the assumption that codebooks have i.i.d. $\mathcal{CN}(0,P)$ entries.
      Condition~2 is established based on the single-user random access converse result in~\cite[Theorem~2]{On_joint} with a properly selected auxiliary distribution (motivated by~\cite{Yang_Beta_beta}).
      See Appendix~\ref{proof_converse_noCSI_noKa} for the complete proof.
    \end{IEEEproof}
  \end{Theorem}

  In~\cite{GuoDN}, a Fano type converse bound was established for Gaussian massive random access channels under the joint error probability criterion.
  In this case, it was pointed out in~\cite{GuoDN} that Fano's converse bound matches the achievability result well in terms of the message-length capacity, and the capacity penalty due to unknown user activities on each of the $K_a$ active users is $H_2(p_a)/p_a$ in the asymptotic regime with infinite number of users.
  In this work, under the assumption of Gaussian codebooks, we extend the Fano type converse result in~\cite{GuoDN} to the multiple-receive-antenna fading channels under the PUPE criterion.
  Moreover, based on the result in~\cite{On_joint}, we establish a finite-blocklength converse bound for the single-user random access problem in multiple-receive-antenna fading channels with unknown user activity, which can also be regarded as a converse bound for the massive random access problem.

\subsubsection{Asymptotic analysis} \label{section3_sub2_subsub4}
  On the basis of the achievability bound in Theorem~\ref{Theorem_noCSI_achi} and the converse bound in Theorem~\ref{prop_converse_noCSI}, we establish scaling laws of the number of reliably served users in Theorem~\ref{Theorem_scalinglaw_noCSI} for a special case in which all users are assumed to be active.
  \begin{Theorem}\label{Theorem_scalinglaw_noCSI}
    Assume that all users are active, i.e. $K_a=K$.
    Each user is equipped with a single antenna and the number of BS antennas is $L$.
    The channel is assumed to be Rayleigh distributed.
    Each user has an individual codebook with size $M$ and length $n$ satisfying the maximum power constraint in~\eqref{power_constraint}.
    Let $ n, L \to \infty$ and $M=\Theta(1)$.
    In the case of no-CSI, when the number of BS antennas is in the order of $L = \Theta \left(n^2\right)$ and the power satisfies $P = \Theta\left(\frac{1}{n^2}\right)$, one can reliably serve up to $K = \mathcal{O}(n^2)$ users.
    A matching converse result is established assuming codebooks have i.i.d. Gaussian entries.
    % one can satisfy the PUPE requirement in~\eqref{PUPE} if and only if the maximum number of users is $K = \Theta(n^2)$.
    \begin{IEEEproof}
        See Appendix \ref{Proof_scalinglaw_noCSI}.
    \end{IEEEproof}
  \end{Theorem}

  In order to obtain the scaling law on the achievability side, both the activity detection problem considered in~\cite{Caire1} and the data detection problem of interest in this work can be formulated as similar sparse support recovery problems.
  This is because one can immediately obtain a data detection scheme from an activity detection scheme by assigning to each user a unique set of codewords, such that a user can transmit the codeword corresponding to its information message.
  Thus, by expanding the number of users from $K$ to $KM$ and expanding the number of active users from $K_a$ to $K$, the scaling law of the activity detection problem in~\cite{Caire1} can be extended to that of the data detection problem as presented in Table~\ref{table:scalinglaw}: under the joint error probability criterion, with blocklength $n\to \infty$ and a sufficient number of BS antennas $L = \Theta\left(n^2 \ln n\right)$, one can reliably serve up to $K = \mathcal{O}\left(n^2\right)$ users when the payload is $J=\Theta(1)$ and the power is $P=\Theta\left( \frac{1}{n^2} \right)$. %(see Appendix~\ref{Proof_scalinglaw_noCSI} for more details).
  Notably, there are some differences between this result and our scaling law in Theorem~\ref{Theorem_scalinglaw_noCSI}.
  First, the joint error probability criterion is used in~\cite{Caire1}, but we utilize the PUPE criterion in this work, which is more appropriate for massive access channels~\cite{A_perspective_on}.
  We point out that the required number of BS antennas can be reduced from $L = \Theta\left(n^2 \ln n\right)$ to $L = \Theta\left(n^2\right)$ when we change from the joint error probability criterion to the PUPE criterion.
  Second, the result in~\cite{Caire1} is on the achievability side; Theorem~\ref{Theorem_scalinglaw_noCSI} is proved from both the achievability and converse sides, in which the converse result relies on the assumption that the codebooks have i.i.d. Gaussian entries.
  Notably, in our regime, it is satisfied that $n^2P=\Theta(1)$, i.e., the energy-per-bit goes to $0$, which is attractive for IoT settings with stringent energy constraints.

  In this subsection, without assuming \emph{a priori} CSI at the receiver, we focus on the regime of $K = \mathcal{O}(n^2)$, because this is the maximum number of users that can be reliably served in the sparse support recovery problem to the best of our knowledge.
  Theorem~\ref{Theorem_scalinglaw_noCSI} shows that, when the power is $P=\Theta\left( \frac{1}{n^2} \right)$, one can reliably serve up to $K = \mathcal{O}\left(n^2\right)$ users with $L = \Theta\left(n^2\right)$ BS antennas.
  However, it is still unknown how the number of reliably served users increases as the number of BS antennas further increases.

\subsection{Pilot-assisted scheme in the no-CSI case} \label{section3_nocsi_pilot}

  The pilot-assisted coded access scheme is widely used in practical wireless systems when there is no \emph{a priori} CSI at the receiver.
  This scheme consists of two stages:
  1)~users transmit dedicated pilots for channel estimation;
  2)~users transmit codewords, and the receiver utilizes the channel estimate obtained in the first stage to decode.
  This methodology falls into the general framework of the mismatched decoder~\cite{mismatch_Asyhari}.
  From an information-theoretic perspective, channel estimation can be simply viewed as a specific form of coding in the no-CSI case as explained in the introduction.
  In this subsection, we only consider a special case where all users are active for simplicity, and establish an upper bound on the PUPE in Theorem~\ref{Theorem_noCSI_achi_pilot}.
  In essence, the achievability bound for the case where all users are active is equivalent to that with knowledge of the active user set.

  \begin{Theorem} \label{Theorem_noCSI_achi_pilot}
    Assume that all users are active, i.e. $K_a=K$.
    Each user is equipped with a single antenna and the number of BS antennas is $L$.
    Assume each user has a dedicated pilot with length $n_p\leq \min\left\{n,K\right\}$ and power $n_p P_p \leq nP$.
    The matrix $\mathbf{B} = \left[ \mathbf{b}_1, \ldots, \mathbf{b}_{K} \right]\in\mathbb{C}^{n_p\times K }$ comprises of pilots of all users,
    which are drawn uniformly at random on an $n_p$-dimensional sphere of radius $\sqrt{n_p P_p}$.
    Each user also has an individual codebook with size $M=2^J$ and length $n_d = n-n_p$, satisfying that the power of each codeword is no more than $nP - n_p P_p$.
    For the pilot-assisted coded access scheme in MIMO quasi-static Rayleigh fading channels, the PUPE can be upper-bounded as
    \begin{equation} \label{eq_noCSI_pilot_PUPE}
      P_e \leq \min_{0< P'< P}  \left\{ p_0 + \sum_{t=1}^{K } \frac{ t}{K } \min \left\{ 1, p_{t} \right\} \right\} ,
    \end{equation}
    where
    \begin{equation} \label{eq_noCSI_pilot_p0}
      p_0 = K \left( 1 - \frac{\gamma\left( n_d , \frac{ n P - n_p P_p }{{P}^{\prime}} \right)}{\Gamma\left( n_d \right)} \right) ,
    \end{equation}
    \begin{equation} \label{eq_noCSI_pilot_pt}
      p_t = \min_{ 0 \leq \nu }
       \left\{ q_{1,t}\left( \nu\right)
       + q_{2,t}\left( \nu\right) \right\}  ,
    \end{equation}
    \begin{equation}\label{eq_noCSI_pilot_q1t}
      q_{1,t}\left( \nu\right) = {\binom {K } {t}}  M^t  \mathbb{E}_{ \tilde{\mathbf{A}}_{\mathcal{K}}, \tilde{\mathbf{A}}_{S_1}, \tilde{\mathbf{A}}^{'}_{S_1},\mathbf{B}} \left[
      \min_{{u\geq 0,r\geq 0,\lambda_{\min}\left(\mathbf{D}\right) > 0}}
      \exp\left\{ r n_d L\nu -\frac{L}{2} \ln \left| \mathbf{D} \right| \right\}  \right] ,
    \end{equation}
    \begin{align}\label{eq_noCSI_pilot_q2t}
      q_{2,t}\left( \nu\right)  =  \min  & \left\{ \mathbb{E}_{ \tilde{\mathbf{A}}_{\mathcal{K}},\mathbf{B}} \left[ \min_{0\leq \delta < \frac{1}{1+\lambda_{max}\left( \tilde{\mathbf{A}}_{\mathcal{K}} \tilde{\mathbf{\Sigma}} \tilde{\mathbf{A}}_{\mathcal{K}}^H \right) } }
      \exp \left\{ - \delta n_d L \nu  \right\}
      \left| \left( 1-\delta \right) \mathbf{I}_{n_d} - \delta \tilde{\mathbf{A}}_{\mathcal{K}} \tilde{\mathbf{\Sigma}} \tilde{\mathbf{A}}_{\mathcal{K}}^H \right|^{-L}   \right] , \right. \notag \\
      &  \;\;\; \left. \min_{0\leq\eta\leq \nu} \left\{ 2 - \frac{\gamma\left( n_dL, n_d L \eta \right)}{\Gamma\left( n_d L \right)}
      - \mathbb{E}_{ \tilde{\mathbf{A}}_{\mathcal{K}},\mathbf{B}} \left[ \frac{\gamma\left( Ln^{*}, \frac{ n_d L \left(\nu - \eta \right)}{ \lambda_{max}\left( \tilde{\mathbf{A}}_{\mathcal{K}} \tilde{\mathbf{\Sigma}} \tilde{\mathbf{A}}_{\mathcal{K}}^H \right) } \!\right)}{\Gamma\left( Ln^{*} \right)} \right]
      \right\} \!\right\} ,
    \end{align}
    \begin{equation}\label{eq_noCSI_pilot_D}
      \mathbf{D} = \left( 1+r \right) \mathbf{I}_{2n_d} + u \left( 1-u+r \right)\bar{\boldsymbol{\Sigma}}_1 + r \bar{\boldsymbol{\Sigma}}_2 - u\left( u-r \right) \bar{\boldsymbol{\Sigma}}_1 \bar{\boldsymbol{\Sigma}}_2,
    \end{equation}
    \begin{equation}
      \bar{\boldsymbol{\Sigma}}_1 =  \begin{bmatrix} \Re\left( \left( \tilde{\mathbf{A}}_{S_1} \!-\! \tilde{\mathbf{A}}^{'}_{S_1} \right) \hat{\mathbf{\Sigma}} \left( \tilde{\mathbf{A}}_{S_1} \!-\! \tilde{\mathbf{A}}^{'}_{S_1} \right)^H \right) & -\Im\left( \left( \tilde{\mathbf{A}}_{S_1} \!-\! \tilde{\mathbf{A}}^{'}_{S_1} \right) \hat{\mathbf{\Sigma}} \left( \tilde{\mathbf{A}}_{S_1} \!-\! \tilde{\mathbf{A}}^{'}_{S_1} \right)^H \right) \\ \Im\left( \left( \tilde{\mathbf{A}}_{S_1} \!-\! \tilde{\mathbf{A}}^{'}_{S_1} \right) \hat{\mathbf{\Sigma}} \left( \tilde{\mathbf{A}}_{S_1} \!-\! \tilde{\mathbf{A}}^{'}_{S_1} \right)^H \right) & \Re\left( \left( \tilde{\mathbf{A}}_{S_1} \!-\! \tilde{\mathbf{A}}^{'}_{S_1} \right) \hat{\mathbf{\Sigma}} \left( \tilde{\mathbf{A}}_{S_1} \!-\! \tilde{\mathbf{A}}^{'}_{S_1} \right)^H \right) \end {bmatrix},
    \end{equation}
    \begin{equation}
      \bar{\boldsymbol{\Sigma}}_2 =  \begin{bmatrix} \Re\left( \tilde{\mathbf{A}}_{\mathcal{K}} \tilde{\mathbf{\Sigma}} \tilde{\mathbf{A}}_{\mathcal{K}}^H  \right) & -\Im\left( \tilde{\mathbf{A}}_{\mathcal{K}} \tilde{\mathbf{\Sigma}} \tilde{\mathbf{A}}_{\mathcal{K}}^H  \right) \\ \Im\left( \tilde{\mathbf{A}}_{\mathcal{K}} \tilde{\mathbf{\Sigma}} \tilde{\mathbf{A}}_{\mathcal{K}}^H  \right) & \Re\left( \tilde{\mathbf{A}}_{\mathcal{K}} \tilde{\mathbf{\Sigma}} \tilde{\mathbf{A}}_{\mathcal{K}}^H  \right) \end {bmatrix} ,
    \end{equation}
    \begin{equation}\label{eq_noCSI_pilot_Sigmahat}
      \hat{\mathbf{\Sigma}} = \mathbf{I}_{K } - \left( \mathbf{I}_{K} + {\mathbf{B}}^H {\mathbf{B}} \right)^{-1} ,
    \end{equation}
    \begin{equation}\label{eq_noCSI_pilot_Sigmatilde}
      \tilde{\mathbf{\Sigma}} = \left( \mathbf{I}_{K } + {\mathbf{B}}^H {\mathbf{B}} \right)^{-1} ,
    \end{equation}
    \begin{equation}\label{eq_noCSI_pilot_rank}
      n^{*} = \min\left\{ K,n_d \right\} .
    \end{equation}
    Here, in a special case where pilots are orthogonal with $n_p = K $, $\hat{\mathbf{\Sigma}}$ in~\eqref{eq_noCSI_pilot_Sigmahat} and $\tilde{\mathbf{\Sigma}}$ in~\eqref{eq_noCSI_pilot_Sigmatilde} reduce to $\hat{\mathbf{\Sigma}} = \frac{n_pP_p}{1+n_pP_p} \mathbf{I}_{K} $ and $\tilde{\mathbf{\Sigma}} = \frac{1}{1+n_pP_p} \mathbf{I}_{K }$, respectively;
    we have $\tilde{\mathbf{A}}_{S_1} = \mathbf{A} \boldsymbol{\Phi}_{ {S}_1} $, $\tilde{\mathbf{A}}^{'}_{S_1} = \mathbf{A} \boldsymbol{\Phi}^{'}_{ {S}_1} $, and $\tilde{\mathbf{A}}_{\mathcal{K}} = \mathbf{A} \boldsymbol{\Phi}_{\mathcal{K} }$;
    the matrix $\mathbf{A}\in\mathbb{C}^{n_d\times MK}$ is the concatenation of codebooks of the $K$ users without power constraint, which has i.i.d. $\mathcal{CN}\left(0, P'\right)$ entries;
    $S_{1}$ is an arbitrary $t$-subset of $\mathcal{K}$;
    the binary selection matrix $\boldsymbol{\Phi}_{ {S}_1} \in\{0,1\}^{MK\times K}$ indicates which codewords are transmitted by users in the set $ {S}_1$, where $\left[\boldsymbol{\Phi}_{ {S}_1}\right]_{(k-1)M + W_k,k} = 1$ if user $k$ in the set $ {S}_1$ is active and the $W_{k}$-th codeword is transmitted, and $\left[\boldsymbol{\Phi}_{ {S}_1}\right]_{(k-1)M + W_k,k} = 0$ otherwise;
    and similarly, $\boldsymbol{\Phi}^{'}_{ {S}_1} \in\{0,1\}^{MK \times K }$ indicates which codewords are not transmitted but decoded for users in the set $ {S}_1$.
  \begin{IEEEproof}[Proof sketch]
    The power of each pilot is $n_pP_p$ and the power of each codeword is no more than $nP-n_pP_p$, thereby satisfying the power constraint in~\eqref{power_constraint}.
    In the pilot transmission phase, users transmit dedicated pilots and the receiver estimates channels based on the MMSE criterion.
    In the data transmission phase, we use the random coding scheme and assume that users transmit codewords uniformly selected from their own codebooks.
    For the pilot-assisted scheme, the decoder has an incorrect estimate of the channel but uses the estimate as if it were perfect, which is different from the case of CSIR.
    Due to the channel estimation error, bounding $\mathbb{P}\left[ \mathcal{F}_t \right]$ is more involved for the pilot-assisted scheme than in the case of CSIR.
    Thus, in this subsection, we only utilize the bounding technique proposed by Fano in~\cite{1961} and simplify the ``good region'' designed in~\eqref{eq_good region_selection} with $\omega=0$.
    See Appendix~\ref{section6} for the complete proof.
  \end{IEEEproof}
  \end{Theorem}

  The following corollary of Theorem~\ref{Theorem_noCSI_achi_pilot} provides an achievability bound on the minimum required energy-per-bit for the pilot-assisted coded access scheme.
  \begin{Corollary} \label{Theorem_noCSI_achi_pilot_energyperbit}
    Assume that all users are active.
    Each user is equipped with a single antenna and the number of BS antennas is $L$.
    Assume each user has a dedicated pilot with length $n_p<n$ and power $n_p P_p < nP$.
    Each user also has an individual codebook with size $M=2^J$ and length $n_d = n-n_p$ satisfying that the power of each codeword is no more than $nP - n_p P_p$.
    For the pilot-assisted scheme in MIMO quasi-static Rayleigh fading channels, the minimum energy-per-bit $E^{*}_{b,\text{no-CSI},K_a}(n,M,\epsilon)$ for satisfying the PUPE requirement in~\eqref{PUPE} can be upper-bounded as
      \begin{equation}
        E^{*}_{b,\text{no-CSI},K_a}(n,M,\epsilon) \leq \inf \frac{n {P}}{J},
      \end{equation}
      where the $\inf$ is taken over all $P > 0$ satisfying that
      \begin{equation} \label{eq_noCSI_epsilon_pilot}
        \epsilon \geq \min_{0< P'< P} \left\{ p_0 + \sum_{t=1}^{K } \frac{ t}{K } \min \left\{ 1, p_{t} \right\} \right\}.
      \end{equation}
      Here, $p_0$ and $p_{t}$ are the same as those in Theorem~\ref{Theorem_noCSI_achi_pilot}.
  \end{Corollary}

  In Corollary~\ref{Theorem_noCSI_achi_pilot_energyperbit}, we derive an achievability bound on the minimum required energy-per-bit for the pilot-assisted transmission scheme.
  As we can see from the result, there exists a tradeoff between the accuracy of the estimated CSI and the blocklength available for data transmission.
  That is, a longer pilot is beneficial to improve the channel estimation performance, but at the price of  reducing the number of channel uses available for data transmission.
  More results on this can be found in Section~\ref{sec_simulation}.

\subsection{Generalizations} \label{section3_generalizations}

  In this subsection, we introduce several possible generalizations of the results in this paper.

  First, we have focused on MIMO quasi-static Rayleigh fading channels in this work. Note that the results can be extended to other types of fading channels, such as Rician fading.
  Specifically, for the CSIR case, the derivations of the achievability bound based on Gallager's $\rho$-trick and the converse bound are independent of the fading distribution (i.e., these bounds can be general).
  The fading distribution only kicks in when evaluating them numerically, and the Rayleigh distribution assumption could simplify the computation.
  In both CSIR and no-CSI cases, the ``good region''-based achievability bounds for Rayleigh fading channels can be extended to Rician fading channels because the main techniques used to derive them, such as Fano's bounding technique, the union bound, Chernoff bound, and moment generating function of quadratic forms, are also applicable when channels are subject to Rician fading.
  In addition, in the case of no-CSI, converse bounds derived in~\cite{sparsity_pattern} are applicable to a general fading model in the single-receive-antenna setting.
  Applying similar ideas in~\cite{sparsity_pattern}, we can extend the converse bound for Rayleigh fading to various types of fading in MIMO channels.

  Second, we have considered the joint activity and data detection problem in MIMO quasi-static Rayleigh fading channels in this work, where each user is assumed to have an individual codebook.
  Note that the results can be extended to the framework of a common codebook. A similar extension with AWGN channels can be found in~\cite{A_perspective_on,indivicommon}.

\section{Numerical Results} \label{sec_simulation}
  In this section, we validate our theoretical results in Section~\ref{section3} through numerical simulations.
  We consider quasi-static fading channels with $L$ BS antennas. The channel between each transmit-receive antenna pair is independently Rayleigh-distributed.
  We assume the blocklength is $n=1000$, payload is $J=100$ bits, and target PUPE is $\epsilon=0.001$.
  The required memory space to compute the bounds is $\mathcal{O}(W^2)$ with $W = \max\{n,K\}$.
  In Section~\ref{sec_simulation1}, we present the number of reliably served active users versus the energy-per-bit when the number of BS antennas is given.
  In Section~\ref{sec_simulation2}, we present the spectral efficiency versus the number of BS antennas for fixed energy-per-bit.
  We use the Monte Carlo method with $500$ samples to evaluate expectations in the converse bounds.
  For the achievability bounds, the parameters outside the expectations are optimized by sampling and exhaustively searching, with the expectations therein evaluated by the Monte Carlo method using $500$ samples; once these parameters are determined, we generate $10000$ samples to obtain ultimate achievability bounds.

\subsection{The number of users versus the energy-per-bit} \label{sec_simulation1}
\begin{figure}
    \centering
    \includegraphics[width=0.6\linewidth]{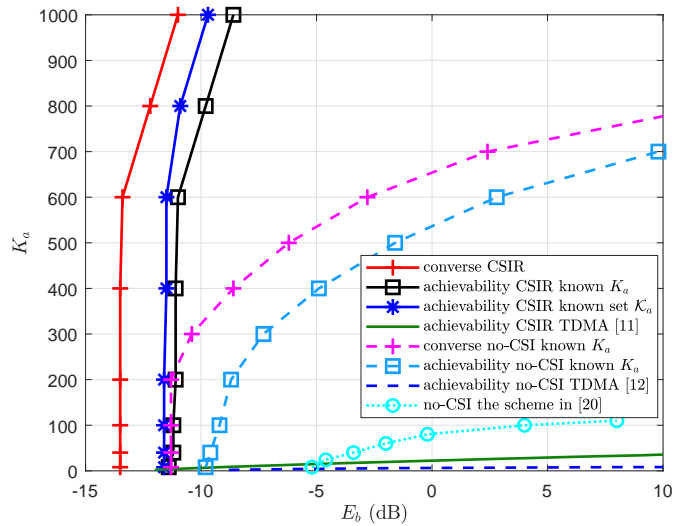}\\
    \caption{The number $K_a$ of active users versus the energy-per-bit $E_b$ with $n=1000$, $J = 100$~bits, $K_a = 0.4K$, $\epsilon = 0.001$, and $L=32$.}
  \label{fig:L32}
  \end{figure}

  In Fig. \ref{fig:L32}, we present our achievability and converse bounds on the minimum required energy-per-bit with known $K_a$, together with the achievability bounds on the orthogonalization scheme time division multiple access~(TDMA)~\cite{Channel_coding_rate,TDMA_yangwei} and the performance of the scheme proposed in~\cite{Caire1}.
  We assume there are $L \!=\! 32$ BS antennas.
  Next, we explain how each curve is obtained:
  \begin{enumerate}
    \item The achievability bound for the case of CSIR with knowledge of the active user set $\mathcal{K}_a$ is based on Corollary~\ref{Theorem_CSIR_achi_knownUE},
        where only Gallager's $\rho$-trick bound $\tilde{p}_{2,t}$ in~\eqref{eq_CSIR_p1t_knownUE} is utilized because it is tighter than the ``good region''-based bound $\tilde{p}_{1,t}$ in our considered regime.
    \item The achievability bound for the case of CSIR with known $K_a$ but unknown $\mathcal{K}_a$ is based on the ``good region'' bound $p_{1,t}$ in Corollary~\ref{Theorem_CSIR_achi_energyperbit}.
        We set $u=\frac{1+r}{2}$ to reduce searching complexity, which is optimal when $\omega=0$.
        Gallager's $\rho$-trick bound $p_{2,t}$ in Corollary~\ref{Theorem_CSIR_achi_energyperbit} is not used because we observe from numerical simulation that it requires an extremely large number of samples to get a good estimate for the massive random access problem.
    \item The converse bound for the case of CSIR is Theorem~\ref{prop_converse_CSIR}.
    \item The achievability bound for the no-CSI case with known $K_a$ is Corollary~\ref{Theorem_noCSI_achi_energyperbit}, where the ``good region''-based bound $p_t$ is provided in Theorem~\ref{Theorem_noCSI_achi}.
        To reduce simulation complexity, we set the parameter $u$ in $p_t$ to be $u=\frac{1+r}{2}$.
        In this case, the term inside the expectation in~\eqref{eq_noCSI_q1t} is a convex function of $r$, which is optimized by Newton's method.
    \item The converse bound for the case of no-CSI with known $K_a$ is Theorem~\ref{prop_converse_noCSI}.
    \item For TDMA, to achieve the spectral efficiency $S_e = \frac{K_aJ}{n}$, we compute the smallest $P$ ensuring the access of an active user with rate $\frac{KJ}{n}$, blocklength $\frac{n}{K}$, target PUPE $\epsilon$, and $L$ BS antennas.
        Specifically, we utilize the $\kappa\beta$ bound
        \cite[Th. 25]{Channel_coding_rate} for the case of CSIR and the bound in~\cite[Eq. (67)]{TDMA_yangwei} for the case of no-CSI, respectively.
    \item For comparison, we present the joint activity and data detection performance of the scheme proposed in~\cite{Caire1} for the case of no-CSI.
        We follow the concatenated coding scheme in~\cite[Section~V]{Caire1}, suitably adapted to our case.
        Specifically, we equally divide a coherence block with length $n = 1000$ into $D=10$ slots. Let each user transmit $J_D=10$ bits over a slot with $n_D = 100$ dimensions, yielding an overall payload $J=100$.
        In each slot, we choose the columns of each coding matrix uniformly i.i.d. from the sphere with radius $P n_D$.
        For the inner code, we assume user $k$ sends the $i_{k,d}$-th column of the coding matrix, where $i_{k,d}\in\left[2^{J_D}\right]$ denotes the message produced by user $k$ in slot $d$.
        For the inner decoder, we use the non-Bayesian approach in~\cite[Algorithm~1]{Caire1}, which is proposed for the unsourced random access model (i.e., the framework of a common codebook).
        To cater for the framework of individual codebooks, we utilize a hard decision on the support of the estimated vector $\hat{\boldsymbol{\gamma}}$ with the threshold $0.08$, and importantly, we restrict that at most one codeword can be decoded in each codebook.
        Moreover, since each user has unique codebook known at the receiver in advance, the decoded messages across different slots can be stitched based on this prior knowledge. Thus, there is no need to utilize the tree code as the outer code.
        % Under the constraint that both the misdetection error probability and the false-alarm error probability are less than $\epsilon$, i.e., $ \max\left\{ P_{e,\mathrm{MD}} , P_{e,\mathrm{FA}} \right\} \leq \epsilon$, we plot the minimum required energy-per-bit for different numbers of active users.
        We obtain the average of the misdetection error probability and the false-alarm error probability, i.e., $P_e = \left( P_{e,\mathrm{MD}} + P_{e,\mathrm{FA}} \right)/2$, and plot the minimum required energy-per-bit to satisfy $P_e \leq \epsilon$ for different numbers of active users.
  \end{enumerate}
  As shown in Fig. \ref{fig:L32}, the gap between our achievability and converse bounds is less than $2.5$~dB in all $K_a$ regimes for the CSIR case and less than $4$~dB for $K_a$ less than $500$ in the case of no-CSI with known $K_a$.
  Thus, our non-asymptotic bounds provide relatively accurate theoretical benchmarks to evaluate practical transmission schemes, which are of considerable importance in massive random access systems.
  In the case of CSIR with known $K_a$, we can observe that the lack of knowledge of the active user set entails a penalty less than $1.2$~dB in terms of energy efficiency.
  %{\blue Although the bound following from Gallager's $\rho$-trick is difficult to evaluate for massive random access in MIMO quasi-static Rayleigh fading channels, it is tighter than the ``good region''-based one in a special case where the set of active users is known at the receiver.}
  As expected, it is more costly to communicate in the no-CSI case than in the CSIR case, especially for a large number of active users.
  Additionally, similar to AWGN channels~\cite{improved_bound} and single-receive-antenna quasi-static fading channels~\cite{finite_payloads_fading}, the almost perfect MUI cancellation effect is observed in multiple-receive-antenna quasi-static Rayleigh fading channels.
  Specifically, when the number of active users is below a critical threshold, the minimum required energy-per-bit is almost a constant in the case of CSIR, although there is a slow growth of the energy-per-bit as $K_a$ increases within this range for the no-CSI case.
  Moreover, we observe that the scheme in~\cite{Caire1} is inferior to the achievability bound in the case of no-CSI, especially when $K_a>80$.
  This is because, although the concatenated coding scheme in~\cite{Caire1} contributes to the manageability of the coding matrix with the dimension as small as $100 \times 1024 K$, it leads to a performance loss since the dimension of a slot is greatly reduced.
  In addition, the orthogonalization scheme TDMA does not have the perfect MUI cancellation effect.
  TDMA is shown to be energy-inefficient for large user densities when user activity is known~\cite{finite_payloads_fading}, and it becomes more energy-inefficient for the random access model since some resources allocated for inactive users are not utilized.
  \begin{figure}
  \centering
  \includegraphics[width=0.6\linewidth]{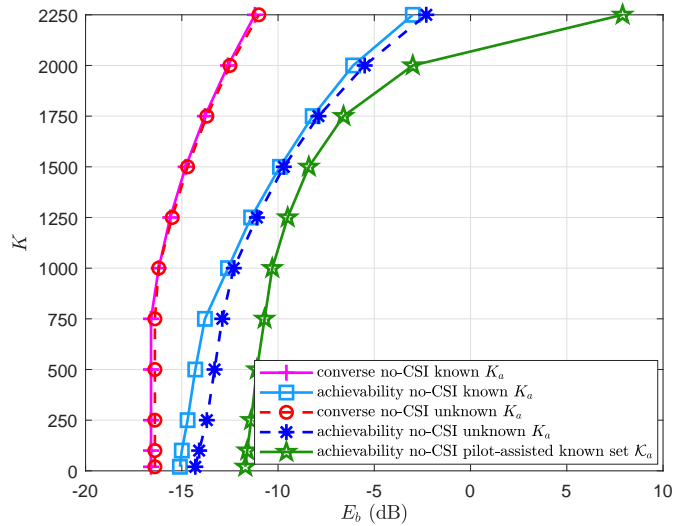}\\
  \caption%{\red The number $K_a$ (resp. the average number $\bar{K}_a$) of reliably served active users versus the energy-per-bit $E_b$ with $n=1000$, $J = 100$~bits, $\epsilon = 0.001$, and $L=128$ in the case of known (unknown) $K_a$. In the case of known $K_a$, we assume $K_a$ is fixed among $K=K_a/0.4$ potential users; in the case of unknown $K_a$, we assume $K_a$ is random and distributed as $K_a\sim \text{Binom}(K,p_a)$ with mean $\bar{K}_a = p_a K$ and $p_a=0.4$.}
  {The number $K$ of potential users versus the energy-per-bit $E_b$ with $n=1000$, $J = 100$~bits, $\epsilon = 0.001$, and $L=128$ in two cases:
  1)~the number of active users is $K_a=0.4K$, which is fixed and known in advance;
  2)~the number of active users is random and unknown, and its distributed $K_a\sim \text{Binom}(K,p_a)$ is known \emph{a priori} with $p_a=0.4$ and mean $\bar{K}_a=0.4K$.}
  \label{fig:L128}
  \end{figure}

  In Fig.~\ref{fig:L128}, we compare the achievability and converse bounds on the minimum required energy-per-bit in the following two settings with no-CSI:
  1)~the number of active users is $K_a=0.4K$, which is fixed and known in advance;
  2)~the number of active users is random and unknown, and its distribution $K_a\sim \text{Binom}(K,0.4)$ is known \emph{a priori}.
  In the case of unknown $K_a$, it is required that $( \epsilon_{\rm MD} + \epsilon_{\rm FA} ) / 2 = \epsilon$.
  Moreover, the achievability bound for a pilot-assisted scheme is also computed.
  Next, we explain how each curve is obtained:
  \begin{enumerate}
    \item The achievability bounds for the no-CSI case are presented for the settings with and without the knowledge of the number $K_a$ of active users at the receiver.
        The bound with known $K_a$ is based on Corollary~\ref{Theorem_noCSI_achi_energyperbit},
        which is computed in a similar way to that in Fig.~\ref{fig:L32}.
        The bound for the setting with unknown $K_a$ is based on Theorem~\ref{Theorem_noCSI_noKa_achi_energyperbit},
        where the decoding radius $r'$ is determined by brute-force searching from the set $\{0,1,\ldots,25\}$.
%        {\red
%        Specifically, we assume $K_a = 0.4K$ users are active.
%        However, the receiver does not know $K_a$ in advance, and it only knows $K_a\in [K_l : K_u]$ with $K_l = 1$ and $K_u = K$.
%        The decoding radius $r'$ is determined by brute-force searching from the set $\{0,1,\ldots,20\}$.
%        Moreover, we require the average of the per-user probabilities of misdetection and false-alarm satisfies $\left( P_{e,\mathrm{MD}} + P_{e,\mathrm{FA}} \right)/2 \leq \epsilon$.}
        % {\red The receiver does not know $K_a$ in advance, and it only knows $K_a\in [K_l : K_u]$ with $K_l = \lfloor 0.35K \rfloor$ and $K_u = \lceil 0.45K \rceil$.
        % We require that both the per-user probability of misdetection and the per-user probability of false-alarm are less than $\epsilon$, i.e., $\epsilon_{e,\mathrm{MD}} = \epsilon_{e,\mathrm{FA}}  = \epsilon$.}
    \item The converse bound for the setting with and without the knowledge of the number $K_a$ of active users is based on Theorem~\ref{prop_converse_noCSI} and Theorem~\ref{prop_converse_noCSI_noKa}, respectively.
%        The converse bounds for the no-CSI case are presented for the settings with and without known $K_a$.
%        The converse bound with known $K_a$ is based on Theorem~\ref{prop_converse_noCSI}, which is computed in a similar way to that in Fig.~\ref{fig:L32}.
%        The bound for the setting with unknown $K_a$ is based on Theorem~\ref{prop_converse_noCSI_noKa},
    \item The achievability bound for the pilot-assisted coded access scheme is based on Corollary~\ref{Theorem_noCSI_achi_pilot_energyperbit} under the assumption that the active user set $\mathcal{K}_a$ is known \emph{a priori}, wherein the power allocation between the pilot and data symbols is optimized and orthogonal pilots of length $n_p=K_a$ are utilized.
  \end{enumerate}
  Our results reveal that the pilot-assisted coded access scheme is suboptimal in the no-CSI case, even if the power allocation between the pilot and data symbols is optimized.
  Specifically, the gap between the achievability bounds of the pilot-assisted scheme and the scheme without explicit channel estimation is less than $3.5$~dB when the number of users is less than $800$ but sees a dramatic increase when the number of users exceeds this.
  Moreover, from the achievability and converse bounds with and without the knowledge of the number $K_a$ of active users at the receiver, we can observe that once the distribution $K_a\sim \text{Binom}(K,p_a)$ is known in advance,
  % there is a small penalty when only the distribution of $K_a$ rather than its exact value is known in advance.
  % there is a slight penalty when the exact number $K_a$ of active users is unknown, under the condition that its distribution is known in advance,
  the uncertainty of the exact value of $K_a$ entails only a small penalty in terms of energy efficiency, with the extra required energy-per-bit less than $0.3$~dB on the converse side and less than $1.1$~dB on the achievability side.
  % the performance loss caused by the uncertainty of user activities is small once the distribution of $K_a$ is known in advance

  \begin{figure}
  \centering
  \includegraphics[width=0.6\linewidth]{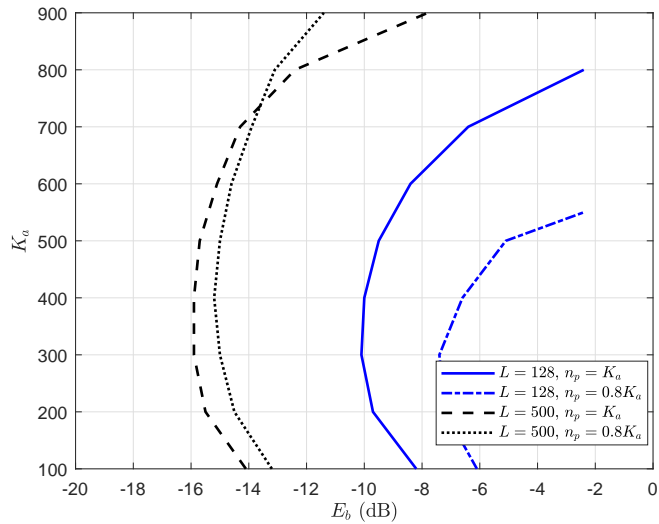}\\
  \caption{The number $K_a$ of active users versus the energy-per-bit $E_b$ for the pilot-assisted scheme with $n=1000$, $J = 100$, $\epsilon = 0.001$, $n_p\in\{K_a,0.8K_a\}$, $L\in\{128,500\}$, $P_p = P$, and $P_d\leq P$.}
  \label{RFig:pilot}
  \end{figure}
  In Fig.~\ref{RFig:pilot}, considering the setup with blocklength $n=1000$, payload $J=100$~bits, PUPE requirement $\epsilon=0.001$, $L\in\{128,500\}$ BS antennas, and known active user set $\mathcal{K}_a$,
  we compare the non-orthogonal-pilot-based scheme with pilot length $n_p=0.8K_a$ (i.e. $n_d = n-0.8K_a$ channel uses for data transmission) and the orthogonal-pilot-based scheme with $n_p=K_a$ (i.e. $n_d = n-K_a$ channel uses for data transmission).
  The non-orthogonal pilots are generated using a sub-sampled discrete Fourier transform matrix.
  As opposed to Fig.~\ref{fig:L128}, the power allocation between the pilot and data symbols is not optimized in Fig.~\ref{RFig:pilot} due to simulation complexity.
  Specifically, we assume the transmitting power of the pilot per channel use is $P_p=P$ and the transmitting power of the data per channel use is $P_d\leq P$ to satisfy the maximum power constraint in~\eqref{power_constraint}.
  As shown in Fig.~\ref{RFig:pilot}, the achievability bound for the scheme based on non-orthogonal pilots is inferior to the orthogonal-pilot-based one in the setup with $L=128$ BS antennas.
  However, when the number of BS antennas increases to $L=500$ and the number of users is above $800$, the scheme based on non-orthogonal pilots of length $n_p=0.8K_a$ outperforms the orthogonal-pilot-based one.
  As a result, for the pilot-assisted scheme, there exists a tradeoff between the channel estimation performance and the blocklength used for data transmission.
  In particular, for a fixed blocklength~$n$, when the numbers of BS antennas and users are large, it is more reasonable to use non-orthogonal pilots to set aside more channel uses for data transmission, instead of allocating an orthogonal pilot to each user.
  This is because when the number of users is large, allocating orthogonal pilots results in little time left for data transmission;
  meanwhile, a large number of BS antennas can mitigate the effects of noise and fast fading, which allows us to reduce the length of pilots.

\subsection{The spectral efficiency versus the number of BS antennas} \label{sec_simulation2}

  \begin{figure*}[!t]
  \centering
  \subfloat[]{\includegraphics[width=0.49\linewidth]{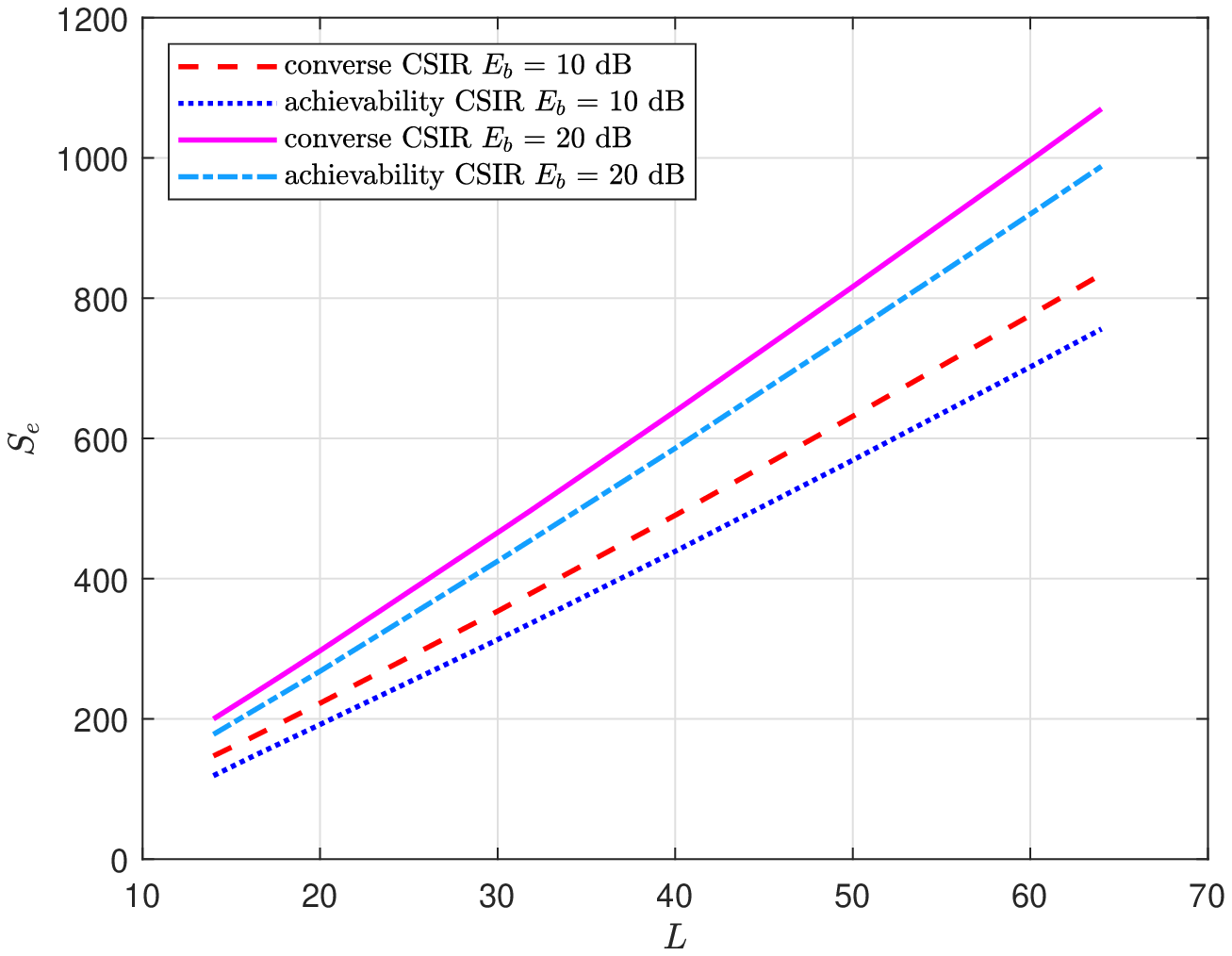} \label{fig:SE_CSIR}}
  \hfil
  \subfloat[]{\includegraphics[width=0.49\linewidth]{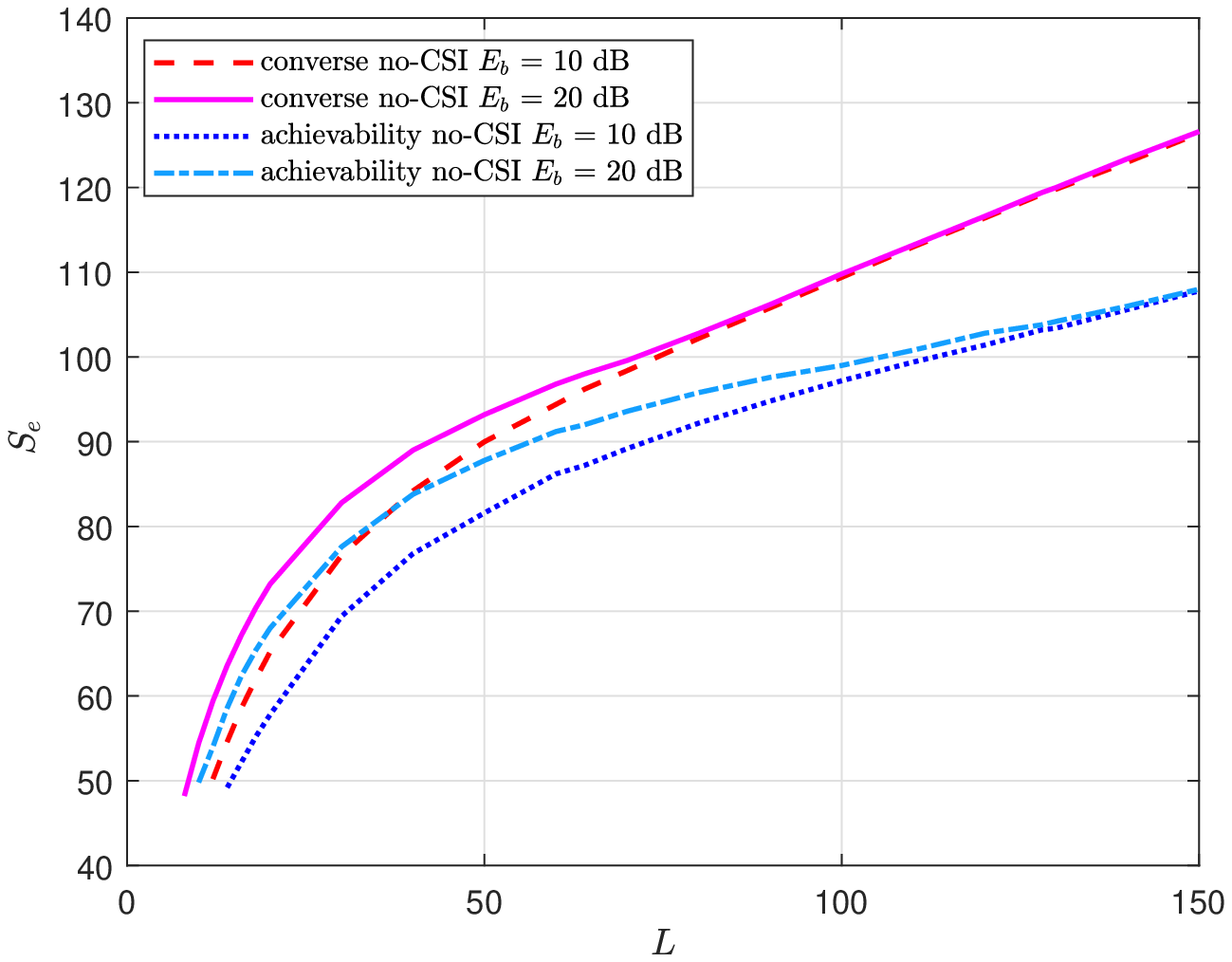}
  \label{fig:SE_noCSI}}
  \caption{The spectral efficiency $S_e$ versus the number $L$ of BS antennas with $n=1000$, $J = 100$~bits, $K_a = 0.4K$, and $\epsilon = 0.001$: (a) CSIR; (b) no-CSI.}
  \label{fig:SE}
  \end{figure*}

  %% 图片内容+参数
  As illustrated in Fig.~\ref{fig:SE}, we present bounds on the maximum spectral efficiency $S_e$ against the number $L$ of BS antennas.
  Specifically, the looser converse bounds \eqref{P_tot_conv_CSIR_bound} in Theorem~\ref{prop_converse_CSIR} and \eqref{P_tot_conv_noCSI_bound} in Theorem~\ref{prop_converse_noCSI} are utilized in the case of CSIR and no-CSI, respectively.
  For the achievability bound in the CSIR case, we utilize the ``good region'' bound $p_{1,t}$ in Corollary~\ref{Theorem_CSIR_achi_energyperbit}, where $\omega$ is set to be $0$ to reduce simulation complexity.
  In this case, the optimal value of $u$ is given by $u=\frac{1+r}{2}$ and the term inside the expectation in~\eqref{eq_CSIR_q1t} becomes a convex function of $r$.
  Thus, the optimal solution of $r$ can be generated by Newton's method.
  The achievability bound in the no-CSI case is Corollary~\ref{Theorem_noCSI_achi_energyperbit}, which is computed similar to that in Fig.~\ref{fig:L32}.
  %% 可以超过码长
  We observe from Fig.~\ref{fig:SE} that, as $L$ increases, the spectral efficiency $S_e$ can exceed $100$, i.e., the number of active users that are reliably served can exceed the blocklength $n$, regardless of whether CSIR is available or not.
  %% 线性+速率降低
  In the case of CSIR, the spectral efficiency increases with $L$ at an approximately constant speed, whereas the increasing speed gradually reduces in the no-CSI case due to the increased channel uncertainty.
  %% 干扰受限+功率受限
  Additionally, as shown in Fig.~\ref{fig:SE_CSIR}, in the case of CSIR, increasing energy-per-bit $E_b$ is beneficial for different values of BS antennas, where the gap between the spectral efficiency for $E_b=10$~dB and $E_b=20$~dB increases as $L$ increases.
  However, as observed in Fig.~\ref{fig:SE_noCSI}, in the case of no-CSI, increasing energy-per-bit contributes only when $K_a$ (or $S_e$) is small, in line with the results in Fig.~\ref{fig:L32}.
  For both the achievability and the converse bounds, the gap between the spectral efficiency for $E_b=10$~dB and $E_b=20$~dB vanishes to zero as $K_a$ grows large, suffering from channel uncertainty in such a worse interference environment.
  In the case of no-CSI, the gap between achievability and converse bounds on the spectral efficiency per antenna is less than $0.13$~bit/s/Hz, regardless of whether $E_b=10$~dB or $E_b=20$~dB.

\section{Conclusion} \label{section7}
  Supporting the transmission of short packets under stringent latency and energy constraints is critically required for next-generation wireless communication networks.
  In this paper, we have considered such a communication system with finite blocklength and payload size.
  Under the PUPE criterion, we have established non-asymptotic achievability and converse bounds on the minimum required energy-per-bit for massive random access in MIMO quasi-static Rayleigh fading channels, with and without \emph{a priori} CSI at the receiver.
  In the case of no-CSI, we consider both the settings with and without the knowledge of the number $K_a$ of active users at the receiver.
  % {\blue Further, we have extended the achievability and converse results in the no-CSI case with known $K_a$ to a general setting where $K_a$ is random and unknown in advance.}
  One key ingredient of the achievability bounds is the design of an appropriate ``good region'', conditioned on which the union bound is applied.
  %% simulation results: tight
  Numerical results demonstrate the tightness of our bounds. Specifically, the gap between the achievability and converse bounds is less than $2.5$~dB for the CSIR case and less than $4$~dB for the no-CSI case in most considered regimes.
  % The no-CSI achievability bounds with and without known $K_a$ show that the extra required energy-per-bit due to the lack of knowledge of $K_a$ is about $5$~dB when $K_a$ is small, and this gap reduces to less than $1.5$~dB as $K_a$ increases.
  The no-CSI achievability and converse bounds show that the extra required energy-per-bit due to the uncertainty of the exact value of $K_a$ is small in the considered regime, under the condition that the distribution of $K_a$ is known \emph{a priori}.
  %% MUI
  The almost perfect MUI cancellation effect for the number of active users below a certain threshold, which was previously observed in AWGN channels~\cite{improved_bound} and single-receive-antenna quasi-static fading channels~\cite{finite_payloads_fading}, is prominent in multiple-receive-antenna quasi-static Rayleigh fading channels with CSIR, although there is a slow growth of the energy-per-bit as the number of active users increases within this range in the no-CSI case.
  %For the no-CSI case, this effect reduces to a slow growth of energy-per-bit as the number of active users increases within a range below a certain threshold.
  %% SE linear nonlinear
  Additionally, in our considered regime, the spectral efficiency grows approximately linearly with the number of BS antennas in the CSIR case, but the lack of CSI at the receiver causes a slowdown in the growth rate.
  %% noncoherent pilot-assisted
  Furthermore, we have evaluated the performance of a pilot-assisted scheme, and numerical results show that it is suboptimal especially when there are many users.
  Overall, we believe our non-asymptotic bounds provide theoretical benchmarks to evaluate practical transmission schemes, and are of considerable importance in massive random access systems.

  %% asymptotic
  Building on these non-asymptotic bounds, assuming $n\to\infty$ and $J=\Theta(1)$, we have obtained scaling laws of the number of reliably served users for a special case where all users are active.
  For the CSIR case, assuming $K\to\infty$, $\ln K = o(n)$, and $KP=\Omega\left(1\right)$, the PUPE requirement is satisfied if and only if $\frac{nL\ln KP}{K}=\Omega\left(1\right)$,
  i.e., if and only if one of the following two relations is satisfied: 1)~$\frac{nL}{K}=\Omega\left(1\right)$ and $KP =\Theta\left(1\right)$; 2)~$\frac{nL\ln KP}{K}=\Omega\left(1\right)$ and $KP \to \infty$.
  The first regime is power-limited and the second regime is degrees-of-freedom-limited.
  The condition $\frac{nL\ln KP}{K}=\Omega\left(1\right)$ shows the great potential of multiple receive antennas to considerably increase the number of reliably served users and reduce the required power $P$ and blocklength $n$.
  For the no-CSI case, we observe a significant difference in the required number of BS antennas between utilizing the PUPE criterion and the joint error probability criterion.
  Specifically, in order to reliably serve $K = \mathcal{O}(n^2)$ users with power $P=\Theta\left(\frac{1}{n^2}\right)$, the required number of BS antennas is reduced from $L = \Theta \left(n^2\ln n\right)$ to $L = \Theta \left(n^2\right)$ when we change from the joint error probability criterion to the PUPE criterion.
  Notably, as presented in Table~\ref{table:scalinglaw}, our scaling laws consider the regime in which the energy-per-bit is finite or goes to $0$, which are crucial in practical communication systems with stringent energy constraints.
  % Moreover, our results are based on the assumption of individual codebooks, which can be extended to the framework of a common codebook.

%%%%%%%%%%%%%%%%%%%%%%%%%%%%%%%%%%%%%%%%%%%%%%%%%%%%%%%%%%%%%%%%%%%%%%%
%%%%%%%%%%%%%%%%%%%%%%%%%%%%%%%%%%%%%%%%%%%%%%%%%%%%%%%%%%%%%%%%%%%%%%%
\appendices
\section{ % A general ``good region''-based achievability bound
         A general upper bound on the PUPE based on Fano's bounding technique} \label{Proof_achi_CSIR_noCSI}

  In this appendix, we provide a general upper bound on the PUPE applying Fano's ``good region'' technique, which is applicable for both CSIR and no-CSI cases.
  This bound is derived under the assumption that $K_a$ is known at the receiver beforehand, and it can be extended to the case without known $K_a$ as introduced in Appendix~\ref{proof_achi_noCSI_noKa_weight}.

  We use a random coding scheme. Specifically, we generate a Gaussian codebook of size $M$ and length $n$ for each user independently.
  Let $\mathcal{C}_k = \left\{\mathbf{c}_{k,1}, \mathbf{c}_{k,2},\ldots, \mathbf{c}_{k,M}\right\} $ denote the codebook of user $k$ without power constraint, where $\mathbf{c}_{k,m}\stackrel{\mathrm{i.i.d.}}{\sim}  \mathcal{CN}\left(0,P'\mathbf{I}_{n}\right)$ for $m\in[M]$ and $k\in\mathcal{K}$.
  We choose $P' < P$ to ensure that we can control the maximum power constraint violation events.
  Let $\mathbf{A} \in \mathbb{C}^{n\times MK}$ denote the concatenation of codebooks of the $K$ users without power constraint.
  If user $k$ is active, let its transmitted codeword be $\mathbf{x}_{(k)} = \mathbf{c}_{(k)} 1 \left\{ \left\|\mathbf{c}_{(k)}\right\|_{2}^{2} \leq n P\right\}$, where $\mathbf{c}_{(k)} = \mathbf{c}_{k,W_{k}}$ with the message $ W_{k} \in [M]$ chosen uniformly at random;
  if user $k$ is inactive, let $\mathbf{x}_{(k)} = \mathbf{c}_{(k)} = \mathbf{0}$.

  The decoder aims to find the estimated set $\hat{\mathcal{K}}_a$ of active users, and find the estimate $\hat{\mathbf{c}}_{(k)}$ of ${\mathbf{c}}_{(k)}$ and corresponding message $\hat{W}_k$ of ${W}_k$ for $ k \in \hat{\mathcal{K}}_a $.
  Let $\hat{\mathbf{c}}_{[ {\hat{\mathcal{K}}}_a]} = \left\{ \hat{\mathbf{c}}_{(k)} \in \mathcal{C}_k:  k\in\hat{\mathcal{K}}_a \right\}$.
  The outputs of the decoder are given by
      \begin{equation}\label{decoder_CSIR_noCSI}
        \left[ \hat{\mathcal{K}}_{a }, \hat{\mathbf{c}}_{[\hat{\mathcal{K}}_a]} \right]
        = \arg\min_{{ \hat{\mathcal{K}}_{a } \subset \mathcal{K}}, {\left| \hat{\mathcal{K}}_{a }\right| = K_{a }} }
        \;\min_{ \left( \hat{\mathbf{c}}_{(k)} \in \mathcal{C}_k \right)_{ k\in\hat{\mathcal{K}}_{a } } }
        \;g\left(  \mathbf{Y} , \hat{\mathbf{c}}_{[\hat{\mathcal{K}}_a]}  \right) ,
      \end{equation}
      \begin{equation}\label{decoder_CSIR_noCSI2}
        \hat{W}_k =  \emph{f}_{\text{en},k}^{-1}\left(\hat{\mathbf{c}}_{(k)}\right) , \;\; k \in \hat{\mathcal{K}}_{a} ,
      \end{equation}
  where $g\left(  \mathbf{Y} ,   \hat{\mathbf{c}}_{[ {\hat{\mathcal{K}}}_a]}  \right)$ denotes the decoding metric.
  We have $\hat{W}_k = 0$ and $\hat{\mathbf{c}}_{(k)}=\boldsymbol{0}$ for $ k \notin \hat{\mathcal{K}}_a $.

  The PUPE in~\eqref{PUPE} can be upper-bounded as
     \begin{align}
       P_{e}& \leq  p_0
       + \mathbb{E}\left[\frac{1}{K_a} \!\sum_{k\in {\mathcal{K}_a}} 1 \left[ W_{k} \neq \hat{W}_{k} \right] \right]_{\text{no power constraint}} \label{eq_PUPE_upper_CSIR_noCSI1} \\
       & = p_0 + \sum_{t=1}^{K_a} \frac{ t}{K_a} \mathbb{P} \left[ \mathcal{F}_t \right]_{\text{no power constraint}} \label{eq_PUPE_upper_CSIR_noCSI},
     \end{align}
  where \eqref{eq_PUPE_upper_CSIR_noCSI1} follows because we change the measure $\mathbf{x}_{(k)} = \mathbf{c}_{(k)} 1 \left\{ \left\|\mathbf{c}_{(k)}\right\|_{2}^{2} \leq n P\right\}$ with power constraint to $\mathbf{x}_{(k)} = \mathbf{c}_{(k)}$ without power constraint by adding a total variation distance upper-bounded by $p_0$ \cite{finite_payloads_fading}.
  Here, $p_0$ is given by
     \begin{equation}
        p_0 = K_a \; \mathbb{P}\left[  \left\|\mathbf{c}_{(k)}\right\|_{2}^{2} > n P  \right]
        = K_a \left( 1 - \frac{\gamma\left(n, \frac{nP}{{P}^{\prime}} \right)}{\Gamma\left(n\right)} \right) ,
     \end{equation}
  which holds because $\left\|\mathbf{c}_{(k)}\right\|_{2}^{2} \sim \frac{P'}{2}\chi^2(2n)$;
  $\mathcal{F}_t = \left\{ \sum_{k\in {\mathcal{K}_a}}  1 \left\{ W_{k} \neq \hat{W}_{k} \right\} = t \right\}$ indicates the event that there are exactly $t$ misdecoded users.
  In what follows, we omit the subscript ``no power constraint'' for simplicity and upper-bound $\mathbb{P} \left[ \mathcal{F}_t  \right]$ applying Fano's ``good region'' technique~\cite{1961}.

  Let the set $S_1 \subset \mathcal{K}_a$ of size $t$ denote the set of users whose codewords are misdecoded.
  Let the set $S_2 \subset \mathcal{K}\backslash\mathcal{K}_a\cup S_1$ of size $t$ denote the set of detected users with false alarm codewords.
  For the sake of simplicity, we rewrite ``$\bigcup_{{S_{1} \subset \mathcal{K}_a,} {\left| S_{1} \right| = t }}  $'' to ``$\bigcup_{S_{1}}$'' and ``$\bigcup_{{S_{2} \subset \mathcal{K} \backslash \mathcal{K}_a\cup S_{1},} {\left| S_{2} \right| = t}}$'' to ``$\bigcup_{S_{2}}$''; and similarly for $\sum$ and $\bigcap$.
  We use $\mathbf{c}_{[S ]} = \left\{ \mathbf{c}_{(k)}: k \in S \right\}$ to denote the set of transmitted codewords corresponding to users in the set $S\subset\mathcal{K}_a$,
  and use $\mathbf{c}^{'}_{[S_2]} \!=\! \left\{ \mathbf{c}'_{(k)} \in \mathcal{C}_k: k \in S_{2}, \mathbf{c}'_{(k)} \neq \mathbf{c}_{(k)} \right\}$ to denote the set of false alarm codewords corresponding to users in the set $S_2\subset\mathcal{K}$.
  Recall that for massive random access in MIMO fading channels, the ``good region'' $\mathcal{R}_{t,S_1}$ is given in~\eqref{eq_good region_selection} for any subset $S_1 \subset {\mathcal{K}}_{a} $ of size $t$.
  We define the event $\mathcal{G}_{\omega,\nu} = \bigcap_{S_1} \left\{ \mathbf{Y} \in \mathcal{R}_{t,S_1} \right\}$. Then, we obtain
    \begin{align}
      \mathbb{P} \left[ \mathcal{F}_t  \right]
      &\leq \mathbb{P} \left[
      \bigcup_{S_{1}} \bigcup_{S_{2}}
      \bigcup_{\mathbf{c}^{'}_{[S_2]}}
      \left\{ g \left(  \mathbf{Y} ,
      \mathbf{c}_{[ {\mathcal{K}}_a \backslash S_1]} \cup \mathbf{c}^{'}_{[S_2]}   \right)
      \leq  g \left(  \mathbf{Y} ,
      \mathbf{c}_{[ {\mathcal{K}}_a ]}   \right) \right\}
      \right] \\
      &\leq \!\min_{ 0 \leq \omega \leq 1 , \nu\geq0 }
      \! \left\{ \! \mathbb{P} \!\left[
      \bigcup_{S_{1}} \bigcup_{S_{2}}
      \bigcup_{\mathbf{c}^{'}_{[S_2]}}
      \left\{ g \!\left(  \mathbf{Y} ,
      \mathbf{c}_{[ {\mathcal{K}}_a \backslash S_1]} \cup \mathbf{c}^{'}_{[S_2]}   \right)
      \leq  g \!\left(  \mathbf{Y} ,
      \mathbf{c}_{[ {\mathcal{K}}_a ]}   \right) \right\} \bigcap \mathcal{G}_{\omega,\nu} \right]
      \!+ \mathbb{P}  \left[ \mathcal{G}_{\omega,\nu}^c \right] \right\} \label{eq_pft_goodregion_CSIR_noCSI},
    \end{align}
  where~\eqref{eq_pft_goodregion_CSIR_noCSI} follows from Fano's bounding technique given in~\eqref{eq_good region_1961}.

  The first probability on the RHS of~\eqref{eq_pft_goodregion_CSIR_noCSI} can be upper-bounded as
    \begin{align}
      & \mathbb{P} \!\left[
      \bigcup_{S_{1}} \bigcup_{S_{2}}
      \bigcup_{\mathbf{c}^{'}_{[S_2]}}
      \left\{ g \!\left(  \mathbf{Y} ,
      \mathbf{c}_{[ {\mathcal{K}}_a \backslash S_1]} \cup \mathbf{c}^{'}_{[S_2]}   \right)
      \leq  g \!\left(  \mathbf{Y} ,
      \mathbf{c}_{[ {\mathcal{K}}_a ]}   \right) \right\} \bigcap \mathcal{G}_{\omega,\nu} \right]\notag \\
      & \leq \! \sum_{S_{1}} \sum_{S_{2}} \sum_{\mathbf{c}^{'}_{[S_2]}}
      \mathbb{E}_{ \mathbf{c}_{[ {\mathcal{K}}_a ]}, \mathbf{c}_{[ {\mathcal{K}}_a \backslash S_1]},  \mathbf{c}^{'}_{[S_2]} }
      \!\!\left[
      \min_{u\geq 0,r\geq 0}  \mathbb{P} \left[ (u-r)g \left(  \mathbf{Y} ,
      \mathbf{c}_{[ {\mathcal{K}}_a ]}   \right)
      - u g \left(  \mathbf{Y} ,
      \mathbf{c}_{[ {\mathcal{K}}_a \backslash S_1]} \cup \mathbf{c}^{'}_{[S_2]}   \right) \right. \right.
      \notag \\
      & \;\;\;\;\;\;\;\;\;\;\;\;\;\;\;\;\;\;\;\;\;\;\;\;\;\;\;\;\;
      \;\;\;\;\;\;\;\;\;\;\;\;\;\;\;\;\;\;
      + r\omega g  \left(  \mathbf{Y} ,
      \mathbf{c}_{[ {\mathcal{K}}_a \backslash S_1]}   \right)
      + r\nu nL \geq 0
      \left| \mathbf{c}_{[ {\mathcal{K}}_a ]}, \mathbf{c}_{[ {\mathcal{K}}_a \backslash S_1]},  \mathbf{c}^{'}_{[S_2]} \right. \Big]\bigg] \label{eq_pft_union_CSIR_noCSI} \\
      & \leq \! \sum_{S_{1}} \sum_{S_{2}} \sum_{\mathbf{c}^{'}_{[S_2]}}
      \mathbb{E}_{ \mathbf{c}_{[ {\mathcal{K}}_a ]}, \mathbf{c}_{[ {\mathcal{K}}_a \backslash S_1]},  \mathbf{c}^{'}_{[S_2]} }
      \!\!\left[ \min_{u\geq 0,r\geq 0}
      \exp\left\{ r\nu nL \right\}
      \!\mathbb{E}_{\mathbf{H},\mathbf{Z}} \Big[
      \exp\!\Big\{ (u-r)g \!\left(  \mathbf{Y} ,
      \mathbf{c}_{[ {\mathcal{K}}_a ]}   \right)
      \right.\notag\\
      & \;\;\;\;\;\;\;\;\;\;\;\;\;\;\;\;\;\;\;\;\;\;\;\;\;\;
      \left.
      - u g \!\left(  \mathbf{Y} ,
      \mathbf{c}_{[ {\mathcal{K}}_a \backslash S_1]} \cup \mathbf{c}^{'}_{[S_2]}   \right)
      + r\omega g  \left(  \mathbf{Y} ,
      \mathbf{c}_{[ {\mathcal{K}}_a \backslash S_1]}   \right)
      \Big\} \right| \mathbf{c}_{[ {\mathcal{K}}_a ]}, \mathbf{c}_{[ {\mathcal{K}}_a \backslash S_1]},  \mathbf{c}^{'}_{[S_2]}
      \Big]\bigg], \label{eq_pft_Chernoff_CSIR_noCSI}
    \end{align}
  where \eqref{eq_pft_union_CSIR_noCSI} follows from the union bound and the fact that $ \mathbb{P}\!\left[ \{a\geq0\} \!\cap\! \{b\geq0\} \right] \leq \mathbb{P}\!\left[ a+b\geq0 \right]$;
  \eqref{eq_pft_Chernoff_CSIR_noCSI} follows by applying the Chernoff bound in Lemma~\ref{Chernoff_bound} shown below to the conditional probability in~\eqref{eq_pft_union_CSIR_noCSI}.

\begin{Lemma}[Section 3.2.4 in~\cite{goodregion}]\label{Chernoff_bound}
Let $Z$ and $W$ be any random variables. Then we have
\begin{equation}
  \mathbb{P}\left[Z \geq 0, W \leq 0\right] \leq \mathbb{E}\left[\exp\left\{s Z-r W\right\}\right], \quad \forall s \geq 0, \quad r \geq 0,
\end{equation}
and
\begin{equation}
  \mathbb{P}\left[W>0\right] \leq \mathbb{E}\left[ \exp\left\{sW\right\} \right], \quad \forall s \geq 0.
\end{equation}
\end{Lemma}

  We obtain an upper bound on $\mathbb{P} \left[ \mathcal{F}_t  \right]$ by substituting \eqref{eq_pft_Chernoff_CSIR_noCSI} into \eqref{eq_pft_goodregion_CSIR_noCSI}.
  Together with~\eqref{eq_PUPE_upper_CSIR_noCSI}, we obtain a general upper bound on the PUPE.
  Note that in both cases of CSIR and no-CSI, the expectations in~\eqref{eq_pft_Chernoff_CSIR_noCSI} and the probability $\mathbb{P}  \left[ \mathcal{G}_{\omega,\nu}^c \right]$ on the RHS of~\eqref{eq_pft_goodregion_CSIR_noCSI} can be further bounded.

\section{ Proof of Theorem \ref{Theorem_CSIR_achi}  } \label{Proof_achiCSIR}
  In this appendix, we prove Theorem~\ref{Theorem_CSIR_achi} to derive an upper bound on the PUPE with known~$K_a$ and CSIR.
  Based on the notation in Appendix~\ref{Proof_achi_CSIR_noCSI}, the ML decoding metric in this case is given~by
  \begin{equation}\label{eq_g_CSIR}
    g\left(  \mathbf{Y} ,   \hat{\mathbf{c}}_{[ {\hat{\mathcal{K}}}_a]}  \right)
    = \sum_{l=1}^{L}\left\|\mathbf{y}_l- \sum_{k\in  \hat{\mathcal{K}}_{a }}h_{k,l}   \hat{\mathbf{c}}_{(k)} \right\|_{2}^{2} .
  \end{equation}
  As introduced in Appendix~\ref{Proof_achi_CSIR_noCSI}, the PUPE can be upper-bounded by~\eqref{eq_PUPE_upper_CSIR_noCSI}. The probability $\mathbb{P}\left[ \mathcal{F}_t \right]$ therein, i.e. the probability of the event that there are exactly $t$ misdecoded users, is upper-bounded in~\eqref{eq_pft_goodregion_CSIR_noCSI} applying Fano's ``good region'' technique.
  In the following Appendix~\ref{Proof_achiCSIR_2}, we particularize the ``good region''-based bound on $\mathbb{P}\left[ \mathcal{F}_t \right]$ given in Appendix~\ref{Proof_achi_CSIR_noCSI} to the case of CSIR;
  then, in Appendix \ref{Proof_achiCSIR_1}, we derive another upper bound on $\mathbb{P} \left[ \mathcal{F}_t \right]$ applying Gallager's error exponent analysis~\cite{1965}.
  The two upper bounds are denoted as $p_{1,t}$ and $p_{2,t}$, respectively.

\subsection{Upper-bounding \texorpdfstring{$\mathbb{P} \left[ \mathcal{F}_t \right]$}{P[Ft]} based on Fano's bounding technique} \label{Proof_achiCSIR_2}
  In this subsection, we particularize the ``good region''-based bound on $\mathbb{P}\left[ \mathcal{F}_t \right]$ in~\eqref{eq_pft_goodregion_CSIR_noCSI} to the CSIR case, followed by further manipulations on the two probabilities on the RHS of~\eqref{eq_pft_goodregion_CSIR_noCSI}.

  Based on the notation in Appendix~\ref{Proof_achi_CSIR_noCSI}, we have $\left|S_1 \cap S_2\right| = t_0 \in [0,t]$.
  Let the binary selection matrix $\boldsymbol{\Phi}_{S_1} \in\{0,1\}^{MK\times K}$ indicate which codewords are transmitted by users in the set $S_1\subset\mathcal{K}_a$. Let the binary selection matrix $\boldsymbol{\Phi}^{'}_{S_2} \in\{0,1\}^{MK\times K}$ indicate which codewords are not transmitted but decoded for users in the set $S_2\subset\mathcal{K}\backslash \mathcal{K}_a \cup S_1$.
  It is satisfied that $\left[\boldsymbol{\Phi}_{S_1}\right]_{(k-1)M + W_k,k} = 1$ if user~$k\in S_1$ is active and the $W_{k}$-th codeword is transmitted by it, and $\left[\boldsymbol{\Phi}_{S_1}\right]_{(k-1)M + W_k,k} = 0$ otherwise; and similarly for $\boldsymbol{\Phi}^{'}_{S_2}$.
  Let $\tilde{\mathbf{A}}_{S_1} = {\mathbf{A}} \boldsymbol{\Phi}_{S_1} \in \mathbb{C}^{n\times K}$ and $\tilde{\mathbf{A}}^{'}_{S_2} = {\mathbf{A}} \boldsymbol{\Phi}^{'}_{S_2}  \in \mathbb{C}^{n\times K}$.
  The conditional expectation in~\eqref{eq_pft_Chernoff_CSIR_noCSI} can be written as
  \begin{align}
     & \mathbb{E}_{\mathbf{H},\mathbf{Z}} \!\! \left[ \left.
      \exp \!\left\{ (u-r)g \!\left(  \mathbf{Y} \!,
      \mathbf{c}_{[ {\mathcal{K}}_a ]}   \right)
      - u g \!\left(  \mathbf{Y} \!,
      \mathbf{c}_{[ {\mathcal{K}}_a \backslash S_1]} \!\cup\! \mathbf{c}^{'}_{[S_2]}   \right)
      + r\omega g \!\left(  \mathbf{Y} \!,
      \mathbf{c}_{[ {\mathcal{K}}_a \backslash S_1]}   \right)
      \right\} \right| \mathbf{c}_{[ {\mathcal{K}}_a ]}, \mathbf{c}_{[ {\mathcal{K}}_a \backslash S_1]},  \mathbf{c}^{'}_{[S_2]}
      \right] \notag \\
     & =  \mathbb{E}_{\mathbf{H},\mathbf{Z}} \!\!\left[ \left. \exp \!\left\{ \!\left( u-r \right) \left\| \mathbf{Z} \right\|_F^2  \!-\!  u \left\| \mathbf{Z} \!+\! \left( \tilde{\mathbf{A}}_{S_1} \!-\! \tilde{\mathbf{A}}^{'}_{S_2} \right) \mathbf{H} \right\|_F^2
     \!+ r\omega \left\| \mathbf{Z} \!+\! \tilde{\mathbf{A}}_{S_1} \mathbf{H} \right\|_F^2
      \right\} \right|  \tilde{\mathbf{A}}_{S_1}, \tilde{\mathbf{A}}^{'}_{S_2} \right]   \\
%     & =  \mathbb{E}_{\mathbf{H},\mathbf{Z}} \! \left[  \prod_{l = 1}^{L} \exp\!\left\{\!
%      - r \left( 1-\omega \right) \left\| \mathbf{z}_l +  \left( \frac{u-r\omega}{r(1-\omega)} \tilde{\mathbf{A}}_{S_1} - \frac{u}{r(1-\omega)} \tilde{\mathbf{A}}^{'}_{S_2} \right) \mathbf{h}_l \right\|_2^2
%      - u \left\| \left( \tilde{\mathbf{A}}_{S_1} - \tilde{\mathbf{A}}^{'}_{S_2} \right) \mathbf{h}_{l} \right\|_2^2
%      \right. \right. \notag\\
%     & \;\;\;\;\;\;\;\;\;\;\;  \left.\left.\left.
%      + r\omega \left\| \tilde{\mathbf{A}}_{S_1} \mathbf{h}_{l} \right\|_2^2
%      + r(1-\omega) \left\| \left( \frac{u-r\omega}{r(1-\omega)} \tilde{\mathbf{A}}_{S_1} - \frac{u}{r(1-\omega)} \tilde{\mathbf{A}}^{'}_{S_2} \right) \mathbf{h}_l \right\|_2^2
%      \right\} \right| \tilde{\mathbf{A}}_{S_1}, \tilde{\mathbf{A}}^{'}_{S_2} \right]   \\
     & =  \left( 1+r \left( 1-\omega \right) \right)^{-nL}
      \mathbb{E}_{ \mathbf{H}} \left[ \exp \left\{
      \frac{ 1 }{ 1+r\left( 1-\omega \right) }  \left\|  \left( (u-r\omega)  \tilde{\mathbf{A}}_{S_1} -  {u} \tilde{\mathbf{A}}^{'}_{S_2} \right) \mathbf{H}  \right\|_F^2 \right.  \right. \notag \\
     &  \left.  \left. \left. \;\;\;\;\;\;\;\;\;\;\;\;\;\;\;\;\;\;\;\;\;\;\;\;\;\;\;\;\;\;\;\;\;\;\;\;\;\;\;\;\;\;\;\;\;\;
     \;\;\;  - u \left\| \left( \tilde{\mathbf{A}}_{S_1} - \tilde{\mathbf{A}}^{'}_{S_2} \right) \mathbf{H} \right\|_F^2
      + r\omega \left\| \tilde{\mathbf{A}}_{S_1} \mathbf{H} \right\|_F^2
      \right\}  \right| \tilde{\mathbf{A}}_{S_1}, \tilde{\mathbf{A}}^{'}_{S_2} \right] \label{eq_q1t_exp_z} \\
     & =  \left( 1+r \left( 1-\omega \right) \right)^{-nL}
      \exp \left\{-L \ln \left| \mathbf{I}_{K} + \tilde{\mathbf{B}} \right| \right\}. \label{eq_q1t_exp_h}
  \end{align}
  Here, \eqref{eq_q1t_exp_z} follows from Lemma~\ref{expectation_bound} provided below by taking the expectation over $\mathbf{Z}$;
  \eqref{eq_q1t_exp_h} also follows from Lemma~\ref{expectation_bound} by taking the expectation over $\mathbf{H}$, under the condition that the minimum eigenvalue of $\tilde{\mathbf{B}}$ satisfies $\lambda_{\min}\left({\tilde{\mathbf{B}}}\right) > -1$, where $\tilde{\mathbf{B}}$ is given by
      \begin{equation}
        \tilde{\mathbf{B}} = \frac{ (1+r-u)(u-r\omega) }{ 1+r\left( 1-\omega \right) }  \left( \tilde{\mathbf{A}}_{S_1} -\! \frac{u}{u-r\omega} \tilde{\mathbf{A}}^{'}_{S_2} \!\right)^H   \!\left( \tilde{\mathbf{A}}_{S_1} \!- \frac{u}{u-r\omega} \tilde{\mathbf{A}}^{'}_{S_2} \right)
      - \frac{r\omega u}{u-r\omega}  \left(\tilde{\mathbf{A}}^{'}_{S_2}\right)^{\!H}  \!\tilde{\mathbf{A}}^{'}_{S_2}.
      \end{equation}

\begin{Lemma}[Corollary 3.2a.2 in \cite{quadratic_form1} and Result 4.4.1 in \cite{quadratic_form2}] \label{expectation_bound}
Let $\mathbf{x} \sim \mathcal{N}\left( \boldsymbol{\mu}, \boldsymbol{\Sigma} \right)$ with $\boldsymbol{\mu} \in \mathbb{R}^{p\times 1} $ and $\boldsymbol{\Sigma} \in \mathbb{R}^{p\times p}$.
Let $\mathbf{D}\in\mathbb{R}^{p\times p}$ be a symmetric matrix.
For any $\gamma$, if the eigenvalues of the matrix $\mathbf{I}_{p}-2 \gamma \boldsymbol{\Sigma}\mathbf{D}$ are positive, the expectation $\mathbb{E}\left[ \exp\left\{ \gamma \mathbf{x}^{T}\mathbf{D}\mathbf{x} \right\} \right]$ is given by
\begin{equation} \label{eq:Lemma_exp1}
\mathbb{E}\left[ \exp\left\{ \gamma \mathbf{x}^{T}\mathbf{D}\mathbf{x} \right\} \right] = \left|\mathbf{I}_{p}-2 \gamma \boldsymbol{\Sigma}\mathbf{D}\right|^{-\frac{1}{2}}
\exp \left\{ \gamma \boldsymbol{\mu}^{T} \mathbf{D} { \left( \mathbf{I}_{p}-2 \gamma \boldsymbol{\Sigma} \mathbf{D} \right)^{-1} } \boldsymbol{\mu}   \right\}.
\end{equation}
In particular, if $\mathbf{x} \sim \mathcal{N}\left( \mathbf{0}, \boldsymbol{\Sigma} \right)$, we have
\begin{equation}\label{eq:Lemma_exp2}
\mathbb{E}\left[ \exp\left\{ \gamma \mathbf{x}^{T}\mathbf{D}\mathbf{x} \right\} \right] = \left|\mathbf{I}_{p}-2 \gamma \boldsymbol{\Sigma}\mathbf{D}\right|^{-\frac{1}{2}} .
\end{equation}
Let $\bar{\mathbf{x}} \in \mathbb{C}^{p\times 1}$ be a complex random vector distributed as $\bar{\mathbf{x}} \sim \mathcal{CN}\left( \mathbf{0}, \bar{\boldsymbol{\Sigma}} \right)$.
Let $\mathbf{B}\in\mathbb{C}^{p\times p}$ be a Hermitian matrix.
For any $\gamma$, if the eigenvalues of the matrix $\mathbf{I}_{p} - \gamma \bar{\boldsymbol{\Sigma}}\mathbf{B}$ are positive, the expectation $\mathbb{E}\left[ \exp\left\{ \gamma \bar{\mathbf{x}}^{H}\mathbf{B}\bar{\mathbf{x}} \right\} \right]$ is given by
\begin{equation}\label{eq:Lemma_exp3}
\mathbb{E}\left[ \exp \left\{ \gamma \bar{\mathbf{x}}^{H}\mathbf{B}\bar{\mathbf{x}} \right\} \right] = \left|\mathbf{I}_{p} - \gamma \bar{\boldsymbol{\Sigma}} \mathbf{B}\right|^{-1} .
\end{equation}
If $\bar{\mathbf{x}} \sim \mathcal{CN}\left( \bar{\boldsymbol{\mu}}, \mathbf{I}_{p}\right)$ with $\bar{\boldsymbol{\mu}} \in \mathbb{C}^{p\times 1} $ and $\gamma<1$, the expectation $\mathbb{E}\left[ \exp\left\{ \gamma \bar{\mathbf{x}}^{H} \bar{\mathbf{x}} \right\} \right]$ is given by
\begin{equation}\label{eq:Lemma_exp4}
\mathbb{E}\left[ \exp \left\{ \gamma \bar{\mathbf{x}}^{H} \bar{\mathbf{x}} \right\} \right] = \left(1 - \gamma \right)^{-p}
\exp \left\{ \frac{\gamma}{1-\gamma} \bar{\boldsymbol{\mu}}^{H}  \bar{\boldsymbol{\mu}}   \right\}.
\end{equation}
\end{Lemma}

  Substituting \eqref{eq_q1t_exp_h} into \eqref{eq_pft_Chernoff_CSIR_noCSI}, we have
    \begin{align}
      & \mathbb{P} \! \left[
      \bigcup_{S_{1}} \bigcup_{S_{2}}
      \bigcup_{\mathbf{c}^{'}_{[S_2]}}
      \left\{ g \!\left(  \mathbf{Y} ,
      \mathbf{c}_{[ {\mathcal{K}}_a \backslash S_1]} \cup \mathbf{c}^{'}_{[S_2]}   \right)
      \leq  g \!\left(  \mathbf{Y} ,
      \mathbf{c}_{[ {\mathcal{K}}_a ]}   \right) \right\} \bigcap \mathcal{G}_{\omega,\nu} \right]\notag \\
      %& \leq  \sum_{S_{1}} \sum_{S_{2}} \sum_{\mathbf{c}^{'}_{[S_2]}}
%      \mathbb{E}_{ \tilde{\mathbf{A}}_{S_1}, \tilde{\mathbf{A}}^{'}_{S_2} }
%      \!\!\left[ \min_{{\substack {u\geq 0,r\geq 0,\\\lambda_{\min}\left({\tilde{\mathbf{B}}}\right) > -1}}}
%      \!\!\left( 1+r \left( 1-\omega \right) \right)^{-nL}
%      \!\exp \!\left\{ r\nu nL -L \ln \!\left| \mathbf{I}_{K} \!+ \tilde{\mathbf{B}} \right| \right\}
%      \right] \label{eq_pft_Chernoff_CSIR1} \\
      & \leq \sum_{t_0 = 0}^{t} C_{t_0,t}\;
      \mathbb{E}_{ \tilde{\mathbf{A}}_{S_1}, \tilde{\mathbf{A}}^{'}_{S_2} }
      \!\!\left[ \min_{{\substack {u\geq 0,r\geq 0,\\\lambda_{\min}\left({\tilde{\mathbf{B}}}\right) > -1}}}
      \!\!\left( 1+r \left( 1-\omega \right) \right)^{-nL}
      \exp \left\{ r\nu nL -L \ln \!\left| \mathbf{I}_{K} + \tilde{\mathbf{B}} \right| \right\}
      \right], \label{eq_pft_Chernoff_CSIR}
    \end{align}
  where $C_{t_0,t} = {\binom {K_a}{t}}  {\binom {t}{t_0}}  {\binom {K-K_a}{t-t_0}} (M-1)^{t_0}  M^{t-t_0}$.
  Here, \eqref{eq_pft_Chernoff_CSIR} follows because the expectation over $\tilde{\mathbf{A}}_{S_1}$ and $\tilde{\mathbf{A}}^{'}_{S_2}$ is unchanged for different $S_1$, $S_2$, and $\mathbf{c}^{'}_{[S_2]}$ once $t_0$ and $t$ are fixed, considering that the codebook matrix $\mathbf{A}$ has i.i.d. $\mathcal{CN}(0,P')$ entries.
  As a result, the first probability on the RHS of~\eqref{eq_pft_goodregion_CSIR_noCSI} is upper-bounded by~\eqref{eq_pft_Chernoff_CSIR}, which is denoted as  $q_{1,t} \left(\omega,\nu \right)$ as presented in~\eqref{eq_CSIR_q1t}.

  In the following, we derive an upper bound $q_{2,t}\left(\omega,\nu \right)$ on the second term $\mathbb{P} \left[ \mathcal{G}_{\omega,\nu}^c \right]$ on the RHS of~\eqref{eq_pft_goodregion_CSIR_noCSI}.
  To obtain $q_{2,t}(\omega,\nu)$, we upper-bound $\mathbb{P}  \left[ \mathcal{G}_{\omega,\nu}^c \right]$ for the case of $t < n$ and $\omega \in (0,1]$, the case of $t \geq n$ and $\omega \in (0,1]$, and the case of $\omega=0$, respectively.

{\textbf{Case 1:}} $t < n, \omega \in (0,1]$.

  Let $\tilde{\mathbf{y}}_{l} = \mathbf{z}_l + \tilde{\mathbf{A}}_{S_1} \mathbf{h}_{l} $.
  Define the event $\mathcal{G}_{\eta} = \bigcap_{S_{1}} \left\{ \sum_{l = 1}^{L} \left\| \mathcal{P}_{\mathbf{c}_{[S_1]}} \mathbf{z}_l \right\|_2^2 \leq tL(1+\eta) \right\}$ for $\eta\geq 0$.
  We can bound $\mathbb{P}  \left[ \mathcal{G}_{\omega,\nu}^c \right]$ as
    \begin{align}
      \mathbb{P}  \left[ \mathcal{G}_{\omega,\nu}^c \right]
      & = \mathbb{P} \left[ \bigcup_{S_{1}} \left\{ \sum_{l = 1}^{L} \left\| \mathbf{z}_l \right\|_2^2
      >  \omega \sum_{l = 1}^{L} \left\| \tilde{\mathbf{y}}_{l} \right\|_2^2
      + nL\nu  \right\} \right]  \\
      & = \mathbb{P} \left[  \bigcup_{S_{1}} \left\{ \sum_{l = 1}^{L} \left\| \mathcal{P}_{\mathbf{c}_{[S_1]}} \mathbf{z}_l \right\|_2^2 + \sum_{l = 1}^{L} \left\| \mathcal{P}_{\mathbf{c}_{[S_1]}}^{\bot} \tilde{\mathbf{y}}_{l} \right\|_2^2
      >  \omega \sum_{l = 1}^{L} \left\| \tilde{\mathbf{y}}_{l} \right\|_2^2
      + nL\nu  \right\} \right]  \label{eq_q2t_proj}\\
      & \leq \sum_{S_{1}} \mathbb{P} \left[  \sum_{l = 1}^{L} \tilde{\mathbf{y}}_{l}^H \left( \mathcal{P}_{\mathbf{c}_{[S_1]}}^{\bot}  - \omega \mathbf{I}_n \right) \tilde{\mathbf{y}}_{l}
      > nL\nu - tL(1+\eta) \right]
      +\mathbb{P} \left[ \mathcal{G}_{\eta}^c \right] \label{eq_q2t_cap},
    \end{align}
  where \eqref{eq_q2t_proj} follows because $\left\| \mathbf{z}_l \right\|_2^2 = \left\| \mathcal{P}_{\mathbf{c}_{[S_1]}} \mathbf{z}_l \right\|_2^2 + \left\| \mathcal{P}_{\mathbf{c}_{[S_1]}}^{\bot} \mathbf{z}_l \right\|_2^2 $ and $\mathcal{P}_{\mathbf{c}_{[S_1]}}^{\bot} \tilde{\mathbf{y}}_{l} = \mathcal{P}_{\mathbf{c}_{[S_1]}}^{\bot} \mathbf{z}_{l} $;
  \eqref{eq_q2t_cap} follows from the bounding technique in~\eqref{eq_good region_1961} and the union bound.

  Next, we focus on two terms on the RHS of \eqref{eq_q2t_cap}.
  We have $\mathcal{P}_{\mathbf{c}_{[S_1]}} \mathbf{z}_l \stackrel{i.i.d.}{\sim} \mathcal{CN}\left( \mathbf{0}, \mathcal{P}_{\mathbf{c}_{[S_1]}} \right)$ for $l\in[L]$ conditioned on $\mathbf{c}_{[S_1]}$.
  %The rank of $\mathcal{P}_{\mathbf{c}_{[S_1]}}$ is $t$ because the vectors in ${\mathbf{c}_{[S_1]}}$ are linearly independent almost surely.
  Let $\mathbf{U}$ be a unitary matrix satisfying $\mathbf{U} \mathcal{P}_{\mathbf{c}_{[S_1]}} \mathbf{U}^H = \mathbf{I}_{(t)}$.
  Conditioned on $\mathbf{c}_{[S_1]}$, we have $\mathbf{U} \mathcal{P}_{\mathbf{c}_{[S_1]}} \mathbf{z}_l \sim \mathcal{CN}\left( \mathbf{0}, \mathbf{I}_{(t)} \right)$, which implies that $\left\| \mathcal{P}_{\mathbf{c}_{[S_1]}} \mathbf{z}_l \right\|_2^2 = \left\| \mathbf{U} \mathcal{P}_{\mathbf{c}_{[S_1]}} \mathbf{z}_l \right\|_2^2 \sim \frac{1}{2} \chi^2\left(2t \right) $ and $\sum_{l=1}^{L}\left\| \mathcal{P}_{\mathbf{c}_{[S_1]}} \mathbf{z}_l \right\|_2^2 \sim \frac{1}{2} \chi^2\left(2tL \right)$.
  Hence, we can obtain
    \begin{equation} \label{eq_q2t_eta}
      \mathbb{P} \left[ \mathcal{G}_{\eta}^c \right]
      \leq \sum_{S_{1}} \mathbb{P}  \left[ \sum_{l = 1}^{L} \left\| \mathcal{P}_{\mathbf{c}_{[S_1]}} \mathbf{z}_l \right\|_2^2 > tL(1+\eta) \right]
      = {\binom {K_a} {t}} \left( 1 - \frac{\gamma\left( tL, tL \left( 1+\eta \right)\right)}{\Gamma\left( tL \right)} \right).
    \end{equation}

  We can bound the first term on the RHS of \eqref{eq_q2t_cap} as follows.
  Let $\mathbf{F}_{S_1} = \mathbf{I}_n + \tilde{\mathbf{A}}_{S_1} \tilde{\mathbf{A}}_{S_1}^H$ which can be decomposed as $\mathbf{F}_{S_1} = \mathbf{F}_{S_1}^{\frac{1}{2}} \mathbf{F}_{S_1}^{\frac{H}{2}}$.
  Conditioned on $\tilde{\mathbf{A}}_{S_1}$, we have $\tilde{\mathbf{y}}_{l} =  \mathbf{F}_{S_1}^{\frac{1}{2}} \tilde{\mathbf{w}}_{l} \sim \mathcal{CN}\left( \mathbf{0}, \mathbf{F}_{S_1} \right)$ where $\tilde{\mathbf{w}}_{l} \sim \mathcal{CN}\left( \mathbf{0}, \mathbf{I}_n \right)$.
  Define the event $\mathcal{G}_{\delta} = \left\{ \sum_{l = 1}^{L} \tilde{\mathbf{w}}_{l}^H   \tilde{\mathbf{w}}_{l} < (1+\delta)nL \right\}$ for $\delta\geq 0$.
  Applying the bounding technique in~\eqref{eq_good region_1961}, we can bound the first probability on the RHS of \eqref{eq_q2t_cap} as
    \begin{align}
      q_{3,t} \left( \omega ,\nu\right)
      & = \mathbb{P} \left[   \sum_{l = 1}^{L} \tilde{\mathbf{w}}_{l}^H \mathbf{F}_{S_1}^{\frac{H}{2}} \left( \mathcal{P}_{\mathbf{c}_{[S_1]}}^{\bot} - \omega \mathbf{I}_n \right) \mathbf{F}_{S_1}^{\frac{1}{2}} \tilde{\mathbf{w}}_{l}
      > nL\nu - tL(1+\eta) \right]  \\
      & \leq \mathbb{P} \left[ \left\{ \sum_{l = 1}^{L} \tilde{\mathbf{w}}_{l}^H \mathbf{F}_{S_1}^{\frac{H}{2}} \!\left( \mathcal{P}_{\mathbf{c}_{[S_1]}}^{\bot} \!- \omega \mathbf{I}_n \right) \mathbf{F}_{S_1}^{\frac{1}{2}} \tilde{\mathbf{w}}_{l}
      > nL\nu - tL(1+\eta) \right\}
      \!\cap  \mathcal{G}_{\delta}  \right]
      \!+ \mathbb{P} \left[ \mathcal{G}_{\delta}^{c} \right] \label{eq_q3t_goodregion} \\
      & = q_{4,t} \left( \omega ,\nu \right)  +  1 - \frac{\gamma\left( nL, nL \left( 1+\delta \right)\right)}{\Gamma\left( nL \right)} \label{eq_q3t_chi}.
    \end{align}

  The probability $q_{4,t} \left( \omega ,\nu \right)$ in \eqref{eq_q3t_chi} can be further upper-bounded as
    \begin{align}
      & q_{4,t} \left( \omega ,\nu \right) \notag \\
      & = \!\mathbb{E}_{ \tilde{\mathbf{A}}_{S_1}} \!\!\!\left[ \mathbb{P} \!\left[ \left. \! \left\{\!\sum_{l = 1}^{L} \!\tilde{\mathbf{w}}_{l}^H \!\left( \!(1-\omega)\mathbf{I}_n \!- \! \mathcal{P}_{\mathbf{c}_{[S_1]}}\! -\! \omega \tilde{\mathbf{A}}_{S_1}  \tilde{\mathbf{A}}_{S_1}^{H} \right) \!\tilde{\mathbf{w}}_{l}
      \!>\!  nL\nu - tL(1+\eta) \!\right\}
      \!\cap   \mathcal{G}_{\delta}  \right| \!  \tilde{\mathbf{A}}_{S_1} \!\right] \right] \label{eq_q4t_condition} \\
      & \leq \!\mathbb{E}_{ \tilde{\mathbf{A}}_{S_1}} \!\!\! \left[ \mathbb{P} \! \left[ \left. \sum_{l=1}^{L} \!\tilde{\mathbf{w}}_{l}^H \!\! \left( \! \mathcal{P}_{\mathbf{c}_{[S_1]}}  \!+\! \omega \tilde{\mathbf{A}}_{S_1} \tilde{\mathbf{A}}_{S_1}^{H} \!\right) \tilde{\mathbf{w}}_{l}
      < L( t(1+\eta) \!-\! n\nu + (1+\delta)n(1-\omega) ) \right| \! \tilde{\mathbf{A}}_{S_1} \!\right] \right] \label{eq_q4t_cap}\! ,\!
    \end{align}
  where \eqref{eq_q4t_condition} holds because the eigenvalues of $(1-\omega)\mathbf{I}_n - \mathcal{P}_{\mathbf{c}_{[S_1]}} - \omega \tilde{\mathbf{A}}_{S_1}  \tilde{\mathbf{A}}_{S_1}^{H}$ are the same as that of $\mathbf{F}_{S_1}^{\frac{H}{2}} \left( \mathcal{P}_{\mathbf{c}_{[S_1]}}^{\bot} - \omega \mathbf{I}_n \right) \mathbf{F}_{S_1}^{\frac{1}{2}}$.
  The conditional probability in \eqref{eq_q4t_cap} can be upper-bounded as
  \begin{align}
      & \mathbb{P} \! \left[ \left. \sum_{l=1}^{L} \tilde{\mathbf{w}}_{l}^H \left( \mathcal{P}_{\mathbf{c}_{[S_1]}} \!+ \omega \tilde{\mathbf{A}}_{S_1} \tilde{\mathbf{A}}_{S_1}^{H} \right) \tilde{\mathbf{w}}_{l}
      < L( t(1+\eta) - n\nu + (1+\delta)n(1-\omega) ) \right| \! \tilde{\mathbf{A}}_{S_1} \right]  \\
      & = \mathbb{P} \left[ \left. \sum_{i = 1}^{t} \frac{ \lambda_i \chi_{i}^{2}(2L) }{2L}
      < t(1+\eta) - n\nu + (1+\delta)n(1-\omega)\right| \tilde{\mathbf{A}}_{S_1} \right] \label{eq_q4t_lambda} \\
      & \leq \mathbb{P} \left[ \left. \frac{\chi^{2}(2tL)}{2tL}
      < \frac{ t(1+\eta) - n\nu + n(1+\delta)(1-\omega) } {t \prod_{i = 1}^{t} \lambda_i^{\frac{1}{t}}}\right| \tilde{\mathbf{A}}_{S_1} \right]  \label{eq_p4t_prod}  \\
      & = \frac{\gamma\left( tL, L \left( { t(1+\eta) - n\nu + n(1+\delta)(1-\omega) }  \right) \left| \mathbf{I}_n + \omega \tilde{\mathbf{A}}_{S_1} \tilde{\mathbf{A}}_{S_1}^{H} \right|^{-\frac{1}{t}} \right)}{\Gamma\left( tL \right)} \label{eq_q4t_chi},
    \end{align}
  where $\lambda_1, \lambda_2, \ldots, \lambda_t$ are non-zero eigenvalues of $\mathcal{P}_{\mathbf{c}_{[S_1]}} \!+ \omega \tilde{\mathbf{A}}_{S_1} \tilde{\mathbf{A}}_{S_1}^{H}$.
  Here, \eqref{eq_q4t_lambda} holds because conditioned on $\tilde{\mathbf{A}}_{S_1}$, the random vectors $\mathcal{U} \tilde{\mathbf{w}}_{l}$ and $\tilde{\mathbf{w}}_{l}$ have the same distribution as $\mathcal{CN} \left( \mathbf{0}, \mathbf{I}_{n} \right)$ for a unitary matrix $\mathcal{U}$ satisfying $\mathcal{U}^H \!\left( \mathcal{P}_{\mathbf{c}_{[S_1]}} \!+\! \omega \tilde{\mathbf{A}}_{S_1} \tilde{\mathbf{A}}_{S_1}^{H} \right) \mathcal{U} = \operatorname{diag}\!\left\{ \lambda_1, \ldots, \lambda_t, 0,\ldots,0 \right\} \in \mathbb{R}^{n\times n}$;
  \eqref{eq_p4t_prod} follows from Lemma~\ref{lemma_chisumprod} shown below;
  \eqref{eq_q4t_chi} holds because $\prod_{i = 1}^{t} \lambda_i = \left| \mathbf{I}_n + \omega \tilde{\mathbf{A}}_{S_1} \tilde{\mathbf{A}}_{S_1}^{H} \right|$.
  Together with~\eqref{eq_q4t_cap}, we can obtain
  \begin{equation}\label{eq_q4t_chi2}
    q_{4,t} \left( \omega ,\nu \right) \leq
    \mathbb{E}_{ \tilde{\mathbf{A}}_{S_1}} \!\! \left[ \frac{\gamma\!\left( tL, L \left( { t(1+\eta) - n\nu + n(1+\delta)(1-\omega) }  \right) \left| \mathbf{I}_n + \omega \tilde{\mathbf{A}}_{S_1} \tilde{\mathbf{A}}_{S_1}^{H} \right|^{-\frac{1}{t}} \right)}{\Gamma\left( tL \right)} \right].
  \end{equation}

  \begin{Lemma}[\cite{chisumprod}] \label{lemma_chisumprod}
    Assume $x_1,\ldots, x_s$ are independently distributed chi-square variables with $m$ degrees of freedom. Assume $\tilde{x} \sim \chi^{2}\left(sm\right)$.
    Let the constant $\gamma_j>0$. Then, for every constant~$c$,
    \begin{equation}
      \mathbb{P} \left({\sum}_{j=1}^{s} \gamma_{j} x_{j} <c\right) \leq \mathbb{P} \left( {\prod}_{j=1}^{s} \gamma_{j}^{\frac{1}{s}} \tilde{x} <c\right).
    \end{equation}
  \end{Lemma}

  Substituting \eqref{eq_q2t_eta}, \eqref{eq_q3t_chi}, and \eqref{eq_q4t_chi2} into \eqref{eq_q2t_cap}, we can obtain an upper bound on $\mathbb{P}  \left[ \mathcal{G}_{\omega,\nu}^c \right]$ in the case of $t < n $ and $\omega \in (0,1]$ as presented in~\eqref{eq_CSIR_q2t}.

{\textbf{Case 2:}} $t \geq n, \omega \in (0,1]$.

  Next, we upper-bound $\mathbb{P}  \left[ \mathcal{G}_{\omega,\nu}^c \right]$ when $t \geq n$ and $\omega\in (0,1]$.
  Recall that $\tilde{\mathbf{y}}_{l} = \mathbf{z}_l + \tilde{\mathbf{A}}_{S_1} \mathbf{h}_{l}$.
  % and $\tilde{\mathbf{A}}_1 = {\mathbf{A}} \boldsymbol{\Phi}_{S_1}$, where $\boldsymbol{\Phi}_{S_1} \in\{0,1\}^{MK\times K}$ denotes which codewords are transmitted by users belonging to the set $S_1\subset\mathcal{K}_a$.
  We define the event $\mathcal{G}_{\eta} = \left\{ \sum_{l = 1}^{L} \left\| \mathbf{z}_l \right\|_2^2 \leq nL(1+\eta) \right\}$ for $\eta\geq 0$.
  We can bound $\mathbb{P}  \left[ \mathcal{G}_{\omega,\nu}^c \right]$ as
    \begin{align}
      \mathbb{P}  \left[ \mathcal{G}_{\omega,\nu}^c \right]
      & \leq \mathbb{P} \left[ \bigcup_{S_{1}} \left\{ \sum_{l = 1}^{L} \left\| \mathbf{z}_l \right\|_2^2
      >  \omega \sum_{l = 1}^{L} \left\| \tilde{\mathbf{y}}_{l} \right\|_2^2
      + nL\nu \right\}
      \bigcap \mathcal{G}_{\eta}  \right]
      +\mathbb{P} \left[ \mathcal{G}_{\eta}^c \right] \label{eq_q2t_goodregion_case2} \\
      & \leq \mathbb{P} \left[ \bigcup_{S_{1}} \left\{\sum_{l = 1}^{L} \left\| \tilde{\mathbf{y}}_{l} \right\|_2^2
      < nL C_{\omega,\nu,\eta} \right\} \right]
      + 1 - \frac{\gamma\left( nL, nL \left( 1+\eta \right)\right)}{\Gamma\left( nL \right)} \\
      & \! \leq {\binom {K_a} {t}}
      \mathbb{E}_{\tilde{\mathbf{A}}_{S_1} } \!\!\left[ \frac{\gamma\!\left( nL, nLC_{\omega,\nu,\eta}  \left| \mathbf{I}_n \!+\! \tilde{\mathbf{A}}_{S_1} {\tilde{\mathbf{A}}}_{S_1}^{H} \right|^{-\frac{1}{n}} \right)}{\Gamma\left( nL \right)} \right]
      \!+ 1 - \frac{\gamma\left( nL, nL \left( 1+\eta \right)\right)}{\Gamma\left( nL \right)} \label{eq_q2t_cap_case2},
    \end{align}
  where $C_{\omega,\nu,\eta} = \frac{ 1+\eta-\nu }{\omega } $.
  Here, \eqref{eq_q2t_cap_case2} follows by applying Lemma \ref{lemma_chisumprod} and from the fact that $\tilde{\mathbf{y}}_{l} \sim \mathcal{CN}\left( \mathbf{0}, \mathbf{I}_n + \tilde{\mathbf{A}}_{S_1} \tilde{\mathbf{A}}_{S_1}^H \right)$ conditioned on $\tilde{\mathbf{A}}_{S_1}$.

{\textbf{Case 3:}} $\omega=0$.

  In the case of $\omega=0$, we have
    \begin{equation}
      \mathbb{P}  \left[ \mathcal{G}_{\omega,\nu}^c \right]
      = \mathbb{P} \left[ \sum_{l = 1}^{L} \left\| \mathbf{z}_l \right\|_2^2
      >  nL\nu \right]
      = 1 - \frac{\gamma\left( nL, nL \nu \right)}{\Gamma\left( nL \right)} .
    \end{equation}

  In conclusion, based on Fano's bounding technique, we have obtained $q_{1,t}\left( \omega,\nu \right)$ in~\eqref{eq_CSIR_q1t} (i.e. an upper bound on the first term in~\eqref{eq_pft_goodregion_CSIR_noCSI}) and $q_{2,t}\left(\omega,\nu \right)$ in~\eqref{eq_CSIR_q2t} (i.e. an upper bound on the second term in~\eqref{eq_pft_goodregion_CSIR_noCSI}),
  which contributes to an upper bound $p_{1,t}$  in~\eqref{eq_CSIR_p2t} on the probability $\mathbb{P} \left[ \mathcal{F}_t \right]$.

\subsection{ Upper-bounding \texorpdfstring{$\mathbb{P} \left[ \mathcal{F}_t \right]$}{P[Ft]} based on Gallager's bounding technique}\label{Proof_achiCSIR_1}

  Let $\mathbf{H}_1$ and $\mathbf{H}_2$ be $t\times L$ submatrices of $\mathbf{H}$ formed by rows corresponding to the support of $S_1$ and $S_2$, respectively.
  Let $\mathbf{A}_{S_1}$ and $\mathbf{A}'_{S_2}$ be $n\times t$ submatrices of $\mathbf{A}$ formed by columns corresponding to the codewords transmitted by users in the set $S_1$ and the codewords not transmitted but decoded for users in the set $S_2$, respectively.
  Then, we have
    \begin{align}
      & \mathbb{P} \left[ \left.  \mathcal{F}_t \right| \mathbf{Z}, \mathbf{H}_1, \mathbf{H}_2, \mathbf{A}_{S_1} \right] \notag \\
      &\leq \mathbb{P} \!\left[ \left.
      \bigcup_{S_{1}} \bigcup_{S_{2}}
      \bigcup_{\mathbf{c}^{'}_{[S_2]}}
      \!\left\{ \sum_{l \in [L]} \left\| \mathbf{z}_l + \!\sum_{k\in S_{1}}\!h_{k,l} \mathbf{c}_{(k)} - \!\sum_{k\in S_{2}}\!h_{k,l} \mathbf{c}'_{(k)} \right\|_2^2
      \!\leq\!  \sum_{l \in [L]} \left\| \mathbf{z}_l \right\|_2^2 \right\}
      \right| \mathbf{Z}, \mathbf{H}_1, \mathbf{H}_2, \mathbf{A}_{S_1} \right] \label{eq_Gallager_rho_a0} \\
      & \leq \sum_{S_{1}} \sum_{S_{2}}
      M^{\rho t} \left( \mathbb{P} \left[ \left.  \left\| \mathbf{Z} + \mathbf{A}_{S_1} \mathbf{H}_1 - \mathbf{A}'_{S_2} \mathbf{H}_2  \right\|_F^2
      \leq \left\| \mathbf{Z} \right\|_F^2 \right| \mathbf{Z}, \mathbf{H}_1, \mathbf{H}_2, \mathbf{A}_{S_1} \right] \right)^{\rho}  \label{eq_Gallager_rho_a},
    \end{align}
    where \eqref{eq_Gallager_rho_a} follows by applying Gallager's $\rho$-trick, i.e., $\mathbb{P}\left[\cup_{j} B_{j}\right] \leq\left(\sum_{j} \mathbb{P}\left[B_{j}\right]\right)^{\rho}$ for any $\rho\in[0,1]$~\cite{1965}, \cite[Section 5.6]{Gallager}.

    The probability on the RHS of \eqref{eq_Gallager_rho_a} can be upper-bounded as
    \begin{align}
      & \mathbb{P} \left[ \left.  \left\| \mathbf{Z} + \mathbf{A}_{S_1} \mathbf{H}_1 - \mathbf{A}'_{S_2} \mathbf{H}_2  \right\|_F^2
      \leq \left\| \mathbf{Z} \right\|_F^2 \right| \mathbf{Z}, \mathbf{H}_1, \mathbf{H}_2,  \mathbf{A}_{S_1} \right] \notag \\
      &  \leq \exp\left\{  \beta  \left\| \mathbf{Z} \right\|_F^2 \right\}
      \mathbb{E}_{ \mathbf{A}_{S_2}^{'}} \left[ \left. \exp\left\{ -\beta  \left\| \mathbf{Z} + \mathbf{A}_{S_1} \mathbf{H}_1 - \mathbf{A}'_{S_2} \mathbf{H}_2  \right\|_F^2 \right\} \right| \mathbf{Z}, \mathbf{H}_1, \mathbf{H}_2,  \mathbf{A}_{S_1} \right]  \label{eq_Chernoff_b} \\
      & =  \exp\left\{ \beta \left\| \mathbf{Z} \right\|_F^2 \right\}
      \left| \mathbf{I}_{2L} + \beta \tilde{\boldsymbol{\Sigma}}_2 \right|^{-\frac{ n}{2}}
      \prod_{i =1}^{n} \exp\left\{ - \beta \tilde{\boldsymbol{\mu}}_i^{T} \left( \mathbf{I}_{2L} + \beta \tilde{\boldsymbol{\Sigma}}_2 \right)^{-1}\tilde{\boldsymbol{\mu}}_i\right\} , \label{eq_expectation_c}
    \end{align}
    where \eqref{eq_Chernoff_b} follows from the Chernoff bound in Lemma~\ref{Chernoff_bound} with $\beta\geq0$,
    and~\eqref{eq_expectation_c} is obtained as follows.
    Let $\boldsymbol{\mu}_i = \left( [\mathbf{Z}]_{i,:} + \left[\mathbf{A}_{S_1}\right]_{i,:} \mathbf{H}_1 \right)^H$ and $\boldsymbol{\nu}_i = \boldsymbol{\mu}_i - \left(  \left[\mathbf{A}'_{S_2}\right]_{i,:} \mathbf{H}_2 \right)^H $.
    Conditioned on $\mathbf{Z}, \mathbf{H}_1, \mathbf{H}_2$, and $\mathbf{A}_{S_1}$, we have $\boldsymbol{\nu}_i \sim \mathcal{CN} \left(\boldsymbol{\mu}_i,  P'\mathbf{H}_2^H \mathbf{H}_2 \right)$ for $i\in[n]$.
    Let $\tilde{\boldsymbol{\nu}}_i = \begin{bmatrix} \Re\left( \boldsymbol{\nu}_i \right)\\ \Im\left(\boldsymbol{\nu}_i\right) \end {bmatrix} $,
    $\tilde{\boldsymbol{\mu}}_i = \begin{bmatrix} \Re\left(\boldsymbol{\mu}_i  \right)\\ \Im\left(\boldsymbol{\mu}_i \right) \end {bmatrix}$,
    and $ \tilde{\boldsymbol{\Sigma}}_2 =  \begin{bmatrix} \Re(P'\mathbf{H}_2^H \mathbf{H}_2) & -\Im(P'\mathbf{H}_2^H \mathbf{H}_2) \\ \Im(P'\mathbf{H}_2^H \mathbf{H}_2) & \Re(P'\mathbf{H}_2^H \mathbf{H}_2)\end {bmatrix}  $.
    We have $\tilde{\boldsymbol{\nu}}_i \sim \mathcal{N}\left(\tilde{\boldsymbol{\mu}}_i, \frac{1}{2}\tilde{\boldsymbol{\Sigma}}_2  \right)$ conditioned on $\mathbf{Z}, \mathbf{H}_1, \mathbf{H}_2$, and $\mathbf{A}_{S_1}$.
    Then, applying Lemma~\ref{expectation_bound}, we can obtain
    \begin{align}
      \mathbb{E}_{ \mathbf{A}_{S_2}^{'}}\!\! \left[ \left. \exp\!\left\{ - \beta \boldsymbol{\nu}_i^{H}\boldsymbol{\nu}_i \right\} \right| \mathbf{Z}, \mathbf{H}_1, \mathbf{H}_2,  \mathbf{A}_{S_1} \right]
      & = \mathbb{E}_{ \mathbf{A}_{S_2}^{'}} \left[ \left. \exp\left\{  -\beta \tilde{\boldsymbol{\nu}}_i^{T} \tilde{\boldsymbol{\nu}}_i \right\}
      \right| \mathbf{Z}, \mathbf{H}_1, \mathbf{H}_2,  \mathbf{A}_{S_1}  \right]  \label{eq_expectation_review} \\
      & = \left| \mathbf{I}_{2L} \!+\! \beta \tilde{\boldsymbol{\Sigma}}_2 \right|^{-\frac{1}{2}}     \!\exp\!\left\{ \!-\beta \tilde{\boldsymbol{\mu}}_i^{T} \!\left( \mathbf{I}_{2L} \!+\! \beta \tilde{\boldsymbol{\Sigma}}_2 \right)^{-1}\!\tilde{\boldsymbol{\mu}}_i\right\},
    \end{align}
    which yields \eqref{eq_expectation_c}.

    Substituting \eqref{eq_expectation_c} into \eqref{eq_Gallager_rho_a} and taking the expectation over $\mathbf{A}_{S_1}$ and $\mathbf{Z}$, we can obtain
    \begin{align}
      & \mathbb{P} \left[ \left. \mathcal{F}_t \right| \mathbf{H}_1, \mathbf{H}_2 \right]
      \notag \\
      &  \leq \sum_{S_{1}} \sum_{S_{2}}
      M^{\rho t}
      \underbrace{ \left| \mathbf{I}_{2L} \!+\! \beta \tilde{\boldsymbol{\Sigma}}_2 \right|^{-\frac{\rho n}{2}}
      \left|
      \begin{bmatrix} \mathbf{I}_{2L} \!+\! \rho \beta  \left( \mathbf{I}_{2L} \!+\! \beta \tilde{\boldsymbol{\Sigma}}_2\right)^{-1} \left(\mathbf{I}_{2L} \!+\!  \tilde{\boldsymbol{\Sigma}}_1 \right) & \rho\beta\!\left( \mathbf{I}_{2L} \!+\! \beta \tilde{\boldsymbol{\Sigma}}_2\right)^{-1}
      \\ -\rho \beta\mathbf{I}_{2L} & \left( 1 - \rho \beta \right)\mathbf{I}_{2L} \end{bmatrix}  \right|^{-\frac{n}{2}} }_{C_{S_{1},S_{2}}}
      \label{eq_pft_H1},
    \end{align}
    where $\tilde{\boldsymbol{\Sigma}}_1 =  \begin{bmatrix} \Re(P'\mathbf{H}_1^H \mathbf{H}_1)  & -\Im(P'\mathbf{H}_1^H \mathbf{H}_1) \\ \Im(P'\mathbf{H}_1^H \mathbf{H}_1)  &  \Re(P'\mathbf{H}_1^H \mathbf{H}_1)\end {bmatrix}$.
    Here, \eqref{eq_pft_H1} follows by applying Lemma~\ref{expectation_bound} and Sylvester's determinant theorem under the condition that $0\leq\beta<1/{\rho}$.
    Then, we have
    \begin{align}
      C_{S_{1},S_{2}}  &
      = \left| \left( 1 - \rho \beta \right)\mathbf{I}_{2L} \right|^{-\frac{n}{2}} \notag \\
      & \;\;\;\; \cdot \left| \left( \mathbf{I}_{2L} \!+\! \beta \tilde{\boldsymbol{\Sigma}}_2 \right)^{ \!\rho} + \rho \beta  \left( \mathbf{I}_{2L} \!+\! \beta \tilde{\boldsymbol{\Sigma}}_2\right)^{\!\rho-1} \!\left(\mathbf{I}_{2L} \!+\!  \tilde{\boldsymbol{\Sigma}}_1 \right)
      + \frac{\rho^2\beta^2}{1-\rho \beta} \left( \mathbf{I}_{2L} \!+\! \beta \tilde{\boldsymbol{\Sigma}}_2\right)^{\!\rho-1} \right|^{-\frac{n}{2}}  \label{eq_proof_prod1} \\
      & =
      \left| \mathbf{I}_{2L} + \beta \tilde{\boldsymbol{\Sigma}}_2 \right|^{-\frac{(-1+\rho)n}{2}}
      \left| \mathbf{I}_{2L}  + \rho \beta \left(1-\rho\beta \right)  \tilde{\boldsymbol{\Sigma}}_1  + \beta  \left(1-\rho\beta \right) \tilde{\boldsymbol{\Sigma}}_2 \right|^{-\frac{n}{2}}   \\
      & =
      \left| \mathbf{I}_{ L} \!+\! \beta P' \mathbf{H}_2^H \mathbf{H}_2 \right|^{ (1-\rho) n }
      \left| \mathbf{I}_{ L}  + \rho \beta \left(1\!-\!\rho\beta \right) P' \mathbf{H}_1^H \mathbf{H}_1  + \beta  \left(1\!-\!\rho\beta \right) P' \mathbf{H}_2^H \mathbf{H}_2 \right|^{- {n} } \label{eq_proof_prod2}\!,
    \end{align}
    where \eqref{eq_proof_prod1} holds because $\left| \begin{bmatrix} \mathbf{A} & \mathbf{B} \\ \mathbf{C} & \mathbf{D} \end{bmatrix}  \right|=\left|\mathbf{D}\right| \left|\mathbf{A}-\mathbf{B} \mathbf{D}^{-1} \mathbf{C}\right|$ when $\mathbf{D}$ is nonsingular, and \eqref{eq_proof_prod2} holds because $\left|\mathbf{D}\right|^2 = \left| \left[\begin{array}{cc} \Re{\left(\mathbf{D}\right)} & -\Im{\left(\mathbf{D}\right)} \\ \Im{\left(\mathbf{D}\right)} & \Re{\left(\mathbf{D}\right)} \end{array} \right] \right| $.

    Substituting \eqref{eq_proof_prod2} into \eqref{eq_pft_H1} and taking the expectation over $\mathbf{H}_1$ and $\mathbf{H}_2$, we have
    \begin{align}
      \mathbb{P} \left[ \mathcal{F}_t \right]
%      & \leq \sum_{S_{1}} \sum_{S_{2}}  M^{\rho t}
%      \mathbb{E}_{\mathbf{H}_1,\mathbf{H}_2}  \left[ \exp \left\{ (1-\rho) n \ln \left| \mathbf{I}_{ L} + \beta P' \mathbf{H}_2^H \mathbf{H}_2 \right| \right.\right.  \notag\\
%      & \;\;\;\;\;\;\;\;\;\;\;\;\;\;\;\;\;\;\;\;\;\;\;\;\;\;\;\;\;\; \left.\left. - n \ln\left| \mathbf{I}_{ L}  +  \beta \left(1-\rho\beta \right) P' \left(\rho \mathbf{H}_1^H \mathbf{H}_1  + \mathbf{H}_2^H \mathbf{H}_2 \right) \right| \right\} \right]  \\
      & \leq \sum_{t_0=0}^{t}  {\binom {K_a} {t}}   {\binom {t} {t_0}}  {\binom {K-K_a} {t-t_0}}  M^{\rho t}\;
      \mathbb{E}_{\mathbf{H}_1,\mathbf{H}_2}  \left[  \exp \left\{ (1-\rho) n \ln \left| \mathbf{I}_{ L} + \beta P' \mathbf{H}_2^H \mathbf{H}_2 \right| \right. \right. \notag\\
      & \;\;\;\;\;\;\;\;\;\;\;\;\;\;\;  \;\;\;\;\;\;\;\;\;\;\;\;\;\;\;\;\;\;\;\;\;\;\;\;\;\;\;\;\;\; \left.\left. - n \ln\left| \mathbf{I}_{ L}  +  \beta \left(1-\rho\beta \right) P' \left(\rho \mathbf{H}_1^H \mathbf{H}_1  + \mathbf{H}_2^H \mathbf{H}_2 \right) \right| \right\} \right], \label{eq_pft_H}
    \end{align}
    where \eqref{eq_pft_H} holds because the expectation is unchanged for different $S_1$ and $S_2$ once $t_0$ and $t$ are fixed, considering that channel coefficients are i.i.d. for different users.
    Taking the minimum over $\rho$ and $\beta$ on the RHS of~\eqref{eq_pft_H}, we obtain an upper bound on $\mathbb{P}\left[ \mathcal{F}_t \right]$ based on Gallager's $\rho$-trick, which is denoted as $p_{2,t}$ in \eqref{eq_CSIR_p1t}.
    This completes the proof of Theorem~\ref{Theorem_CSIR_achi}.

\section{Proof of Corollary \ref{Theorem_CSIR_achi_knownUE}} \label{proof_eq_CSIR_p1t_knownUE}
  In a special case where all users are active (i.e. $K_a=K$), the set $ {S}_1$ of misdecoded users is the same as the set $ {S}_2$ including detected users with false alarm codewords.
  Thus, $\tilde{p}_{1,t}$ can be easily obtained from Theorem~\ref{Theorem_CSIR_achi} and its proof is omitted here for the sake of brevity.
  Moreover, the proof of $\tilde{p}_{2,t}$ is provided in Appendix \ref{proof_eq_CSIR_p1t_knownUE1}; the upper bound $\tilde{p}_{2,t}^{\text{u}}$ on $\tilde{p}_{2,t}$ is derived in Appendix~\ref{proof_eq_CSIR_p1t_knownUE2}.

\subsection{Proof of \texorpdfstring{\eqref{eq_CSIR_p1t_knownUE}}{(23)} }  \label{proof_eq_CSIR_p1t_knownUE1}
  In a special case where all users are active, i.e., $K_a=K$, $\mathbf{H}_1 = \mathbf{H}_2$, and $S_1 = S_2$, it is easy to see that the optimum value of $\beta$ minimizing \eqref{eq_pft_H} is given by $\beta^{*} = 1/(1+\rho)$. Then, we have
    \begin{equation}
      \mathbb{P} \left[ \left. \mathcal{F}_t \right| \mathbf{H}_1 \right]
      \leq \sum_{S_{1}}  M^{\rho t}
      \exp \left\{  -\rho n \ln \left| \mathbf{I}_{ L} + \frac{P'}{1+\rho} \mathbf{H}_1^H \mathbf{H}_1 \right| \right\}.
    \end{equation}
    Taking the expectation over $\mathbf{H}_1$, we have
    \begin{align}
      \mathbb{P} \left[  \mathcal{F}_t  \right]
      & \leq \min_{0\leq\rho\leq 1} \sum_{S_{1}}  M^{\rho t} \; \mathbb{E}_{ \mathbf{H}_1 } \! \left[
      \exp \left\{ -\rho n \ln \left| \mathbf{I}_{ L} + \frac{P'}{1+\rho} \mathbf{H}_1^H \mathbf{H}_1 \right| \right\} \right] \\
      & = \min_{0\leq\rho\leq 1}  {\binom {K } {t}} M^{\rho t} \; \mathbb{E}_{ \mathbf{G} } \! \left[
      \exp \left\{ - L \ln \left| \mathbf{I}_{t} + \frac{P'}{1+\rho} {\mathbf{G}} {\mathbf{G}}^H \right| \right\} \right] \label{eq_HG} ,
    \end{align}
    where \eqref{eq_HG} holds when $\rho n$ is an integer and each element of $\mathbf{H}_1 \in \mathbb{C}^{t\times L}$ and $\mathbf{G} \in \mathbb{C}^{t\times \rho n}$ is i.i.d. $\mathcal{CN}\left( 0,1 \right)$ distributed. This is because
    \begin{align}
      \mathbb{E}_{ \mathbf{H}_1, \mathbf{G} } \!\! \left[
      \exp \!\left\{ -\frac{P'}{1+\rho} \left\| \mathbf{G}^H \mathbf{H}_1 \right\|_F^2 \right\} \right]\!
      & = \mathbb{E}_{ \mathbf{H}_1 } \!\! \left[ \prod_{i=1}^{\rho n}
      \mathbb{E} \!\left[ \left. \exp \!\left\{ -\frac{P'}{1+\rho}
      \left\| \left( [\mathbf{G}]_{:,i}\right)^H \!\mathbf{H}_1 \right\|_2^2 \right\}
      \right| \mathbf{H}_1 \right] \right] \\
      & = \mathbb{E}_{ \mathbf{H}_1 } \! \left[
      \exp \left\{ - \rho n \ln \left| \mathbf{I}_{L} + \frac{P'}{1+\rho} {\mathbf{H}}_1^H {\mathbf{H}}_1 \right| \right\} \right] \label{eq_HG1}\\
      & = \mathbb{E}_{ \mathbf{G} } \!  \left[ \prod_{l=1}^{L}
      \mathbb{E}  \left[ \left. \exp \left\{ -\frac{P'}{1+\rho}
      \left\| \mathbf{G}^H \left[\mathbf{H}_1\right]_{:,l} \right\|_2^2 \right\}
      \right| \mathbf{G} \right] \right] \\
      & = \mathbb{E}_{ \mathbf{G} } \!  \left[
      \exp \left\{ - L \ln \left| \mathbf{I}_{t} + \frac{P'}{1+\rho} {\mathbf{G}} {\mathbf{G}}^H \right| \right\} \right] \label{eq_HG2},
    \end{align}
  where \eqref{eq_HG1} and \eqref{eq_HG2} follows from Lemma \ref{expectation_bound}.
  Denote the RHS of~\eqref{eq_HG} as $\tilde{p}_{2,t}$. This completes the proof of \eqref{eq_CSIR_p1t_knownUE}.

\subsection{Proof of the upper bound \texorpdfstring{$\tilde{p}_{2,t}^{\text{u}}$}{p2tu} on \texorpdfstring{$\tilde{p}_{2,t}$}{p2t}} \label{proof_eq_CSIR_p1t_knownUE2}
  Recall that each element of $\mathbf{G} \in \mathbb{C}^{t\times \rho n}$ is i.i.d. $\mathcal{CN}\left( 0,1 \right)$ distributed.
  In the case of $\rho n \geq t + L$, the expectation in \eqref{eq_CSIR_p1t_knownUE} can be upper-bounded as
    \begin{align}
      \mathbb{E}_{ \mathbf{G} } \! \left[ \left| \mathbf{I}_{t} + \frac{P'}{1+\rho} {\mathbf{G}} {\mathbf{G}}^H \right|^{-L} \right]
      & \leq \left( \frac{P'}{ 1+\rho } \right)^{-Lt} \mathbb{E}_{ \mathbf{G} } \! \left[ \left|   {\mathbf{G}} {\mathbf{G}}^H \right|^{-L} \right] \label{eq_CSIR_p1t_knownUE_bound_proof1_1}\\
      & = \left( \frac{P'}{ 1+\rho } \right)^{-Lt}
      \mathbb{E} \left[ \prod_{i=\rho n - t + 1}^{\rho n} \left( \frac{\chi^2(2i)}{2} \right)^{-L} \right]  \label{eq_CSIR_p1t_knownUE_bound_proof1_2} \\
      & = \left( \frac{P'}{ 1+\rho } \right)^{-Lt}
      \prod_{i=\rho n - t + 1}^{\rho n}  \frac{\Gamma(i-L)}{\Gamma(i)} \label{eq_CSIR_p1t_knownUE_bound_proof1_3},
    \end{align}
    where \eqref{eq_CSIR_p1t_knownUE_bound_proof1_1} follows because $\left| \mathbf{I}+\mathbf{A}\right| \geq \left| \mathbf{A} \right| $ when $\mathbf{A}$ is a positive semidefinite matrix;
    \eqref{eq_CSIR_p1t_knownUE_bound_proof1_2} follows because the determinant of the Wishart matrix $\left|   {\mathbf{G}} {\mathbf{G}}^H \right|$ has the same distribution as the product of independent random variables with chi-square distributions, i.e., $\prod_{i=\rho n - t + 1}^{\rho n} \frac{\chi^2(2i)}{2} $~\cite[Section 3.5]{wishart_deter};
    \eqref{eq_CSIR_p1t_knownUE_bound_proof1_3} follows from the moments of chi-square random variables.

    Applying similar ideas, we can upper-bound this term in the case of $\rho n \leq t - L$ as follows:
    \begin{equation}
      \mathbb{E}_{ \mathbf{G} } \! \left[ \left| \mathbf{I}_{t} + \frac{P'}{1+\rho} {\mathbf{G}} {\mathbf{G}}^H \right|^{-L} \right]
      \leq \left( \frac{P'}{1+\rho} \right)^{-L\rho n }
      \prod_{i=t - \rho n + 1}^{t}  \frac{\Gamma(i-L)}{\Gamma(i)} . \label{eq_CSIR_p1t_knownUE_bound_proof2_3}
    \end{equation}

    Substituting \eqref{eq_CSIR_p1t_knownUE_bound_proof1_3} and \eqref{eq_CSIR_p1t_knownUE_bound_proof2_3} into \eqref{eq_CSIR_p1t_knownUE} and considering $\tilde{p}_{2,t} \leq 1$, we can obtain \eqref{eq_CSIR_p1t_knownUE_bound1}.

\section{Proof of Theorem \ref{prop_converse_CSIR}}\label{proof_converse_CSIR}
  In this appendix, we prove Theorem \ref{prop_converse_CSIR} to establish a converse bound on the minimum required energy-per-bit for the CSIR case.
    %Let $\mathcal{W}_{\mathcal{K}_a} = \left\{W_k: k\in {\mathcal{K}_a} \right\}$ denote the set of transmitted messages of active users,
%    $\mathcal{X}_{\mathcal{K}_a} = \left\{\mathbf{x}_{(k)}: k \in {\mathcal{K}_a} \right\}$ denote corresponding codewords,
%    $\mathbf{Y} \in \mathbb{C}^{n\times L}$ be the received signal, and $\hat{\mathcal{W}}_{\mathcal{K}_a} = \left\{\hat{W}_k: k\in{\hat{\mathcal{K}}_a} \right\}$ be the set of decoded messages.
%    We have the Markov chain: $\mathcal{W}_{\mathcal{K}_a} \to \mathcal{X}_{\mathcal{K}_a} \to \mathbf{Y} \to \hat{\mathcal{W}}_{\mathcal{K}_a} $.
    We assume a genie $G$ reveals the set $\mathcal{K}_a$ of active users and a subset $S_1 \subset \mathcal{K}_a$ for messages $\mathcal{W}_{S_1} = \left\{W_k: k\in{S_1} \right\}$ and corresponding fading coefficients to the decoder.
    It is evident that a converse bound in the genie case is a converse bound for the problem without genie.
    Let $S_2= \mathcal{K}_{a}\backslash S_1$ of size $t$.
    Let $\boldsymbol{\Phi}_{S_2} \in\{0,1\}^{MK\times K}$ denote which codewords are transmitted by users in the set $S_2\subset\mathcal{K}_a$, where $\left[\boldsymbol{\Phi}_{S_2}\right]_{(k-1)M + W_k,k} = 1$ if the $k$-th user belonging to the set $S_2$ is active and the $W_{k}$-th codeword is transmitted, and $\left[\boldsymbol{\Phi}_{S_2}\right]_{(k-1)M + W_k,k} = 0$ otherwise.
    The equivalent received signal of the $l$-th antenna at the BS is given by
    \begin{equation}\label{receive_y_G}
      \mathbf{y}_l^G = \sum_{k\in{S_2}}{h}_{k,l}\mathbf{x}_{(k)}+\mathbf{z}_l
      = \mathbf{X}{\boldsymbol{\Phi}}_{S_2}\mathbf{h}_{l} +\mathbf{z}_l  \in \mathbb{C}^{n}.
    \end{equation}
    The equivalent received message over all antennas is given by
    \begin{equation}
      \mathbf{Y}^G = \mathbf{X}{\boldsymbol{\Phi}}_{S_2}\mathbf{H} +\mathbf{Z},
    \end{equation}
    where $\mathbf{Y}^G = [\mathbf{y}_1^G,\mathbf{y}_2^G,\ldots, \mathbf{y}_L^G ] \in \mathbb{C}^{n\times L}$, and $\mathbf{H}$ and $\mathbf{Z}$ are defined in Section~\ref{section2}.
    Denote the decoded signal for the $k$-th user with genie as $\hat{W}_{k}^{G}$.
    Let $\mathcal{M}_k =1\left[ {W}_{k}\neq\hat{W}_{k}^{G}\right]$ and $P_{e,k}^G = \mathbb{E}\left[ \mathcal{M}_k \right]$. We have $P_{e,k}^G = 0$ for $k\in S_1$.
    The averaged PUPE is $P_{e}^G=\frac{1}{K_a}\sum_{k\in {S_2}} P_{e,k}^G \leq \epsilon$.

    Based on the Fano inequality, we have
    \begin{equation}\label{eq_fano}
      \frac{t}{K_a}J - P_{e}^G \log_2\left(2^{J}-1\right) - \frac{1}{K_a}\sum_{k\in S_2}h_2\left( P_{e,k}^G \right) \leq \frac{1}{K_a}\sum_{k\in S_2}I_2\left( W_k;\hat{W}_k^{G} \right).
    \end{equation}
    Considering the concavity of $h_2(\cdot)$ and the inequality that $P_{e}^G \leq \epsilon \leq 1-\frac{1}{2^J}$, we have
    \begin{equation} \label{eq_hinf}
      P_{e}^G \log_2\left(2^J-1\right) + \frac{1}{K_a}\sum_{k\in S_2}h_2\left( P_{e,k}^G \right)  \leq \epsilon J + h_2\left( \epsilon \right).
    \end{equation}
    Denote $\mathcal{W}_{S_2} = \left\{W_k: k\in{S_2} \right\}$,
    $\mathcal{X}_{S_2} = \left\{\mathbf{x}_{(k)}: k\in{S_2} \right\}$, and $\hat{\mathcal{W}}_{S_2}^G = \left\{\hat{W}_k^G: k\in{S_2} \right\}$.
    The matrix $ \mathbf{H}_t $ is a $t\times L$ submatrix of $\mathbf{H}$ corresponding to fading coefficients of users in the set $S_2$.
    We can upper-bound $\sum_{k\in S_2} I_2\left(W_k; \hat{W}_k^{G}  \right)$ as
    \begin{align}
      \sum_{k\in S_2} I_2\left(W_k; \hat{W}_k^{G}  \right)
      &= H_2\left( \mathcal{W}_{S_2} \right) - \sum_{k\in S_2} H_2\left(W_k \left| \hat{W}_k^{G} \right. \right)  \label{eq_conv_IH1} \\
      & \leq H_2 \left( \mathcal{W}_{S_2} \right) - H_2\left(\mathcal{W}_{S_2} \left| \hat{\mathcal{W}}_{S_2}^{G} \right. \right)  \label{eq_conv_H2}  \\
      &= I_2\left(\mathcal{W}_{S_2} ; \hat{\mathcal{W}}_{S_2}^{G}  \right) \label{eq_conv_IH2}  \\
      &\leq I_2\left(\mathcal{X}_{S_2} ; \mathbf{Y}^{G}  \right)  \label{eq_mutualinf1} \\
      &\leq n \; \mathbb{E}_{\mathbf{H}_t} \! \left[\log_2  \left| \mathbf{I}_L + P \mathbf{H}_t^H \mathbf{H}_t \right|  \right],\label{eq_mutualinf}
    \end{align}
    where $H_2(x)$ denotes the entropy of a random variable $x$. Here, \eqref{eq_conv_H2} follows because
    \begin{equation}\label{eq_conv_HH}
      H_2\left(\mathcal{W}_{S_2} \left| \hat{\mathcal{W}}_{S_2}^{G} \right. \right)
      = \sum_{k\in S_2} H_2\left(W_k \left| \hat{\mathcal{W}}_{S_2}^{G}, {W}_{1}, \ldots, {W}_{k-1} \right. \right)
      \leq \sum_{k\in S_2} H_2\left(W_k \left| \hat{W}_{k}^{G} \right. \right),
    \end{equation}
    \eqref{eq_mutualinf1} follows due to the data processing inequality and the Markov chain: $\mathcal{W}_{S_2} \to \mathcal{X}_{S_2} \to \mathbf{Y}^{G} \to \hat{\mathcal{W}}_{S_2}^{G} $,
    and~\eqref{eq_mutualinf} holds because both $\mathbf{X}$ and $\mathbf{Z}$ are independent for $n$ channel uses and
    the normal distribution of codewords maximizes the entropy for a given variance~\cite[Theorem 8.6.5]{elements_IT}.

    Substituting \eqref{eq_hinf} and \eqref{eq_mutualinf} into \eqref{eq_fano}, we can obtain \eqref{P_tot_conv_CSIR} in Theorem \ref{prop_converse_CSIR}.
    Then, applying the concavity of $\log_2\left| \cdot \right|$ function, we obtain \eqref{P_tot_conv_CSIR_bound}, which completes the proof of Theorem \ref{prop_converse_CSIR}.

\section{ Proof of Theorem \ref{Theorem_scalinglaw_CSIR}  } \label{Proof_scalinglaw_CSIR}
  To prove Theorem \ref{Theorem_scalinglaw_CSIR}, we first establish an achievability result in Appendix \ref{Proof_scalinglaw_CSIR_achi} and then prove a converse result in Appendix \ref{Proof_scalinglaw_CSIR_conv} for the CSIR case assuming all users are active, i.e. $K=K_a$.
\subsection{ Achievability  } \label{Proof_scalinglaw_CSIR_achi}
  In a special case where all users are active, the PUPE can be upper-bounded as
  \begin{align}
    P_{e}& \leq \mathbb{E}\left[\frac{1}{K } \sum_{k\in {\mathcal{K} }} 1 \left[ W_{k} \neq \hat{W}_{k} \right] \right]_{\text{no power constraint}} + \tilde{p}_0 \\
    & \leq \epsilon_1 + \mathbb{P}\left[\frac{1}{K}  \sum_{k\in {\mathcal{K} }} 1 \left[ W_{k} \neq \hat{W}_{k} \right] \geq \epsilon_1 \right]_{\text{no power constraint}}
    + \tilde{p}_0 \\
    & = \epsilon_1 + \sum_{t = \lceil \epsilon_1 K \rceil}^{K } \mathbb{P} \left[ \mathcal{F}_t \right]_{\text{no power constraint}}
    + \tilde{p}_0 \label{Proof_eq_PUPE_upper_CSIR},
  \end{align}
  where $\epsilon_1$ is a positive constant less than $\epsilon$; $\mathcal{F}_t = \left\{ \sum_{k\in {\mathcal{K} }} 1 \left\{ W_{k} \neq \hat{W}_{k} \right\} = t \right\}$ denotes the event that there are exactly $t$ misdecoded users;
  $\tilde{p}_0$ upper-bounds the power constraint violation probability given by
  \begin{equation}
    \tilde{p}_0 = K  \; \mathbb{P} \left[\frac{x}{2 n}>\frac{P}{{P}^{\prime}} \right]
    \leq K \exp\left\{ -n \left( \frac{P}{{P}^{\prime}} - \sqrt{2\frac{P}{{P}^{\prime}}-1} \right) \right\} ,\;\; x\sim\chi^2(2n) \label{Proof_eq_CSIR_p0},
  \end{equation}
  which follows from Lemma~\ref{chi_square_bound1} presented below.
  It is easy to see that $c_P = \frac{P}{{P}^{\prime}} - \sqrt{2\frac{P}{{P}^{\prime}}-1} $ is a positive finite constant, provided that $\frac{P}{P'}-1$ is a positive finite constant.
  In the case of $\ln K = o(n)$, we have $\tilde{p}_0 \leq \exp \left\{ o(n) - c_P n \right\} \to 0$ as $n \to \infty$.

  \begin{Lemma}[\cite{chi_square_upperbound}]\label{chi_square_bound1}
    Let $x\sim \chi^{2}\left(  m \right) $ be a central chi-square distributed variable with $m$ degrees of freedom. For $\forall a>0$,
    \begin{equation}
      \mathbb{P}\left[ x - m \geq a \right]
      \leq \exp\left\{-\frac{1}{2}\left(a+m -\sqrt{m } \sqrt{2 a+m }\right)\right\}.
    \end{equation}
  \end{Lemma}

  An upper bound $\tilde{p}_{1,t}$ on $\mathbb{P} \left[ \mathcal{F}_t \right]_{\text{no power constraint}}$ is given in Corollary~\ref{Theorem_CSIR_achi_knownUE}.
  Next, we pay attention to upper-bounding $\tilde{p}_{1,t}$, thereby finding the condition under which $\sum_{t = \lceil \epsilon_1 K \rceil}^{K } \tilde{p}_{1,t} \to 0$ and thus $P_e\leq \epsilon$.
  In the case of $\epsilon_1 K \geq n +L$, for $t=\lceil \epsilon_1 K \rceil, \lceil \epsilon_1 K \rceil + 1, \ldots, K $, we have
    \begin{align}
      \tilde{p}_{1,t}
      & \leq {\binom {K } {t}}
      M^{t}  \left( \frac{P'}{2} \right)^{-Ln }
      \prod_{i=t - n + 1}^{t}  \frac{\Gamma(i-L)}{\Gamma(i)} \label{Proof_eq_CSIR_p1t_knownUE_bound1} \\
      & \leq {\binom {K } {t}}
      M^{t}  \left( \frac{P'\left( t - n + 1 - L \right)}{2} \right)^{-Ln }
      \label{Proof_eq_CSIR_p1t_knownUE_bound2},
    \end{align}
  where \eqref{Proof_eq_CSIR_p1t_knownUE_bound1} follows from \eqref{eq_CSIR_p1t_knownUE_bound1} and \eqref{eq_CSIR_p1t_knownUE_bound} by allowing $\rho=1$, and \eqref{Proof_eq_CSIR_p1t_knownUE_bound2} follows from the equality that $\Gamma(x)=(x-1)!$ for any positive integer $x$.

  Let $t = \theta K $ with $\theta \in S_{\theta} = \left\{ \frac{1}{K }, \frac{2}{K }, \ldots, 1 \right\} \cap \left[ \epsilon_1,1\right]$. We have
  \begin{align}
      \sum_{t = \lceil \epsilon_1 K \rceil}^{K } \tilde{p}_{1,t}
      & \leq \sum_{t = \lceil \epsilon_1 K \rceil}^{K } {\binom {K } {t}}
      M^{t}  \left( \frac{P'\left( t - n + 1 - L \right)}{2} \right)^{-Ln }
      \label{Proof_eq_CSIR_sump1t_knownUE_bound1}\\
      & \leq \exp \! \left\{ o\!\left(K \right)  + \! K  \max_{\theta \in S_{\theta}  }
      \left\{  h(\theta) + \theta \ln M - \!\frac{Ln}{K }\ln \left( \frac{P' \!\left( \theta K  -  n + 1 - L \right) }{2} \right)  \right\} \right\} \label{Proof_eq_CSIR_sump1t_knownUE_bound2}\\
      & \leq  \exp \! \left\{ o\!\left(K \right)  + \! K
      \!\left( h \left(\frac{1}{2}\right) + \ln M - \frac{Ln}{K }\ln  \left(  \frac{P' \left( \epsilon_1 K  -  n + 1 - L \right) }{2} \right)  \right) \right\}, \label{Proof_eq_CSIR_sump1t_knownUE_bound3}
    \end{align}
  where \eqref{Proof_eq_CSIR_sump1t_knownUE_bound2} follows from the inequality that~\cite[Example 11.1.3]{elements_IT}
  \begin{equation} \label{Proof_eq_CSIR_Kt}
    \binom {K } {t} \leq \exp\left\{ K h(\theta)\right\}.
  \end{equation}
  Therefore, in the case of $K\to \infty$, we have $\sum_{t = \lceil \epsilon_1 K \rceil}^{K } \tilde{p}_{1,t} \to 0$ provided that the finite constant
  \begin{equation} \label{Proof_eq_CSIR_scalinglaw_cond1}
    c_1 = \frac{Ln}{K }\ln \left( \frac{P' \left( \epsilon_1 K -  n + 1 - L \right) }{2} \right) - h \left(\frac{1}{2}\right) - \ln M > 0.
  \end{equation}

  Assume $M = \Theta(1)$, $K$ and $n\to \infty$, and $K \gg \frac{n+L-1}{\epsilon_1}$. In the case of $KP'=\Omega(1)$, we can obtain that~\eqref{Proof_eq_CSIR_scalinglaw_cond1} is satisfied if and only if $\frac{nL\ln KP'}{K}=\Omega\left(1\right)$, which can be divided into the following two relations:
  \begin{enumerate}
    \item We assume $P'K$ is a finite positive constant satisfying $ P'K > \frac{2}{\epsilon_1}$. In this case, we have
  \begin{equation}
      c_1 = c_3 \frac{nL}{K} - c_2 , \label{Proof_eq_CSIR_cond_knownUE_bound1_regime1}
  \end{equation}
  where $c_2 = h \left(\frac{1}{2}\right) + \ln M $ and $c_3$ is a finite positive constant.
  In order to satisfy the condition in~\eqref{Proof_eq_CSIR_scalinglaw_cond1}, it is possible to choose $\frac{nL}{K}=\Omega\left(1\right)$ and $P'K =\Theta\left(1\right)$.
  An example for this case is that the number of BS antennas satisfies $L=\Theta\left( n \right)$, the power satisfies $P'=\Theta\left(\frac{1}{n^2}\right)$ and the number of users satisfies $K = \Theta (n^2)$.
    \item In the case of $P'K\to\infty$, we have
  \begin{equation}
      c_1 =  {Ln} \frac{\ln KP' }{K }
      - {Ln} \frac{\mathcal{O}(1)}{K } - c_2, \label{Proof_eq_CSIR_cond_knownUE_bound1}
  \end{equation}
  where $c_2 = h \left(\frac{1}{2}\right) + \ln M $. Applying~\eqref{Proof_eq_CSIR_cond_knownUE_bound1}, in order to satisfy the condition in~\eqref{Proof_eq_CSIR_scalinglaw_cond1}, it is possible to choose $\frac{nL\ln KP'}{K}=\Omega\left(1\right)$ with $KP' \to \infty$.
  An example for this case satisfies $L=\Theta\left( \frac{n}{\ln n} \right)$, $P'=\Theta\left(\frac{1}{n}\right)$, and $K = \Theta (n^2)$.
  \end{enumerate}

  Combining \eqref{Proof_eq_CSIR_p0} and \eqref{Proof_eq_CSIR_scalinglaw_cond1}, we conclude that assuming $K, n \to \infty$, $\ln K = o(n)$, $KP=\Omega(1)$, $M=\Theta(1)$, and $K \geq \frac{n+L-1}{\epsilon_1}$, the PUPE requirement $P_e \leq \epsilon$ is satisfied provided that $\frac{nL\ln KP}{K}=\Omega\left(1\right)$.
  In particular, the PUPE requirement is satisfied for $K = \frac{n+L-1}{\epsilon_1}$ users when $KP=\Omega(1)$ (the condition $\frac{nL\ln KP}{K}=\Omega\left(1\right)$ is satisfied directly in this case).
  It was proved in~\cite[Appendix~A-C]{finite_payloads_fading} that, if one can achieve a certain PUPE for $K$ users, it will also be possible to achieve the same PUPE for less than $K$ users.
  Thus, one can reliably serve $K \leq \frac{n+L-1}{\epsilon_1}$ users provided that $\frac{n+L-1}{\epsilon_1}P=\Omega(1)$, or under a stricter condition that $KP=\Omega(1)$.
  As a result, assuming $K, n \to \infty$, $\ln K = o(n)$, $KP=\Omega(1)$, and $M=\Theta(1)$, the PUPE requirement $P_e \leq \epsilon$ is satisfied if $\frac{nL\ln KP}{K}=\Omega\left(1\right)$.
  That is, it is possible to choose the following two regimes: $\frac{nL}{K}=\Omega\left(1\right)$ and $KP =\Theta\left(1\right)$;
  $\frac{nL\ln KP}{K}=\Omega\left(1\right)$ and $KP \to \infty$.
  In particular, when the number of BS antennas is $L=\Theta\left( n \right)$ (resp. $L=\Theta\left( \frac{n}{\ln n} \right)$) and the power satisfies $P =\Theta\left(\frac{1}{n^2}\right)$ (resp. $P =\Theta\left(\frac{1}{n}\right)$), we can reliably serve $K = \mathcal{O}(n^2)$ users.

\subsection{ Converse  } \label{Proof_scalinglaw_CSIR_conv}
  Assume that $t = \theta K$ with $\theta \in S'_{\theta} = \left\{ \frac{1}{K }, \frac{2}{K}, \ldots, 1 \right\}$.
  We consider the case where $\epsilon$ and $J$ are finite positive constants.
  Following from Theorem~\ref{prop_converse_CSIR}, the minimum required energy-per-bit is larger than $\inf \frac{nP}{J}$, where the infimum is taken over all $P>0$ satisfying that
  \begin{equation}\label{Proof_P_tot_conv_CSIR_bound}
    \left( \theta - \epsilon \right) J - h_2 \left( \epsilon \right) \leq
    \frac{nL}{K }  \log_2\left( 1 + \theta PK  \right),   \forall \theta \in S'_{\theta}.
  \end{equation}
  When $\theta \leq \theta' = \frac{h_2 \left( \epsilon \right)}{J} + \epsilon$, \eqref{Proof_P_tot_conv_CSIR_bound} is satisfied for any positive $P,n,L,$ and $K$.

  Next, assuming $n,K\to\infty$, $KP=\Omega\left(1\right)$, and $J=\Theta(1)$, the inequality in~\eqref{Proof_P_tot_conv_CSIR_bound} holds for any $\theta \in S'_{\theta} \cap \left( \theta', 1\right]$ if and only if $\frac{nL\ln PK  }{K }=\Omega(1)$, which can be divided into the following two relations:
  1)~$KP=\Theta\left(1\right)$ and $\frac{nL}{K} = \Omega(1)$; 2)~$KP\to\infty$ and $\frac{nL\ln KP}{K} = \Omega(1)$.
  Moreover, since the RHS of~\eqref{Proof_P_tot_conv_CSIR_bound} is a monotonically decreasing function of $K$, when the number of BS antennas is $L=\Theta\left( n \right)$ (resp. $L=\Theta\left( \frac{n}{\ln n} \right)$) and the power satisfies $P =\Theta\left(\frac{1}{n^2}\right)$ (resp. $P =\Theta\left(\frac{1}{n}\right)$), $K$ users can be reliably served only if $K=\mathcal{O}(n^2)$.
  %the number of users that can be reliably served should be no more than $K=\Theta(n^2)$.

  Together with the case of $\theta \leq \theta'$, assuming $n,K\to\infty$, $KP=\Omega\left(1\right)$, and $J=\Theta(1)$, the inequality in~\eqref{Proof_P_tot_conv_CSIR_bound} holds for any $\theta \in S'_{\theta}$ if and only if $\frac{nL\ln PK }{K }=\Omega(1)$, i.e., if and only if one of the two relations mentioned above is satisfied.

  Together with the achievability result in Appendix~\ref{Proof_scalinglaw_CSIR_achi}, we conclude that assuming $n,K\to\infty$, $\ln K = o(n)$, $KP=\Omega\left(1\right)$, and $J=\Theta(1)$, the PUPE requirement $P_e \leq \epsilon$ is satisfied if and only if $\frac{nL\ln PK }{K }=\Omega(1)$, i.e., if and only if one of the following two relations is satisfied: 1)~$\frac{nL}{K}=\Omega\left(1\right)$ and $KP =\Theta\left(1\right)$; 2)~$\frac{nL\ln KP}{K}=\Omega\left(1\right)$ and $KP \to \infty$.
  In particular, when the number of BS antennas is $L=\Theta\left( n \right)$ (resp. $L=\Theta\left( \frac{n}{\ln n} \right)$) and the power satisfies $P =\Theta\left(\frac{1}{n^2}\right)$ (resp. $P =\Theta\left(\frac{1}{n}\right)$), the number of users that can be reliably served is in the order of $K = \mathcal{O}(n^2)$.

\section{ Proof of Theorem \ref{Theorem_noCSI_achi}  } \label{section5}
  In this appendix, we prove Theorem~\ref{Theorem_noCSI_achi} to establish an achievability bound on the PUPE in the case of no-CSI with known $K_a$.
  As introduced in Appendix~\ref{Proof_achi_CSIR_noCSI}, the PUPE can be upper-bounded by~\eqref{eq_PUPE_upper_CSIR_noCSI}. The probability $\mathbb{P}\left[ \mathcal{F}_t \right]$ therein, i.e. the probability of the event that there are exactly $t$ misdecoded users, is upper-bounded  in~\eqref{eq_pft_goodregion_CSIR_noCSI} applying Fano's ``good region'' technique~\cite{1961}.
  In the following, we particularize the ``good region''-based bound on $\mathbb{P}\left[ \mathcal{F}_t \right]$ given in Appendix~\ref{Proof_achi_CSIR_noCSI} to the no-CSI case,
  followed by further manipulations on the two probabilities on the RHS of~\eqref{eq_pft_goodregion_CSIR_noCSI}.

  Based on the notation introduced in Appendix~\ref{Proof_achi_CSIR_noCSI}, the ML decoding metric $g\left(  \mathbf{Y} , \hat{\mathbf{c}}_{[ {\hat{\mathcal{K}}}_a]}  \right)$ in the case of no-CSI is given by~\cite{Caire1}
  \begin{align}
    g\left(  \mathbf{Y} ,   \hat{\mathbf{c}}_{[ {\hat{\mathcal{K}}}_a]}  \right)
    & =  L \ln \left| \mathbf{I}_n + {\sum}_{k\in  \hat{\mathcal{K}}_{a }} \hat{\mathbf{c}}_{(k)} \hat{\mathbf{c}}_{(k)}^{H}  \right|
    + \operatorname{tr}\left( \left( \mathbf{I}_n+ {\sum}_{k\in  \hat{\mathcal{K}}_{a }} \hat{\mathbf{c}}_{(k)} \hat{\mathbf{c}}_{(k)}^{H} \right)^{-1} \mathbf{Y} \mathbf{Y}^H \right) \\
    & = L \ln \left| \mathbf{I}_n+ \mathbf{A} {\boldsymbol{\Gamma}}^{'}_{\hat{\mathcal{K}}_a}\mathbf{A}^H \right|
    + \operatorname{tr}\left( \left( \mathbf{I}_n+ \mathbf{A} {\boldsymbol{\Gamma}}^{'}_{\hat{\mathcal{K}}_a} \mathbf{A}^H \right)^{-1} \mathbf{Y} \mathbf{Y}^H \right). \label{eq_g_noCSI}
  \end{align}
  Here, the matrix $\mathbf{A} \in \mathbb{C}^{n\times MK}$ denotes the concatenation of codebooks of the $K$ users, which has i.i.d. $\mathcal{CN}\left(0,P'\right)$ entries;
  the matrix ${\boldsymbol{\Gamma}}^{'}_{S } = \operatorname{diag} \left\{ {\boldsymbol{\gamma}}^{'}_{S } \right\} \in \left\{ 0,1 \right\}^{KM\times KM} $, where $\left[ {\boldsymbol{\gamma}}^{'}_{S } \right]_{(k-1)M + W_k} = 1$ if $k\in S $ and the $W_{k}$-th codeword is decoded for this user, and $\left[ {\boldsymbol{\gamma}}^{'}_{S} \right]_{(k-1)M + W_k} = 0$ otherwise.
  Similarly, let ${\boldsymbol{\Gamma}}_{S} = \operatorname{diag} \left\{ {\boldsymbol{\gamma}}_{S} \right\}\in \left\{ 0,1 \right\}^{KM\times KM}$ be a diagonal matrix, where $\left[ {\boldsymbol{\gamma}}_{S} \right]_{(k-1)M + W_k} = 1$ if $k\in S$ and the $W_{k}$-th codeword is transmitted by this user, and $\left[ {\boldsymbol{\gamma}}_{S} \right]_{(k-1)M + W_k} = 0$ otherwise.
  In the following, we denote $g\left( \mathbf{Y} , \hat{\mathbf{c}}_{[ {\hat{\mathcal{K}}}_a]}  \right)$ as $g\left( {\boldsymbol{\Gamma}}^{'}_{\hat{\mathcal{K}}_a} \right)$ for simplicity.

  Let $\mathbf{A}_{S} \in \mathbb{C}^{n \times |S|}$ denote the concatenation of transmitted codewords of active users in the set $S\subset\mathcal{K}_a$ and
  let $\mathbf{A}^{'}_{S_2} \in \mathbb{C}^{n \times |S_2|}$ denote the concatenation of false-alarm codewords for users in the set $S_2\subset\mathcal{K}\backslash\mathcal{K}_a\cup S_1$.
  Denote ${\mathbf{A}}_{all} = \left\{ {\mathbf{A}}_{ \mathcal{K}_a },  {\mathbf{A}}_{ \mathcal{K}_a \backslash S_1 },   {\mathbf{A}}^{'}_{  S_2} \right\}$.
  Define $\mathbf{F}$, $\mathbf{F}^{'}$, and $\mathbf{F}_1$ as in~\eqref{eq_noCSI_F}, \eqref{eq_noCSI_Fprime}, and \eqref{eq_noCSI_F1}, respectively.
  The conditional expectation in~\eqref{eq_pft_Chernoff_CSIR_noCSI} can be written as
  \begin{align}
      & \mathbb{E}_{\mathbf{H},\mathbf{Z}} \left[ \left.
      \exp \left\{ (u-r) g\left(\boldsymbol{\Gamma}_{\mathcal{K}_a} \right)
      - u g\left( {\boldsymbol{\Gamma}}^{'}_{ \mathcal{K}_a \backslash S_1 \cup S_2}   \right)
      + r\omega g\left(  {\boldsymbol{\Gamma}}_{ \mathcal{K}_a \backslash S_1 } \right)
      \right\} \right| \mathbf{c}_{[ {\mathcal{K}}_a ]}, \mathbf{c}_{[ {\mathcal{K}}_a \backslash S_1]},  \mathbf{c}^{'}_{[S_2]}
      \right] \notag \\
      & =  \exp \left\{
      (u-r) L \ln\left| \mathbf{F} \right|  - u L \ln\left| \mathbf{F}^{'} \right|  + r\omega L \ln\left| \mathbf{F}_1 \right|   \right\}    \notag\\
      & \;\;\;\;  \cdot \mathbb{E}_{ \mathbf{H},\mathbf{Z} } \left[ \left. \exp \left\{
      \operatorname{tr}\left( \! \mathbf{Y}^H \! \left( (u-r) \mathbf{F}^{-1}
      - u \left( \mathbf{F}^{'} \right)^{-1}
      + r\omega \mathbf{F}_{1}^{-1} \right)\mathbf{Y} \right)
      \right\} \right| \mathbf{A}_{all}  \right]  \\
      & = \exp \left\{ L \left(  (u-r) \ln \left|\mathbf{F}\right|
      - u \ln \left| {\mathbf{F}^{'}} \right|
      + r\omega \ln \left| \mathbf{F}_{1} \right|
      - \ln \left| \mathbf{B} \right| \right)
      \right\}   \label{eq_noCSI_q1t_exp_y}.
  \end{align}
  Here, \eqref{eq_noCSI_q1t_exp_y} follows from Lemma \ref{expectation_bound} by taking the expectation over $\mathbf{H}$ and $\mathbf{Z}$ provided that the minimum eigenvalue of $\mathbf{B}$ satisfies  $\lambda_{\min}\left(\mathbf{B}\right) > 0$, where the matrix $\mathbf{B}$ is given by
      \begin{equation}
        \mathbf{B} = (1-u+r) \mathbf{I}_n
      + u \left( \mathbf{F}' \right)^{-1} \mathbf{F}
      - r\omega \mathbf{F}_{ 1}^{-1} \mathbf{F}.
      \end{equation}

  Then, we have
    \begin{align}
      & \mathbb{P} \! \left[
      \bigcup_{S_{1}} \bigcup_{S_{2}}
      \bigcup_{\mathbf{c}^{'}_{[S_2]}}
      \left\{ g \left( {\boldsymbol{\Gamma}}^{'}_{ \mathcal{K}_a \backslash S_1 \cup S_2}   \right)
      \leq  g \left(\boldsymbol{\Gamma}_{\mathcal{K}_a} \right) \right\} \bigcap \mathcal{G}_{\omega,\nu} \right]\notag \\
%      & \leq \! \sum_{S_{1}} \sum_{S_{2}} \sum_{\mathbf{c}^{'}_{[S_2]}}
%      \mathbb{E}_{\mathbf{A}_{all}} \!\! \left[ \min_{{\substack{u\geq 0,r\geq 0, \\ \lambda_{\min}\left({ {\mathbf{B}}}\right) > 0}}}
%      \!\exp \!\left\{ L \!\left( rn\nu + (u-r) \ln \left|\mathbf{F}\right|
%      - u \ln \left| {\mathbf{F}^{'}} \right|
%      + r\omega \ln \left| \mathbf{F}_{1} \right|
%      - \ln \left| \mathbf{B} \right| \right)
%      \right\}  \right]  \label{eq_pft_Chernoff_noCSI1} \\
      & \leq  C_{t}\;
      \mathbb{E}_{ \mathbf{A}_{all} } \!\! \left[ \min_{{\substack{u\geq 0,r\geq 0, \\ \lambda_{\min}\left({ {\mathbf{B}}}\right) > 0}}}
      \!\exp \!\left\{ L \!\left( rn\nu + (u-r) \ln \left|\mathbf{F}\right|
      - u \ln \left| {\mathbf{F}^{'}} \right|
      + r\omega \ln \left| \mathbf{F}_{1} \right|
      - \ln \left| \mathbf{B} \right| \right)
      \right\}  \right]\!, \label{eq_pft_Chernoff_noCSI}
    \end{align}
  where $C_{t} = {\binom {K_a}{t}} {\binom {K-K_a+t}{t}} M^{t}$.
  Here, \eqref{eq_pft_Chernoff_noCSI} follows by substituting \eqref{eq_noCSI_q1t_exp_y} into \eqref{eq_pft_Chernoff_CSIR_noCSI} and follows from the fact that the expectation in~\eqref{eq_pft_Chernoff_noCSI} is unchanged for different $S_1$, $S_2$, and $\mathbf{c}^{'}_{[S_2]}$ once $t$ is fixed, considering that the codebook matrix $\mathbf{A}$ has i.i.d. $\mathcal{CN}(0,P')$ entries.
  As a result, the first probability on the RHS of~\eqref{eq_pft_goodregion_CSIR_noCSI} is upper-bounded by~\eqref{eq_pft_Chernoff_noCSI}, denoted as  $q_{1,t} \left(\omega,\nu \right)$ in~\eqref{eq_noCSI_q1t}.

  Next, we proceed to upper-bound the second term $\mathbb{P} \left[ \mathcal{G}_{\omega,\nu}^c \right]$ on the RHS of~\eqref{eq_pft_goodregion_CSIR_noCSI}.
  Denote $\mathbf{A}_{all} = \left\{ \mathbf{A}_{ \mathcal{K}_a }, \mathbf{A}_{ \mathcal{K}_a \backslash S_1 } \right\}$ and define the event $\mathcal{G}_\delta \!=\! \left\{ \sum_{i=1}^{n} \!\! \frac{\chi_i^2(2L)}{2} \!\leq\! nL(1+\delta) \right\}$ for $\delta\geq0$.
  We have
    \begin{align}
      \mathbb{P} \! \left[ \mathcal{G}_{\omega,\nu}^c \right]
      & = \mathbb{P} \left[ \bigcup_{S_1} \left\{ g\left(\boldsymbol{\Gamma}_{\mathcal{K}_a} \right) > \omega g\left(  {\boldsymbol{\Gamma}}_{ \mathcal{K}_a \backslash S_1 } \right) + nL\nu \right\} \right] \\
      & \leq \sum_{S_1}  \mathbb{E}_{ \mathbf{A}_{all} } \! \left[ \mathbb{P} \left[   \left. \sum_{l=1}^{L} \left( \tilde{\mathbf{y}}_l^H \left(\mathbf{I}_n  - \omega \mathbf{F}^{\frac{H}{2}}\mathbf{F}_1^{-1}\mathbf{F}^{\frac{1}{2}} \right) \tilde{\mathbf{y}}_l \right)
      > C_F
      \right| \mathbf{A}_{all} \right] \right]  \label{eq_q2t_tildey} \\
      & \leq \!\min_{\delta\geq 0} \sum_{S_1} \!\!\left\{ \!\mathbb{E}_{ \mathbf{A}_{all} } \! \! \left[  \mathbb{P} \!\left[ \left. \!\left\{ \sum_{i=1}^{n} \!\left(1-\omega-\omega \lambda_i\right) \frac{\chi_i^2(2L)}{2}
      \!>   C_F \!\right\}
      \bigcap \mathcal{G}_\delta \right| \mathbf{A}_{all}
      \right] \right]
      \!+ \! \mathbb{P} \left[ \mathcal{G}_\delta^c \right] \right\} \label{eq_noCSI_q2t_goodregion}\\
      & = \min_{\delta\geq 0} \left\{ \sum_{S_1} q_{3,t}\left( \omega, \nu \right)
      + \binom {K_a} {t} \left(1 - \frac{\gamma\left( nL, nL \left( 1+\delta \right)\right)}{\Gamma\left( nL \right)} \right) \right\} \label{eq_noCSI_q2t_chi},
    \end{align}
  where $C_F \!= \!\omega L \ln\left|\mathbf{F}_1\right| - L \ln\left|\mathbf{F}\right|  + nL\nu$;  $\lambda_1,\ldots, \lambda_n$ are eigenvalues of $\mathbf{F}_1^{-1} \mathbf{A}_{S_1} \mathbf{A}_{S_1}^H$ in decreasing order with the first $m=\min\left\{ n,t \right\}$ eigenvalues being positive and all of the rest being $0$.
%  and thus $1-\omega-\omega\lambda_1, \ldots, 1-\omega-\omega\lambda_n$ are the eigenvalues of $(1-\omega)\mathbf{I}_n  - \omega\mathbf{F}_1^{-1} \mathbf{A}\boldsymbol{\Gamma}_{S_1}\mathbf{A}^H $, i.e., the eigenvalues of $\mathbf{I}_n  - \omega \mathbf{F}^{\frac{H}{2}}\mathbf{F}_1^{-1}\mathbf{F}^{\frac{1}{2}}$, in increasing order.
  Here, \eqref{eq_q2t_tildey} follows from the union bound and the fact that $\mathbf{y}_{l} =  \mathbf{F}^{\frac{1}{2}} \tilde{\mathbf{y}}_{l} \stackrel{i.i.d.}{\sim} \mathcal{CN}\left( \mathbf{0}, \mathbf{F} \right)$ conditioned on $\mathbf{A}_{ \mathcal{K}_a }$, where $\tilde{\mathbf{y}}_{l} \stackrel{i.i.d.}{\sim} \mathcal{CN}\left( \mathbf{0}, \mathbf{I}_n \right)$ for $l\in[L]$;
  \eqref{eq_noCSI_q2t_goodregion} follows from the bounding technique in~\eqref{eq_good region_1961}, and the fact that
  conditioned on $\mathbf{A}_{ all }$,
  $\mathcal{U} \tilde{\mathbf{y}}_{l}$ and $\tilde{\mathbf{y}}_{l} $ have the same distribution as $\mathcal{CN} \left( \mathbf{0}, \mathbf{I}_{n} \right)$ for the unitary matrix $\mathcal{U}$ satisfying $\mathcal{U}^H \left(\mathbf{I}_n  - \omega \mathbf{F}^{\frac{H}{2}}\mathbf{F}_1^{-1}\mathbf{F}^{\frac{1}{2}} \right) \mathcal{U} = \boldsymbol{\Lambda} $,
  where $\boldsymbol{\Lambda} = \operatorname{diag} \left\{ 1-\omega-\omega\lambda_1, \ldots, 1-\omega-\omega\lambda_n \right\}$;
  \eqref{eq_noCSI_q2t_chi} holds because $\sum_{i=1}^{n} \chi_i^2(2L)$ has the same distribution as $\chi^2(2nL)$ considering that $\chi_i^2(2L)$, $i=1,\ldots,n$, are independent.

  The first term on the RHS of \eqref{eq_noCSI_q2t_chi} can be bounded as
    \begin{align}
      \sum_{S_1} q_{3,t}\left( \omega, \nu \right)
      & \leq \sum_{S_1} \mathbb{E}_{ \mathbf{A}_{all} } \! \left[  \mathbb{P} \left[ \left.  \sum_{i=1}^{n} \lambda_i  \frac{\chi_i^2(2L)}{2} < \frac{nL(1+\delta)(1-\omega) - C_F}{\omega}  \right| \mathbf{A}_{all}
      \right] \right] \label{eq_noCSI_q3t_cap} \\
%      & \leq \sum_{S_1} \mathbb{E}_{\blue \mathbf{A}_{all} } \! \left[ \mathbb{P} \left[ \left. \prod_{i=1}^{m}  \lambda_i^{\frac{1}{m}}  \frac{\chi^2(2mL)}{2}  <
%      \frac{ nL(1+\delta)(1-\omega) - C_F } {\omega}
%      \right| \mathbf{A}_{all} \right] \right] \label{eq_noCSI_q2t_prod}\\
      & \leq \binom {K_a} {t} \mathbb{E}_{ \mathbf{A}_{all} } \! \left[
      \frac{\gamma\left( Lm, L \prod_{i=1}^{m}\lambda_i^{-\frac{1}{m}} \frac{ n(1+\delta)(1-\omega) - \omega\ln\left|\mathbf{F}_1\right| + \ln\left|\mathbf{F}\right|  - n\nu }
      {\omega}  \right)}{\Gamma\left( Lm \right)}  \right] \label{eq_noCSI_q3t_chi},
    \end{align}
  where \eqref{eq_noCSI_q3t_chi} follows from Lemma \ref{lemma_chisumprod} and the fact that the number of non-zero eigenvalues of $\mathbf{F}_1^{-1} \mathbf{A}_{S_1}\mathbf{A}_{S_1}^H$ is $m=\min\left\{ n,t \right\}$, which are denoted as $\lambda_1,\ldots, \lambda_m$ in decreasing order as aforementioned.
  %\eqref{eq_noCSI_q3t_chi} follows because the expectation term is unchanged for different $S_1$ with fixed size $t$, considering that the codebook matrix has i.i.d. $\mathcal{CN}(0,P')$ entries.
  Substituting \eqref{eq_noCSI_q3t_chi} into \eqref{eq_noCSI_q2t_chi}, we can obtain an upper bound on $\mathbb{P}  \left[ \mathcal{G}_{\omega,\nu}^c \right]$, which is denoted as $q_{2,t}\left( \omega, \nu \right) $ in \eqref{eq_noCSI_q2t}.

   In conclusion, based on Fano's bounding technique, we have obtained $q_{1,t}\left( \omega,\nu \right)$ in~\eqref{eq_noCSI_q1t} (i.e. an upper bound on the first probability in~\eqref{eq_pft_goodregion_CSIR_noCSI}) and $q_{2,t}\left(\omega,\nu \right)$ in~\eqref{eq_noCSI_q2t} (i.e. an upper bound on the second probability in~\eqref{eq_pft_goodregion_CSIR_noCSI}),
   which contributes to an upper bound $p_{t}$ given in~\eqref{eq_noCSI_pt} on the probability $\mathbb{P} \left[ \mathcal{F}_t \right]$.
   This completes the proof of Theorem~\ref{Theorem_noCSI_achi}.

%%%%%%%%%%%%%%%%%%%%%%%%%%%%%%%%%%%%%%%%%%%%%%%%%%%%%%%%%%%%
%%%%%%%%%%%%%%%%%%%%%%%%%%%%%%%%%%%%%%%%%%%%%%%%%%%%%%%%%%%%
\section{Proof of Theorem \ref{Theorem_noCSI_noKa_achi_energyperbit}}
\label{proof_achi_noCSI_noKa_weight}
  In this appendix, we prove Theorem~\ref{Theorem_noCSI_noKa_achi_energyperbit} to establish an achievability bound on the PUPE for the scenario in which the number $K_a$ of active users is random and unknown.
%  Specifically, we assume that $K_a$ follows a distribution with PMF $P_{K_a}(\cdot)$ and the range of $K_a$ is denoted as $[{\rm{K}}_l:{\rm{K}}_u]$ with $0\leq {\rm{K}}_l \leq {\rm{K}}_u \leq K$.
  In this case, the decoder first obtains an estimate ${\rm{K}}'_a$ of $K_a$ via an energy-based estimator.
  Then, given ${\rm{K}}'_a$, the decoder produces a set of decoded codewords, which is denoted as $\hat{\mathbf{c}}_{[\hat{\mathcal{K}}_a]}$.
  The number of codewords in the set $\hat{\mathbf{c}}_{[\hat{\mathcal{K}}_a]}$ belongs to an interval around ${\rm{K}}'_a $,
  i.e., it is satisfied that $ | \hat{\mathcal{K}}_a  | \in [{\rm{K}}_{a,l}^{'},{\rm{K}}_{a,u}^{'}]$,
  where ${\rm{K}}'_{a,l} = \max\left\{ 0 , {\rm{K}}'_{a} -r' \right\}$, ${\rm{K}}'_{a,u} = \min\left\{ K , {\rm{K}}'_{a} +r' \right\}$, and $r'$ denotes a nonnegative integer referred to as the decoding radius.
  Based on the notation introduced in Appendix~\ref{Proof_achi_CSIR_noCSI}, the per-user probability of misdetection in~\eqref{eq:MD} can be upper-bounded as
  \begin{align}
    P_{e,\mathrm{MD}}
    & = \mathbb{E} \left[ \frac{1}{K_a} \sum_{k\in {\mathcal{K}_a}} 1\left[  W_{k} \neq \hat{W}_{k}  \right] \right] \\
    % & \leq \mathbb{E}\left[ \frac{1}{K_a} \sum_{k\in {\mathcal{K}_a}} 1\left[  W_{k} \neq \hat{W}_{k}  \right] \right]_{\text{no power constraint}}  +  p_0 \label{eq_PUPE_MD_upper_noCSI_noKa1_r} \\
    % & = \sum_{{K}_a^{'}=K_l}^{K_u} \mathbb{P}\left[ K_a \to {K}_a^{'} \right] \mathbb{E}\left[ \left. \frac{1}{K_a} \sum_{k\in % {\mathcal{K}_a}} 1\left[  W_{k} \neq \hat{W}_{k}  \right]  \right| K_a \to {K}_a^{'} \right]_{\text{no power constraint}}  +  p_0 \\
    & \leq  \! \sum_{{\rm{K}}_a=1}^{K} \!\!P_{K_a}({\rm{K}}_a)\!
    \sum_{{\rm{K}}'_a=0}^{K}
    \sum_{t \in \mathcal{T}_{{\rm{K}}'_a} }
    \!\frac{t+({\rm{K}}_a-{\rm{K}}'_{a,u})^{+}}{{\rm{K}}_a}
    \mathbb{P} \left[  \mathcal{F}_{t} \cap \left\{ {\rm{K}}_a  \!\to\! {\rm{K}}'_{a} \right\} \right]_{\text{no power constraint}} +  p_0. \label{eq_PUPE_MD_upper_noCSI_noKa_r_weight}
  \end{align}
  Here, the integer $t$ takes value in $\mathcal{T}_{{\rm{K}}'_a}$ defined in~\eqref{eq_noCSI_noKa_estimate_Tset1} because the number of misdetected codewords, given by $t+({\rm{K}}_a-{\rm{K}}'_{a,u})^{+}$, is lower-bounded by $({\rm{K}}_a-{\rm{K}}'_{a,u})^{+}$ and upper-bounded by the total number ${\rm{K}}_a$ of transmitted messages;
  $\mathcal{F}_{t} %= \left\{ \sum_{k\in {\mathcal{K}_a}} 1 \left\{ W_{k} \neq \hat{W}_{k} \right\} = t+(K_a-{K}_{a,u}^{'})^{+} \right\}
  $ denotes the event that there are exactly $t+({\rm{K}}_a-{\rm{K}}'_{a,u})^{+}$ misdetected codewords;
  $\left\{{\rm{K}}_a \to {\rm{K}}_{a}^{'}\right\}$ denotes the event that the estimation of $K_a$ results in ${\rm{K}}'_a$;
  $p_0$ denotes an upper bound on the total variation distance between the measures with and without power constraint given by
  \begin{equation}
    p_0 = \mathbb{E}[K_a] \left( 1 - \frac{\gamma\left(n, \frac{nP}{{P}^{\prime}} \right)}{\Gamma\left(n\right)} \right).
  \end{equation}
  Likewise, the per-user probability of false-alarm in~\eqref{eq:FA} can be upper-bounded as
  \begin{align}
    P_{e,\mathrm{FA}}
    & = \mathbb{E} \left[ \frac{1}{ | \hat{\mathcal{K}}_a  | } \sum_{k\in {\hat{\mathcal{K}}_a}} 1\left[  \hat{W}_{k} \neq {W}_{k}  \right] \right] \\
    & \leq \!  \sum_{{\rm{K}}_a=0}^{K} \! P_{K_a}({\rm{K}}_a) \!\! \sum_{{\rm{K}}'_a=0}^{K}
    \sum_{t \in \mathcal{T}_{{\rm{K}}'_a} }
    \sum_{t' \in \mathcal{T}_{{\rm{K}}'_a,t} }
    \!\!\!\frac{t' +  ( {\rm{K}}_{a,l}^{'} \!-\! {\rm{K}}_a )^{+}}{ \hat{\rm{K}}_a  }
    \mathbb{P} \! \left[  \mathcal{F}_{t,t'} \!\cap\! \left\{ {\rm{K}}_a \to {\rm{K}}'_{a} \right\} \right]_{\text{no power constraint}}
    +  p_0 ,\label{eq_PUPE_FA_upper_noCSI_noKa_r_weight}
  \end{align}
  where $\hat{\rm{K}}_a$ denotes the number of detected codewords as given in~\eqref{eq_noCSI_noKa_estimate_Kahat};
  $\mathcal{F}_{t,t'}$ denotes the event that there are exactly $t+({\rm{K}}_a-{\rm{K}}'_{a,u})^{+}$ misdetected codewords and $t' + ( {\rm{K}}'_{a,l} - {\rm{K}}_a )^{+}$ falsely alarmed codewords;
  the integer $t'$ takes value in $\mathcal{T}_{{\rm{K}}'_a}$ defined in~\eqref{eq_noCSI_noKa_estimate_Tset2} because:
  i)~$\hat{\rm{K}}_a$ must be in $[{\rm{K}}'_{a,l} : {\rm{K}}'_{a,u} ]$;
  ii)~the number of falsely alarmed codewords is lower-bounded by $( {\rm{K}}'_{a,l} - {\rm{K}}_a )^{+}$;
  iii)~there exist falsely alarmed codewords only when $\hat{\rm{K}}_a\geq 1$.
%  $\mathcal{F}_{t,t'} = \left\{ \sum_{k\in {\mathcal{K}_a}} 1 \left\{ W_{k} \neq \hat{W}_{k} \right\} = t+(K_a-{K}_{a,u}^{'})^{+} \right\}
%  \cap \left\{ \sum_{k\in{\hat{\mathcal{K}}_a}} 1 \left\{ \hat{W}_{k} \neq W_{k} \right\} = t' + ( K_{a,l}^{'} - K_a )^{+} \right\} $

  Next, we omit the subscript ``no power constraint'' for the sake of brevity.
  The probability in the RHS of~\ref{eq_PUPE_MD_upper_noCSI_noKa_r_weight} can be bounded as
  \begin{align}
    \mathbb{P} \left[  \mathcal{F}_{t} \cap \left\{ {\rm{K}}_a \to {\rm{K}}'_{a} \right\} \right]
    & = \mathbb{P} \left[  \mathcal{F}_{t}
    \cap \left\{ |\hat{\mathcal{K}}_{a}| \in [{\rm{K}}'_{a,l} , {\rm{K}}'_{a,u} ] \right\}
    \cap \left\{ {\rm{K}}_a \to {\rm{K}}'_{a} \right\} \right] \label{eq_PUPE_MD_FA_upper_noCSI_noKa1_r_weight} \\
    & \leq \min\left\{ \mathbb{P} \left[  \mathcal{F}_{t}
    \cap \left\{ |\hat{\mathcal{K}}_{a}| \in [{\rm{K}}'_{a,l} , {\rm{K}}'_{a,u} ] \right\} \right]
    , \mathbb{P} \left[ {\rm{K}}_a \to {\rm{K}}'_{a} \right] \right\} \label{eq_PUPE_MD_FA_upper_noCSI_noKa2_r_weight}\\
%    & \leq \min\left\{ \mathbb{P} \left[  \mathcal{F}_{t}
%    \left|  |\hat{\mathcal{K}}_{a}| \in [{K}_{a,l}^{'},{K}_{a,u}^{'}] \right.\right]
%    , \mathbb{P} \left[ K_a \to {K}_{a}^{'} \right] \right\} \\
    & \leq \min\left\{ \sum_{t' \in \bar{\mathcal{T}}_{{\rm{K}}'_a,t} } \mathbb{P} \left[  \mathcal{F}_{t,t'}
    \left|  |\hat{\mathcal{K}}_{a}| \in [{\rm{K}}'_{a,l},{\rm{K}}'_{a,u} ] \right.\right]
    , \mathbb{P} \left[ {\rm{K}}_a \to {\rm{K}}'_{a}  \right] \right\}.
  \end{align}
  Here, $\bar{\mathcal{T}}_{{\rm{K}}'_a,t}$ is defined in \eqref{eq_noCSI_noKa_estimate_Tset2bar}, which is obtained similar to $\mathcal{T}_{{\rm{K}}'_a,t}$ with the difference that the number $\hat{\rm{K}}_a$ of detected codewords can be $0$;
  \eqref{eq_PUPE_MD_FA_upper_noCSI_noKa1_r_weight} follows because the event ${\rm{K}}_a \to {\rm{K}}'_{a}$ implies that $ |\hat{\mathcal{K}}_{a}| \in [{\rm{K}}'_{a,l} , {\rm{K}}'_{a,u} ]$~\cite{noKa};
  % , i.e., the decoder aims to find the estimated set $\hat{\mathcal{K}}_a$ of active users with size belonging to $[{K}_{a,l}^{'},{K}_{a,u}^{'}]$, and decode the transmitted message for each user in the set $\hat{\mathcal{K}}_a$~\cite{noKa};
  \eqref{eq_PUPE_MD_FA_upper_noCSI_noKa2_r_weight} follows from the fact that the joint probability is upper-bounded by each of the individual probabilities.
  Similarly, the probability in the RHS of~\ref{eq_PUPE_FA_upper_noCSI_noKa_r_weight} can be bounded as
%  \begin{align}
%    \mathbb{P} \left[  \mathcal{F}_{t,t'} \cap \left\{ K_a \to {K}_{a}^{'} \right\} \right]
%    & = \mathbb{P} \left[  \mathcal{F}_{t,t'}
%    \cap \left\{ |\hat{\mathcal{K}}_{a}| \in [{K}_{a,l}^{'},{K}_{a,u}^{'}] \right\}
%    \cap \left\{ K_a \to {K}_{a}^{'} \right\} \right] \\
%    & \leq \min\left\{ \mathbb{P} \left[  \mathcal{F}_{t,t'}
%    \cap \left\{ |\hat{\mathcal{K}}_{a}| \in [{K}_{a,l}^{'},{K}_{a,u}^{'}] \right\} \right]
%    , \mathbb{P} \left[ K_a \to {K}_{a}^{'} \right] \right\}  \\
%    & \leq \min\left\{ \mathbb{P} \left[  \mathcal{F}_{t,t'}
%    \left|  |\hat{\mathcal{K}}_{a}| \in [{K}_{a,l}^{'},{K}_{a,u}^{'}] \right.\right]
%    , \mathbb{P} \left[ K_a \to {K}_{a}^{'} \right] \right\} . \label{eq_PUPE_MD_FA_upper_noCSI_noKa_r}
%  \end{align}
  \begin{equation}
    \mathbb{P} \left[  \mathcal{F}_{t,t'} \cap \left\{ {\rm{K}}_a \to {\rm{K}}'_{a} \right\} \right]
    \leq \min\left\{ \mathbb{P} \left[  \mathcal{F}_{t,t'}
    \left|  |\hat{\mathcal{K}}_{a}| \in [ {\rm{K}}'_{a,l} , {\rm{K}}'_{a,u} ] \right.\right]
    , \mathbb{P} \left[ {\rm{K}}_a \to {\rm{K}}'_{a} \right] \right\} . \label{eq_PUPE_MD_FA_upper_noCSI_noKa_r}
  \end{equation}
   Then, we proceed to bound $\mathbb{P} \left[  \mathcal{F}_{t,t'}
    \left|  |\hat{\mathcal{K}}_{a}| \in [ {\rm{K}}'_{a,l} , {\rm{K}}'_{a,u} ] \right.\right]$ and $\mathbb{P} \left[ {\rm{K}}_a \to {\rm{K}}'_{a} \right]$, respectively,
   which are two ingredients of upper-bounding $P_{e,\mathrm{MD}}$ and $P_{e,\mathrm{FA}}$.

\subsection{Upper-bounding \texorpdfstring{$\mathbb{P}\! \left[ {\rm{K}}_a \!\to\! {\rm{K}}'_{a} \right]$}{pKa}}

  % Since the receiver does not know $K_a$, we let it estimate $K_a$ from $\mathbf{Y}$.
  Given the channel output $\mathbf{Y}$, the receiver estimates $K_a$ as
  \begin{equation}\label{eq:proof_noCSI_noKa_Kaestimate}
    {\rm{K}}'_{a} = \arg \min_{\tilde{{\rm{K}}}_a \in\left[{\rm{K}}_{l}, {\rm{K}}_{u}\right] } m(\mathbf{Y}, \tilde{{\rm{K}}}_a),
  \end{equation}
  where $m(\mathbf{Y}, \tilde{\rm{K}}_a)$ denotes the energy-based estimation metric given by
  \begin{equation}
    m \left( \mathbf{Y} , \tilde{\rm{K}}_a \right)
    = \left| \left\| \mathbf{Y} \right\|_{F}^{2}
    - nL \left( 1 + \tilde{\rm{K}}_a P^{\prime} \right) \right|.
  \end{equation}
  Denote $C_{{\rm{K}}'_a,\tilde{\rm{K}}_a} = \frac{{\rm{K}}'_a+\tilde{\rm{K}}_a}{2}$.
  In the case of $\tilde{\rm{K}}_a \neq {\rm{K}}'_a$, the event $m\left( \mathbf{Y}, {\rm{K}}'_a \right)
    \leq m(\mathbf{Y}, \tilde{\rm{K}}_a )$ is equivalent to
  \begin{equation}
    \begin{cases}
      \left\| \mathbf{Y} \right\|_{F}^{2} \leq nL\left(1+C_{{\rm{K}}'_a,\tilde{\rm{K}}_a} P^{\prime}\right), & \text { if } {\rm{K}}'_a < \tilde{\rm{K}}_a \\
      \left\| \mathbf{Y} \right\|_{F}^{2} \geq nL\left(1+C_{{\rm{K}}'_a,\tilde{\rm{K}}_a} P^{\prime}\right), & \text { if } {\rm{K}}'_a > \tilde{\rm{K}}_a
    \end{cases}.
  \end{equation}
  As a result, the probability of the event that $K_a$ is estimated as ${\rm{K}}'_a$ is upper-bounded as
  \begin{align}
    \mathbb{P} \left[ {\rm{K}}_a \!\to\! {\rm{K}}'_{a} \right]\!
    & \leq \mathbb{P}\left[ m\left( \mathbf{Y}, {\rm{K}}'_a \right)
    \leq m( \mathbf{Y}, \tilde{\rm{K}}_a ), \forall \tilde{\rm{K}}_a \neq {\rm{K}}'_a \right] \\
    & \leq \min_{ \tilde{\rm{K}}_a \in \left[0:K\right] , \tilde{\rm{K}}_a \neq {\rm{K}}'_a }
    \mathbb{P}\left[m\left( \mathbf{Y}, {\rm{K}}'_a \right)
    \leq m(\mathbf{Y}, \tilde{\rm{K}}_a )\right] \\
    & = \min_{ \tilde{\rm{K}}_a \in \left[0:K\right], \tilde{\rm{K}}_a \neq {\rm{K}}'_a }
    1 \!\left[ {\rm{K}}'_a \!<\! \tilde{\rm{K}}_a \right]
    \mathbb{P}\!\left[ \left\| \mathbf{Y} \right\|_{F}^{2} \leq nL\!\left(\!1+C_{{\rm{K}}'_a,\tilde{\rm{K}}_a} P^{\prime}\right) \right] \notag \\
    & \;\;\;\;\;\;\;\;\;\;\;\;\;\;\;\;\;
    \;\;\;\;\;\;\;
    +  1 \!\left[ {\rm{K}}'_a \!>\! \tilde{\rm{K}}_a \right]
    \mathbb{P}\!\left[ \left\| \mathbf{Y} \right\|_{F}^{2} \geq nL\!\left(\!1+C_{{\rm{K}}'_a,\tilde{\rm{K}}_a} P^{\prime}\right) \right]. \label{eq:proof_noCSI_noKa_Ka_Kahat}
  \end{align}

  The probability $ \mathbb{P} \!\left[ \left\| \mathbf{Y} \right\|_{F}^{2} \leq nL \left( 1+C_{{\rm{K}}'_a,\tilde{\rm{K}}_a} P^{\prime}\right) \right] $ on the RHS of~\eqref{eq:proof_noCSI_noKa_Ka_Kahat} can~be bounded following two approaches.
  First, applying the Chernoff bound and Lemma~\ref{expectation_bound}, we~have
  %% part1---Chernoff
  \begin{equation}
    \mathbb{P} \!\left[ \left\| \mathbf{Y} \right\|_{F}^{2} \leq nL \left( 1+C_{{\rm{K}}'_a,\tilde{\rm{K}}_a} P^{\prime}\right) \right]
    \leq \mathbb{E}_{\mathbf{A}_{\mathcal{K}_a}} \!\!\left[ \min_{\rho\geq0} \exp\left\{ \rho nL\left(1+C_{{\rm{K}}'_a,\tilde{\rm{K}}_a} P^{\prime}\right)
    - L \ln \left| \mathbf{I}_n   +  \rho\mathbf{F} \right| \right\} \right] \label{eq:proof_noCSI_noKa_Ka_Kahat11},
  \end{equation}
  where $\mathbf{A}_{\mathcal{K}_a} \in \mathbb{C}^{n\times {\rm{K}}_a}$ denotes the concatenation of the transmitted codewords of ${\rm{K}}_a$ active users and $\mathbf{F} = \mathbf{I}_n + \mathbf{A}_{\mathcal{K}_a} \mathbf{A}_{\mathcal{K}_a}^{H}$.
  %% part1---eigenvalue
  Define the event $\mathcal{G}_{\eta} = \left\{ \sum_{i=1}^{n} \frac{\chi_i^2(2L)}{2} \geq nL\eta \right\}$ for $\eta\geq0$.
  Then, we can obtain another upper bound as follows:
  \begin{align}
    & \mathbb{P}\left[ \left\|\mathbf{Y}\right\|_F^{2} \leq nL\left(1+ C_{{\rm{K}}'_a,\tilde{\rm{K}}_a} P^{\prime}\right) \right] \notag \\
    % & = \mathbb{E}\left[ \mathbb{P}\left[ \left. \sum_{l=1}^{L}\left\|\mathbf{y}_l\right\|_2^{2} \leq nL\left(1+\frac{\hat{K}_a+\tilde{K}_a}{2} P^{\prime}\right) \right| \mathbf{A}_{\mathcal{K}_a} \right] \right] \\
    & = \mathbb{E}_{\mathbf{A}_{\mathcal{K}_a}} \!\left[ \mathbb{P}\left[ \left.
    \sum_{l=1}^{L}  \tilde{\mathbf{y}}_l^H
    \left(\mathbf{I}_n + \mathbf{A}_{\mathcal{K}_a}\mathbf{A}_{\mathcal{K}_a}^H \right) \tilde{\mathbf{y}}_l
    \leq nL\left(1+C_{{\rm{K}}'_a,\tilde{\rm{K}}_a} P^{\prime}\right) \right| \mathbf{A}_{\mathcal{K}_a} \right] \right] \label{eq:proof_noCSI_noKa_Ka_Kahat12_tildey} \\
    % & = \mathbb{E}\left[ \mathbb{P}\left[ \left.
    % \sum_{i=1}^{n} (1+\lambda_i) \frac{\chi^2_i(2L)}{2}
    % \leq nL\left(1+\frac{\hat{K}_a+\tilde{K}_a}{2} P^{\prime}\right) \right| \mathbf{A}_{\mathcal{K}_a} \right] \right] \\
    & \leq \min_{\eta>0} \!\left\{ \mathbb{E}_{\mathbf{A}_{\mathcal{K}_a}} \!\!\left[ \mathbb{P}\!\left[ \left.\!
    \left\{ \sum_{i=1}^{n} (1 + \lambda'_i) \frac{\chi^2_i(2L)}{2}
    \!\leq\! nL\!\left(1\!+\!C_{{\rm{K}}'_a,\tilde{\rm{K}}_a} P^{\prime}\right) \right\}
    \cap \mathcal{G}_\eta \right| \mathbf{A}_{\mathcal{K}_a} \right] \right]
    \!+  \mathbb{P} \left[ \mathcal{G}_\eta^c \right]
    \right\} \label{eq:proof_noCSI_noKa_Ka_Kahat12_gr}\\
    % & \leq \min_{\eta>0} \mathbb{E}\left[ \mathbb{P}\left[ \left.
    % \sum_{i=1}^{n} \lambda_i \frac{\chi^2_i(2L)}{2}
    % \leq nL\left( 1 + \frac{\hat{K}_a+\tilde{K}_a}{2} P^{\prime} - \eta \right) \right| \mathbf{A}_{\mathcal{K}_a} \right] \right]
    % + \mathbb{P} \left[ \mathcal{G}_\eta^c \right] \\
    & \leq \min_{\eta>0} \left\{ \mathbb{E}_{\mathbf{A}_{\mathcal{K}_a}} \!\!\left[
    \mathbb{P} \left[ \left.
    \prod_{i=1}^{m'} {(\lambda^{'}_i)}^{\frac{1}{m'}} \frac{\chi^2(2Lm')}{2}
    \leq  nL \left( 1 + C_{{\rm{K}}'_a,\tilde{\rm{K}}_a} P^{\prime} - \eta \right) \right| \mathbf{A}_{\mathcal{K}_a} \right]
    \right]
    + \mathbb{P} \left[ \mathcal{G}_\eta^c \right]  \right\} \label{eq:proof_noCSI_noKa_Ka_Kahat12_prod} \\
    & = \min_{\eta>0} \left\{\mathbb{E}_{\mathbf{A}_{\mathcal{K}_a}} \!\!\left[
      \frac{\gamma\left( Lm', \prod_{i=1}^{m'} {(\lambda^{'}_i)}^{-\frac{1}{m'}}
      nL\left( 1 + C_{{\rm{K}}'_a,\tilde{\rm{K}}_a} P^{\prime} - \eta \right)
      \right)}{\Gamma\left( Lm' \right)}  \right]
    + \frac{\gamma\left( nL, nL\eta \right)}{\Gamma\left( nL \right)} \right\}, \label{eq:proof_noCSI_noKa_Ka_Kahat12}
  \end{align}
  where $\lambda'_1,\ldots, \lambda'_n$ are eigenvalues of $\mathbf{A}_{\mathcal{K}_a} \mathbf{A}_{\mathcal{K}_a}^H$ in decreasing order with the first $m' = \min\left\{ n,{\rm{K}}_a \right\}$ eigenvalues being positive and others being $0$.
  Here, \eqref{eq:proof_noCSI_noKa_Ka_Kahat12_tildey} holds because $\mathbf{y}_{l} =  \mathbf{F}^{\frac{1}{2}} \tilde{\mathbf{y}}_{l} \sim \mathcal{CN}\left( \mathbf{0}, \mathbf{F} \right)$ conditioned on $\mathbf{A}_{ \mathcal{K}_a }$, where $\tilde{\mathbf{y}}_{l} \sim \mathcal{CN}\left( \mathbf{0}, \mathbf{I}_n \right)$;
  \eqref{eq:proof_noCSI_noKa_Ka_Kahat12_gr} follows from the ``good region'' technique in~\eqref{eq_good region_1961};
  \eqref{eq:proof_noCSI_noKa_Ka_Kahat12_prod} follows from Lemma~\ref{lemma_chisumprod}.
  Taking the minimum of \eqref{eq:proof_noCSI_noKa_Ka_Kahat11} and \eqref{eq:proof_noCSI_noKa_Ka_Kahat12}, we obtain the ultimate upper bound on $\mathbb{P}\left[ \left\|\mathbf{Y}\right\|_F^{2} \leq nL\left(1+C_{{\rm{K}}'_a,\tilde{\rm{K}}_a} P^{\prime}\right) \right]$ denoted as $p_{{\rm{K}}_a\to {\rm{K}}'_a,1}$ in~\eqref{eq_noCSI_noKa_pKa_Kahat1}.

  Likewise, we can derive two upper bounds on $\mathbb{P}\left[ \left\|\mathbf{Y}\right\|_F^{2} \geq nL\left(1+C_{{\rm{K}}'_a,\tilde{\rm{K}}_a} P^{\prime}\right) \right]$.
  Taking the minimum value of them, we obtain the ultimate upper bound on it, denoted as $p_{{\rm{K}}_a\to {\rm{K}}'_a,2}$ in~\eqref{eq_noCSI_noKa_pKa_Kahat2}.
%  \begin{equation}
%    p_{K_a\to K'_a,2,1} = \mathbb{E}_{\mathbf{A}_{\mathcal{K}_a}}
%    \left[ \min_{0\leq \rho < 1/(1+\lambda'_1)} \exp\left\{ -\rho nL\left(1+C_{K'_a,\tilde{K}_a} P^{\prime}\right)
%    -L \ln \left|  \mathbf{I}_n - \rho\mathbf{F} \right|  \right\} \right], \label{eq:proof_noCSI_noKa_Ka_Kahat21}
%  \end{equation}
%  %% part2---eigenvalue
%  \begin{equation}
%    p_{K_a\to K'_a,2,2} = \min_{\eta>0}
%    \left\{ 2 - \mathbb{E}_{\mathbf{A}_{\mathcal{K}_a}} \left[
%      \frac{\gamma\left( Lm, \frac{nL}{\lambda_1}
%      \left( 1 + C_{K'_a,\tilde{K}_a} P^{\prime} - \eta \right)
%      \right)}{\Gamma\left( Lm \right)}  \right]
%    - \frac{\gamma\left( nL, nL\eta \right)}{\Gamma\left( nL \right)}    \right\}, \label{eq:proof_noCSI_noKa_Ka_Kahat22}
%  \end{equation}
%  where \eqref{eq:proof_noCSI_noKa_Ka_Kahat21} follows from the Chernoff bound and \eqref{eq:proof_noCSI_noKa_Ka_Kahat22} follows from the ``good region'' technique.
%  Details are omitted for brevity.

\subsection{Upper-bounding \texorpdfstring{$\mathbb{P} \left[  \mathcal{F}_{t,t'}
    \left|  |\hat{\mathcal{K}}_{a}| \in [{\rm{K}}'_{a,l} , {\rm{K}}'_{a,u} ] \right.\right]$}{pFt}}
  In this subsection, we utilize the MAP decoder to upper-bound $\mathbb{P} \! \left[  \mathcal{F}_{t,t'}
    \left|  |\hat{\mathcal{K}}_{a}| \in [{\rm{K}}'_{a,l} , {\rm{K}}'_{a,u}] \right.\right]$.
  Under the condition that $|\hat{\mathcal{K}}_{a}| \in [{\rm{K}}'_{a,l} , {\rm{K}}'_{a,u}]$, the outputs of the decoder are given by
      \begin{equation}\label{decoder_CSIR_noCSI_noKa}
        \left[ \hat{\mathcal{K}}_{a }, \hat{\mathbf{c}}_{[\hat{\mathcal{K}}_a]} \right]
        = \arg\min_{{ \hat{\mathcal{K}}_{a } \subset \mathcal{K}}, { | \hat{\mathcal{K}}_{a } | \in [{\rm{K}}'_{a,l} , {\rm{K}}'_{a,u}] } }
        \;\min_{ \left( \hat{\mathbf{c}}_{(k)} \in \mathcal{C}_k \right)_{ k\in\hat{\mathcal{K}}_{a } } }
        \;g\left(  \mathbf{Y} , \hat{\mathbf{c}}_{[\hat{\mathcal{K}}_a]}  \right) ,
      \end{equation}
      \begin{equation}\label{decoder_CSIR_noCSI2_noKa}
        \hat{W}_k =  \emph{f}_{\text{en},k}^{-1}\left(\hat{\mathbf{c}}_{(k)}\right) , \;\; k \in \hat{\mathcal{K}}_{a} ,
      \end{equation}
  where the MAP decoding metric $g\left(  \mathbf{Y} ,   \hat{\mathbf{c}}_{[ {\hat{\mathcal{K}}}_a ]}  \right)$ is given by
  \begin{equation}
    g\!\left(  \mathbf{Y} ,   \hat{\mathbf{c}}_{[ {\hat{\mathcal{K}}}_a]}  \right)
    = L \ln \! \left| \mathbf{I}_n \!+\! \mathbf{A} {\boldsymbol{\Gamma}}'_{\hat{\mathcal{K}}_a}\mathbf{A}^H \right|
    + \operatorname{tr}\!\left( \!\left( \mathbf{I}_n \!+\! \mathbf{A} {\boldsymbol{\Gamma}}'_{\hat{\mathcal{K}}_a} \mathbf{A}^H \right)^{-1} \! \mathbf{Y} \mathbf{Y}^H \right)
    - \ln \! \left( P_{K_a}(|{\hat{\mathcal{K}}}_a|) M^{-|{\hat{\mathcal{K}}}_a|} \right) . \label{eq_g_noCSI_weight}
  \end{equation}
  Here, ${\boldsymbol{\Gamma}}^{'}_{ S }$ is defined in Appendix~\ref{section5}.
  In the following, $g\left(  \mathbf{Y} ,   \hat{\mathbf{c}}_{[ {\hat{\mathcal{K}}}_a]}  \right)$ is denoted as $g\left( \hat{\boldsymbol{\Gamma}}_{\hat{\mathcal{K}}_a} \right)$ for simplicity.

  Let the set $S_1 \subset \mathcal{K}_a$ of size $t+({\rm{K}}_a-{\rm{K}}'_{a,u})^{+}$ denote the set of users whose codewords are misdecoded.
  The set $S_1$ can be divided into two subsets $S_{1,1}$ and $S_{1,2}$ of size $({\rm{K}}_a-{\rm{K}}'_{a,u})^{+}$ and $t$, respectively.
  Let the set $S_2 \subset \mathcal{K}\backslash\mathcal{K}_a\cup S_1$ of size $t'+({\rm{K}}'_{a,l}-{\rm{K}}_a)^{+}$ denote the set of detected users with false-alarm codewords.
  % {\blue Let the set $S_3 \subset \mathcal{K}\backslash\mathcal{K}_a\backslash S_2$ denote a set of users of size $({K}^{'}_{a,l}-K_a)^{+}$.}
  Let $S_{2,1}$ denote an arbitrary subset of $S_2$ of size $({\rm{K}}'_{a,l}-{\rm{K}}_a)^{+}$.
  For the sake of simplicity, we rewrite ``$\bigcup_{{S_{1} \subset \mathcal{K}_a,} {\left| S_{1} \right| = t+({\rm{K}}_a-{\rm{K}}'_{a,u})^{+} }}  $'' to ``$\bigcup_{S_{1}}$''
  and ``$\bigcup_{{S_{2} \subset \mathcal{K} \backslash \mathcal{K}_a\cup S_{1}, \left| S_{2} \right| = t'+({\rm{K}}'_{a,l}-{\rm{K}}_a)^{+} } }$'' to ``$\bigcup_{S_2}$''; similarly for $\sum$ and $\bigcap$.
  We rewrite $\left\{ \mathbf{c}'_{(k)} \in \mathcal{C}_k: k \in S_{2}, \mathbf{c}'_{(k)} \neq \mathbf{c}_{(k)} \right\}$ to $\mathbf{c}^{'}_{[S_2]}$ for short, which denotes the set of false alarm codewords corresponding to users in the set $S_2$.
  Define $\mathbf{A}_{S}$, $\mathbf{A}^{'}_{S}$, ${\boldsymbol{\Gamma}}_{ S }$ and ${\boldsymbol{\Gamma}}^{'}_{ S }$ as in Appendix~\ref{section5}.
  % Let $\mathbf{A}_{S} \in \mathbb{C}^{n \times |S|}$ denote the concatenation of transmitted codewords of active users in the set $S\subset\mathcal{K}_a$. Let $\mathbf{A}^{'}_{S} \in \mathbb{C}^{n \times |S|}$ denote the concatenation of false-alarm codewords for users in the set $S\subset\mathcal{K}$.
  Define the event $\mathcal{G}_{\omega,\nu} = \bigcap_{S_1} \left\{ \mathbf{Y} \in \mathcal{R}_{t,S_1} \right\}$ as in Appendix~\ref{Proof_achi_CSIR_noCSI}.
  Following similar ideas in~\eqref{eq_pft_goodregion_CSIR_noCSI}, we have
  \begin{align}
      & \mathbb{P} \left[  \mathcal{F}_{t,t'} \left|  |\hat{\mathcal{K}}_{a}| \in [{\rm{K}}'_{a,l} , {\rm{K}}'_{a,u} ] \right.\right] \notag \\
%      & \leq \mathbb{P} \left[ \left.
%      {\bigcup}_{S_{1}} {\bigcup}_{S_{2}}
%      {\bigcup}_{\mathbf{c}^{'}_{[S_2]}}
%      \left\{ g \left( {\boldsymbol{\Gamma}}^{'}_{ \mathcal{K}_a \backslash S_1 \cup S_2 }   \right)
%      \leq g \left( \boldsymbol{\Gamma}^{'}_{ \mathcal{K}_a \backslash S_{1,1} \cup S_{2,1} } \right)  \right\}
%      \right| |\hat{\mathcal{K}}_{a}| \in [{K}_{a,l}^{'},{K}_{a,u}^{'}] \right] \\
      & \leq \!\min_{ 0 \leq \omega \leq 1 , \nu\geq0 }
      \! \left\{ \! \mathbb{P} \!\left[ \left.
      {\bigcup}_{S_{1}} {\bigcup}_{S_{2}}
      {\bigcup}_{\mathbf{c}^{'}_{[S_2]}}
      \!\!\left\{ g \!\left( {\boldsymbol{\Gamma}}^{'}_{ \mathcal{K}_a \backslash S_1 \cup S_2 }   \right)
      \leq g \!\left( \boldsymbol{\Gamma}^{'}_{ \mathcal{K}_a \backslash S_{1,1} \cup S_{2,1} } \right)  \right\} \cap \mathcal{G}_{\omega,\nu}
      \right| \!|\hat{\mathcal{K}}_{a}| \!\in\! [ {\rm{K}}'_{a,l} , {\rm{K}}'_{a,u} ] \right]
      \right. \notag \\
      & \;\;\;\;\;\;\;\;\;\;\;\;\;\;\;\;\;\;\;  \left.
      \!+ \mathbb{P}\!  \left[ \left. \mathcal{G}_{\omega,\nu}^c \right| |\hat{\mathcal{K}}_{a}| \in [{\rm{K}}'_{a,l} , {\rm{K}}'_{a,u} ] \right] \right\}. \label{eq:proof_noCSI_noKa_pftt}
  \end{align}

  Similar to \eqref{eq_pft_Chernoff_CSIR_noCSI} and \eqref{eq_pft_Chernoff_noCSI}, we can obtain an upper bound on the first probability on the RHS of~\eqref{eq:proof_noCSI_noKa_pftt},
  which is denoted as $q_{1,{\rm K}'_a,t,t'}$ in~\eqref{eq_noCSI_noKa_estimation_q1t}.
%  \begin{align}
%      & \mathbb{P} \!\left[ \left.
%      {\bigcup}_{S_{1}} {\bigcup}_{S_{2}}
%      {\bigcup}_{\mathbf{c}^{'}_{[S_2]}}
%      \!\!\left\{ g \!\left( {\boldsymbol{\Gamma}}^{'}_{ \mathcal{K}_a \backslash S_1 \cup S_2 }   \right)
%      \leq g \!\left( \boldsymbol{\Gamma}^{'}_{ \mathcal{K}_a \backslash S_{1,1} \cup S_{2,1} } \right)  \right\} \cap \mathcal{G}_{\omega,\nu}
%      \right| \!|\hat{\mathcal{K}}_{a}| \!\in\! [{K}_{a,l}^{'},{K}_{a,u}^{'}] \right]  \notag \\
%      & \leq  C_{K_a^{'},t,t'}
%      \mathbb{E}_{ \mathbf{A}_{all} } \!\! \left[ \min_{{\substack{u\geq 0,r\geq 0, \\ \lambda_{\min}\left({ {\mathbf{B}}}\right) > 0}}}
%      \!\!\exp \!\left\{ L \!\left( rn\nu + u \ln \!\left|\mathbf{F}''\right| - r \ln \! \left|\mathbf{F}\right|
%      - u \ln \!\left| {\mathbf{F}'} \right|
%      + r\omega \ln \!\left| \mathbf{F}_{1} \right|
%      - \ln \!\left| \mathbf{B} \right| \right)
%      \right\}  \right], \label{eq:proof_noCSI_noKa_pftt1}
%  \end{align}
%  where ${\mathbf{A}}_{all} = \left\{ {\mathbf{A}}_{ \mathcal{K}_a },  {\mathbf{A}}_{ S_1 },  {\mathbf{A}}_{ S_{1,1} },   {\mathbf{A}}^{'}_{ S_2},   {\mathbf{A}}^{'}_{ S_{2,1}} \right\}$;
%  $C_{K_a^{'},t,t'}$,
%  $\mathbf{F}$,
%  $\mathbf{F}^{'}$,
%  $\mathbf{F}^{''}$,
%  $\mathbf{F}_1$,
%  and $\mathbf{B}$ are given in Theorem~\ref{Theorem_noCSI_noKa_achi_energyperbit}.
%  The RHS of~\eqref{eq:proof_noCSI_noKa_pftt1} is denoted as $q_{1,K'_a,t,t'}$ in~\eqref{eq_noCSI_noKa_estimation_q1t}.
  In the case of $t+({\rm K}_a-{\rm K}'_{a,u})^{+} > 0$, the second probability on the RHS of~\eqref{eq:proof_noCSI_noKa_pftt} can be bounded as in Appendix~\ref{section5}.
  When $t+({\rm K}_a-{\rm K}'_{a,u})^{+} = 0$, we have
  \begin{align}
    \mathbb{P}  \left[ \mathcal{G}_{\omega,\nu}^c \right]
    % & = \mathbb{P} \left[ \bigcup_{S_1} \left\{ g\left(\boldsymbol{\Gamma}_{\mathcal{K}_a} \right) > \omega g\left(  {\boldsymbol{\Gamma}}_{ \mathcal{K}_a \backslash S_1 } \right) + nL\nu \right\} \right] \\
    & = \mathbb{P} \left[ g\left(\boldsymbol{\Gamma}_{\mathcal{K}_a} \right) >  \frac{nL\nu}{1-\omega} \right] \\
    % & = \mathbb{P} \left[ \operatorname{tr}\left( \mathbf{Y}^H \mathbf{F}^{-1} \mathbf{Y} \right) > \frac{nL\nu}{1-\omega} - L \ln \left| \mathbf{F} \right| \right] \\
%    & = \mathbb{E}_{\mathbf{A}_{\mathcal{K}_a}} \left[ \mathbb{P} \left[ \left.
%    \frac{\chi^2(2nL)}{2}
%    > \frac{nL\nu}{1-\omega} - L \ln \left| \mathbf{F} \right| \right| \mathbf{A}_{\mathcal{K}_a}
%    \right] \right] \\
    & = \mathbb{E}_{\mathbf{A}_{\mathcal{K}_a}} \left[ 1 - \frac{\gamma\left( nL, \frac{nL\nu}{1-\omega} - L \ln \left| \mathbf{F} \right|  +  b \right)}{\Gamma\left( nL \right)}  \right] , \label{eq:proof_noCSI_noKa_pft_t0_rbig}
  \end{align}
  where the constant $b$ is given in~\eqref{eq_noCSI_noKa_estimation_b}.
  The RHS of \eqref{eq:proof_noCSI_noKa_pft_t0_rbig} is denoted as $q_{2,{\rm K}'_a,t,0}$ in~\eqref{eq_noCSI_noKa_estimation_q2t0}.
  This concludes the proof of Theorem~\ref{Theorem_noCSI_noKa_achi_energyperbit}.

\section{Proof of Theorem \ref{prop_converse_noCSI}} \label{proof_converse_noCSI}
  In this appendix, we prove Theorem \ref{prop_converse_noCSI} to establish a Fano type converse bound on the minimum required energy-per-bit for the no-CSI case.
  Let $\bar{\mathbf{y}} = \left[\mathbf{y}_1^T, \mathbf{y}_2^T, \ldots, \mathbf{y}_L^T \right]^T \in \mathbb{C}^{nL\times 1}$ be a vector obtained by concatenating the received signals of $L$ antennas at the BS.
  Let $\bar{\mathbf{X}}_{K_aM}$ be an $n\times K_{a} M$ submatrix of ${\mathbf{X}}$ including codebooks of $K_a$ active users and denote $\bar{\mathbf{X}}  = \operatorname{diag} \left\{\bar{\mathbf{X}}_{K_aM}, \ldots, \bar{\mathbf{X}}_{K_aM} \right\} \in \mathbb{C}^{nL\times K_aML} $.
  %Assume $\bar{\mathbf{X}}_{K_aM}$ has i.i.d. $\mathcal{CN}(0,P)$ entries.
    Let $\bar{\mathbf{H}}_{l} \in \mathbb{C}^{K_a M \times K_a M} $ be a block diagonal matrix, where block $k$ is a diagonal $M \times M$ matrix with  all diagonal entries equal to $h_{k,l} \sim \mathcal{CN}(0,1) $.
    Let $\bar{\mathbf{H}}  = \left[\bar{\mathbf{H}}_{1}, \ldots , \bar{\mathbf{H}}_{L} \right]^T\in \mathbb{C}^{K_a ML \times K_a M} $.
    The vector $\bar{\boldsymbol{\beta}}  \in \left\{ 0,1 \right\}^{K_aM}$ includes $K_a$ blocks, where each block is of size $M$ and includes one 1; we have $\left[ \bar{\boldsymbol{\beta}} \right]_{(k-1)M + W_k} = 1$ if the $W_{k}$-th codeword is transmitted by user $k$, and $\left[ \bar{\boldsymbol{\beta}} \right]_{(k-1)M + W_k} = 0$ otherwise.
    Then, we can model the communication system~as
    \begin{equation}
      \bar{\mathbf{y}} = \bar{\mathbf{X}}  \bar{\mathbf{H}}  \bar{\boldsymbol{\beta}} + \bar{\mathbf{z}},
    \end{equation}
    where $\bar{\mathbf{z}} \in \mathbb{C}^{nL\times 1} $ with each entry i.i.d. from $\mathcal{CN}(0,1)$.

    We assume a genie reveals the set of active users.
    Similar to the analysis in Appendix \ref{proof_converse_CSIR}, we have~\cite{finite_payloads_fading}
    \begin{equation} \label{eq_conv_fano_nocsi}
      \left( 1-\epsilon \right) J - h_2 \left(\epsilon\right)
      \leq \frac{1}{K_a} I_2 \left(\left.\bar{\boldsymbol{\beta}};\bar{\mathbf{y}} \right| \bar{\mathbf{X}}\right).
    \end{equation}
    Based on the chain rule of the mutual information, we have
    \begin{align}
      I_2\left( \left.\bar{\boldsymbol{\beta}}, \bar{\mathbf{H}} \bar{\boldsymbol{\beta}}; \bar{\mathbf{y}} \right| \bar{\mathbf{X}}\right)
      & = I_2\left(\left.\bar{\boldsymbol{\beta}};\bar{\mathbf{y}} \right|\bar{\mathbf{X}}\right) + I_2\left(\left.\bar{\mathbf{H}}\bar{\boldsymbol{\beta}};\bar{\mathbf{y}} \right|\bar{\boldsymbol{\beta}}, \bar{\mathbf{X}}\right)  \\
      & = I_2\left(\left. \bar{\mathbf{H}}\bar{\boldsymbol{\beta}};\bar{\mathbf{y}} \right|\bar{\mathbf{X}}\right) + I_2\left(\left.\bar{\boldsymbol{\beta}};\bar{\mathbf{y}} \right|\bar{\mathbf{H}}\bar{\boldsymbol{\beta}}, \bar{\mathbf{X}}\right).
    \end{align}
    Since $\bar{\boldsymbol{\beta}} \to \bar{\mathbf{H}}\bar{\boldsymbol{\beta}} \to  (\bar{\mathbf{y}},\bar{\mathbf{X}})$ forms a Markov chain, the mutual information $I_2 \left(\left.\bar{\boldsymbol{\beta}};\bar{\mathbf{y}} \right| \bar{\mathbf{H}}\bar{\boldsymbol{\beta}}, \bar{\mathbf{X}}\right) = 0$. Hence, we have~\cite[Eq. (78)]{sparsity_pattern}
    \begin{equation} \label{eq_conv_inf_equality}
      I_2 \left(\left. \bar{\boldsymbol{\beta}};\bar{\mathbf{y}} \right| \bar{\mathbf{X}}\right)
    = I_2 \left(\left. \bar{\mathbf{H}}\bar{\boldsymbol{\beta}}; \bar{\mathbf{y}}  \right| \bar{\mathbf{X}}\right) - I_2 \left( \left.\bar{\mathbf{H}}\bar{\boldsymbol{\beta}}; \bar{\mathbf{y}}  \right| \bar{\boldsymbol{\beta}}, \bar{\mathbf{X}}\right).
    \end{equation}

    Next, we focus on the two terms on the RHS of~\eqref{eq_conv_inf_equality}.
    We have
    \begin{align}
      I_2\left(\left. \bar{\mathbf{H}}\bar{\boldsymbol{\beta}};\bar{\mathbf{y}} \right|\bar{\mathbf{X}} = \bar{\mathbf{X}}^r\right)
      & =  I_2\left(\bar{\mathbf{H}}\bar{\boldsymbol{\beta}}; \bar{\mathbf{X}}^r\bar{\mathbf{H}}\bar{\boldsymbol{\beta}}+ \bar{\mathbf{z}} \right)  \\
      & \leq \sup_{\mathbf{u}} I_2\left(\mathbf{u}; \bar{\mathbf{X}}^r\mathbf{u}+\bar{\mathbf{z}} \right)   \label{eq_conv_inf_sup}\\
      & = \log_2 \left| \mathbf{I}_{nL} + \frac{1}{M}\bar{\mathbf{X}}^r \left(\bar{\mathbf{X}}^r\right)^{H} \right|   \label{eq_conv_inf_sup2}\\
      &  = L \log_2 \left| \mathbf{I}_{n } + \frac{1}{M}\bar{\mathbf{X}}_{K_aM}^{r} \left(\bar{\mathbf{X}}_{K_aM}^{r}\right)^{H} \right|,\label{eq_conv_inf_sup3}
    \end{align}
    where $\bar{\mathbf{X}}_{K_aM}^r$ is a realization of $\bar{\mathbf{X}}_{K_aM}$ and $\bar{\mathbf{X}}^r  \!=\! \operatorname{diag} \!\left\{ \bar{\mathbf{X}}_{K_aM}^r, \ldots, \bar{\mathbf{X}}_{K_aM}^r \right\}$ is a realization of $\bar{\mathbf{X}}$.
    The supremum in \eqref{eq_conv_inf_sup} is over
    $\mathbf{u}$ with $\mathbb{E} \!\left[\mathbf{u} \right] \!=\! \mathbf{0}$ and $\mathbb{E} \!\left[\mathbf{u} \mathbf{u}^{H} \right] \!=\! \mathbb{E} \!\left[ \left(\bar{\mathbf{H}}\bar{\boldsymbol{\beta}} \right) \!\left(\bar{\mathbf{H}}\bar{\boldsymbol{\beta}} \right)^{ H} \right] \!=\! \frac{1}{M}\mathbf{I}_{K_aML}$.
    The supremum is achieved when $\mathbf{u} \!\sim\! \mathcal{CN} \!\left( \mathbf{0}, \frac{1}{M}\mathbf{I}_{K_aML} \right)$ \cite{sparsity_pattern}, which implies \eqref{eq_conv_inf_sup2}.
    Then, we~have
    \begin{equation}
      I_2 \left( \left. \bar{\mathbf{H}}\bar{\boldsymbol{\beta}}; \bar{\mathbf{y} } \right| \bar{\mathbf{X}} \right)
      \leq L \mathbb{E} \left[  \log_2 \left| \mathbf{I}_{n} + \frac{1}{M} \bar{\mathbf{X}}_{K_aM} \bar{\mathbf{X}}_{K_aM}^{ H}  \right|  \right]. \label{eq_conv_inf_term1_y}
    \end{equation}
    Under the assumption that the entries of codebooks are i.i.d. with mean zero and variance $P$, the expectation on the RHS of~\eqref{eq_conv_inf_term1_y} can be upper-bounded as
    \begin{equation}
      \mathbb{E} \left[  \log_2 \left| \mathbf{I}_{n} + \frac{1}{M} \bar{\mathbf{X}}_{K_aM} \bar{\mathbf{X}}_{K_aM}^{ H}  \right|  \right]
      \leq \min \left\{ n \log_2 \left( 1 + K_aP \right),  K_a M  \log_2 \left( 1 + \frac{1}{M}nP \right) \right\},\label{eq_conv_inf_term1}
    \end{equation}
    where \eqref{eq_conv_inf_term1} follows from the concavity of the $\log_2 \left|\cdot\right|$ function.
    We denote the RHS of~\eqref{eq_conv_inf_term1} as $C$ for simplicity.

    A lower bound on $I \left( \left.\bar{\mathbf{H}}\bar{\boldsymbol{\beta}}; \bar{\mathbf{y}}  \right| \bar{\boldsymbol{\beta}}, \bar{\mathbf{X}}\right)$ can be derived as follows. Let $\tilde{\mathbf{X}}_{K_a} \in \mathbb{C}^{n\times K_{a}} $ be a submatrix of ${\mathbf{X}}$ formed by columns corresponding to the support of $\bar{\boldsymbol{\beta}}$.
    Let $\tilde{\mathbf{H}}_{K_a}$ be a $K_{a}\times L$ submatrix of ${\mathbf{H}}$ including fading coefficients between $K_{a}$ active users and $L$ antennas of the receiver.
    Then, the received signal given in \eqref{eq_y} can be rewritten as
    \begin{equation}
      \mathbf{Y} = \tilde{\mathbf{X}}_{K_a} \tilde{\mathbf{H}}_{K_a}  + {\mathbf{Z}}.
    \end{equation}
    We have
    \begin{align}
      I_2\left(\left.\bar{\mathbf{H}}\bar{\boldsymbol{\beta}};\bar{\mathbf{y}} \right|\bar{\boldsymbol{\beta}}= \bar{\boldsymbol{\beta}}^r, \bar{\mathbf{X}} = \bar{\mathbf{X}}^r \right)
      &= I_2\left(\bar{\mathbf{H}}\bar{\boldsymbol{\beta}}^r; \bar{\mathbf{X}}^r\bar{\mathbf{H}}\bar{\boldsymbol{\beta}}^r + \bar{\mathbf{z}}  \right)   \\
      &= I_2\left(\tilde{\mathbf{H}}_{K_a}; \tilde{\mathbf{X}}_{K_a}^r \tilde{\mathbf{H}}_{K_a}  + {\mathbf{Z}}  \right)   \\
      %&= H_2 \left( \bar{\mathbf{X}}^r\bar{\mathbf{H}}\bar{\boldsymbol{\beta}}^r + \bar{\mathbf{z}} \right)
%      - H_2\left( \bar{\mathbf{z}}  \right)    \\
%      &= H_2\left( \tilde{\mathbf{X}}_{K_a}^r \tilde{\mathbf{H}}_{K_a}  + {\mathbf{Z}}  \right)
%      - H_2\left( \mathbf{Z}  \right)    \\
      &= L\log_2 \left| \mathbf{I}_{n} + \tilde{\mathbf{X}}_{K_a}^r \left(\tilde{\mathbf{X}}_{K_a}^r\right)^{H} \right|,
    \end{align}
    where $\bar{\boldsymbol{\beta}}^r$ is a realization of $\bar{\boldsymbol{\beta}}$ and $\tilde{\mathbf{X}}_{K_a}^r$ is a realization of $\tilde{\mathbf{X}}_{K_a}$. Hence, applying Sylvester's determinant theorem, we have
    \begin{equation}  \label{eq_conv_inf_term2}
      I_2 \left( \left.\bar{\mathbf{H}}\bar{\boldsymbol{\beta}};\bar{\mathbf{y}} \right| \bar{\boldsymbol{\beta}},  \bar{\mathbf{X}} \right)
      = L \mathbb{E} \left[ \log_2  \left| \mathbf{I}_{n} + \tilde{\mathbf{X}}_{K_a} \tilde{\mathbf{X}}_{K_a}^{ H} \right|  \right]
      = L \mathbb{E} \left[ \log_2  \left| \mathbf{I}_{K_a} + \tilde{\mathbf{X}}_{K_a}^{ H}  \tilde{\mathbf{X}}_{K_a} \right|  \right] .
    \end{equation}

  Substituting \eqref{eq_conv_inf_term1} and \eqref{eq_conv_inf_term2} into \eqref{eq_conv_inf_equality}, we obtain an upper bound on $I \left(\left. \bar{\boldsymbol{\beta}};\bar{\mathbf{y}} \right| \bar{\mathbf{X}}\right)$.
  Substituting this bound into~\eqref{eq_conv_fano_nocsi}, the proof of \eqref{P_tot_conv_noCSI} in Theorem \ref{prop_converse_noCSI} is completed.

  Under the assumption that $K$ users generate their codebooks independently with each entry i.i.d. from $\mathcal{CN}(0,P)$, we further lower-bound $\mathbb{E} \left[ \log_2  \left| \mathbf{I}_{K_a} + \tilde{\mathbf{X}}_{K_a}^{ H} \tilde{\mathbf{X}}_{K_a} \right|  \right]$ in~\eqref{eq_conv_inf_term2} in the remainder of this appendix.
  In the case of $K_a > n$, let $\alpha_i \sim \frac{\chi^2\left( 2 \left(K_a-i+1 \right)\right)}{2} $ for $i=1,\ldots,n$. We have
  \begin{align}
      \mathbb{E}\! \left[ \log_2  \left| \mathbf{I}_{K_a} + \tilde{\mathbf{X}}_{K_a}^{ H} \tilde{\mathbf{X}}_{K_a} \right|  \right]
      & \geq \sum_{i=1}^{n} \mathbb{E} \left[  \log_2  \left( 1 + P \alpha_i\right)  \right] \label{Proof_noCSI_logdet_IWishart_lbound}\\
      & =  \sum_{i=1}^{n} \mathbb{E} \left[ \log_2  \alpha_i  \right]
      + \sum_{i=1}^{n} \mathbb{E} \!\left[ \log_2  \!\left( {P + \frac{1}{\alpha_i }} \right)  \right]  \\
      % & = \log_2 e \! \sum_{i=1}^{n }\! \psi(K_a-i+1) + \sum_{i=1}^{n} \mathbb{E} \!\left[ \log_2  \!\left( {P + \frac{1}{\alpha_i }} \right)  \right] \label{Proof_noCSI_logdet_IWishart} \\
      & \geq \log_2 e \! \sum_{i=1}^{n }\! \psi(K_a-i+1) + \sum_{i=1}^{n} \log_2  \!\left( {P + \frac{1}{ K_a -i + 1 }} \right), \label{Proof_noCSI_logdet_IWishart_lbound2}
    \end{align}
  where \eqref{Proof_noCSI_logdet_IWishart_lbound} follows from Lemma~\ref{Lemma_logdet_Iwishart_lb} shown below;
  \eqref{Proof_noCSI_logdet_IWishart_lbound2} follows because $\mathbb{E}\!\left[ \ln \frac{\chi^2\left( 2b \right)}{2 } \right]\!=\!\psi(b)$ with $\psi(x)$ denoting Euler's digamma function, and follows from Jensen's inequality considering $\log_2\left(P+\frac{1}{x}\right)$ is a convex function of $x$.
  Let $b_1=J{\left(1 - \epsilon\right)} - h_2 \left( \epsilon \right)$.
  Substituting \eqref{Proof_noCSI_logdet_IWishart_lbound2} into \eqref{P_tot_conv_noCSI}, we have
  \begin{equation}
    b_1  \leq \frac{LC}{K_a}
    - \frac{L}{K_a} \sum_{i=1}^{n} \left( \psi(K_a-i+1) \log_2 e +  \log_2  \!\left( {P + \frac{1}{ K_a -i + 1 }} \right) \right) \label{Proof_P_tot_conv_noCSI_loose_nless}.
  \end{equation}

  \begin{Lemma}[Section 4.1.1 in~\cite{logdet_IWishart}] \label{Lemma_logdet_Iwishart_lb}
      For $b>0$. A central complex Wishart matrix $\mathbf{W} \sim \mathcal{W}_{m}(n, \mathbf{I})$, with $n\geq m$, satisfies
      \begin{equation}
        \mathbb{E}\left[\log_{2} \left|\mathbf{I}_m + b \mathbf{W}\right| \right] >
        \sum_{i=n-m+1}^{n} \mathbb{E}\left[ \log_2 \left(1+ b \frac{\chi^2(2i)}{2} \right)\right] ,
      \end{equation}
      where $\chi^2(2i)$ is a chi-square variate with $2i$ degrees of freedom.
  \end{Lemma}

  Likewise, when $K_a \leq n$, we have
  \begin{equation}
      \mathbb{E}\! \left[ \log_2  \left| \mathbf{I}_{K_a} + \tilde{\mathbf{X}}_{K_a}^{ H} \tilde{\mathbf{X}}_{K_a} \right|  \right]
      \geq \log_2 e \sum_{i=1}^{K_a } \psi(n-i+1) + \sum_{i=1}^{K_a} \log_2  \!\left( {P + \frac{1}{ n -i + 1 }} \right). \label{Proof_noCSI_logdet_IWishart_lbound_Kaless}
    \end{equation}
  Substituting \eqref{Proof_noCSI_logdet_IWishart_lbound_Kaless} into \eqref{P_tot_conv_noCSI}, when $K_a \leq n$, we have   \begin{equation}
    b_1 \leq \frac{LC}{K_a}
    - \frac{L}{K_a} \sum_{i=1}^{K_a} \left( \psi(n-i+1) \log_2 e +  \log_2  \!\left( {P + \frac{1}{ n -i + 1 }} \right) \right) \label{Proof_P_tot_conv_noCSI_loose_Kaless}   .
  \end{equation}
  Together with \eqref{Proof_P_tot_conv_noCSI_loose_nless}, the proof of \eqref{P_tot_conv_noCSI_bound} is completed, which concludes the proof of Theorem \ref{prop_converse_noCSI}.

%%%%%%%%%%%%%%%%%%%%%%%%%%%%%%%%%%%%%%%%%%%%%%%%%%%%%%%%%%%%%%%%%%%%%%%%
%%%%%%%%%%%%%%%%%%%%%%%%%%%%%%%%%%%%%%%%%%%%%%%%%%%%%%%%%%%%%%%%%%%%%%%%
\section{Proof of Theorem \ref{prop_converse_noCSI_noKa}} \label{proof_converse_noCSI_noKa}

  In this appendix, we prove Theorem \ref{prop_converse_noCSI_noKa} to establish a converse bound on the minimum required energy-per-bit for the case in which there is no CSI at the receiver and the number $K_a$ of active users is random and unknown.
  In Appendix~\ref{proof_converse_noCSI_noKa_multiple_user}, we establish a converse bound for the scenario with multiple users;
  in Appendix~\ref{proof_converse_noCSI_noKa_single_user}, we establish a converse bound for the scenario with knowledge of the activities of $K-1$ potential users and the transmitted codewords and channel coefficients of active users among them, which is also a converse bound for the massive random access problem.

\subsection{Multiple-user random access converse bound}\label{proof_converse_noCSI_noKa_multiple_user}
  In this part, we use the Fano inequality to derive a converse bound on the minimum required energy-per-bit for the multiple-user case when $K_a$ is random and unknown.
  Define $\bar{\mathbf{y}}$, $\bar{\mathbf{X}}$, $\bar{\mathbf{X}}_{KM}$, $\bar{\mathbf{H}}_{l}$, and $\bar{\mathbf{H}}$ as in Appendix~\ref{proof_converse_noCSI}.
  Let the vector $\bar{\boldsymbol{\beta}} \in \left\{ 0,1 \right\}^{KM}$ indicate which codewords are transmitted by active users, which includes $K$ blocks with each block of size $M$ and including at most one 1.
  Specifically, according to the random access model described in Section~\ref{section2}, for $m\in[M]$ and $k\in[K]$, we have $\mathbb{P}\left[ \left[ \bar{\boldsymbol{\beta}} \right]_{(k-1)M + m} = 0\right] = 1-\frac{p_a}{M}$ and $\mathbb{P}\left[ \left[ \bar{\boldsymbol{\beta}} \right]_{(k-1)M + m} = 1\right] = \frac{p_a}{M}$.
  Then, we can model the communication system as
    \begin{equation}
      \bar{\mathbf{y}} = \bar{\mathbf{X}}  \bar{\mathbf{H}}  \bar{\boldsymbol{\beta}} + \bar{\mathbf{z}},
    \end{equation}
  where $\bar{\mathbf{z}} \in \mathbb{C}^{nL\times 1} $ with each entry i.i.d. from $\mathcal{CN}(0,1)$.

%  Let $\bar{\mathbf{y}} = \left[\mathbf{y}_1^T, \ldots, \mathbf{y}_L^T \right]^T \in \mathbb{C}^{nL\times 1}$ be a vector obtained by concatenating the received signals of $L$ antennas at the BS.
%  Let $\bar{\mathbf{X}}_{KM}$ be an $n\times KM$ submatrix of $\mathbf{X}$ including codebooks of $K$ potential users and denote $\bar{\mathbf{X}}  = \operatorname{diag} \left\{\bar{\mathbf{X}}_{KM}, \ldots, \bar{\mathbf{X}}_{KM} \right\} \in \mathbb{C}^{nL\times KML}$.
%  Let $\bar{\mathbf{H}}_{l} \in \mathbb{C}^{K M \times K M} $ be a block diagonal matrix, where block $k$ is a diagonal $M \times M$ matrix with all diagonal entries equal to $h_{k,l} \sim \mathcal{CN}(0,1) $.
%  Let $\bar{\mathbf{H}}  = \left[\bar{\mathbf{H}}_{1}, \ldots , \bar{\mathbf{H}}_{L} \right]^T\in \mathbb{C}^{K ML \times K M}$.
%
%  Specifically, we assume $K_a\sim \text{Binom}(K,p_a)$ with mean $\bar{K}_a=p_aK$.

  Let $\mathcal{M}_k =1\left[ {W}_{k}\neq\hat{W}_{k} \right]$ and $P_{e,k} = \mathbb{E}\left[ \mathcal{M}_k \right]$.
  The error requirements in~\eqref{eq:MD} and~\eqref{eq:FA} can be loosened to
  \begin{equation}
    P_e = \frac{1}{K} \sum_{k\in \mathcal{K}} P_{e,k} \leq  \epsilon_{\rm MD} + \epsilon_{\rm FA}.
  \end{equation}
  For $k\in\mathcal{K}$, a Fano type argument gives
    \begin{equation}
      (1-P_{e,k}) H_2(W_k|\bar{\mathbf{X}})
      - h_2( P_{e,k} )
      \leq I_2(W_k;\hat{W}_k|\bar{\mathbf{X}}) , \label{eq_fano_noKa4}
    \end{equation}
    %\begin{align}
%      H_2(W_k|\bar{\mathbf{X}})
%      % & = H_2(W_k|\hat{W}_k,\bar{\mathbf{X}}) + I (W_k;\hat{W}_k|\bar{\mathbf{X}}) \label{eq_fano_noKa1} \\
%      & = H_2( W_k,\mathcal{M}_k |\hat{W}_k,\bar{\mathbf{X}} ) + I_2 (W_k;\hat{W}_k|\bar{\mathbf{X}}) \label{eq_fano_noKa2} \\
%      & = H_2( \mathcal{M}_k |\hat{W}_k,\bar{\mathbf{X}} ) + H_2( W_k | \mathcal{M}_k , \hat{W}_k,\bar{\mathbf{X}} ) + I_2 (W_k;\hat{W}_k|\bar{\mathbf{X}})  \label{eq_fano_noKa3} \\
%      & \leq h_2( P_{e,k} ) + P_{e,k} \;\! H_2( W_k | \bar{\mathbf{X}} ) +
%      I_2(W_k;\hat{W}_k|\bar{\mathbf{X}}) , \label{eq_fano_noKa4}
%    \end{align}
  where $H_2(x)$ denotes the entropy of a random variable $x$ and $h_2( \cdot )$ denotes the binary entropy function.
%  Here, \eqref{eq_fano_noKa2} holds because the error indicator $\mathcal{M}_k$ is determined by $W_k$ and $\hat{W}_k$;
%  \eqref{eq_fano_noKa4} follows from the fact that the condition can reduce the entropy and the inequality that
%    \begin{align}
%      & H_2( W_k | \mathcal{M}_k , \hat{W}_k ,\bar{\mathbf{X}} ) \notag \\
%      & = \mathbb{P}[\mathcal{M}_k\!=\!1] \;\! H_2( W_k | \mathcal{M}_k \!=\! 1 , \hat{W}_k ,\bar{\mathbf{X}} )
%      + \mathbb{P}[\mathcal{M}_k\!=\!0] \;\! H_2( W_k | \mathcal{M}_k \!=\! 0 , \hat{W}_k ,\bar{\mathbf{X}} ) \\
%      % & = P_{e,k} \;\! H_2( W_k | \mathcal{M}_k = 1 , \hat{W}_k ,\bar{\mathbf{X}} ) \\
%      & \leq P_{e,k} \;\! H_2( W_k | \bar{\mathbf{X}} ).
%    \end{align}
  The entropy $H_2(W_k| \bar{\mathbf{X}})$ can be computed as
    \begin{equation}\label{eq_fano_entropy}
      H_2(W_k| \bar{\mathbf{X}})
      = H_2(W_k) = -(1-p_a)\log_2(1-p_a) - p_a\log_2\frac{p_a}{M}
      = h_2(p_a) + p_a J.
    \end{equation}
  Substituting \eqref{eq_fano_entropy} into \eqref{eq_fano_noKa4} and taking the summation over $k\in\mathcal{K}$ on both sides of \eqref{eq_fano_noKa4}, we have
    \begin{equation}\label{eq_fano_noKa5}
      K ( 1 - P_{e} ) \;\! ( h_2(p_a) + p_a J )
      - \sum_{k\in\mathcal{K}} h_2( P_{e,k} )
      \leq \sum_{k\in\mathcal{K}} I_2(W_k;\hat{W}_k \;\! | \;\! \bar{\mathbf{X}} ) .
    \end{equation}
  Considering the concavity of $h_2(\cdot)$ and the inequality that $P_{e} \leq \epsilon_{\rm MD} + \epsilon_{\rm FA} \leq 1-\frac{1}{1+2^{h_2(p_a) + p_a J}}$, we have
    \begin{equation} \label{eq_hinf_noKa}
      P_{e}\;\! ( h_2(p_a) + p_a J ) + \frac{1}{K} \sum_{k\in\mathcal{K}} h_2( P_{e,k} )
      \leq (\epsilon_{\rm MD} + \epsilon_{\rm FA})\;\! ( h_2(p_a) + p_a J ) +  h_2( \epsilon_{\rm MD} + \epsilon_{\rm FA} ) .
    \end{equation}
  %Similar to \eqref{eq_mutualinf1},
  Moreover, following from~\eqref{eq_mutualinf1}, we have $ \sum_{k\in\mathcal{K}} I_2(W_k;\hat{W}_k\;\!\! | \;\!\! \bar{\mathbf{X}}) \leq I_2(W_{\mathcal{K}};\bar{\mathbf{y}}\;\!\!  | \; \!\! \bar{\mathbf{X}}) = I_2( \bar{\boldsymbol{\beta}};\bar{\mathbf{y}} \;\!\! | \;\!\! \bar{\mathbf{X}} )$.
  Together with~\eqref{eq_conv_inf_equality},~\eqref{eq_fano_noKa5}, and~\eqref{eq_hinf_noKa}, we can obtain
    \begin{equation}
      K (1-\epsilon_{\rm MD} - \epsilon_{\rm FA}) \left( h_2(p_a) + p_a J \right)
      - K h_2( \epsilon_{\rm MD} + \epsilon_{\rm FA} )
      \leq
      I_2 \left(\left. \bar{\mathbf{H}}\bar{\boldsymbol{\beta}}; \bar{\mathbf{y}}  \right| \bar{\mathbf{X}}\right) - I_2 \left( \left.\bar{\mathbf{H}}\bar{\boldsymbol{\beta}}; \bar{\mathbf{y}}  \right| \bar{\boldsymbol{\beta}}, \bar{\mathbf{X}}\right) .  \label{eq_fano_noKa6}
    \end{equation}

    Next, we focus on the two terms on the RHS of~\eqref{eq_fano_noKa6}.
    Following from similar ideas used in~\eqref{eq_conv_inf_sup3} and \eqref{eq_conv_inf_term1_y} with the difference that $\mathbb{E} \!\left[ \left(\bar{\mathbf{H}}\bar{\boldsymbol{\beta}} \right) \!\left(\bar{\mathbf{H}}\bar{\boldsymbol{\beta}} \right)^{ H} \right] \!=\! \frac{p_a}{M}\mathbf{I}_{KML}$,
    we have
    \begin{equation}
      I_2 \left( \left. \bar{\mathbf{H}}\bar{\boldsymbol{\beta}}; \bar{\mathbf{y} } \right| \bar{\mathbf{X}} \right)
      \leq L \mathbb{E} \left[  \log_2 \left| \mathbf{I}_{n} + \frac{p_a}{M} \bar{\mathbf{X}}_{KM} \bar{\mathbf{X}}_{KM}^{ H}  \right|  \right]. \label{eq_conv_inf_term1_y_noKa}
    \end{equation}
    Under the assumption that the entries of codebooks are i.i.d. with mean zero and variance $P$, the expectation on the RHS of~\eqref{eq_conv_inf_term1_y_noKa} can be upper-bounded as
    \begin{equation}
      \mathbb{E} \left[  \log_2 \left| \mathbf{I}_{n} + \frac{p_a}{M} \bar{\mathbf{X}}_{KM} \bar{\mathbf{X}}_{KM}^{ H}  \right|  \right]
      \leq \min \left\{ n \log_2 \left( 1 + p_aKP \right),  K M  \log_2 \left( 1 + \frac{p_a}{M}nP \right) \right\}. \label{eq_conv_inf_term1_noKa}
    \end{equation}
    Moreover, $I_2 \left( \left.\bar{\mathbf{H}}\bar{\boldsymbol{\beta}}; \bar{\mathbf{y}}  \right| \bar{\boldsymbol{\beta}}, \bar{\mathbf{X}}\right)$ can be computed as
    \begin{align}
      I_2 \left( \left.\bar{\mathbf{H}}\bar{\boldsymbol{\beta}};\bar{\mathbf{y}} \right| \bar{\boldsymbol{\beta}},  \bar{\mathbf{X}} \right)
      & = L \; \mathbb{E} \left[ \log_2  \left| \mathbf{I}_{n} + \tilde{\mathbf{X}}_{K_a} \tilde{\mathbf{X}}_{K_a}^{ H} \right|  \right] \label{eq_conv_inf_term2_noKa1} \\
      & = L \sum_{{\rm{K}}_a=0}^{K} \left( P_{K_a}({\rm{K}}_a)
      \mathbb{E} \left[ \left. \log_2  \left| \mathbf{I}_{n} + \tilde{\mathbf{X}}_{K_a}  \tilde{\mathbf{X}}_{K_a}^{ H} \right|  \right| K_a = {\rm{K}}_a \right] \right),\label{eq_conv_inf_term2_noKa}
    \end{align}
    where $\tilde{\mathbf{X}}_{K_a} \in \mathbb{C}^{n\times K_{a}} $ denotes a submatrix of ${\mathbf{X}}$ formed by columns corresponding to the support of $\bar{\boldsymbol{\beta}}$;
    $P_{K_a}({\rm{K}}_a)$ denotes the probability of the event that there are exactly ${\rm K}_a$ active users given in~\eqref{eq_conv_pKa_noKa};
    \eqref{eq_conv_inf_term2_noKa1} follows from \eqref{eq_conv_inf_term2}.
    Combining \eqref{eq_fano_noKa6}, \eqref{eq_conv_inf_term1_y_noKa}, and \eqref{eq_conv_inf_term2_noKa}, the proof of the converse bound for the multiple-user case is completed.

\subsection{Single-user random access converse bound} \label{proof_converse_noCSI_noKa_single_user}

  The converse bound for the scenario with knowledge of the activities of $K-1$ potential users and the transmitted codewords and channel coefficients of active users among them, can be regarded as a converse bound for the massive random access problem.
  In this case, it is equivalent to assume that there is a single user in the system with active probability $p_a$.
  If this user is active, it equiprobably selects a message $W$ from $\left\{1,2,\ldots,M \right\}$, and the corresponding codeword is denoted as $\mathbf{x}_W\in\mathbb{C}^{n}$ satisfying the maximum power constraint
  \begin{equation}\label{Req:equal_power_constraint}
    \left\| \mathbf{x}_W \right\|_2^2 \leq n P.
  \end{equation}
  Let $\mathcal{F} \subset \mathbb{C}^n$ be a set of permissible channel inputs as specified by~\eqref{Req:equal_power_constraint}.
  If this user is inactive, we assume $W=0$ and $\mathbf{x}_W = \mathbf{0}$.
  The received signal is given by
  \begin{equation}
    \mathbf{Y} = \mathbf{x}_W \mathbf{h}^{T} + \mathbf{Z} \in \mathbb{C}^{n\times L},
  \end{equation}
  where the vector $\mathbf{h}\sim \mathcal{CN}(\mathbf{0},\mathbf{I}_{L})$ includes channel fading coefficients between the user and $L$ antennas at the BS and the noise matrix $\mathbf{Z}\in \mathbb{C}^{n\times L}$ has i.i.d. $\mathcal{CN}(0,1)$ entries.

  Denote the decoded message as $\hat{W} \in \{0,1,\ldots,M\}$.
  We define three types of error probabilities as follows:
  the probability of the event that the receiver detects the presence of a message even though the user is inactive is given by
    \begin{equation} \label{eqR:PUPE_singleUE_FA}
      P_{e,1}
      = \mathbb{P} \left[ \hat{W} \neq 0 | W = 0 \right] , % \leq \epsilon_{\rm{FA}} ,
    \end{equation}
  the probability of the event that the receiver does not decode correctly a transmitted message is given by
    \begin{equation} \label{eqR:PUPE_singleUE_IE}
      P_{e,2}
      = \frac{1}{ M } \sum_{m \in [M]} \mathbb{P} \left[ \hat{W} \neq m | W = m \right] , %\leq \epsilon_{\rm{IE}} ,
    \end{equation}
  and the probability of the event that the receiver erroneously decides that the user is inactive is given by
    \begin{equation} \label{eqR:PUPE_singleUE_MD}
      P_{e,3}
      = \frac{1}{ M } \sum_{m \in [M]} \mathbb{P} \left[ \hat{W} = 0 | W = m \right] , % \leq \epsilon_{\rm{MD}} ,
    \end{equation}
  where $P_{e,3}\leq P_{e,2}$.
  Then, the error requirements in~\eqref{eq:MD} and~\eqref{eq:FA} can be rewritten as
    \begin{equation} \label{eqR:MD}
      P_{e,{\rm{MD}}} = p_a P_{e,2} \leq \epsilon_{\rm{MD}},
    \end{equation}
    \begin{equation} \label{eqR:FA}
      P_{e,{\rm{FA}}} = (1-p_a) P_{e,1} \leq \epsilon_{\rm{FA}}.
    \end{equation}

  An upper bound on the number of codewords that are compatible with the requirement that $P_{e,1}$, $P_{e,2}$, and $P_{e,3}$ do not exceed $\epsilon_{1}$, $\epsilon_{2}$, and $\epsilon_{3}$, respectively, is provided in~\cite[Theorem~2]{On_joint}.
  By changing the error requirement in~\cite{On_joint} to~\eqref{eqR:MD} and~\eqref{eqR:FA} and by considering the multiple-receive-antenna setting, we obtain the following meta-converse result:
  \begin{prop} \label{R_Theorem_AWGN_singleUE_conv}
    Consider the single-user setup, where the user is active with probability $p_a$.
%    It becomes active with probability $p_a$ and keeps silent otherwise.
%    If it is active, it equiprobably selects a message from $\left\{1,2,\ldots,M \right\}$ and transmits it to the BS.
%    The number of BS antennas is assumed to be $L$.
%    The user has a codebook of size $M=2^J$ and length $n$, in which each codeword satisfies the maximum power constraint in~\eqref{Req:equal_power_constraint}.
    Let $Q_{ Y^{n\times L} }$ be an arbitrary distribution on $\mathcal{Y}^{n\times L}$.
    When both the CSI and the user activity are unknown, every $(n,M,\epsilon_{\rm{MD}},\epsilon_{\rm{FA}},P)_{\text{no-CSI,no-}K_a}$ code satisfies
      \begin{equation} \label{eqR:beta_conv_AWGN_singleUE_1}
        M \leq \sup_{ \substack{ {P_{X^{n}}: \mathbf{x} \in \mathcal{F} }\\{\epsilon_{1},\epsilon_{2},\epsilon_{3} \in [0,1]} } }
        \frac{ 1 - \beta_{1-\epsilon_{1}}  \left( P_{ Y^{n\times L} | X^{n} = \mathbf{0} } , Q_{ Y^{n\times L} } \right) }{ \beta_{ 1-\epsilon_{2} }
        \left( P_{ X^{n} } P_{Y^{n\times L}|X^{n}} , P_{X^{n}} Q_{Y^{n\times L}} \right)  } ,
      \end{equation}
    where
      \begin{equation} \label{eqR:beta_conv_AWGN_singleUE_2}
        \beta_{ 1-\epsilon_{3} } \left( P_{ Y^{n\times L} } , Q_{ Y^{n\times L} } \right) \leq
        1 - \beta_{1-\epsilon_{1}}  \left( P_{ Y^{n\times L} | X^{n} = \mathbf{0} } , Q_{ Y^{n\times L} } \right) ,
      \end{equation}
      \begin{equation} \label{eqR:conv_MD_IE}
        \epsilon_{3} \leq \epsilon_{2} ,
      \end{equation}
      \begin{equation} \label{eqR:conv_MD}
        p_a \epsilon_{2} = \epsilon_{\rm{MD}} ,
      \end{equation}
      \begin{equation} \label{eqR:conv_FA}
        (1-p_a) \epsilon_{1} = \epsilon_{\rm{FA}} .
      \end{equation}
    % Here, the auxiliary distribution is chosen as $Q_{Y^{n}} = \mathcal{N}(0,\mathbf{I}_n)$.
  \end{prop}

  Proposition~\ref{R_Theorem_AWGN_singleUE_conv} presents a meta-converse bound for the single-user random access problem.
  However, evaluating this bound is numerically intractable because it involves an optimization over all possible input distributions.
  Next, we proceed to loosen Proposition~\ref{R_Theorem_AWGN_singleUE_conv} and obtain an easy-to-evaluate bound as provided in Theorem~\ref{prop_converse_noCSI_noKa}.

  Following from the inequality that $M_m(n,\epsilon,P) \leq M_e(n+1,\epsilon,P)$~\cite[Lemma~39]{Channel_coding_rate}, which relates the numbers of codewords under maximum power constraint and equal power constraint,
  the condition in~\eqref{eqR:beta_conv_AWGN_singleUE_1} can be loosened to
  \begin{align}
    M & \leq \sup_{ \substack{ {P_{X^{n+1}}: \mathbf{x} \in \mathcal{F}^{n+1} }\\{\epsilon_{1},\epsilon_{2},\epsilon_{3} \in [0,1]} } }
    \frac{ 1 - \beta_{1-\epsilon_{1}}  \left( P_{ Y^{(n+1)\times L} | X^{n+1} = \mathbf{0} } , Q_{ Y^{(n+1)\times L} } \right) }{ \beta_{ 1-\epsilon_{2} }
    \left( P_{ X^{n+1} } P_{Y^{(n+1)\times L}|X^{n+1}} , P_{X^{n+1}} Q_{Y^{(n+1)\times L}} \right)  } \label{eqR:beta_conv_AWGN_singleUE_1e1} \\
    & = \sup_{ \epsilon_{1},\epsilon_{2},\epsilon_{3} \in [0,1] }
    \frac{ \epsilon_{1} }{ \beta_{ 1-\epsilon_{2} }
    \left( P_{Y^{(n+1)\times L}|X^{n+1}=\mathbf{x}_1} ,   Q_{Y^{(n+1)\times L}} \right)  } ,
    \label{eqR:beta_conv_AWGN_singleUE_1e2}
%    & = \sup_{ \epsilon_{\rm{FA}},\epsilon_{\rm{MD}},\epsilon_{\rm{IE}} \in [0,1] }
%    \frac{ \epsilon_{\rm{FA}} }{ Q \left(  Q^{-1} \left( 1-\epsilon_{\rm{IE}} \right) + \sqrt{(n+1)P'}  \right)  } ,
%    \label{eqR:beta_conv_AWGN_singleUE_1e3}
  \end{align}
  where $\mathcal{F}^{n+1} \!=\! \left\{ \mathbf{x}\in\mathbb{C}^{n+1} \!: \!\left\| \mathbf{x} \right\|_2^2 = \!(n+1)P\right\}$,
  the auxiliary distribution is chosen as $Q_{Y^{(n+1)\times L}} = P_{ Y^{(n+1)\times L} | X^{n+1} = \mathbf{0} }  = \prod_{l=1}^{L} \mathcal{CN}(0,\mathbf{I}_{n+1})$~\cite{Yang_Beta_beta},
  and \eqref{eqR:beta_conv_AWGN_singleUE_1e2} follows from~\cite[Lemma~29]{Channel_coding_rate} for any input $\mathbf{x}_1 \in \mathcal{F}^{n+1}$.
  Meanwhile, under $Q_{Y^{(n+1)\times L}} = \prod_{l=1}^{L} \mathcal{CN}(0,\mathbf{I}_{n+1})$, the condition in~\eqref{eqR:beta_conv_AWGN_singleUE_2} becomes
      \begin{equation} \label{eqR:beta_conv_AWGN_singleUE_e2}
        \beta_{ 1-\epsilon_{3} } \left( P_{ Y^{(n+1)\times L} } , Q_{ Y^{(n+1)\times L} } \right) \leq
        \epsilon_{1}.
      \end{equation}
  Since $\alpha \mapsto \beta_{ \alpha } \left( P_{ Y^{(n+1)\times L} } , Q_{ Y^{(n+1)\times L} } \right) $ is monotonically nondecreasing, we can combine \eqref{eqR:beta_conv_AWGN_singleUE_e2} and \eqref{eqR:conv_MD_IE} as
  \begin{equation} \label{eqR:beta_conv_AWGN_singleUE_2_proof}
    \beta_{ 1-\epsilon_{2} } \left( P_{ Y^{(n+1)\times L} } , Q_{ Y^{(n+1)\times L} } \right) \leq
    \epsilon_{1} .
  \end{equation}
  Following from~\cite[Lemma~6]{Yang_Beta_beta}, we can obtain that
  \begin{equation}
    \beta_{ 1-\epsilon_{2} } \left( P_{ Y^{(n+1)\times L} } , Q_{ Y^{(n+1)\times L} } \right) \leq
    M \beta_{ 1-\epsilon_{2} } \left( P_{ Y^{(n+1)\times L} | X^{n+1} = \mathbf{x}_1 } , Q_{ Y^{(n+1)\times L} } \right) .
  \end{equation}
  Together with \eqref{eqR:beta_conv_AWGN_singleUE_1e2} and \eqref{eqR:beta_conv_AWGN_singleUE_2_proof}, we observe that the condition in \eqref{eqR:beta_conv_AWGN_singleUE_2_proof} is satisfied once \eqref{eqR:beta_conv_AWGN_singleUE_1e2} is satisfied.

  Next, we proceed to compute $\beta_{ 1-\epsilon_{2} }
  \left( P_{Y^{(n+1)\times L}|X^{n+1}=\mathbf{x}_1} ,   Q_{Y^{(n+1)\times L}} \right)$.
  We have
  \begin{equation} \label{Req:proof_fenmu_HG}
    \ln \frac{ d P_{ Y^{(n+1)\times L} | X^{n+1} = \mathbf{x}_1 } }{ d Q_{ Y^{(n+1)\times L} } }
    = -L \ln(1+(n+1)P) + \sum_{l=1}^{L} \mathbf{y}_l^H \left( \mathbf{I}_{n+1} - ( \mathbf{I}_{n+1}+\mathbf{x}_1 \mathbf{x}_1^H )^{-1} \right) \mathbf{y}_l .
  \end{equation}
  % It is convenient to define independent Gaussian vectors $\mathbf{z}_l \sim \mathcal{CN}(0,\mathbf{I}_n)$.
  Under $P_{ Y^{(n+1)\times L} | X^{n+1} = \mathbf{x}_1 }$, \eqref{Req:proof_fenmu_HG} is distributed the same as
  \begin{equation} \label{Req:proof_fenmu_H}
    H % = -L \ln(1+nP) + \sum_{l=1}^{L} \mathbf{z}_l^H \mathbf{x}_1 \mathbf{x}_1^H \mathbf{z}_l
    = -L \ln(1+(n+1)P) + (n+1)P \frac{\chi^2(2L)}{2}  .
  \end{equation}
  Under $Q_{ Y^{(n+1)\times L} }$, \eqref{Req:proof_fenmu_HG} is distributed the same as
  \begin{equation} \label{Req:proof_fenmu_G}
    G % = -L \ln(1+nP) + \frac{1}{1+nP'} \sum_{l=1}^{L} \mathbf{z}_l^H \mathbf{x}_1 \mathbf{x}_1^H \mathbf{z}_l
    = -L \ln(1+(n+1)P) + \frac{(n+1)P}{1+(n+1)P} \frac{\chi^2(2L)}{2} .
  \end{equation}
  Thus, we have
  \begin{equation}\label{Req:proof_fenmu1}
    \beta_{1-\epsilon_{2}}  \left( P_{ Y^{(n+1)\times L} | X^{n+1} = \mathbf{x}_1 } , Q_{ Y^{(n+1)\times L} } \right)
    = \mathbb{P} \left[
    G \geq \bar{r}
    \right]
    = \mathbb{P} \left[ \chi^2(2L) \geq (1+(n+1)P)r
    \right] ,
  \end{equation}
  where $\bar{r} $ and $r $ are chosen to satisfy
  \begin{equation}\label{Req:proof_fenmu2}
    \mathbb{P} \left[
    H \leq \bar{r}
    \right]
    = \mathbb{P} \left[ \chi^2(2L) \leq r
    \right] = \epsilon_{2} .
  \end{equation}
  Thus, the single-user random access bound is obtained.
  It completes the proof of Theorem~\ref{prop_converse_noCSI_noKa}.

%%%%%%%%%%%%%%%%%%%%%%%%%%%%%%%%%%%%%%%%%%%%%%%%%%%%%%%%%%%%%%%%%%%%%%%%%%%%%%%%%%%%%%%%%%
%%%%%%%%%%%%%%%%%%%%%%%%%%%%%%%%%%%%%%%%%%%%%%%%%%%%%%%%%%%%%%%%%%%%%%%%%%%%%%%%%%%%%%%%%%
\section{ Proof of Theorem \ref{Theorem_scalinglaw_noCSI}  } \label{Proof_scalinglaw_noCSI}
  In this appendix, we prove Theorem \ref{Theorem_scalinglaw_noCSI} to establish a scaling law for the no-CSI case under the PUPE criterion and the assumption that all users are active.
  The achievability and converse scaling laws are established in Appendix~\ref{Proof_scalinglaw_noCSI_achi} and Appendix~\ref{Proof_scalinglaw_noCSI_conv}, respectively.

%  To obtain the scaling law on the achievability side, both the activity detection problem considered in~\cite{Caire1} (whose proof relies on \cite{support_recovery}) and the data detection problem of interest in this work can be formulated as similar sparse support recovery problems.
%  This is because one can immediately obtain a data detection scheme from an activity detection scheme by assigning to each user a unique set of codewords, such that a user can transmit the codeword corresponding to its information message.
%  Thus, the achievability scaling law in~\cite{Caire1}, which is established for the activity detection problem under the joint error probability criterion, can be extended to the data detection problem as follows: under the joint error probability criterion, with blocklength $n\to \infty$ and a sufficient number of BS antennas $L = \Theta\left(n^2 \ln n\right)$, one can reliably serve $K = \mathcal{O}\left(n^2\right)$ users when the payload $J = \Theta(1)$ and the power $P=\Theta\left(\frac{1}{n^2}\right)$.
%  In Appendix~\ref{Proof_scalinglaw_noCSI_achi}, we further establish an achievability scaling law for the data detection problem under the PUPE criterion.
%  As will become clear shortly, the required numbers of BS antennas are significantly different under the two criteria.
%  Then, we derive a matching converse result in Appendix~\ref{Proof_scalinglaw_noCSI_conv} under the PUPE criterion and the assumption of i.i.d. Gaussian codebooks.
%

\subsection{ Achievability  } \label{Proof_scalinglaw_noCSI_achi}
  Assume that the matrix $\mathbf{A} \in \mathbb{C}^{n\times KM}$ consists of codewords of all users, with columns drawn uniformly i.i.d. from the sphere of radius $\sqrt{nP}$. The power constraint in~\eqref{power_constraint} is fulfilled in this case. Then, the PUPE can be upper-bounded as
  \begin{equation}
    P_{e} \leq \epsilon_1 + \mathbb{P}\left[\frac{1}{K}  \sum_{k\in {\mathcal{K} }} 1 \left[ W_{k} \neq \hat{W}_{k} \right] \geq \epsilon_1 \right]
    = \epsilon_1 + \mathbb{P} \left[ {\bigcup}_{t = \lceil \epsilon_1 K \rceil}^{K } \mathcal{F}_t \right] \label{eq_PUPE_upper_noCSI_scalinglaw},
  \end{equation}
  where the positive constant $\epsilon_1<\epsilon$ and $\mathcal{F}_t$ denotes the event that there are exactly $t$ misdecoded users.
  Denote the set of codewords of $K$ users as $S_{all}$ of size $KM$ and the set of the transmitted codewords of $K$ users as $S_{\mathcal{K}}$ of size $K$.
  Let ${\boldsymbol{\Gamma}}_{S} \!= \!\operatorname{diag} \!\left\{ {\boldsymbol{\gamma}}_{S} \right\} \!\in\! \left\{ 0,1 \right\}^{KM\times KM}$, where $\left[ {\boldsymbol{\gamma}}_{S} \right]_{i} \!=\! 1$ if the $i$-th codeword in the set $S$ is transmitted by a user, and $\left[ {\boldsymbol{\gamma}}_{S} \right]_{i} = 0$ otherwise. Similarly, let ${\boldsymbol{\Gamma}}^{'}_{S} \!=\! \operatorname{diag} \!\left\{ {\boldsymbol{\gamma}}^{'}_{S} \right\}$, where $\left[ {\boldsymbol{\gamma}}^{'}_{S} \right]_{i} \!=\! 1$ if the $i$-th codeword in the set $S$ is decoded for a user, and $\left[ {\boldsymbol{\gamma}}^{'}_{S} \right]_{i} \!=\! 0$ otherwise.
  Applying the decoding metric given in Appendix~\ref{section5}, we can bound $\mathbb{P} \left[ \mathcal{F}_t  \right]$~as
  \begin{align}
      \mathbb{P} \left[ \mathcal{F}_t  \right]
      &\leq \mathbb{P} \left[
      \bigcup_{ S_1\subset S_{\mathcal{K}}, |S_{1}|=t } \bigcup_{S_2\subset S_{all}\backslash S_{\mathcal{K}}, |S_2|=t}
      \left\{ g\left( {\boldsymbol{\Gamma}}^{'}_{ S_{\mathcal{K}} \backslash S_1 \cup S_2}   \right)
      \leq g \left(\boldsymbol{\Gamma}_{S_{\mathcal{K}}} \right)  \right\}
       \right] \label{eq_Pft_upper_noCSI_scalinglaw00} \\
      &\leq \binom {K} {t} \binom {KM-K} {t} \mathbb{P} \left[ g\left( {\boldsymbol{\Gamma}}^{'}_{ S_{\mathcal{K}} \backslash S_1 \cup S_2}   \right)
      \leq g \left(\boldsymbol{\Gamma}_{S_{\mathcal{K}}} \right)
      \right] \label{eq_Pft_upper_noCSI_scalinglaw0} \\
      &\leq \exp\left\{ t \ln \left( \frac{e^2K^2M}{t^2} \right) \right\}
      \mathbb{P} \left[ g\left( {\boldsymbol{\Gamma}}^{'}_{ S_{\mathcal{K}} \backslash S_1 \cup S_2}   \right)
      \leq g \left(\boldsymbol{\Gamma}_{S_{\mathcal{K}}} \right)
      \right] \label{eq_Pft_upper_noCSI_scalinglaw1},
  \end{align}
  where \eqref{eq_Pft_upper_noCSI_scalinglaw0} follows from the union bound and \eqref{eq_Pft_upper_noCSI_scalinglaw1} holds because $\binom {a } {b} \leq \left( \frac{ea}{b} \right)^{b}$ for $a\geq b>0$.
  Denote ${\mathbf{A}}_{all} = \left\{ {\mathbf{A}} ,  \boldsymbol{\Gamma}_{S_{\mathcal{K}}} ,   {\boldsymbol{\Gamma}}^{'}_{ S_{\mathcal{K}} \backslash S_1 \cup S_2} \right\}$.
  We can upper-bound the probability on the RHS of~\eqref{eq_Pft_upper_noCSI_scalinglaw1}~as
  \begin{align}
    \mathbb{P} \! \left[ g\!\left( {\boldsymbol{\Gamma}}^{'}_{ S_{\mathcal{K}} \backslash S_1 \cup S_2}   \right)
    \!\leq\! g \!\left(\boldsymbol{\Gamma}_{S_{\mathcal{K}}} \right) \right]
    &\!\leq\! \mathbb{E}_{  {\mathbf{A}}_{all} }  \!\!
    \left[ \min_{ u\geq 0 }
    \mathbb{E}_{ \mathbf{H},\mathbf{Z} }\! \! \left[ \left. \exp \left\{ u g \! \left(\boldsymbol{\Gamma}_{S_{\mathcal{K}}} \right)
    \!-\! u g\! \left( {\boldsymbol{\Gamma}}^{'}_{ S_{\mathcal{K}} \backslash S_1 \cup S_2}   \right)
    \right\} \right| {\mathbf{A}}_{all} \right] \right] \label{eq_Pft_upper_noCSI_Chernoff1} \\
    &\!=\! \mathbb{E}_{  {\mathbf{A}} } \! \left[\exp \left\{ \!-L  \left(
    \!-\frac{1}{2} \ln \!\left| \mathbf{F} \right| - \frac{1}{2} \ln \!\left| \mathbf{F}^{'} \right| + \ln \!\left| \frac{1}{2}\mathbf{F} + \frac{1}{2}\mathbf{F}^{'} \right| \right) \right\} \right] \!, \label{eq_Pft_upper_noCSI_Chernoff3}
  \end{align}
  where $\mathbf{F} = \mathbf{I}_n+ \mathbf{A} \boldsymbol{\Gamma}_{S_\mathcal{K}} \mathbf{A}^H$ and
  $\mathbf{F}^{'}  = \mathbf{I}_n+ \mathbf{A} {\boldsymbol{\Gamma}}^{'}_{ S_{\mathcal{K}} \backslash S_1 \cup S_2 }  \mathbf{A}^H$; \eqref{eq_Pft_upper_noCSI_Chernoff1} follows by applying Lemma~\ref{Chernoff_bound} conditioned on ${\mathbf{A}}_{all}$;
  \eqref{eq_Pft_upper_noCSI_Chernoff3} follows from Lemma~\ref{expectation_bound} by allowing $u=\frac{1}{2}$ and taking the expectation over $\mathbf{H}$ and $\mathbf{Z}$, and from the fact that the expectation is unchanged for different $\boldsymbol{\Gamma}_{S_{\mathcal{K}}}$ and $  {\boldsymbol{\Gamma}}^{'}_{ S_{\mathcal{K}} \backslash S_1 \cup S_2}$.
  Substituting~\eqref{eq_Pft_upper_noCSI_Chernoff3} into~\eqref{eq_Pft_upper_noCSI_scalinglaw1}, we obtain an upper bound on $\mathbb{P} \left[ \mathcal{F}_t  \right]$, which is a special case of the upper bound on $\mathbb{P} \left[ \mathcal{F}_t  \right]$ in Appendix~\ref{section5} by allowing $\nu \to \infty$, $\omega=1$, $r=0$, and $u=\frac{1}{2}$.

  Then, we aim to lower-bound $f\!\left( \mathbf{F},\mathbf{F}^{'} \right) \!=\! -\frac{1}{2} \ln \!\left| \mathbf{F} \right| - \frac{1}{2} \ln \!\left| \mathbf{F}^{'} \right| + \ln \!\left| \frac{1}{2}\mathbf{F} \!+\! \frac{1}{2}\mathbf{F}^{'} \right|$ in~\eqref{eq_Pft_upper_noCSI_Chernoff3}. We~have
  \begin{align}
    f\left( \mathbf{F},\mathbf{F}^{'} \right)
    &\geq \frac{1}{8} \operatorname{tr} \left( \left( \mathbf{F}-\mathbf{F}^{'} \right) \left( \frac{\mathbf{F}+\mathbf{F}^{'}}{2} \right)^{-1} \left( \mathbf{F}-\mathbf{F}^{'} \right) \left( \frac{\mathbf{F}+\mathbf{F}^{'}}{2} \right)^{-1} \right) \label{eq_Dkl_lower_noCSI_scalinglaw0} \\
    % &\geq \frac{1}{8} \sigma_{min} \left( \left( \frac{\mathbf{F}+\mathbf{F}^{'}}{2} \right)^{-1} \right) \operatorname{tr} \left( \left( \mathbf{F}-\mathbf{F}^{'} \right) \left( \frac{\mathbf{F}+\mathbf{F}^{'}}{2} \right)^{-1} \left( \mathbf{F}-\mathbf{F}^{'} \right)  \right) \label{eq_Dkl_lower_noCSI_scalinglaw1} \\
    &\geq \frac{1}{8} \sigma_{min}^2 \left( \left( \frac{\mathbf{F}+\mathbf{F}^{'}}{2} \right)^{-1} \right) \operatorname{tr} \left( \left( \mathbf{F}-\mathbf{F}^{'} \right)  \left( \mathbf{F}-\mathbf{F}^{'} \right)  \right) \label{eq_Dkl_lower_noCSI_scalinglaw2} \\
    &= \frac{ \left\| \mathbf{F}-\mathbf{F}^{'} \right\|_F^2 }
    { 8 \sigma_{max}^2\left( \frac{\mathbf{F}+\mathbf{F}^{'}}{2} \right) } , \label{eq_Dkl_lower_noCSI_scalinglaw3}
  \end{align}
  where $\sigma_{min}\left( \mathbf{A} \right)$ (resp., $\sigma_{max}\left( \mathbf{A} \right)$) denotes the minimum (resp., maximum) singular value of $\mathbf{A}$;
  \eqref{eq_Dkl_lower_noCSI_scalinglaw0} follows from Lemma~\ref{Lemma_Dkl_lb} shown below; \eqref{eq_Dkl_lower_noCSI_scalinglaw2} follows by applying the inequality $\operatorname{tr}\left( \mathbf{A}\mathbf{B}\right) \geq \sigma_{min}\left( \mathbf{A} \right) \operatorname{tr}\left( \mathbf{B}\right)$ twice for positive semi-definite matrices $\mathbf{A}$ and $\mathbf{B}$, and from the cyclic property of trace;
  \eqref{eq_Dkl_lower_noCSI_scalinglaw3} follows because the matrix $\frac{\mathbf{F}+\mathbf{F}^{'}}{2}$ is positive definite, $\sigma_{min}\left( \mathbf{A}^{-1}\right) = \frac{1}{\sigma_{max}\left( \mathbf{A} \right)}$ for positive definite matrix $\mathbf{A}$, the matrix $\mathbf{F}-\mathbf{F}^{'}$ is Hermitian, and $\operatorname{tr}\left( \mathbf{B}^{H} \mathbf{B} \right) = \left\| \mathbf{B} \right\|_F^2$.

  \begin{Lemma}[Proposition 1 in \cite{support_recovery}] \label{Lemma_Dkl_lb}
    Let $p_1$ and $p_2$ be two multivariate Gaussian distributions with zero mean and positive definite covariance matrices $\boldsymbol{\Sigma}_1$ and $\boldsymbol{\Sigma}_2$, respectively. Then, the $\frac{1}{2}$-R$\acute{\rm{e}}$nyi divergence between $p_1$ and $p_2$ is bounded as
      \begin{align}
        D_{\frac{1}{2}}\left(p_{1}, p_{2}\right)
        & = -\frac{1}{2} \ln \left| \boldsymbol{\Sigma}_{1} \right| - \frac{1}{2} \ln \left| \boldsymbol{\Sigma}_{2} \right| + \ln \left| \frac{1}{2}\boldsymbol{\Sigma}_{1} + \frac{1}{2} \boldsymbol{\Sigma}_{2} \right|  \\
        & \geq \frac{1}{2} \operatorname{tr}
        \left(\left(\boldsymbol{\Sigma}_{1}-\boldsymbol{\Sigma}_{2}\right)\left(\boldsymbol{\Sigma}_{1}+\boldsymbol{\Sigma}_{2}\right)^{-1}
        \left(\boldsymbol{\Sigma}_{1}-\boldsymbol{\Sigma}_{2}\right)\left(\boldsymbol{\Sigma}_{1}+\boldsymbol{\Sigma}_{2}\right)^{-1}\right).
      \end{align}
  \end{Lemma}

  In the following, we follow similar ideas in~\cite{Caire1} to lower-bound $\left\| \mathbf{F}-\mathbf{F}^{'} \right\|_F^2$ and upper-bound $\sigma_{max}^2\left( \frac{\mathbf{F}+\mathbf{F}^{'}}{2} \right)$, respectively.
  Following from the Restricted Isometry Property (RIP) results in~\cite[Theorems~2 and~5 and Appendix~A]{Caire1}, under the condition that
  \begin{equation}\label{eq_K_cond_noCSI_scalinglaw}
    \frac{2n(n-1)}{M} \leq K \leq \min\left\{ \frac{c_1 n(n-1)}{2\ln^2\left( \frac{eM}{2c_1} \right)},
    \frac{\exp\left\{ \frac{\sqrt{2n(n-1)}}{c_2} \right\}}{4M} \right\},
  \end{equation}
  with probability exceeding $1-\exp\left\{ -c_{\delta} \sqrt{n(n-1)} \right\}$ on a draw of concatenated codebooks of $K$ users, we have
  \begin{equation}
    \left\| \mathbf{F}-\mathbf{F}^{'} \right\|_F^2 \geq (1-\delta) n^2 P^2 \epsilon_1 K, \label{eq_RIP_lower_noCSI_scalinglaw}
  \end{equation}
  where $0<c_1<1$, $c_2>0$, $c_\delta>0$, and $0<\delta<1$ are universal constants.

  Following from the large deviation result~\cite[Theorem~4.6.1]{High_Dimensional_Prob}, an upper bound on $\sigma_{max} \left( \frac{\mathbf{F}+\mathbf{F}^{'}}{2} \right)$ can be derived as
  \begin{equation}
    \sigma_{\max } \left( \frac{\mathbf{F}+\mathbf{F}^{'}}{2} \right)
    \leq P C^{\prime}
    \left( 2\ln\left(\frac{eM}{2}\right) + \frac{\ln(2/\epsilon_2)}{\max\left\{ K,n\right\}} \right)
    \max \left\{ K, n \right\} + 1, \label{eq_daviation_lower_noCSI_scalinglaw}
  \end{equation}
  with probability at least $1-\exp\left(-\beta\max\left(K,n\right)\right)$ for constants $C'>0$ and $\beta>0$.
  Following from~\cite[Appendix~A]{Caire1}, \eqref{eq_daviation_lower_noCSI_scalinglaw} is satisfied independent of transmitted codewords and decoded codewords with probability at least $1-\epsilon_2$ where the positive constant $\epsilon_2$ is less than $\epsilon$.

  Denote by $\mathcal{G}$ the event that \eqref{eq_RIP_lower_noCSI_scalinglaw} and \eqref{eq_daviation_lower_noCSI_scalinglaw} hold for all possible sets of transmitted codewords and decoded codewords.
  If the event $\mathcal{G}$ occurs, \eqref{eq_Dkl_lower_noCSI_scalinglaw3} can be lower-bounded as
  \begin{equation}
      -\frac{1}{2} \ln \left| \mathbf{F} \right| - \frac{1}{2} \ln \left| \mathbf{F}^{'} \right| + \ln \left| \frac{1}{2}\mathbf{F} + \frac{1}{2}\mathbf{F}^{'} \right|
      \geq \frac{m^{*}\epsilon_1 K}{4}  \label{eq_Dt_bound_noCSI_scalinglaw}.
  \end{equation}
  where
  \begin{equation}\label{eq_m_cond_noCSI_scalinglaw}
    m^{*} \geq \frac{ 1-\delta }{2\left( C' \left( 2\ln\left(\frac{eM}{2}\right) + \frac{\ln(2/\epsilon_2)}{\max\left\{ K,n\right\}} \right) \max\left\{ \frac{K}{n}, 1\right\} + \frac{1}{nP} \right)^2} .
  \end{equation}

  Therefore, we have
  \begin{align}
    \mathbb{P}\left[ \bigcup_{t = \lceil \epsilon_1 K \rceil}^{K } \mathcal{F}_t \cap \mathcal{G}\right]
    & \leq \sum_{t = \lceil \epsilon_1 K \rceil}^{K }  \mathbb{P}\left[ \left. \mathcal{F}_t \right| \mathcal{G}\right] \label{eq_pftG_upper_noCSI_scalinglaw1} \\
    & \leq \sum_{t = \lceil \epsilon_1 K \rceil}^{K } \exp\left\{ t \ln \left( \frac{e^2K^2M}{t^2} \right) \right\}
    \mathbb{P} \left[ \left. g\left( {\boldsymbol{\Gamma}}^{'}_{ S_{\mathcal{K}} \backslash S_1 \cup S_2}   \right) \leq g \left(\boldsymbol{\Gamma}_{S_{\mathcal{K}}} \right) \right| \mathcal{G}
    \right]  \label{eq_pftG_upper_noCSI_scalinglaw2}   \\
    & \leq \sum_{t = \lceil \epsilon_1 K \rceil}^{K } \exp \left\{  t \ln \left( \frac{e^2K^2M}{t^2} \right)
    -L \frac{m^{*}\epsilon_1 K}{4}  \right\}  \label{eq_pftG_upper_noCSI_scalinglaw3} \\
    & \leq (1-\epsilon_1)K  \exp \left\{  K \left( \ln \left( \frac{e^2 M}{\epsilon_1^2} \right)
    - \frac{m^{*}L\epsilon_1}{4}  \right)  \right\} \label{eq_PUPE_upper_noCSI_scalinglaw2},
  \end{align}
  where \eqref{eq_pftG_upper_noCSI_scalinglaw1} follows from the union bound and the inequality that $\mathbb{P}\left[ \mathcal{G}_1 \cap \mathcal{G}_2 \right] \!=\! \mathbb{P}\left[ \mathcal{G}_2  \right] \mathbb{P}\left[ \left. \mathcal{G}_1 \right| \mathcal{G}_2 \right] \!\leq\! \mathbb{P}\left[ \left. \mathcal{G}_1 \right| \mathcal{G}_2 \right]$ for events $\mathcal{G}_1$ and $\mathcal{G}_2$;
  \eqref{eq_pftG_upper_noCSI_scalinglaw2} follows from~\eqref{eq_Pft_upper_noCSI_scalinglaw1};
  \eqref{eq_pftG_upper_noCSI_scalinglaw3} follows from~\eqref{eq_Pft_upper_noCSI_Chernoff3} and~\eqref{eq_Dt_bound_noCSI_scalinglaw}.
  % Since~\eqref{eq_Pft_upper_noCSI_Chernoff3} is obtained by applying the Chernoff bound with respect to the noise matrix, the probability in~\eqref{eq_pftG_upper_noCSI_scalinglaw2} conditioned on the event $\mathcal{G}$ can be bounded similarly.
  In the case of
  \begin{equation}\label{eq_L_cond_noCSI_scalinglaw}
    c = \frac{m^{*}L\epsilon_1}{4} - \ln \left( \frac{e^2 M}{\epsilon_1^2} \right) > 0,
  \end{equation}
  we have $\mathbb{P}\left[ \bigcup_{t = \lceil \epsilon_1 K \rceil}^{K } \mathcal{F}_t \cap \mathcal{G}\right] \leq \exp\left\{ o(K) - c   K \right\}$.
  Together with \eqref{eq_PUPE_upper_noCSI_scalinglaw}, we have
  \begin{align}
    P_{e} & \leq \epsilon_1 + \mathbb{P}\left[ \bigcup_{t = \lceil \epsilon_1 K \rceil}^{K } \mathcal{F}_t \cap \mathcal{G}\right]  +  \mathbb{P}\left[\mathcal{G}^c\right] \\
    & \leq  \epsilon_1 + \exp\left\{ o(K) - c  K \right\} + \exp\left\{ - c_{\delta} (n-1) \right\} + \epsilon_2 \label{eq_PUPE_upper_noCSI_scalinglaw4},
  \end{align}
  where $\epsilon_1 + \epsilon_2 < \epsilon$, and $c, c_{\delta} > 0$ are universal constants.
  As $n,K\to\infty$, the error requirement $P_e\leq\epsilon$ in \eqref{PUPE} is satisfied.

  Combining \eqref{eq_K_cond_noCSI_scalinglaw}, \eqref{eq_m_cond_noCSI_scalinglaw}, \eqref{eq_L_cond_noCSI_scalinglaw}, and \eqref{eq_PUPE_upper_noCSI_scalinglaw4}, we can obtain the following scaling law.
  Supposing $M=\Theta(1)$ and $n,K,L\to\infty$, it is possible to serve $K = \Theta(n^2)$ users with $L=\Theta\left(n^2\right)$ BS antennas and power $P=\Theta\left(\frac{1}{n^2}\right)$, such that the PUPE constraint is satisfied.
  It was proved in~\cite[Appendix~A-C]{finite_payloads_fading} that, if one can achieve a certain PUPE for $K$ users, it will also be possible to achieve the same PUPE for less than $K$ users.
  As a result, under the PUPE criterion, we can reliably serve $K = \mathcal{O}(n^2)$ users when $L=\Theta\left(n^2\right)$,  $P=\Theta\left(\frac{1}{n^2}\right)$, and $M=\Theta(1)$.

%  The scaling law for the activity detection problem is established in~\cite[Corollary~1]{Caire1}, where the finite SNR regime is considered.
%  We remark that, following from (101) in~\cite{Caire1}, the scaling law in~\cite[Corollary~1]{Caire1} continues to hold if the SNR is in the order of $\Theta\left(\frac{1}{n^2}\right)$, i.e., if $P$ is in the order of $\Theta\left(\frac{1}{n^2}\right)$ when the noise variance $\sigma^2=1$.
%  As a result, the scaling law of the activity detection problem in~\cite{Caire1} can be extended to that of the data detection problem as follows: under the joint error probability criterion, with a coherence block of dimension $n\to \infty$ and a sufficient number of BS antennas $L = \Theta\left(n^2 \ln n\right)$, one can reliably serve up to $K = \mathcal{O}\left(n^2\right)$ users when the payload $J = \Theta(1)$ and the power $P=\Theta\left(\frac{1}{n^2}\right)$ in the case of no-CSI.

\subsection{ Converse  } \label{Proof_scalinglaw_noCSI_conv}
  We assume $n,L\to \infty$ and $\epsilon$ and $M$ are positive finite constants.
  Recall that $b_1=J{\left(1 - \epsilon\right)} - h_2 \left( \epsilon \right)$.
  Following from Theorem \ref{prop_converse_noCSI}, in the case of $K >n$, the minimum required energy-per-bit is larger than $\inf \frac{nP}{J}$, where the infimum is taken over all $P>0$ satisfying
  \begin{equation}
    b_1  \leq \frac{nL}{K } \log_2 \left( 1 + K P \right)
    - \frac{L}{K } \sum_{i=1}^{n} \left( \psi(K -i+1) \log_2 e +  \log_2  \!\left( {P + \frac{1}{ K -i + 1 }} \right) \right) \label{Proof_P_tot_conv_noCSI_scalinglaw_loose_nless} ,
  \end{equation}
  where $\psi(\cdot)$ is Euler's digamma function.
  The RHS of~\eqref{Proof_P_tot_conv_noCSI_scalinglaw_loose_nless} can be further bounded and we have
  \begin{align}
    b_1 & \leq  \frac{L}{K } \sum_{i=1}^{n} \left( \log_2 \left( \frac{1 + K P}{1+(K -i+1)P}\right) + \frac{\log_2 e}{K -i+1}  \right)  \label{Proof_P_tot_conv_noCSI_scalinglaw_loose_nless2_1} \\
    & \leq  \frac{nL}{K } \left( \log_2 \left(  \frac{1 + K P}{1+(K -n+1)P}  \right) + \frac{\log_2 e}{K -n+1}  \right) \\
    & \leq  \frac{nL\log_2 e}{K } \left(  \frac{(n-1) P}{1+(K -n+1)P} + \frac{1}{K -n+1}  \right)  ,\label{Proof_P_tot_conv_noCSI_scalinglaw_loose_nless2}
  \end{align}
  where \eqref{Proof_P_tot_conv_noCSI_scalinglaw_loose_nless2_1} follows by applying the inequality $\psi(x)\geq \ln x - \frac{1}{x}$ for $x>0$ into~\eqref{Proof_P_tot_conv_noCSI_scalinglaw_loose_nless}; \eqref{Proof_P_tot_conv_noCSI_scalinglaw_loose_nless2} follows because $\log_2(1+x)\leq x\log_2 e$ for $x\geq 0$.
  It is evident that the RHS of~\eqref{Proof_P_tot_conv_noCSI_scalinglaw_loose_nless2} is a monotonically decreasing function of $K $.
  In the case of $L=\Theta(n^2)$ and $P=\Theta\left(\frac{1}{n^2}\right)$, the condition in~\eqref{Proof_P_tot_conv_noCSI_scalinglaw_loose_nless2} is satisfied if and only if the number of users satisfies $n < K \leq \Theta(n^2)$.

  In the case of $1\leq K\leq n$, following from Theorem~\ref{prop_converse_noCSI}, the minimum required energy-per-bit is larger than $\inf \frac{nP}{J}$, where the infimum is taken over all $P>0$ satisfying
  \begin{equation}
    b_1  \leq ML \log_2 \left( 1 + \frac{nP}{M} \right)
    - \frac{L}{K}  \sum\limits_{i=0}^{K-1} \left( \psi(n-i) \log_2 e +  \log_2  \left( {P + \frac{1}{ n -i }} \right) \right) \label{Proof_P_tot_conv_noCSI_scalinglaw_loose_Kless}.
  \end{equation}
  Similar to~\eqref{Proof_P_tot_conv_noCSI_scalinglaw_loose_nless2}, the RHS of~\eqref{Proof_P_tot_conv_noCSI_scalinglaw_loose_Kless} can be further upper-bounded and we have
  \begin{equation}
    b_1 \leq ML \log_2 \left( 1 + \frac{nP}{M} \right)
    - L \log_2\left(1 + (n-K+1)P \right)
    + \frac{ L \log_2 e }{n-K+1}. \label{Proof_P_tot_conv_noCSI_scalinglaw_loose_Kaless_scale2}
  \end{equation}
  In the case of $K=1$, \eqref{Proof_P_tot_conv_noCSI_scalinglaw_loose_Kaless_scale2} reduces to
  \begin{equation}
    b_1 \leq ML \log_2 \left( 1 + \frac{nP}{M} \right)
    - L \log_2\left(1 + nP \right)
    + \frac{ L \log_2 e }{n} \label{Proof_P_tot_conv_noCSI_scalinglaw_loose_Kaless_scale2_K1}.
  \end{equation}
  It is evident that $ML \log_2 \left( 1 + \frac{nP}{M} \right) - L \log_2\left(1 + nP \right) \geq 0$.
  Thus, when $L=\Theta(n^2)$ and $P=\Theta\left(\frac{1}{n^2}\right)$, the condition in~\eqref{Proof_P_tot_conv_noCSI_scalinglaw_loose_Kaless_scale2_K1} is satisfied for $K=1$.
  Since the RHS of~\eqref{Proof_P_tot_conv_noCSI_scalinglaw_loose_Kaless_scale2} is a monotonically increasing function of $K$, \eqref{Proof_P_tot_conv_noCSI_scalinglaw_loose_Kaless_scale2} is satisfied for any $1\leq K\leq n$ in this case.

  Taking both the cases of $1\leq K\leq n$ and $K>n$ into consideration, we can draw the conclusion that when $M = \Theta(1)$, $n\to\infty$, $L=\Theta(n^2)$, and $P=\Theta\left(\frac{1}{n^2}\right)$, all users can be reliably served only if $K=\mathcal{O}(n^2)$.

  Together with the achievability result in Appendix~\ref{Proof_scalinglaw_noCSI_achi}, we conclude that assuming $M = \Theta(1)$ and $n\to\infty$, with $L = \Theta \left(n^2\right)$ BS antennas and the power $P=\Theta\left(\frac{1}{n^2}\right)$, one can satisfy the error requirement if and only if the number of users is $K = \mathcal{O}(n^2)$ when all users are active and there is no \emph{a priori} CSI at the receiver.

\section{ Proof of Theorem \ref{Theorem_noCSI_achi_pilot}  } \label{section6}
  In this appendix, we prove Theorem \ref{Theorem_noCSI_achi_pilot} to establish an achievability bound for the pilot-assisted coded access scheme.
  Specifically, we consider a special case where all users are active, i.e. $K_a=K$.
  We use pilots drawn uniformly at random on an $n_p$-dimensional sphere of radius $\sqrt{n_p P_p}$.
  Thus, these pilots, denoted as $\mathbf{b}_{1}, \ldots, \mathbf{b}_{K} $ with length $n_p$, satisfy that $\left\|\mathbf{b}_{k}\right\|_2^2 = n_p P_p$ for $k\in\mathcal{K}$.
  Denote $\mathbf{B} = \left[ \mathbf{b}_1,  \ldots, \mathbf{b}_{K}\right] \in \mathbb{C}^{n_p\times K}$.
  The received signal of the $l$-th antenna at the BS in the pilot transmission phase is given by
  \begin{equation}
    \mathbf{y}_{l,p} = \sum_{k\in{\mathcal{K}}}{h}_{k,l} {\mathbf{b}}_{k}+\mathbf{z}_{l,p} = {\mathbf{B}} \mathbf{h}_l +\mathbf{z}_{l,p}  \in \mathbb{C}^{ n_p },
  \end{equation}
  where ${h}_{k,l} \sim \mathcal{CN}(0,1)$ denotes the fading coefficient between the $k$-th user and the $l$-th antenna of the BS, which is i.i.d. across different users and different BS antennas;
  the vector $ \mathbf{h}_l = \left[ {h}_{1,l}, \ldots,{h}_{K,l} \right]^T \in \mathbb{C}^{K} $;
  the noise vector $\mathbf{z}_{l,p} \sim \mathcal{CN}(\mathbf{0},\mathbf{I}_{n_p})$, which is i.i.d. across $L$ BS antennas.

  The BS performs MMSE channel estimation. The estimated channel for the $l$-th antenna of the BS is given by
  \begin{equation}
    \hat{\mathbf{h}}_l  = \left( \mathbf{I}_{K } + {\mathbf{B}}^H {\mathbf{B}} \right)^{-1} {\mathbf{B}}^H \mathbf{y}_{l,p},
  \end{equation}
  where $\hat{\mathbf{h}}_l = \left[ \hat{h}_{1,l}, \ldots, \hat{h}_{K ,l} \right]^T \in \mathbb{C}^{K }$ is distributed as $\hat{\mathbf{h}}_l \sim \mathcal{CN} \left( \mathbf{0}, \hat{\mathbf{\Sigma}} \right)$ with $\hat{\mathbf{\Sigma}} = \mathbf{I}_{K } - \left( \mathbf{I}_{K } + {\mathbf{B}}^H {\mathbf{B}} \right)^{-1} $.
  From the orthogonality principle of the MMSE estimation, the channel estimation error $ \tilde{\mathbf{h}}_l = {\mathbf{h}}_l - \hat{\mathbf{h}}_l = \left[ \tilde{h}_{1,l}, \ldots, \tilde{h}_{K ,l} \right]^T$ is independent of $\hat{\mathbf{h}}_l$, and is distributed as $\tilde{\mathbf{h}}_l \sim \mathcal{CN} \left( \mathbf{0}, \tilde{\mathbf{\Sigma}} \right)$ with $\tilde{\mathbf{\Sigma}} = \left( \mathbf{I}_{K } + {\mathbf{B}}^H {\mathbf{B}} \right)^{-1} $.
  For fixed pilot matrix $\mathbf{B}$, both the channel estimation $\hat{\mathbf{h}}_l$ and the channel estimation error $ \tilde{\mathbf{h}}_l$ are i.i.d. across $L$ BS antennas.

  Similar to Appendix~\ref{Proof_achi_CSIR_noCSI}, we use a random coding scheme in the data transmission phase by generating Gaussian codebooks of size $M$ and length $n_d = n-n_p$ without power control, which for the $k$-th user is denoted as $\mathcal{C}_k = \left\{\mathbf{c}_{k,1},\ldots, \mathbf{c}_{k,M}\right\}$ with $\mathbf{c}_{k,m}\stackrel{i.i.d.}{\sim} \mathcal{CN}\left(0,P'\mathbf{I}_{n_d}\right)$ for $k\in\mathcal{K}$ and $m\in[M]$.
  We choose $P'<\frac{n P - n_p P_p}{n_d}$ to ensure that we can control the maximum power constraint violation event.
  The matrix $\mathbf{A} \in \mathbb{C}^{n_d\times MK }$ denotes the concatenation of codebooks of the $K $ users.
  Let the transmitted codeword of user $k$ be $\mathbf{x}_{(k)} = \mathbf{c}_{(k)} 1 \left\{ \left\|\mathbf{c}_{(k)}\right\|_{2}^{2} \leq n P - n_p P_p \right\} $, where $\mathbf{c}_{(k)} = \mathbf{c}_{k,W_{k}}$ with the message $ W_{k} \in [M]$ chosen uniformly at random.
  The received signal of the $l$-th antenna in the data transmission phase is given by
  \begin{equation}
    \mathbf{y}_{l,d} = \sum_{k\in{\mathcal{K}}}{h}_{k,l} \mathbf{x}_{(k)} + \mathbf{z}_{l,d}  \in \mathbb{C}^{ n_d },
  \end{equation}
  where ${h}_{k,l}$ is defined as aforementioned, and the noise vector $\mathbf{z}_{l,d} \sim \mathcal{CN}(\mathbf{0},\mathbf{I}_{n_d})$, which are i.i.d. across $L$ antennas.
  Denote the signals received over $L$ antennas as $ \mathbf{Y}_d = \left[ \mathbf{y}_{1,d},\ldots,\mathbf{y}_{L,d} \right] \in \mathbb{C}^{n_d\times L}$.

  The decoder has an incorrect estimate of the channel, but uses the estimate as if it were perfect.
  Based on the notation in Appendix~\ref{Proof_achi_CSIR_noCSI}, the decoding metric in this case is given~by
  \begin{equation}\label{eq_g_noCSI_pilot}
    g\left(  \mathbf{Y}_d , \hat{\mathbf{c}}_{[ \mathcal{K} ]} \right) = \sum_{l=1}^{L}\left\|\mathbf{y}_{l,d}- \sum_{k\in  {\mathcal{K}} } \hat{h}_{k,l} \hat{\mathbf{c}}_{(k)} \right\|_{2}^{2}.
  \end{equation}
  The decoder outputs
      \begin{equation}\label{decoder_noCSI_pilot}
        \hat{\mathbf{c}}_{[ \mathcal{K} ]}
        =  \min_{\left( \hat{\mathbf{c}}_{(k)} \in \mathcal{C}_k \right)_{k\in {\mathcal{K}} }}
        g\left(  \mathbf{Y}_d , \hat{\mathbf{c}}_{[ \mathcal{K} ]} \right) ,
      \end{equation}
      \begin{equation}
        \hat{W}_k =  \emph{f}_{\text{en},k}^{-1}\left(\hat{\mathbf{c}}_{(k)}\right) , \;\; k \in {\mathcal{K}} .
      \end{equation}

  We can upper-bound the PUPE as in~\eqref{eq_PUPE_upper_CSIR_noCSI} by allowing $K_a=K$, where the total variation distance $p_0$ for the pilot-assisted scheme is given by
      \begin{equation}
         p_0 = K\; \mathbb{P}\left[  \left\|\mathbf{c}_{(k)}\right\|_{2}^{2} > n P - n_p P_p \right]
         = K \left( 1 - \frac{\gamma\left( n_d , \left( n P - n_p P_p \right)/{{P}^{\prime}} \right)}{\Gamma\left( n_d \right)} \right) .
      \end{equation}
  In the remainder of this appendix, we upper-bound $\mathbb{P} \left[ \mathcal{F}_t \right]$ in~\eqref{eq_PUPE_upper_CSIR_noCSI} relying on the standard bounding technique proposed by Fano~\cite{1961}.
  Compared with the case of CSIR, upper-bounding $\mathbb{P} \left[ \mathcal{F}_t \right]$ is more involved for the pilot-assisted coded access scheme due to the channel estimation error.
  Hence, we simplify the ``good region'' introduced in Section~\ref{section3_sub_goodregion} by allowing $w=0$ and obtain
  \begin{equation}
    \mathcal{R}  =  \left\{ \mathbf{Y}_d : g\left(  \mathbf{Y}_d, \mathbf{c}_{[\mathcal{K}_a]}  \right)
    \leq  n_d L\nu\right\}.
  \end{equation}
  Define the event $\mathcal{G}_{\nu} = \left\{ \mathbf{Y}_d \in \mathcal{R} \right\}$.
  By replacing $\mathcal{G}_{\omega,\nu}$ with $\mathcal{G}_{\nu}$ and allowing $S_1=S_2$, the upper bound on $\mathbb{P}\left[ \mathcal{F}_t  \right]$ in~\eqref{eq_pft_goodregion_CSIR_noCSI} becomes
    \begin{align}
      \mathbb{P} \left[ \mathcal{F}_t  \right]
      &\leq  \min_{ \nu\geq0 }
      \left\{  \mathbb{P} \left[
      \bigcup_{S_{1}}
      \bigcup_{\mathbf{c}^{'}_{[S_1]}}
      \left\{ g \left(  \mathbf{Y}_d ,
      \mathbf{c}_{[ {\mathcal{K}}_a \backslash S_1]} \cup \mathbf{c}^{'}_{[S_1]}   \right)
      \leq  g \left(  \mathbf{Y}_d ,
      \mathbf{c}_{[ {\mathcal{K}}_a ]}   \right) \right\} \bigcap \mathcal{G}_{ \nu} \right]
      + \mathbb{P}  \left[ \mathcal{G}_{ \nu}^c \right] \right\} \label{eq_pft2_goodregion_pilot} \\
      & = \min_{ \nu\geq 0 }
       \left\{ \mathbb{P}  \left[\mathcal{G}_{e} \cap \mathcal{G}_{\nu} \right]
       + \mathbb{P}  \left[ \mathcal{G}_{\nu}^c \right] \right\} \label{eq_pft2_union_pilot} .
    \end{align}
  In the following, we bound $\mathbb{P}  \left[\mathcal{G}_{e} \cap \mathcal{G}_{\nu} \right]$ and $\mathbb{P}  \left[ \mathcal{G}_{\nu}^c \right]$, respectively.

  Define $\tilde{\mathbf{A}}_{S_1}$, $\tilde{\mathbf{A}}^{'}_{S_1}$, and $\tilde{\mathbf{A}}_{\mathcal{K} }$ as in Theorem~\ref{Theorem_noCSI_achi_pilot}.
  Denote $\mathbf{H} = \left[ {\mathbf{h}}_1,\ldots,{\mathbf{h}}_L \right]\in \mathbb{C}^{K\times L}$,
  $\hat{\mathbf{H}} = \left[ \hat{\mathbf{h}}_1,\ldots,\hat{\mathbf{h}}_L \right]\in \mathbb{C}^{K\times L}$, $\tilde{\mathbf{H}} = \left[ \tilde{\mathbf{h}}_1,\ldots,\tilde{\mathbf{h}}_L \right]\in \mathbb{C}^{K\times L}$,
  and $ \mathbf{Z}_d = \left[ \mathbf{z}_{1,d},\ldots,\mathbf{z}_{L,d} \right]\in \mathbb{C}^{n_d\times L}$.
  Denote ${\mathbf{A}}_{all} = \left\{ \tilde{\mathbf{A}}_{\mathcal{K}}, \tilde{\mathbf{A}}_{S_1}, \tilde{\mathbf{A}}^{'}_{S_1} \right\}$.
  Using the above notation, we can obtain
    \begin{align}
      \mathbb{P}  \left[\mathcal{G}_{e} \cap \mathcal{G}_{\nu} \right]
%      & \leq \sum_{S_{1}} \sum_{\mathbf{c}^{'}_{[S_1]}}
%      \mathbb{P} \left[ \left\{ \left\| \mathbf{Z}_d + \tilde{\mathbf{A}}_{\mathcal{K}} \tilde{\mathbf{H}} + \left( \tilde{\mathbf{A}}_{S_1} - \tilde{\mathbf{A}}^{'}_{S_1} \right) \hat{\mathbf{H}} \right\|_F^2
%      \leq \left\| \mathbf{Z}_d + \tilde{\mathbf{A}}_{\mathcal{K}} \tilde{\mathbf{H}} \right\|_F^2 \right\} \right. \notag\\
%      &  \;\;\;\;\;\;\;\;\;\;\;\;\;\;\;\;\;\;\; \left.
%      \cap  \left\{ \left\| \mathbf{Z}_d + \tilde{\mathbf{A}}_{\mathcal{K}} \tilde{\mathbf{H}} \right\|_F^2
%      \leq n_d L\nu\right\}   \right]  \label{eq_q1t_goodregion_pilot}  \\
      & \leq \sum_{S_{1}} \sum_{\mathbf{c}^{'}_{[S_1]}} \mathbb{E}_{ {\mathbf{A}}_{all} , \mathbf{B} } \left[ \min_{u\geq 0,r\geq 0} \exp\left\{ r n_d L \nu \right\}
      \mathbb{E}_{ \mathbf{Z}_d ,\tilde{\mathbf{H}}, \hat{\mathbf{H}} } \left[ \left. \exp \left\{  \left( u-r \right) \left\| \mathbf{Z}_d + \tilde{\mathbf{A}}_{\mathcal{K}} \tilde{\mathbf{H}} \right\|_F^2   \right. \right. \right. \right. \notag \\
      & \;\;\;\;\;\;\;\;\;\;\;\;\;\;\;\;\;\;\;\;\;\;\;\;\;\;\;\;
      \;\;\;
      \left. \left. \left. \left.
      - u \left\| \mathbf{Z}_d + \tilde{\mathbf{A}}_{\mathcal{K}} \tilde{\mathbf{H}} + \left( \tilde{\mathbf{A}}_{S_1} - \tilde{\mathbf{A}}^{'}_{S_1} \right) \hat{\mathbf{H}}\right\|_F^2 \right\} \right| {\mathbf{A}}_{all} , \mathbf{B} \right]  \right]  \label{eq_q1t_chernoff_pilot} \\
      & \leq {\binom {K } {t}}  M^t  \mathbb{E}_{ {\mathbf{A}}_{all}, \mathbf{B} } \left[
      \min_{{u\geq 0,r\geq 0,\lambda_{\min}\left(\mathbf{D}\right) > 0}}
      \exp\left\{ r n_d L\nu -\frac{L}{2} \ln \left| \mathbf{D} \right| \right\}  \right], \label{eq_q1t_exp_pilot}
    \end{align}
  where \eqref{eq_q1t_chernoff_pilot} follows by applying the Chernoff bound in Lemma~\ref{Chernoff_bound} to the probability $\mathbb{P}  \left[\mathcal{G}_{e} \cap \mathcal{G}_{\nu} \right]$ conditioned on ${\mathbf{A}}_{all}$ and $\mathbf{B}$;
  \eqref{eq_q1t_exp_pilot} follows from Lemma~\ref{expectation_bound} by taking the expectation over $\tilde{\mathbf{H}}, \hat{\mathbf{H}}$, and $\mathbf{Z}_d$ provided that the eigenvalues of $\mathbf{D}$ are positive, with the expression of $\mathbf{D}$ given in~\eqref{eq_noCSI_pilot_D}.
  The term on the RHS of \eqref{eq_q1t_exp_pilot} is denoted as $q_{1,t} \left( \nu \right)$ as presented in \eqref{eq_noCSI_pilot_q1t}.

  In the remainder of this appendix, we upper-bound $\mathbb{P}  \left[ \mathcal{G}_{\nu}^c \right]$ in two ways.
  Let $\lambda_1,\ldots,\lambda_{n_d}$ denote the eigenvalues of $\tilde{\mathbf{A}}_{\mathcal{K}}  \tilde{\mathbf{\Sigma}} \tilde{\mathbf{A}}_{\mathcal{K}}^H$ in decreasing order with rank $n^{*} = \min\left\{ K,n_d \right\}$.
  First, applying the Chernoff bound and Lemma~\ref{expectation_bound}, we have
    \begin{equation}
      \mathbb{P}  \left[ \mathcal{G}_{\nu}^c \right]
      \leq \mathbb{E}_{ \tilde{\mathbf{A}}_{\mathcal{K}} , \mathbf{B} } \! \left[ \min_{0\leq \delta < 1/\left(1+\lambda_{1} \right) }  \exp\left\{ - \delta n_d L \nu  \right\}
      \left| \left( 1-\delta \right) \mathbf{I}_{n_d} - \delta \tilde{\mathbf{A}}_{\mathcal{K}}  \tilde{\mathbf{\Sigma}} \tilde{\mathbf{A}}_{\mathcal{K}} ^H \right|^{-L}   \right] \label{eq_q2t_method1_pilot} .
    \end{equation}
  Define the event $\mathcal{G}_{\eta} = \left\{ \frac{ \chi^2\left(2n_dL\right) }{2} \leq n_dL\eta \right\}$ for $\eta\geq0$.
  Alternatively, we have
    \begin{align}
      \mathbb{P}  \left[ \mathcal{G}_{\nu}^c \right]
      & = \mathbb{E}_{ \tilde{\mathbf{A}}_{\mathcal{K}} , \mathbf{B} } \! \left[ \mathbb{P} \left[ \left. \sum_{l \in [L]} \left\| \mathbf{z}_{l,d} + \tilde{\mathbf{A}}_{\mathcal{K}} \tilde{\mathbf{h}}_l \right\|_2^2
>  n_d L \nu \right| \tilde{\mathbf{A}}_{\mathcal{K}} , \mathbf{B} \right]  \right] \\
      %& = \mathbb{E}_{\blue \tilde{\mathbf{A}}_{\mathcal{K}} , \mathbf{B} } \! \left[ \mathbb{P} \left[ \left. \sum_{i = 1}^{n_d} \left( 1+\lambda_i \right) \frac{ \chi^2_i\left(2L\right) }{2}
%>  n_d L \nu \right| \tilde{\mathbf{A}}_{\mathcal{K}} , \mathbf{B} \right]  \right] \\
      & \leq \!\min_{0\leq\eta\leq \nu} \!\!\left\{ \! \mathbb{E}_{ \tilde{\mathbf{A}}_{\mathcal{K}} , \mathbf{B} } \! \left[ \mathbb{P}\! \left[ \!\left. \left\{ \!\frac{ \chi^2\!\left(2n_dL\right) }{2}
      \!+\! \sum_{i = 1}^{n^{*}} \! \frac{ \lambda_i\chi^2_i\!\left(2L\right) }{2}
      \!>\!  n_d L \nu \!\right\}
\cap \mathcal{G}_{\eta}  \right| \tilde{\mathbf{A}}_{\mathcal{K}}, \mathbf{B}  \right]  \right]
\!+ \!\mathbb{P}\! \left[ \mathcal{G}_{\eta}^c \right] \!\right\} \label{eq_q2t_goodregion_pilot}\\
      & \leq \!\min_{0\leq\eta\leq \nu} \!\!\left\{ \! \mathbb{E}_{ \tilde{\mathbf{A}}_{\mathcal{K}} , \mathbf{B} } \!  \left[ \mathbb{P} \!\left[ \left. \sum_{i = 1}^{n^{*}} \!\frac{ \lambda_i \chi^2_i\left(2L\right) }{2}
\!>\! n_d L (\nu - \eta ) \right| \tilde{\mathbf{A}}_{\mathcal{K}} , \mathbf{B} \right]  \right]
\!+ 1 - \frac{\gamma\left( n_dL,n_d L \eta \right)}{\Gamma\left( n_dL \right)} \!\right\} \label{eq_q2t_cap_pilot}\!,
    \end{align}
  where the conditional probability on the RHS of \eqref{eq_q2t_cap_pilot} can be further upper-bounded as
    \begin{align}
      \mathbb{P} \!\left[ \left. \sum_{i = 1}^{n^{*}} \lambda_i \frac{ \chi^2_i\left(2L\right) }{2}
>  n_d L \left(\nu - \eta \right) \right| \tilde{\mathbf{A}}_{\mathcal{K}} , \mathbf{B} \right]
      & \leq  \mathbb{P} \!\left[ \left. \lambda_1 \frac{ \chi^2\left(2Ln^{*}\right) }{2}
>  n_d L \left(\nu - \eta \right) \right| \tilde{\mathbf{A}}_{\mathcal{K}} , \mathbf{B} \right]  \\
      & =  1 - \frac{\gamma\left( Ln^{*}, \frac{ n_d L \left(\nu - \eta \right)}{\lambda_1} \right)}{\Gamma\left( Ln^{*} \right)} \label{eq_q2t_term1_pilot}.
    \end{align}

  % Substituting \eqref{eq_q2t_term2_pilot} and \eqref{eq_q2t_term1_pilot} into \eqref{eq_q2t_cap_pilot}, we can obtain an upper bound on $\mathbb{P}  \left[ \mathcal{G}_{\nu}^c \right]$.
  Taking the minimum value of \eqref{eq_q2t_method1_pilot} and \eqref{eq_q2t_cap_pilot}, we can obtain the ultimate upper bound on $\mathbb{P}  \left[ \mathcal{G}_{\nu}^c \right]$, which is denoted as $q_{2,t} \left( \nu \right)$ as in~\eqref{eq_noCSI_pilot_q2t}. This concludes the proof of Theorem \ref{Theorem_noCSI_achi_pilot}.


\begin{thebibliography}{00}
% 1
\bibitem{classical_MAC_Liao} H. Liao, ``A coding theorem for multiple access communications,'' in \emph{Proc. IEEE Int. Symp. Inform. Theory (ISIT)}, Pacific Grove, USA, Jan. 1972, pp. 1--5.
% 2
\bibitem{classical_MAC_Ahlswede} R. Ahlswede, ``Multi-way communication channels,'' in \emph{Proc. IEEE Int. Symp. Inform. Theory (ISIT)}, Tsahkadsor, USSR, Sep. 1971, pp. 23--52.
% 3
\bibitem{elements_IT} T. M. Cover and J. A. Thomas, \emph{Elements of Information Theory}, 2nd ed. Hoboken, NJ, USA: John Wiley $\&$ Sons, 2006.
% 4
\bibitem{GuoDN} X. Chen, T.-Y. Chen, and D. Guo, ``Capacity of Gaussian many-access channels,'' \emph{IEEE Trans. Inf. Theory}, vol. 63, no. 6, pp. 3516--3539, Feb. 2017.
% 5
\bibitem{WeiF} F. Wei, Y. Wu, W. Chen, W. Yang, and G. Caire, ``On the fundamental limits of MIMO massive multiple access channels,'' in \emph{Proc. IEEE Int. Conf. Commun. (ICC)}, Shanghai, China, May 2019.
% 6
\bibitem{A_perspective_on} Y. Polyanskiy, ``A perspective on massive random-access,'' in \emph{Proc. IEEE Int. Symp. Inf. Theory (ISIT)}, Aachen, Germany, Jun. 2017, pp. 2523--2527.
% 7
\bibitem{improved_bound} I. Zadik, Y. Polyanskiy, and C. Thrampoulidis, ``Improved bounds on Gaussian MAC and sparse regression via Gaussian inequalities,'' in \emph{Proc. IEEE Int. Symp. Inf. Theory (ISIT)}, Paris, France, Jul. 2019, pp. 430--434.
% 8
\bibitem{finite_payloads_fading} S. S. Kowshik and Y. Polyanskiy, ``Fundamental limits of many-user MAC with finite payloads and fading,'' \emph{IEEE Trans. Inf. Theory}, vol. 67, no. 9, pp. 5853--5884, Sep. 2021.
% 9
\bibitem{GaoJY} J. Gao, Y. Wu and W. Zhang, ``Energy-efficiency of massive random access with individual codebook,'' in \emph{Proc. IEEE Global Commun. (GLOBECOM)}, Dec. 2020.
% 10
\bibitem{wuyp} Y. Wu, X. Gao, S. Zhou, W. Yang, Y. Polyanskiy, and G. Caire, ``Massive access for future wireless communication systems,'' \emph{IEEE Wireless Commun.}, vol. 27, no. 4, pp. 148--156, Aug. 2010.
% 11
\bibitem{Channel_coding_rate} Y. Polyanskiy, H. V. Poor, and S. Verd\'u, ``Channel coding rate in the finite blocklength regime,'' \emph{IEEE Trans. Inf. Theory}, vol. 56, no. 5, pp. 2307--2359, May 2010.
% 12
\bibitem{TDMA_yangwei} W. Yang, G. Durisi, T. Koch, and Y. Polyanskiy, ``Quasi-static multiple-antenna fading channels at finite blocklength,'' \emph{IEEE Trans. Inf. Theory}, vol. 60, no. 7, pp. 4232--4265, Jul. 2014.
% 13
\bibitem{K_MAC} E. MolavianJazi and J. N. Laneman, ``A second-order achievable rate region for Gaussian multi-access channels via a central limit theorem for functions,'' \emph{IEEE Trans. Inf. Theory}, vol. 61, no. 12, pp. 6719--6733, Dec. 2015.
% 14
\bibitem{K_MAC2} R. C. Yavas, V. Kostina, and M. Effros, ``Gaussian multiple and random access channels: Finite-blocklength analysis,'' \emph{IEEE Trans. Inf. Theory}, vol. 67, no. 11, pp. 6983--7009, Nov. 2021.
% 15
\bibitem{indivicommon} R. C. Yavas, V. Kostina, and M. Effros, ``Random access channel coding in the finite blocklength regime,'' \emph{IEEE Trans. Inf. Theory}, vol. 67, no. 4, pp. 2115--2140, Apr. 2021.
% 16
\bibitem{RAC_fading} S. S. Kowshik, K. Andreev, A. Frolov, and Y. Polyanskiy, ``Energy efficient coded random access for the wireless uplink,'' \emph{IEEE Trans. Commun.}, vol. 68, no. 8, pp. 4694--4708, Aug. 2020.
% 17
\bibitem{estimate_Ka} O. L. A. L{\'o}pez, G. Brante, R. D. Souza, M. Juntti, and M. Latva-aho, ``Coordinated pilot transmissions for detecting the signal sparsity level in a massive IoT network under Rayleigh fading,'' May 2022, arxiv:2205.00406. [Online]. Available: https://arxiv.org/abs/2205.00406
\bibitem{On_joint} A. Lancho, J. {\"O}stman, and G. Durisi, ``On joint detection and decoding in short-packet communications,'' in \emph{Proc. IEEE Global Commun. Conf. (GLOBECOM)}, Madrid, Spain, Dec. 2021.
% 17
\bibitem{noKa} K.-H. Ngo, A. Lancho, G. Durisi, and A. Graell i Amat, ``Unsourced multiple access with random user activity,'' Feb. 2022, arxiv:2202.06365. [Online]. Available: https://arxiv.org/abs/2202.06365
% 18
\bibitem{Caire1} A. Fengler, S. Haghighatshoar, P. Jung, and G. Caire, ``Non-bayesian activity detection, large-scale fading coefficient estimation, and unsourced random access with a massive MIMO receiver,'' \emph{IEEE Trans. Inf. Theory}, vol. 67, no. 5, pp.~2925--2951, May 2021.
% 19
\bibitem{Zheng} L. Zheng and D. N. C. Tse, ``Communication on the Grassmann manifold: A geometric approach to the noncoherent multiple-antenna channel,'' \emph{IEEE Trans. Inf. Theory}, vol. 48, no. 2, pp. 359--383, Feb. 2002.
% 20
\bibitem{pilot_coding1} W. Yang, G. Durisi, and E. Riegler, ``On the capacity of large-MIMO block-fading channels,'' \emph{IEEE J. Sel. Areas Commun.}, vol. 31, no. 2, pp. 117--132, Feb. 2013.
% 21
\bibitem{pilot_coding2} A. Lapidoth, ``On the asymptotic capacity of stationary Gaussian fading channels,'' \emph{IEEE Trans. Inf. Theory}, vol. 51, no. 2, pp. 437--446, Feb. 2005.
% 22
\bibitem{pilot_Durisi1} J. {\"O}stman, G. Durisi, E. G. Str{\"o}m, M. C. Coskun, and G. Liva, ``Short packets over block-memoryless fading channels: Pilot-assisted or noncoherent transmission?'' \emph{IEEE Trans. Commun.}, vol. 67, no. 2, pp. 1521--1536, Feb. 2019.
% 23
\bibitem{pilot_Durisi2} J. {\"O}stman, A. Lancho, G. Durisi, and L. Sanguinetti, ``URLLC with massive MIMO: Analysis and design at finite blocklength,''  \emph{IEEE Trans. Wireless Commun.}, vol. 20, no. 10, pp. 6387--6401, Oct. 2021.

\bibitem{Yuwei_active} L. Liu and W. Yu, ``Massive connectivity with massive MIMO--Part I: Device activity detection and channel estimation,'' \emph{IEEE Trans. Signal Process.}, vol. 66, no. 11, pp. 2933--2946, Jun. 2018.
\bibitem{Gao_active} G. Sun, Y. Li, X. Yi, W. Wang, X. Gao, L. Wang, F. Wei, and Y. Chen, ``Massive grant-free OFDMA with timing and frequency offsets,'' \emph{IEEE Trans. Wireless Commun.}, vol. 21, no. 5, pp. 3365--3380, May 2022.

% 24
\bibitem{mismatch_Shamai} H. Weingarten, Y. Steinberg, and S. Shamai (Shitz), ``Gaussian codes and weighted nearest neighbor decoding in fading multiple-antenna channels,'' \emph{IEEE Trans. Inf. Theory}, vol. 50, no. 8, pp. 1665--1686, Aug. 2004.
% 25
\bibitem{mismatch_Asyhari} A. T. Asyhari and A. Guill{\'e}n i F{\`a}bregas, ``Nearest neighbor decoding in MIMO block-fading channels with imperfect CSIR,'' \emph{IEEE Trans. Inf. Theory}, vol. 58, no. 3, pp. 1483--1517, Mar. 2012.
% 26
\bibitem{1961} R. M. Fano, \emph{Transmission of Information}. Jointly published by the MIT Press and John Wiley $\&$ Sons, 1961.
% 27
\bibitem{1965} R. G. Gallager, ``A simple derivation of the coding theorem and some applications,'' \emph{IEEE Trans. Inf. Theory}, vol. 11, no.~1, pp. 3--18, Jan. 1965.
% 28
\bibitem{goodregion} I. Sason and S. Shamai (Shitz), ``Performance analysis of linear codes under maximum-likelihood decoding: A tutorial'', in \emph{Foundations and Trends in Communications and Information Theory.} Delft, The Netherlands: now Publishers, 2006, vol. 3, no. 1--2, pp. 1--222.
% 29
\bibitem{Yang_scalinglaw_singleuser} W. Yang, G. Durisi, and Y. Polyanskiy, ``Minimum energy to send $k$ bits over multiple-antenna fading channels,''  \emph{IEEE Trans. Inf. Theory}, vol. 62, no. 12, pp. 6831--6853, Dec. 2016.
% 30
\bibitem{Ravi} J. Ravi and T. Koch, ``Scaling laws for gaussian random many-access channels,'' \emph{IEEE Trans. Inf. Theory}, vol. 68, no. 4, pp. 2429--2459, Apr. 2022.



% 31
\bibitem{SB} H. Herzberg and G. Poltyrev, ``Techniques for bounding the probability of decoding error for block coded modulations structures,'' \emph{IEEE Trans. Inf. Theory}, vol. 40, no. 3, pp. 903--911, May 1994.
% 32
\bibitem{TB} E. R. Berlekamp, ``The technology of error correction codes,'' \emph{Proc. IEEE}, vol. 68, no. 5, pp. 564--593, May 1980.
% 33
\bibitem{VP} H. V. Poor, \emph{An Introduction to Signal Detection and Estimation}, 2nd ed. New York, NY, USA: Springer, 1994.
% 34
\bibitem{Shamai_Bettesh} I. Bettesh and S. Shamai, ``Outages, expected rates and delays in multiple-users fading channels,'' in \emph{Proc. Conf. Inf. Sci. Syst. (CISS)}, Princeton, USA, Mar. 2000.
% 35
\bibitem{Beta_dis} W. Yang, G. Durisi, T. Koch, and Y. Polyanskiy, ``Quasi-static SIMO fading channels at finite blocklength,'' in \emph{Proc. IEEE Int. Symp. Inf. Theory (ISIT)}, Istanbul, Turkey, Jul. 2013, pp. 1531--1535.

% 36
\bibitem{sparsity_pattern} G. Reeves and M. C. Gastpar, ``Approximate sparsity pattern recovery: Information-theoretic lower bounds,'' \emph{IEEE Trans. Inf. Theory}, vol. 59, no. 6, pp. 3451--3465, Jun. 2013.
\bibitem{noCSI_conv} J. {\"O}stman, W. Yang, G. Durisi, and T. Koch, ``Diversity versus multiplexing at finite blocklength,'' in \emph{Proc. IEEE Int. Symp. Wireless Commun. Syst. (ISWCS)}, Barcelona, Spain, Aug. 2014, pp. 702--706.

\bibitem{Yang_Beta_beta} W. Yang, A. Collins, G. Durisi, Y. Polyanskiy, and H. V. Poor, ``Beta-beta bounds: Finite-blocklength analog of the golden formula,'' \emph{IEEE Trans. Inf. Theory}, vol. 64, no. 9, pp. 6236--6256, Sep. 2018.

% 37
\bibitem{quadratic_form1} A. M. Mathai and B. P. Serge, \emph{Quadratic Forms in Random Variables: Theory and Applications}. New York, NY, USA: Marcel Dekker, 1992.
% 38
\bibitem{quadratic_form2} A. A. Mohsenipour, ``On the distribution of quadratic expressions in various types of random vectors,'' Ph.D. dissertation, UWO, Ontario, Canada, Nov. 2012.
% 39
\bibitem{chisumprod} M. Okamoto, ``An inequality for the weighted sum of $\chi^{2}$ variates,'' \emph{Bulletin Math. Stat.}, vol. 9, no. 2--3, pp. 69--70, Oct. 1960.
% 40
\bibitem{Gallager} R. G. Gallager, \emph{Information Theory and Reliable Communication}. New York, NY, USA: John Wiley $\&$ Sons, 1968.
% 41
\bibitem{wishart_deter} A. Edelman, ``Eigenvalues and condition numbers of random matrices,'' Ph.D. dissertation, Dept. Math., MIT, Cambridge, MA, USA, May 1989.
% 42
\bibitem{chi_square_upperbound} L. Birg\'e, ``An alternative point of view on Lepski's method,'' \emph{Lecture Notes-Monograph Series}, pp. 113--133, 2001.
% 43
\bibitem{logdet_IWishart} G. J. Foschini and M. J. Gans, ``On limits of wireless communications in a fading environment when using multiple antennas,'' \emph{Wireless Personal Commun.}, vol. 6, no. 3, pp. 311--335, Mar. 1998.

% 44
\bibitem{support_recovery} S. Khanna and C. R. Murthy, ``On the support recovery of jointly sparse Gaussian sources via sparse Bayesian learning,'' Mar. 2017, arXiv:1703.04930. [Online]. Available: http://arxiv.org/abs/1703.04930
% 45
\bibitem{High_Dimensional_Prob} R. Vershynin, \emph{High-Dimensional Probability: An Introduction with Applications in Data Science} (Cambridge Series in Statistical and Probabilistic Mathematics). Cambridge, UK: Cambridge University Press, 2018.


\end{thebibliography}
\end{document}